\documentclass[online,final,faculty=firw,department=cws,phddegree=cws]{adsphd}

\title{Semi-Automated Modular Formal Verification of Critical Software}
\subtitle{Liveness and Completeness Thresholds}

\author{Tobias}{Reinhard}

\supervisor{Prof.~dr. B. Jacobs}{}
\supervisor{Prof.~dr. A. Timany}{}
\president{Prof.~dr.~ir. D. Vandermeulen}{}
\jurymember{Prof.~dr.~ir. F. Piessens}{}
\jurymember{Prof.~dr.~ir. T. Schrijvers}{}
\externaljurymember{Prof. dr. M. Huisman}{Universiteit Twente}
\externaljurymember{Prof. dr. P. Müller}{ETH Zürich}

\researchgroup{DistriNet}
\website{reito.eu} % Leave empty to hide
\email{} % Leave empty to hide

\address{Celestijnenlaan 200A box 2402}
\date{February 2024}
\copyyear{2024}
\usepackage{nomencl}   % For nomenclature
\renewcommand{\nomname}{List of Symbols}

\makenomenclature%

\usepackage{glossaries} % For list of abbreviations

\makeglossaries%

\usepackage{textcomp} % nice greek alphabet
\usepackage{pifont}   % Dingbats
\usepackage{booktabs}
\usepackage{amssymb,amsthm}
\usepackage{amsmath}

\usepackage{xparse}
\usepackage{xspace}
\usepackage{thmtools}
\usepackage{hyperref}
\usepackage{mathpartir}
\usepackage{caption}
\usepackage{subcaption}
\usepackage{subfiles}
\usepackage{fancyvrb}
\usepackage{listings}
\usepackage[skins]{tcolorbox}

\usepackage[numbers]{natbib}			% provides \citeauthor
\usepackage{resources}
\usepackage{notationFonts}
\usepackage{misc}
\usepackage{genMath}
\usepackage{collections}
\usepackage{functions}
\usepackage{graphs}
\usepackage{notes}
\usepackage{threadPools}
\usepackage{levels}
\usepackage{obligations}
\usepackage{permissions}
\usepackage{forkPathIDs}
\usepackage{lists}
\usepackage{subfiles}
\usepackage{separationLogic}
\usepackage{hoareTriples}
\usepackage{genMath}
\usepackage{algebra}
\usepackage{substitution}
\usepackage{tuples}

\usepackage{tikz-cd}
\usetikzlibrary{babel}			 % without this package every label on an edge in a tikz-cd graph leads to errors

	\usepackage{abruptExit-syntax}
	\usepackage{abruptExit-ProofRules}
	\usepackage{abruptExit-reduction}
	\usepackage{abruptExit-safety}
	\usepackage{abruptExit-resources}
	\usepackage{abruptExit-variables}

	\usepackage{sigExt-syntax}
	\usepackage{sigExt-ProofRules}
	\usepackage{sigExt-reduction}
	\usepackage{sigExt-programOrderGraph}
	\usepackage{sigExt-universe}

	\usepackage{correctness}
	\usepackage{equivalence}
	\usepackage{gen-CT-defs}

\begin{document}

%%%%%%%%%%%%%%%%%%%%%%%%%%%%%%%%%%%%%%%%%%%%%%%%%%%%%%%%%%%%%%%%%%%%%%
% Definition of commands and environments
% Whenever possible, definitions are merely copied from papers.
% This may lead to redundancies.
% Definitions must be placed in the top file. Otherwise, hard-to-debug
% errors arise such, e.g., when definitions are used within figure 
% captions.

	% ++++++++++++++++++++++++++++++++++++++++++++++++++++++++++++++++++++
	% For all section/papers
	% ++++++++++++++++++++++++++++++++++++++++++++++++++++++++++++++++++++
		\NewDocumentCommand{\citationBox}{m}
			{\begin{tcolorbox}
					[boxrule=-0.07mm, 
					 leftrule=1.5mm, 
					 colback=blue!5!white,
					 colframe=violet!75!black,
%					frame west hidden ,
					 %borderline west={1pt}{0pt}{red}
					]
					#1
			\end{tcolorbox}}
	
		\newtheorem{theorem}{Theorem}[section]
		\newtheorem{definition}[theorem]{Definition}
		\newtheorem{lemma}[theorem]{Lemma}
		\newtheorem{corollary}[theorem]{Corollary}
		\newtheorem{observation}[theorem]{Observation}
		\newtheorem{assumption}[theorem]{Assumption}
		\newtheorem{notation}[theorem]{Notation}
		\newtheorem{workflow}[theorem]{Workflow}
		\newtheorem{remark}[theorem]{Remark}
		\newtheorem{terminology}[theorem]{Terminology}
		
		\NewDocumentCommand{\deleteBlock}{m}{}

	% ++++++++++++++++++++++++++++++++++++++++++++++++++++++++++++++++++++
	% Commands for Abrupt Exit
	% ++++++++++++++++++++++++++++++++++++++++++++++++++++++++++++++++++++
    \NewDocumentCommand{\RRedRuleNameSet}{}{\ensuremath{N_\keywordFont{r}}\xspace}
		%    \NewDocumentCommand{\progOrdGraph}{}{\ensuremath{\mathcal{G}}\xspace}
		\NewDocumentCommand{\progOrdGraphConfig}{}{\ensuremath{g}\xspace}
		\NewDocumentCommand{\progOrdGraphLF}{}
			{\ensuremath{\progOrdGraph_{\fixedPredNameFont{lf}}}\xspace}
		\NewDocumentCommand{\nodesOf}{m}
			{\ensuremath{\fixedFuncNameFont{nodes}(#1)}\xspace}
		\NewDocumentCommand{\edgesOf}{m}
			{\ensuremath{\fixedFuncNameFont{edges}(#1)}\xspace}
		\NewDocumentCommand{\leavesOf}{m}
			{\ensuremath{\fixedFuncNameFont{leaves}(#1)}\xspace}
		
		\NewDocumentCommand{\sumOb}{m}{\ensuremath{S_o(#1)}\xspace}
		\NewDocumentCommand{\sumCred}{m}{\ensuremath{S_c(#1)}\xspace}
		
		\NewDocumentCommand{\aeExProg}{}
			{\ensuremath{\cmdVar_\text{ex2}}\xspace}

	% ++++++++++++++++++++++++++++++++++++++++++++++++++++++++++++++++++++
	% Commands for Ghost Signal paper
	% ++++++++++++++++++++++++++++++++++++++++++++++++++++++++++++++++++++
		\NewDocumentCommand{\m}{}{\progVar{m}}
		\NewDocumentCommand{\x}{}{\progVar{x}}
		\NewDocumentCommand{\y}{}{\progVar{y}}
		
		\NewDocumentCommand{\locX}{}{\metaVar{\hlocVar_\x}}
		\NewDocumentCommand{\locM}{}{\metaVar{\hlocVar_\m}}
		
		\NewDocumentCommand{\invM}{}{\metaVar{\assLockInvVar}}
		\NewDocumentCommand{\levM}{}{0\xspace}
		\NewDocumentCommand{\meMut}{}{\metaVar{mut}} % TODO: remove
		\NewDocumentCommand{\meM}{}{\progVar{mut}}
		
		\NewDocumentCommand{\valX}{}{\metaVar{\valVar_\x}}
		
		% degree of initial permission in termination proof sketch sketch of minimal example
		\NewDocumentCommand{\meStartDeg}{}{1}
		\NewDocumentCommand{\meWaitDeg}{}{0}
		\NewDocumentCommand{\meLevS}{}{1}
		\NewDocumentCommand{\meIdS}{}{\progVar{sig}}%{\metaVar{\idVar_\sigVar}}
		\NewDocumentCommand{\meS}{}{\sigVar}

	% ++++++++++++++++++++++++++++++++++++++++++++++++++++++++++++++++++++
	% Commands for CT paper
	% ++++++++++++++++++++++++++++++++++++++++++++++++++++++++++++++++++++
		\DefineVerbatimEnvironment{code}{BVerbatim}{baseline=t}

		\NewDocumentCommand{\sizeVar}{}{\metaVar{s}}
		\NewDocumentCommand{\sizeLow}{}
			{\metaVar{\sizeVar_-}}
		\NewDocumentCommand{\sizeHigh}{}
			{\metaVar{\sizeVar_+}}
%		\NewDocumentCommand{\cmdVar}{}{\metaVar{c}}
		\NewDocumentCommand{\arrayVar}{}{\metaVar{a}}
		\NewDocumentCommand{\indexVar}{}{\metaVar{i}}
		\NewDocumentCommand{\leftRangeBound}{}{\metaVar{L}}
		\NewDocumentCommand{\rightRangeBound}{}{\metaVar{R}}
		\NewDocumentCommand{\sizeBoundVar}{}{\metaVar{S}}
		\NewDocumentCommand{\depthBoundVar}{}{\metaVar{D}}
		\NewDocumentCommand{\zVar}{}{\metaVar{Z}}
		
		\NewDocumentCommand{\travPattern}{o o o}
			{\ensuremath{
				\mathit{trav}
				_{
					\IfValueTF{#1}{#1}{\leftRangeBound},
					\IfValueTF{#2}{#2}{\rightRangeBound}
				}
				^{
					\IfValueTF{#3}{#3}{\zVar}
				}
			}\xspace}
		\NewDocumentCommand{\travPatternVars}{o o o m}
			{\metaVar{\travPattern[#1][#2][#3](#4)}}
		
		\NewDocumentCommand{\travPatternClass}{}
			{\ensuremath{
				\mathit{Trav}
			}\xspace}

		\NewDocumentCommand{\compProg}{o o o}
			{\ensuremath{
				\mathit{comp}
				_{
					\IfValueTF{#1}{#1}{\leftRangeBound},
					\IfValueTF{#2}{#2}{\rightRangeBound}
				}
				^{
					\IfValueTF{#3}{#3}{\zVar}
				}
			}\xspace}
		\NewDocumentCommand{\compProgVars}{o o o m}
			{\metaVar{\compProg[#1][#2][#3](#4)}}
		
		\NewDocumentCommand{\compProgClass}{}
			{\ensuremath{
				\mathit{Comp}
			}\xspace}

		\NewDocumentCommand{\sumPattern}{o o o}
			{\ensuremath{
				\mathit{sum}
				_{
					\IfValueTF{#1}{#1}{\leftRangeBound},
					\IfValueTF{#2}{#2}{\rightRangeBound}
				}
				^{
					\IfValueTF{#3}{#3}{\zVar}
				}
			}\xspace}
		\NewDocumentCommand{\sumPatternVars}{o o o m}
			{\metaVar{\sumPattern[#1][#2][#3](#4)}}

		\NewDocumentCommand{\ctVar}{}{\metaVar{Q}}
		\NewDocumentCommand{\ctElem}{o}
			{\ensuremath{
				\metaVar{q}
				\IfValueT{#1}{
					_{#1}
				}
			}\xspace}
		\NewDocumentCommand{\ctElemLow}{}{\ctElem[-]}
		\NewDocumentCommand{\ctElemHigh}{}{\ctElem[+]}
		\NewDocumentCommand{\domVar}{}{\metaVar{X}}
		\NewDocumentCommand{\constraintSetVar}{}{\metaVar{K}}
		
		\NewDocumentCommand{\memsafe}{m}
			{\ensuremath{
				\texttt{memsafe}(#1)
			}\xspace}

		\NewDocumentCommand{\vcVar}{}{\metaVar{vc}}
		
		\NewDocumentCommand{\vcFalse}{}
			{\ensuremath{
				\vcVar_{\text{false}}
			}\xspace}
		
		\NewDocumentCommand{\vcTrav}{}
			{\ensuremath{
				\vcVar_{\text{trav}}
			}\xspace}
		
		\NewDocumentCommand{\vcSum}{o}
			{\ensuremath{
				\vcVar_{\text{sum}}
				\IfValueT{#1}{
					^{#1}
				}
			}\xspace}
		
		\NewDocumentCommand{\preVar}{}{\metaVar{A}}
		\NewDocumentCommand{\postVar}{}{\metaVar{B}}
		\NewDocumentCommand{\memAssVar}{}{\metaVar{M}}
		
		\NewDocumentCommand{\arrayPredName}{}
			{\fixedPredNameFont{array}\xspace}
		\NewDocumentCommand{\arrayPred}{O{\arrayVar} O{\sizeVar}}
			{\ensuremath{
				\arrayPredName(#1, #2)
			}\xspace}
		
		\NewDocumentCommand{\unrelStructVar}{}{\metaVar{y}}
		\NewDocumentCommand{\unrelSizeVar}{}
			{\metaVar{k}}
		\NewDocumentCommand{\dataPredName}{}
			{\fixedPredNameFont{complex\_data}}
		\NewDocumentCommand{\unrelDataPred}{O{\unrelStructVar} O{\unrelSizeVar}}
			{\ensuremath{
				\dataPredName(#1, #2)
			}\xspace}
		\NewDocumentCommand{\vcComp}{}{\metaVar{\vcVar_{comp}}}

%		\NewDocumentCommand{\resVar}{}{\metaVar{r}}
		\NewDocumentCommand{\wlpName}{}{\fixedFuncNameFont{wlp}\xspace}
		\NewDocumentCommand{\wlp}{m O{\resVar} m}
			{\ensuremath{
				\wlpName
				(#1, \lambda #2.\ #3)
			}\xspace}
		
		\NewDocumentCommand{\nVar}{}{\metaVar{n}}
		
		\NewDocumentCommand{\vcLine}{m}
			{\ensuremath{
				(\texttt{\##1})
			}\xspace}
		
		\NewDocumentCommand{\vcTransComment}{m}
			{\texttt{// #1}}

		\NewDocumentCommand{\subAssVar}{o}
			{\ensuremath{
				\metaVar{A}
				\IfValueT{#1}{
					_{#1}
				}
			}\xspace}
		
		\NewDocumentCommand{\freeVars}{m}
			{\ensuremath{
				\fixedFuncNameFont{freeVars}(#1)
			}\xspace}

		\definecolor{codegreen}{rgb}{0,0.6,0}
		\definecolor{codegray}{rgb}{0.5,0.5,0.5}
		\definecolor{codepurple}{rgb}{0.58,0,0.82}
		\definecolor{backcolour}{rgb}{0.95,0.95,0.92}
		
		\lstdefinestyle{WHILEstyle}{
			backgroundcolor=\color{backcolour},   
			commentstyle=\color{codegreen},
			keywordstyle=\color{magenta},
			numberstyle=\tiny\color{codegray},
			stringstyle=\color{codepurple},
			basicstyle=\ttfamily\footnotesize,
			breakatwhitespace=false,         
			breaklines=true,                 
			captionpos=b,                    
			keepspaces=true,                 
			numbers=left,                    
			numbersep=5pt,                  
			showspaces=false,                
			showstringspaces=false,
			showtabs=false,                  
			tabsize=2
		}
		\lstset{style=WHILEstyle}
%%%%%%%%%%%%%%%%%%%%%%%%%%%%%%%%%%%%%%%%%%%%%%%%%%%%%%%%%%%%%%%%%%%%%%

%%%%%%%%%%%%%%%%%%%%%%%%%%%%
% general type setting
	\NewDocumentCommand{\caseStudyFont}{}{\tiny}

% arxiv submission
	\NewDocumentCommand{\arxivShift}{}{\hspace{1.4cm}}
	\NewDocumentCommand{\ifArxiv}{m}
			{#1}

%%%%%%%%%%%%%%%%%%%%%%%%%%%%

\makefrontcoverXII

\maketitle

\frontmatter % to get \pagenumbering{roman}

\includepreface{preface}
\includeabstract{abstract}
\includeabstractnl{abstractnl}

% To create a list of abbreviations, there are 2 options
% 1. manual creation of list of abbreviations and inclusion as a chapter
%    \includeabbreviations{abbreviations}
% 2. automatic generation via the glossary package
%    \glossary{name=MD,description=molecular dynamics}
%\myprintglossary

% To create a list of symbols, there are 2 options
% 1. include a manually created nomenclature as a chapter
%    \includenomenclature{nomenclaturechapter}
% 2. automatic generation via the nomencl package
%    \nomenclature[cB]{$c_B(\vec{x})$}{Characteristic function of $B$}
%\myprintnomenclature

\tableofcontents
%\listoffigures
%\listoftables

%%%%%%%%%%%%%%%%%%%%%%%%%%%%%%%%%%%%%%%%%%%%%%%%%%%%%%%%%%%%%%%%%%%%%%

\mainmatter % to get \pagenumbering{arabic}

% Show instructions on a separate page
%\instructionschapters\cleardoublepage

%
  \graphicspath{{\chaptersdir/introduction/image/}}%
  %\cleardoublepage%
  % !TeX root = ../../thesis.tex
\chapter{Introduction}\label{ch:introduction}

Critical software is everywhere and almost every piece of critical infrastructure depends on it, in one way or another.
The electrical power we use every day is produced by power plants whose safety heavily depends on the correctness of a variety of software systems.
For instance, in case of an emergency dedicated \emph{shutdown systems}~\cite{Lawford2012VerificationNuclearSystems, Wassyng2011SoftwareCertificationNuclearIndustry} are responsible to automatically power down reactors and prevent disaster.
But also a power plant's normal operation is heavily software controlled.
Specifically, they rely on sophisticated networks of hardware sensors and software components, called \emph{supervisory control and data acquisition} (SCADA) systems~\cite{Igure2006ScadaSecurityIssues}.
These collect data and make automatic decisions about the plant's operation or assist human decision makers.
Their correctness is critical to any power plant's safety.
Moreover, the complex networks that supply our homes every day with electricity, gas and water are also operated by SCADA systems~\cite{Teixeira2011ScadaSecurityStudy}.
They can typically be accessed remotely,
which makes them vulnerable to cyber attacks exploiting bugs~\cite{Teixeira2011ScadaSecurityStudy, Igure2006ScadaSecurityIssues}.

To a similar extent, we critically rely on software when we travel.
Many of the safety features that cars offer nowadays are provided by software~\cite{Panaroni2008AutomotiveSoftwareOverview}:
ABS (anti-lock braking system) prevents wheels from locking during emergency braking.
ACC (adaptive cruise control) ensures a safe distance to the vehicle ahead.
ESC (electronic stability control) is yet another service helping the car to stay on track by controlled breaking.
Just to name a few.

%\TodoTextNote{Trains}

Aircrafts use onboard software to control about every aspect of their journey.
They navigate via GPS software~\cite{Sathaye2019AttacksAircraftLanding}.
Dedicated software controls the slats and flaps -- the parts of the wing used to alter or keep the altitude~\cite{Chen2015ModelCheckingAircraft}.
The most critical part of any flight, the landing, is also computer guided~\cite{Sathaye2019AttacksAircraftLanding}.
Faults in any of these programs can have critical consequences.
%Sadly, the groundings of two Boeing 737 MAX planes in October 2018 and March 2019 demonstrate what can happen when correctness is not taken seriously~\cite{Herkert2020Boeing737MAXAccidents}.
The two groundings of Boeing 737 MAX planes in October 2018 and March 2019 should serve as a sad reminder of that~\cite{Herkert2020Boeing737MAXAccidents}.~\footnote{
	We should note that the software was only one of multiple flaws that ultimately led to the groundings.
}

\paragraph{Correctness Properties}
It is easy to agree that correctness is essential for critical systems.
Though, what exactly it means for software to be correct differs from case to case.
Two important classes of correctness properties are \emph{safety} and \emph{liveness}~\cite{Lamport1997CorrectnessMulticore, Alpern1985Liveness}.

\paragraph{Safety}
We call a program \emph{safe} if something bad never happens.
For instance, a program is memory safe if it never accesses invalid memory (e.g. unallocated or already freed memory).
This is typically the most basic property that we want for every program, 
%since it is required for any other form of correct behaviour.
since we need it to ensure any other correctness property.
%the most basic correctness that we typically want for every program is memory safety.
A very common class of memory safety errors are array out-of-bounds accesses, where we read or write memory past the array bounds.
This type of error is often exploited in cyber attacks~\cite{Irmak2018OverviewAttackSCADA, Cowan2000BufferOverflowAttacks, Fu2012BufferOverflowExploit}.

Another safety property that we commonly care about in concurrent applications is data race freedom~\cite{Balasundaram1989DataRace}.
%A data race occurs when two threads try to concurrently write to the same memory cell (or any other shared resource) (or when one reads and one writes non-atomically).
A data race occurs when a thread (non-atomically) writes a memory cell (or any other shared resource) concurrently with another thread writing or reading it.
Whenever that happens the outcome is undefined or non-deterministic if the involved operations are atomic.
In the worst case, a concurrent write can corrupt the memory cell.
If the subsequent program depends on the memory cell's value, its behaviour becomes unpredictable.
Even worse, depending on the concrete programming language used, the semantics of the entire program might become undefined.

\paragraph{Liveness}
Liveness is the counterpart to safety.
We call a program \emph{live} if something good eventually happens.
For instance, we often want computations to terminate, because we care about their result.
Non-terminating computations block the entire subsequent program.
Hence, termination is a commonly desired liveness property,
even though the name liveness suggests otherwise.
%Though the name suggests otherwise, termination is commonly desired a liveness property.

Imagine a distributed banking system processing monetary transactions.
For each transaction, it would debit the specified amount from the sending account, send it to the recipient's institution and then add it to their account.
One of the most important properties of such a system is that the transferred money eventually arrives on the recipient's account, which is also a liveness property.

\paragraph{Approaches to Correctness}
Over the years the community came up with different approaches to reduce the number of bugs and thereby improve the correctness of software systems.
It remains, however, very difficult to reason about programs.
Many interesting properties are undecidable, e.g.,
memory~safety~\cite{Rosu2009_Undecidable_MemorySafety},
deadlock~freedom~\cite{Gouda1984_Undecidable_DeadlockFreedom_FiniteStateMachines}~\footnote{
	The cited work concerns communicating finite state machines, but the work applies straightforwardly to deadlock freedom in concurrent programs.
}
and
termination~\cite{Rice1953RicesTheorem}.
In a nutshell this means that for sufficiently complex programs it is not possible to reason precisely and automatically about every possible execution.

\paragraph{Testing}
Test driven development~\cite{Geras2004TestDrivenDevelopment}, i.e.,
% writing tests in before the source code we develop.
writing tests first and implementing the actual functionality later,
has proven valuable in finding bugs early on.
However, as Dijkstra said: 
``Program testing can be used to show the presence of bugs, but never to show their absence!''~\cite{Dijkstra1970NotesOnStructuredProgramming}
Even for relatively simple programs, the number of tests we had to write to cover every possible scenario, would be infeasibly large.
Approaches like property based testing~\cite{Claessen2000QuickCheck} and fuzzing~\cite{Miller1990ReliabilityUnixUtils} try to tackle this issue by automatically generating test cases and thereby increasing coverage.
However, due to the sheer number of possible inputs and runtime behaviours, even automatic testing is not sufficient to guarantee that we did not miss any bug.
The situation gets even worse when we move from sequential, single-threaded software to concurrent, multi-threaded computations.
In the latter case, we not only have to explore all possible inputs but also all possible thread interleavings to ensure that our software works as expected.
By that the number of required test cases explodes exponentially.
Meanwhile, the main advantage of testing is its simplicity.

\paragraph{Bounded Model Checking}
Model checking~\cite{Clarke1982ModelChecking, Queille1982ModelChecking} is an alternative approach to automatically find bugs by exhaustively simulating every possible program execution.
Since programs do not necessarily terminate, neither does model checking them.
Moreover, the number of different states that the program might be in and which we have to check grows exponentially with the complexity of the program, the number of possible inputs and the length of the executions.
This is known as the state space explosion problem~\cite{Browne1989NetworksIdenticalProcesses,
	Clarke2008BmcStateExplosion, 
	Park2000JavaMC, 
	Clarke2000CounterExGuidedRefinement} 
and it makes straight-forward model checking unfeasible, even for terminating programs.

Bounded model checking (BMC)~\cite{Biere1999SymbolicMC} overcomes this issue by bounding both the length of executions and the input size.
That is, when we bounded model check a program sorting lists of integers, we specify an input size bound $S$ and a depth bound $D$.
Then the model checker simulates all executions for lists with up to $S$ elements and it stops the simulation after $D$ program steps.

Compared to testing, BMC is a much more exhaustive approach to ensure program correctness.
Over the years, BMC has become an established technique in industry, though still far less widespread or well-known than testing.
At this point we have industrial grade bounded model checkers like CBMC~\cite{Clarke2004CBMC} and ESBMC~\cite{Cordeiro2011ESBMC}
that are relatively easy to use.
Writing a CBMC proof for instance requires little more than specifying the bounds $S$ and $D$ as well as which correctness properties the model checker should look for.
The rest happens automatically.
Since CBMC simulates concrete executions, it can report a counterexample for every found bug.
Just like with testing, this makes it easy for the developer to understand and fix the problem.

The main selling point of bounded model checkers is that they manage to strike a compromise between usability and exhaustiveness.
It achieves this by under-approximating the program behaviour, i.e., exploring a (small) subset of finite prefixes of concrete program executions.
However, as with all under-approximating approaches it does not suffice to ensure that the program we checked is correct.
Maybe we missed a bug that only shows up for bigger inputs or longer program executions.
To make things worse,  the state space explosion makes it usually only feasible to check small inputs.

\paragraph{Deductive Verification}
The approaches we discussed so far focus on finding bugs.
With every bug we find and fix our program converges a bit more towards correctness.
With these approaches we can, however, never be certain that no bug is hiding in a corner that we didn't explore.
One way to overcome this limitation is deductive verification~\cite{Turing1949ProgramVerification, Hoare1968HoareLogic}.
The underlying idea is that we analyse the program and use mathematical reasoning to rule out every possible error.
When this succeeds, we know for sure that the program is correct.
Hence, it is the most powerful correctness approach among the ones discussed here.

So why do we bother with testing or model checking at all?
For once, to validate the scenarios considered by our proof.
Yet also because proofs are hard!
In order to logically reason about the behaviour of a program we have to understand how it works and -- even worse -- why it should be correct.
That is, we have to manually come up with a correctness argument in the form of specifications, invariants and inductive proofs.
The deductive verifier then checks that every proof step we took is correct and that our argument holds.
This requires not only a detailed understanding of the software's inner workings but also expertise in the logic the deductive verifier is built upon.
This makes deductive verification hard to access for developers without any background in formal logic.
In contrast, testing and BMC allow us to more or less treat the program as a black box.

Another difference is that a deductive verifier never executes the program.
That means that we never look at any concrete program state in an execution.
Instead we deduce correctness from the program's static structure.
Our correctness argument must be general enough to cover all possible executions.
This often forces us to over-generalise in the sense that our argument also covers impossible executions, especially for loops and recursion.
As a result, it is not possible to verify every correct program.

%Nevertheless, deductive verification is the best approach that we have to truly ensure correctness of complex critical systems.
%For some applications such as power plant shutdown systems or aircraft controls failure is not an option and the stakes are high enough to warrant the tedious correctness proofs.
%Over the years, many research tools such as 
%VeriFast~\cite{Jacobs2011Verifast},
%VerCors~\cite{Huisman2017VerCors},
%Viper~\cite{Mueller2016Viper},
%Why3~\cite{Filliatre2013Why3},
%and
%VCC~\cite{Cohen2009VCC}
%have been developed.
%They are occasionally used in industrial projects such as the verification of %FreeRTOS~\cite{website_FreeRTOS}.
%Moreover, a small number of industrial grade tools such as
%Dafny~\cite{Leino2010Dafny},
%Frama-C~\cite{Cuoq2012FramaC}
%and
%AdacCore Spark~\cite{McCormick2015AdaCoreSpark}
%exist with a long track record in critical areas like cryptography and security~\cite{Cook2018SecurityVerificationAWS, Rungta2022BillionSMTQueries, RepoDafnyAWSEncryptionSDK, Haque2020AdaCoreSparkVerificationOfSeparationKernel, Fagin2013AdaCoreSparkSecureDNS}.

Nevertheless, deductive verification (complemented by validation testing) is the best approach that we have to truly ensure correctness of complex critical systems.
For some applications failure is not an option and the stakes are high enough to warrant the tedious correctness proofs.
Over the years, many research tools such as 
VeriFast~\cite{Jacobs2011Verifast},
VerCors~\cite{Huisman2017VerCors},
Viper~\cite{Mueller2016Viper},
Why3~\cite{Filliatre2013Why3},
and
VCC~\cite{Cohen2009VCC}
have been developed.
While their focus is primarily academic, some are also used in industrial projects such as the verification of FreeRTOS~\cite{Repo_ReinhardAWSInternship_FreeRTOSSchedulerVeriFastProof, website_FreeRTOS} or cloud infrastructure~\cite{Cook2018SecurityVerificationAWS}.
Moreover, a small number of industrial grade tools such as
Dafny~\cite{Leino2010Dafny}
and
AdacCore Spark~\cite{McCormick2015AdaCoreSpark}
exist with a long track record in critical areas like cryptography and security~\cite{Cook2018SecurityVerificationAWS, Rungta2022BillionSMTQueries, RepoDafnyAWSEncryptionSDK, Haque2020AdaCoreSparkVerificationOfSeparationKernel, Fagin2013AdaCoreSparkSecureDNS}.
Meanwhile, works like loop invariant synthesis~\cite{Qin2010LoopInvSynth, Leino2005LoopInvRefine} and -- recently --	 the application of large language models~\cite{Yang2023ArxivLeanDojoLLM, Wu2023ArxivLemurLLMForProgramVerification} try to make deductive verification simpler and thereby more accessible to the masses.

\paragraph{Advantages and Disadvantages}
Testing is a good way to find bugs early in the development cycle.
It requires little expertise and the results are easy to understand.
But given the very limited guarantees it entails, testing alone is clearly insufficient to guarantee correctness of any critical system.
Yet, it complements deductive verification well and serves as an easy way to validate proof assumptions.

In contrast, BMC provides much stronger guarantees.
Nonetheless, it succeeds in preserving the main advantages of testing:
It is fairly easy to apply.
Especially because it allows us to treat most of the target program as a black box.
And it yields real, i.e., non-spurious, counterexamples when it fails.
The combination makes it easy to find and fix bugs.
Its main drawback is that BMC proofs are unsound, because they only explore a subset of all inputs and only bounded executions.
Consequently, no matter how big we set the bounds, we can never be certain that no bug hides in some dark corner we did not explore.

Deductive verification, in contrast, is sound and yields unbounded guarantees.
This makes it an appropriate approach for critical systems, where we cannot tolerate any failure.
Yet, it should be complemented by validation, e.g., testing, to ensure the proof considers the right scenarios.
Deductive proofs are hard, though, and require significant expertise.
At the same time, the approach is incomplete.
That is, many programs are impossible to verify even though they are fully correct.
Hence, choosing the right correctness approach is a trade-off between the degree of guarantees we desire and resources we are willing to spend.

\paragraph{Termination of Busy Waiting}
%In this dissertation we take a closer look at liveness and safety verification.
In concurrent programs responsibilities are often distributed across threads.
The same holds for dependencies.
It is very common for a thread $A$ to depend on data produced by another thread $B$.
Both threads can run concurrently until $A$ arrives at a point where it needs to access said data.
Whenever that happens $A$ has to wait until the data arrives.
That is, the thread is blocked until $B$ finishes the computation it is responsible for.
We call this pattern \emph{blocking concurrency}.
One possible implementation that we focus on in this dissertation is for $A$ to loop until the awaited data arrives and to check the status in every iteration.
We call this \emph{busy waiting}~\cite{MellorCrummey1991AlgorithmsFS}.

Such a system can easily get stuck if one thread fails to adhere to its responsibility or if we accidentally end up with cyclic dependencies.
Ensuring that this not the case, i.e., that the system is \emph{live}, is far from trivial.
Even more so to formally prove it live.
Every such proof has to show that $A$'s wait loop eventually terminates.
To achieve that, we can first tie $A$'s busy waiting to the progress of $B$.
In a next step, we can prove that $B$ eventually produces the data $A$ is waiting for.

We see that reasoning about whether or not $A$'s busy waiting eventually terminates, forces us to take $B$'s progress into account.
Hence, reasoning about such liveness properties and dependencies is difficult to achieve modularly, i.e., thread-locally.
In \S~\ref{ch:paper-2-ghostSignals} we present the first work that allows us verify termination of busy waiting modularly.
We present a separation logic~\cite{Reynolds2002SeparationLA, OHearn2001LocalRA} and a new type of ghost resource~\cite{Jung2016HigherorderGS} that we call \emph{ghost signal}. 

Ghost resources are objects that do not exist during the runtime of a program.
Instead we only introduce them for the sake of our proof.
Naturally, we have to make sure that they can be introduced and erased without affecting the program's semantics or control flow.
Intuitively, a ghost signal is a witness of some other thread promising to do something.
For instance, suppose thread~$A$ possesses a ghost signal tied to the data $A$ is waiting for.
Then it knows that another thread promised to produced said data.
Hence, we can use it in a thread-local proof to justify that $A$'s busy waiting eventually terminates.

\paragraph{Completeness Thresholds}
BMC is unsound by construction since there's always a chance that we chose our bounds too small to find all bugs.
While this is obviously a big disadvantage, it is also where BMC's usability comes from.
In general, bounded proofs 
such as the one that we see in BMC 
are easy to automate since they only explore a finite set of finite, concrete program executions.
Also for that reason, failed proofs yield non-spurious counterexamples.
The latter help developers by showing them precisely under which circumstance the bugs occur.
The same does not hold for unbounded proofs like the ones we get with deductive verification.

BMC proofs are often justified by the intuition that if there is a bug in a program, then it also tends to show up for small inputs and early in an execution.
The problem is that we do not know when this intuition is true and when it is not.
If we knew that it was correct for certain programs, it would allow us to turn bounded proofs into unbounded ones.

Though, we know at least that this intuition holds for bounded proofs of specific properties $\phi$ over finite state transition systems $T$~\cite{Biere1999SymbolicMC, Clarke2004CompletenessAC, Bundala2012OnTM}.
To formalise it, the model checking community came up with the concept of \emph{completeness thresholds} (CTs)~\cite{Clarke2004CompletenessAC}:
Intuitively, such a threshold is a number $k$ that is large enough such that if $T$ violates $\phi$, then there is also a violation that can be reached with at most $k$ steps.
In other words, if there is a bug, we can find it by exploring all finite path prefixes up to the length $k$.

Any bounded proof that exceeds the threshold is sound.
From a verification point of view it would make more sense to call it a \emph{soundness threshold}.
%Unfortunately, the terms \emph{sound} and \emph{complete} are used in opposite ways in the model checking and verification communities~\cite{?}.
Unfortunately, the model checking and verification communities use the terms \emph{sound} and \emph{complete} in opposite ways.
%For the verification community a logic is sound if any proof entails the proven property.
For the verification community, soundness means that any proven property indeed holds.
The model checking community calls this completeness, because the proof did not miss any bugs.

In \S~\ref{ch:paper-3-ct4ms} we present the first notion of CTs for program verification.
In particular, we define them in the context of memory safety proofs and show that we can extract them for array traversing programs.
The programs we study do not correspond to finite transition systems.
Hence, we cannot reuse the existing results or approaches to CTs from the model checking literature.
Though, to highlight the connection to previous CT research in the model checking community, we chose to continue the terminology instead of calling them soundness thresholds.
Yet, apart from the term CT we continue to use \emph{sound} and \emph{complete} in the verification sense.

For us, a CT is not a number $k$.
Instead it is a subdomain $\ctVar \subseteq \domVar$ of an input domain~\domVar that is big enough to ensure that a bounded proof only exploring~\ctVar covers all relevant cases.
Suppose we want to prove memory safety of a program that iterates over an array of size~$\sizeVar \in \N$.
How can we know whether it is sufficient to check memory safety for arrays of size~$\sizeVar < 10$?
In \S~\ref{ch:paper-3-ct4ms} we show that we can extract that knowledge from the program's verification condition and thereby reduce unbounded memory safety proofs to bounded ones.

%%%%%%%%%%%%%%%%%%%%%%%%%%%%%%%%%%%%%%%%%%%%%%%%%%
% Keep the following \cleardoublepage at the end of this file, 
% otherwise \includeonly includes empty pages.
\cleardoublepage

% vim: tw=70 nocindent expandtab foldmethod=marker foldmarker={{{}{,}{}}}
%

%\includechapter{manual} % Remove this chapter
%
  \graphicspath{{\chaptersdir/state-of-the-art/image/}}%
  %\cleardoublepage%
  % !TeX root = ../../thesis.tex
\chapter{State of the Art}\label{ch:state-of-the-art}

\section{Deductive Liveness Verification of Concurrent Software}

We can broadly classify major deductive approaches to reason about concurrent software along two dimensions:
(i)~The abstraction level and
(ii)~the foundation.

\paragraph{Abstraction Level}
Most popular approaches~\cite{OHearn2012APO,
	Owicki1976VerifyingPropiertiesGhostVariables,
	Xu1997RelyGuarentee,
	VafeiadisCSLSound,
	Vafeiadis2007RGSep,
	Jung2018IrisGroundUp,
	Lundberg2020HoareStyleLogic}, like the one we propose in \S~\ref{ch:paper-2-ghostSignals}, work directly with the target program and reason about the effects of every single instruction.
A less widespread alternative is contextual refinement.
The idea is to reason about a more abstract model instead of the actual program we want to verify.
In this context, a refinement is a program transformation that preserves the properties we are interested in and at the same time abstracts implementation details away.
Refinement-based works include
\citet{Turon2013LogicalRelationsRefinement},
\citet{Gu2015DeepSpecAbstractionLayers},
\citet{Boussabbeh2016TerminationRefinement},
\citet{Carvalho2018VerificationConditionVariables}.
Works that employ a mixture of both approaches include: 
CaReSL~\cite{Turon2013UnifyingRefinementHoare},
LiLi~\cite{Liang2016LiliAPL, Liang2017LiliProgressOC},
Conditional Contextual Refinement~\cite{Song2023ConditionalContextualRefinement}.
% separation logic
%\citeauthor{Tassarotti2017ConcurrentTerminationRefinement}~\cite{Tassarotti2017ConcurrentTerminationRefinement},

\paragraph{Logical Foundation}
The second dimension is the foundation we choose to base our reasoning on.
While the landscape of logical foundations is vast, most fall into one of two broad categories: 
(i)~Type-based reasoning and (ii)~program logics à la Floyd-Hoare.

%(i) As the name suggests, the key idea behind type-based logics is to express the program behaviour in form of a program type and to use type-theory

(i) The foundation for all type-based approaches is the Curry-Howard correspondence which equates logical propositions $P$ with types $T$ and consequently proofs of $P$ with values of type $T$.
Dependent type theories like the calculus of constructions~\cite{Coquand1988CalculusOfConstructions} form the basis of popular state-of-the-art proof assistants like Coq~\cite{website_Coq} and Lean~\cite{website_Lean, deMoura2015Lean}.
Due to their expressiveness, they mostly serve as a trusted foundation to mechanise and check higher-level reasoning.
For instance, Iris~\cite{Jung2018IrisGroundUp} and Verifiable C/VST-Floyd~\cite{Apple2014VerifiableC, Cao2018VSTFloyd} are both higher-order separation logics mechanised in Coq.
The expressiveness of dependent types comes at the cost of automation.
Meanwhile refinement types are sufficiently restricted such that assertions/typings can be automatically checked by SMT solvers.
One of the most notable projects using refinement types for automatic program verification is Liquid Haskell~\cite{Vazou2014LiquidHaskell}.
More recent works include Flux~\cite{Lehmann2023Flux_LiquidTypesRust} to verify Rust code and CN~\cite{Pulte2023CN_separationLogicRefinement} which combines refinement types and separation logic to verify C programs.

(ii)~The central idea of Hoare-logic is that we specify the behaviour of a program in the form of pre- and post conditions.
This simple paradigm proved very effective to reason about imperative programs.
Since its introduction, many successors have been proposed~\cite{OHearn2012APO,
	Owicki1976VerifyingPropiertiesGhostVariables,
	Xu1997RelyGuarentee,
	VafeiadisCSLSound,
	Vafeiadis2007RGSep,
	Jung2018IrisGroundUp,
	Lundberg2020HoareStyleLogic}.
One notable step along the evolution of Hoare-style logics is the Owicki-Gries method~\cite{Owicki1976VerifyingPropiertiesGhostVariables}, which introduced some of the concepts we still use today, such as ghost variables, critical sections and resource invariants.
Another notable, though not very wide-spread, type of Hoare-style logics is rely-guarantee reasoning~\cite{Xu1997RelyGuarentee} and its successor deny-guarantee reasoning~\cite{Dodds2009DenyGuaranteeReasoning}.

Most notably, separation logic~\cite{Reynolds2002SeparationLA, OHearn2001LocalRA} became the de-facto standard to reason about heap-manipulating programs.
Dynamic frames~\cite{Kassios2006DynamicFrames} and implicit dynamic frames~\cite{Smans2009ImplicitDynamicFrames} are conceptually similar and used in verifiers such as Dafny~\cite{Leino2010Dafny}, though certainly less wide-spread than separation logic.
In particular, concurrent separation logic~\cite{Reynolds2002SeparationLA, OHearn2001LocalRA} is now the de-facto standard to reason about concurrent, imperative programs.
The key idea is fine-grained reasoning about ownership of resources.
This is also the foundation for the approach we present in \S~\ref{ch:paper-2-ghostSignals}.
Over the years, separation logic has been extended by many concepts to address the challenges that arise when we try to reason about concurrent programs.
Higher-order ghost state~\cite{Jung2016HigherorderGS} makes very expressive invariants and specifications possible.
Fractional permissions~\cite{Boyland2003FractionalPermissions, Brookes2016ConcurrentSL} allow us to distribute permissions among threads in a fine-grained manner that for instance differentiates between read and write permissions.

\paragraph{Proving Liveness}
Hoare-style logics -- and in particular separation logic -- are by construction safety logics.
%Separation logic is by construction -- just like all Hoare-style logic -- a safety logic.
There are, however, several extensions that allow us to prove liveness properties by reducing them to safety ones.
Obligations~\cite{Kobayashi2006ANT} 
%(originally not introduced in the context of separation logic) 
allow us to prove finite blocking by reasoning about responsibility distribution among threads.
Obligations are a very versatile concept that can be used to reason about many different forms of blocking, e.g.,
acquisition of locks~\cite{finite-blocking, Leino2010DeadlockFreeCA},
reading from channels~\cite{finite-blocking, Leino2010DeadlockFreeCA},
waiting for a heap cell to be~set~(\S~\ref{ch:paper-2-ghostSignals}).
Call permissions~\cite{Jacobs2018ModularTerminationVerification} support reasoning about termination of loops and function calls by forming a decreasing measure.

\paragraph{Non-Blocking Concurrency}
We can divide the landscape of concurrent programs into blocking and non-blocking ones.
In case of the former, threads' progress depends on that of other threads.
Non-blocking ones are in contrast programs where different threads cannot block each other's progress.
Several previous works such as 
\citet{Moir1995WaitFreeRenaming,
	Hurault2019NonBlockingRenaming},
\citet{Gotsman2009NonBlocking, 
	total-tada}
focus on verifying liveness properties of non-blocking algorithms.
Notably, Total-TaDA~\cite{total-tada} is a total correctness logic that supports proving termination of concurrent, non-blocking programs.
Moreover, it also allows us to prove non-impedance, i.e., that threads cannot prevent each other's progress.

\paragraph{Blocking Concurrency}
\citet{finite-blocking} propose a technique to verify blocking behaviour in non-terminating, concurrent programs.
They consider blocking behaviour that stems from primitive blocking operations:
Reading from built-in channels, acquiring built-in locks, joining another thread, i.e., waiting until it terminates.
Their approach is based on several techniques that we also reuse for the work we present in \S~\ref{ch:paper-2-ghostSignals}.
In particular, they use obligations, decreasing measures and levels to prove that waiting terminates.

A non-primitive form of blocking behaviour is busy-waiting, where a thread waits by looping until the desired event occurs.
\citet{Liang2016LiliAPL, Liang2017LiliProgressOC} propose LiLi, a separation logic to verify liveness of blocking constructs implemented via busy waiting.
Their proofs rely on contextual, liveness preserving refinement.
Ghost signals (\S~\ref{ch:paper-2-ghostSignals}) allow the sound generation of fuel to justify indefinite waiting in a context that provably leads to termination.
Tada Live~\cite{DOsualdo2021TadaLive}, a separation logic published in parallel to our ghost signal paper, also allows to prove termination of busy waiting loops.
We discuss it in more depth in \S~\ref{ch:conclusion}.

Condition variables are another way to implement blocking behaviour.
Threads can register to wait for access to a specific resource and are woken up when the resource becomes available.
In their paper \citet{Carvalho2018VerificationConditionVariables} propose technique to verify liveness of Java programs that synchronise via condition variables.
The liveness property they consider is that each thread which enters a critical section eventually leaves it, which is essentially a termination property.
They reduce termination to reachability in coloured Petri Nets.
The latter can be checked by existing Petri Net analysis tools.
\citet{Hamin2018DeadlockFreeM} show that we can also use obligations to reason about the liveness of code that uses condition variables for synchronisation.

\section{Reducing Unbounded to Bounded Proofs}

	The fundamental idea behind model checking is to express a problem as property of a transition system and then exhaustively check that it holds for all paths.
	This is often unfeasible for two reasons:
	(i)~The system contains infinite paths that can never be fully checked.
	(ii)~The state space explosion problem~\cite{Browne1989NetworksIdenticalProcesses,
		Clarke2008BmcStateExplosion,
		Park2000JavaMC, 
		Clarke2000CounterExGuidedRefinement}.
	Transition systems grow very fast with the complexity of the original problem we care about.
	Since our computational resources are limited, we can often only check a small prefix of the entire system.
	
	The model checking community came up with two fundamentally different approaches to overcome this limitation:
	(i)~Heavily reducing the state space via abstractions and
	(ii)~approximating unbounded proofs by bounded ones.

	\paragraph{Simplification by Abstraction}
	\emph{Predicate abstraction}~\cite{Graf1997PredicateAbstraction} over-approximates concrete states in a finite transition system by abstract states expressed in terms of hand-crafted predicates.
	Coming up with suitable predicates is hard and leads to spurious counterexamples~\cite{Ball2002SLAM}.
	To counter this issue, \citet{Clarke2000CEGAR} proposed \emph{counterexample-guided abstraction refinement}~(CEGAR).
	Each found counterexample is checked against the concrete transition system to determine whether it is spurious.
	When that is the case, we automatically refine our abstractions to ensure that the refined abstract transition system does not permit the spurious counterexample anymore.
	A special form of this approach is \emph{lazy abstraction}~\cite{Henzinger2002LazyAbstraction} where we limit the refinement to states where the previous abstraction yields a spurious counterexample.
	This reduces the size increase inherent to abstraction refinement.
	Another abstraction refinement approach worth mentioning is \emph{property-directed reachability}~(PDR)~\cite{Bradley2011SatMCWithoutUnrolling, 
		Een2011ImplPDR, 
		Hoder2012GeneralizedPDR}.
	While both CEGAR and PDR use counterexamples to refine state abstractions,
	a major difference is that PDR preserves a precise transition relation~\cite{Hoder2012GeneralizedPDR}.

	\paragraph{Approximating Proofs}
	Bounded model checking (BMC), i.e., approximating unbounded proofs by bounded ones, was originally introduced by \citet{Biere1999SymbolicMC} and is now a wide-spread technique.
	State-of-the-art software-model checkers such as CBMC~\cite{Clarke2004CBMC} and ESBMC~\cite{Cordeiro2011ESBMC}
	use this paradigm.
	Instead of aiming for an exhaustive check, they only focus on a small prefix of the transition system.
	One way is to keep the transition system as is, but to set a depth bound that limits the length up to which path prefixes are explored~\cite{Biere1999SymbolicMC}.
	Checkers typically also allow for more fine-grained bounds, such as separate bounds for specific input sizes and loop depths.
	Fine-grained bounds essentially allow the user to trim specific parts of the transition system.

	\paragraph{Completeness Thresholds}
	While this approach is very popular, it only gives us bounded guarantees that can not straight-forwardly be extrapolated to the entire transition system.
%	The question when such an extrapolation is sound has been studied extensively for finite state transition systems~\cite{Biere1999SymbolicMC, Clarke2004CompletenessAC, Kroening2003EfficientCO, Bundala2012OnTM, Abdulaziz2018FormallyVA, Heljanko2005IncrementalAC, Awedh2004ProvingMP, McMillan2003InterpolationAS}.
	However, in the very paper introducing BMC \citet{Biere1999SymbolicMC} already observed that finite, lasso-shaped prefixes (i.e., prefixes ending in a backloop) can represent the behaviour of infinite lasso-shaped paths.
	Thus they can serve as finite witnesses of infinite behaviour.
	This might sound profane at first glance, but it is actually a very fundamental insight.
	As \citeauthor{Biere1999SymbolicMC} showed, it allows us to extract unbounded guarantees from bounded proofs, 
	given the right circumstances.
	%if the checked formula and bound have the right properties.
	The concept was later popularised by \citet{Clarke2004CompletenessAC} under the term \emph{completeness threshold}.
	Consider a finite transition system $T$ and a property of interest $\phi$.
	We call a number $k \in \N$ a \emph{completeness threshold}~(CT) for $T$ and $\phi$ if the following holds: 
	If $\phi$ holds for all path prefixes up to length $k$, then $\phi$ holds for all paths in $T$, i.e., 
	$T \models_k \phi  \,\Rightarrow\,  T \models \phi$.
	
	Notice that the term CT is used inconsistently in the literature.
	While \citet{Clarke2004CompletenessAC} use the definition above, 
	\citet{Kroening2003EfficientCO} refer to a CT as the minimal $k$ with
	$T \models_k \phi  \,\Rightarrow\,  T \models \phi$.
	We stick to the original definition.
	
	Since their introduction, CTs have been studied extensively for finite state transition systems~\cite{Biere1999SymbolicMC, Clarke2004CompletenessAC, Kroening2003EfficientCO, Bundala2012OnTM, Abdulaziz2018FormallyVA, Heljanko2005IncrementalAC, Awedh2004ProvingMP, McMillan2003InterpolationAS, Kroening2005OverApproxBMC}.
	Finding minimal CTs is in general just as hard as checking the unbounded system~\cite{Kroening2011LinearCTs}.
	Moreover, the question when we can soundly extrapolate a bounded proof to the entire system highly depends on the property $\phi$ we are trying to prove.
	Hence, the research focuses on finding over-approximations specific to certain types of formulas.
	Existing works characterise CTs in terms of key properties of the transition system $T$.
	For instance, the diameter is a CT for the simple LTL 
	safety properties~$Gp$~(i.e. $\square p$), where $p$ is local~\cite{Biere1999SymbolicMC, Kroening2011LinearCTs}.
	Meanwhile the recurrence diameter is a CT for simple LTL 
	liveness properties~$Fp$~(i.e. $\lozenge p$)~\cite{Kroening2011LinearCTs}.
	In this context, the diameter of $T$ is the maximal distance between two nodes in $T$ and the recurrence diameter is the length of the longest loop-free path in $T$.
	That is, whenever we want to model check a simple safety or liveness property, it suffices to run a bounded proof up to the diameter or recurrence diameter.
	That way, we can get sound unbounded guarantees from a bounded proof.

	That sounds great, but we should note that both CTs pose very large bounds, which are typically too big to explore in practice.
	Both are often exponential in the number of state variables in $T$~\cite{Kroening2005OverApproxBMC}.
	Moreover, since both are hard to compute, they are typically over-approximated as well~\cite{Kroening2005OverApproxBMC, Pardalos2004ComplexityPathProblems, Bjoerklund2004ApproxLongestPaths}, leading to even larger CTs.
	They are also worse-case tight in terms of the chosen key property~\cite{Kroening2011LinearCTs}.
	Hence, we cannot hope to find smaller CTs for $Gp$ or $Fp$ that are expressible purely in terms of $T$'s diameter or recurrence diameter.
	To make things worse, CTs for more complex LTL formulas have much larger CTs that can be quadratic, super-polynomial and even exponential~\cite{Bundala2012OnTM, Kroening2011LinearCTs}.

	An interesting combination of CTs and CEGAR is
	\citeauthor{Sinha2012SmallShortWorlds}'s
	\emph{Short-Small-World} ($S^2W$) work~\cite{Sinha2012SmallShortWorlds}.
	They use counterexamples to determine which parts of their state should be represented precisely.
	The rest can be abstracted away, which makes the system smaller and easier to model check.
	Finally, they compute an over-approximation of the reduced system's reachability diameter and use it as CT to bounded model check their system.
	If BMC fails, they collect the found counterexample and return to the CEGAR phase.
	The CT guarantees $S^2W$'s soundness.
	In addition, it also guarantees ``short'' counterexamples (relatively speaking), making them easier to process in the CEGAR phase.
	
	All existing, non-abstracting works on CTs focus on finite state transition systems.
	Software that processes arbitrary large data structures, e.g., arrays or lists, corresponds, however, in general to infinite state transition systems.
	Hence, the key properties we typically use to describe CTs, like the (recurrence) diameter, are in general infinite.
	The same holds thereby for the existing CT approximations.
	In other words, we cannot reuse existing results to reason about CTs for software that processes arbitrary large data.
	In \S~\ref{ch:paper-3-ct4ms} we present the first notion of CTs for memory safety proofs and show how to extract them from the program's verification condition.
	In this work we consider array-traversing programs and focus on CTs for input variables, in particular on CTs for the size of the traversed array.
	Hence, for us, CTs are not depth bounds $k \in \N$ but subdomains.
	Let 
	\hoareTriple{\preVar}{\cmdVar}{\postVar}
	be a program specification and let \varVar be an input variable with domain \domVar.
	A subdomain $\ctVar \subseteq \domVar$ is a CT for \varVar in 
	\hoareTriple{\preVar}{\cmdVar}{\postVar} if
	correctness of the specification on \ctVar implies the unbounded specification, i.e.,
  	$$
		\models 
			\forall x \in \ctVar.\
			\hoareTriple{\preVar}{\cmdVar}{\postVar}
		\quad\Rightarrow\quad
		\models
			\forall x \in \domVar.\
			\hoareTriple{\preVar}{\cmdVar}{\postVar}.
	$$
	In \S~\ref{ch:paper-3-ct4ms} we illustrate the concept by characterising CTs for a simple array traversal pattern.
	Moreover, in \S~\ref{ch:paper-3-ct4ms} and \S~\ref{ch:conclusion} we also sketch plans to characterise CTs for complex programs that traverse a data structure (array, list, etc.) in a memory-layout-preserving way.
	We also sketch plans our approach to arbitrary correctness properties.
	
	\paragraph{Reasoning about Coverage}
	Intuitively, a subdomain $\ctVar \subseteq \domVar$ is a CT if it contains the right inputs to reach (and refute) all potentially problematic behaviours of our program.
	In this sense, reasoning about CTs means reasoning about a specific form of coverage.
	Coverage also plays a big role in traditional testing techniques, like fuzzing~\cite{Miller1990ReliabilityUnixUtils} and property-based testing~\cite{Claessen2000QuickCheck}.
	Compared to hand-written tests, automated approaches make it easier to reach a higher coverage due to the bigger number of tests we can generate.
	Just as in the context of BMC and CTs, the higher the coverage the stronger the guarantees that our tests convey.
	The various notions of code coverage that we consider in testing,
	e.g.,
	method coverage,
	statement coverage,
	path coverage~\cite{Ammann2008IntroTesting, Hilton2018TestCoverageEvolution},
	are much weaker than what we need to reason about CTs.
	Furthermore, existing testing works do not care about providing formal guarantees for the coverage they achieve.
	Recently, \citet{Zhou2023CoverageTypes} proposed \emph{coverage types} to reason formally and automatically about coverage of test generators.
	Their approach combines refinement types~\cite{Vazou2014LiquidHaskell} and incorrectness logic~\cite{OHearn2019IncorrectnessLogic} and supports sophisticated structural notions of coverage.

	\paragraph{Cut-Offs, Small Worlds, Small Models}
	All of these terms are equivalent to CTs but are mostly used in a different model checking subcommunity than CTs.
	An exception that proves the rule is the \emph{Small-Short-World} work by \citet{Sinha2012SmallShortWorlds}.
	Many distributed protocols can be represented as a parameterised system
	$P^n = P_1 || \dots || P_n$ 
	consisting of $n$ identical and parallel processes $P_i$ that each can be represented by the same finite state transition system.
	Intuitively, a distributed protocol should not behave wildly different for every possible number of processes $n$.
	In other words, there should be a finite bound $K$ big enough such that $K$ running processes
	$P_1 || \dots || P_K$
	exhibit all interesting properties~\cite{Namjoshi2007SymmetryCompletenessParameterizedSys}.
	\citet{Emerson1995ReasoningAboutRings} formalised this intuition in the context of token rings~\cite{Namjoshi2007SymmetryCompletenessParameterizedSys}.
	Finding a small cut-off $K$ then allows us to represent the finite system
	$P^K = P_1 || \dots || P_K$
	in one finite transition system and use classic model checking to reason about it.
	Note that this setting is fundamentally different from the one considered by classic CTs à la \citet{Biere1999SymbolicMC} and \citet{Clarke2004CompletenessAC}.
	In particular, it differs fundamentally from the setting we consider for our CT work in~\S~\ref{ch:paper-3-ct4ms}.
	
	The technique is also known as \emph{small model} or \emph{small world} approach~\cite{Sinha2012SmallShortWorlds}.
	Since its introduction, it has proven a valuable technique to simplify reasoning about a range of topologies, e.g., token rings, 
	hybrid automata networks~\cite{Johnson2012SmallModelHybridAutomataNetworks},
	parameterised networks of timed automata~\cite{Spalazzi2014ParameterizedMCTimedSystems},
	general composed network topologies~\cite{Clarke2004NetworkDecomposition, Aminof2014ParamMCTokenPassing}.
	Furthermore, it has since been applied to other domains where systems can be represented by parameterised transition systems.
	For instance, several works use cut-offs to simplify the verification of security properties in hypervisors~\cite{Franklin2010ScalableParametricVerificationSecurity,
		Franklin2012ParametricVerificationAddressSeparation, 
		Sinha2012SmallShortWorlds}.
	
	\citet{Franklin2010ScalableParametricVerificationSecurity}
	consider the verification of security properties concerning page table accesses.
	They focus on a restricted setting that yields nice cutoffs by construction:
	A guarded programming language and a restricted LTL for the specification of security properties $\phi(n)$, where $n$ refers to the number of page table entries.
	Page table accesses are modelled via array accesses.
	Neither their programming nor specification language support programs or specifications~$\phi(n)$ that differentiate between array accesses at different indices.
	They show that they can reduce any expressible security property $\phi(n)$ to $\phi(1)$.
	Impressively, even though they consider such a restricted setting, they manage to verify two realistic case studies: 
	The access control mechanisms in the hypervisors SecVisor~\cite{Seshadri2007SecVisor} and sHype~\cite{Sailer2005sHype}.
	\citet{Franklin2012ParametricVerificationAddressSeparation} extend this work to address separation properties over multi-level data structures such as nested page tables.

	\paragraph{$k$-induction}
	Another approach that has gained popularity since its introduction is \emph{$k$-induction}.
	Originally, \citet{Sheeran2000k_induction} proposed k-induction as a SAT-based technique to prove safety properties over finite transition systems.
	Since then, the approach has been adopted to program verification~\cite{Donaldson2011k_induction, 
		Guo2022_k_induction_C, 
		Alhawi2021_k_induction_invInference_C,
		Gadelha2017LoopsC_k_induction} 
	and integrated in the C model checker ESBMC~\cite{Gadelha2019ESBMC_k_induction}.
	As the name suggests, $k$-induction is a generalised form of proofs by induction.
	The approach's special peculiarity is that we replace loop invariants by $k$ unrolled loop iterations.
	This eliminates the difficult task of manually finding or generating loop invariants.
	Once a $k$ has been set, the loop has been unrolled and once the right assertions have been set to represent the inductive step, we can use standard BMC to carry out the proof.
	We see that $k$-induction does not reduce unbounded proofs to bounded ones.
	However, it reduces the task of finding a traditional loop invariant to finding a big enough $k$ such that $k$ unrolled iterations carry enough information to be representative of the loop's general behaviour.
	Both approaches are thus similar in spirit.

%%%%%%%%%%%%%%%%%%%%%%%%%%%%%%%%%%%%%%%%%%%%%%%%%%
% Keep the following \cleardoublepage at the end of this file, 
% otherwise \includeonly includes empty pages.
\cleardoublepage

% vim: tw=70 nocindent expandtab foldmethod=marker foldmarker={{{}{,}{}}}
%

% Insert here your own chapters
% Chapters are expected to be in a tex-file with the given name dot
% tex and in a directory with the given name in the chapters
% directory.

%\includechapter{paper-1-abruptExit}
%
  \graphicspath{{\chaptersdir/paper-2-ghostSignals/image/}}%
  %\cleardoublepage%
  % !TeX root = ../../thesis.tex
\chapter{Ghost Signals: Verifying Termination of Busy Waiting}\label{ch:paper-2-ghostSignals}

\NewDocumentCommand{\gsSecDir}{}
	{chapters/paper-2-ghostSignals/CAV_paper_extended}

This chapter was previously published as:
\citationBox
{
	\textbf{T. Reinhard}, and B. Jacobs. 
	\emph{Ghost signals: Verifying Termination of Busy Waiting.}
	In Proceedings of 33rd 
	International Conference on Computer-Aided Verification (CAV) (2021), 
	vol. 12760 of Lecture Notes in Computer Science, Springer, pp. 27–50.
	\\
	\href
		{https://doi.org/10.1007/978-3-030-81688-9_2}
		{doi.org/10.1007/978-3-030-81688-9\_2}
}

\section{Introduction}
	
    Programs for multiprocessor machines commonly perform busy waiting for synchronization~\cite{Muehlemann1980MethodFR, MellorCrummey1991AlgorithmsFS}. 
    In this chapter, we propose a separation logic~\cite{Reynolds2002SeparationLA, OHearn2001LocalRA} to modularly verify termination of such programs under fair scheduling.
    Specifically, we consider programs where some threads busy-wait for a certain condition $C$ over a shared data structure to hold, e.g., a memory flag being set by other threads.
    By modularly, we mean that we reason about each thread and each function in isolation.
    That is, we do not reason about thread scheduling or interleavings.
    We only consider these issues when proving the soundness of our logic.
    Assuming fair scheduling is necessary since busy-waiting for a condition~$C$ only terminates if the thread responsible for establishing the condition is sufficiently often scheduled to establish $C$.
    
    Busy waiting is an example of \emph{blocking} behaviour, where a thread's progress \emph{requires interference} from other threads. This is not to be confused with \emph{non-blocking} concurrency, where a thread's progress does not rely on---and may in fact be \emph{impeded} by---interference from other threads. Existing proposed approaches for verifying termination of concurrent programs consider only programs that only involve non-blocking concurrent objects \cite{total-tada}, or \emph{primitive blocking constructs} of the programming language, such as acquiring built-in mutexes, receiving from built-in channels, joining threads, or waiting for built-in monitor condition variables \cite{Leino2010DeadlockFreeCA,finite-blocking,Hamin2018DeadlockFreeM}, or both \cite{Jacobs2018ModularTerminationVerification}. Existing techniques that do support busy waiting are not Hoare logics; instead, they verify termination-preserving \emph{contextual refinements} between more concrete and more abstract implementations of busy-waiting concurrent objects \cite{Liang2017LiliProgressOC,Kim2017LayerByLayer}. In contrast, we here propose the first conventional program logic for modular verification of termination of programs involving busy waiting, using Hoare triples as module specifications.

    In order to prove that a busy-waiting loop terminates, we have to prove that it performs only finitely many iterations.
    To do this we introduce a special form of  \emph{ghost resources}~\cite{Jung2016HigherorderGS} which we call \emph{ghost signals}.
    As ghost resources they only exist on the verification level and hence do not affect the program's runtime behaviour.
    Signals are initially unset and come with an obligation to set them.
    Setting a signal does not by definition correspond to any runtime condition.
    So, in order to use a signal \sigVar effectively, anyone using our approach has to prove an invariant stating that \sigVar is set if and only if the condition of interest holds.
    Further, the proof author must prove that every thread discharges all its obligations by performing the corresponding actions, e.g., by setting a signal and establishing the corresponding condition by setting the memory flag.
%    When we verify a program, we must prove that every thread discharges all its obligations by performing the corresponding actions, e.g., by setting a signal and establishing the corresponding condition by setting the memory flag.
    
    In our verification approach we tie every busy-waiting loop to a finite set of ghost signals \SigSetVar that correspond to the set of conditions the loop is waiting for.
    Every iteration that does not terminate the loop must be justified by the proof author proving that some signal $\sigVar \in \SigSetVar$ has indeed not been set, yet.
    This way, we reduce proving termination to proving that no signal is waited for infinitely often.
    
    Our approach ensures that no thread directly or indirectly waits for itself by requiring the proof author
    (i)~to choose a well-founded and partially ordered set of levels \LevelSet and
    (ii)~to assign a level to every signal
    and by
    (iii)~only allowing a thread to wait for a signal if the signal's level is lower than the level of each held obligation.
    This guarantees that every signal is waited for only finitely often and hence that every busy-waiting loop terminates.
    We use this to prove that every program that is verified using our approach indeed terminates.

    We start by gradually introducing the intuition behind our verification approach and the concepts we use.
    In \S~\ref{gs:sec:Nutshell:ThreadSafePhysicalSignals} and \S~\ref{gs:sec:Nutshell:NonThreadSafePhysicalSignals} we present the main aspects of using signals to verify termination.
    We start by treating them as physical thread-safe resources and only consider busy waiting for a signal to be set.
    Then, we drop thread-safety and explain how to prove data-race- and deadlock-freedom.
    In \S~\ref{gs:sec:Nutshell:AbitraryDataStructures} and \S~\ref{gs:sec:Nutshell:SignalErasure} we generalize our approach to busy waiting for arbitrary conditions over arbitrary data structures and then lift signals to the verification level by introducing ghost signals. 

    In \S~\ref{gs:sec:RealisticExample} we sketch the verification of a realistic producer-consumer example involving a bounded FIFO to demonstrate our approach's usability and address fine-grained concurrency in \S~\ref{gs:sec:FineGrainedConcurrency}.
    Further, we describe the available tool support in \S~\ref{gs:sec:ToolSupport} and discuss integrating higher-order features in \S~\ref{gs:sec:HigherOrder}.
    We conclude by comparing our approach to related work and reflecting on it in \S~\ref{gs:sec:RelatedWork} and \S~\ref{gs:sec:Conclusion}.
    
    We formally define our logic and prove its soundness in appendix \S~\ref{ch:appendix-paper-2-ghostSignals}.
    To keep the presentation in this chapter simple, we assume busy-waiting loops to have a certain syntactical form. 
    In our technical report~\cite{Reinhard2020GhostSignalsTR}, included as appendix~\S~\ref{ch:appendix-ghostSignals-TR}, we present a generalised version of our logic and its soundness proof.    
    Further, we verify the realistic example presented in \S~\ref{gs:sec:RealisticExample} in full detail in appendix~\S~\ref{ch:appendix-paper-2-ghostSignals} and in the technical report  (appendix~\S~\ref{ch:appendix-ghostSignals-TR}), using the respective version of our logic.
    We used our tool support to verify C versions of the bounded FIFO example and the CLH lock.
    The tool we used and the annotated .c files can be found at~\cite{verifast2104,ArtifactJacobs2020VerifastGhostSignalConsumerProducerBoundedFifo,ArtifactJacobs2020VerifastCLHLock}.

\section{A Guide on Verifying Termination of Busy Waiting}\label{gs:sec:ApproachInANutshell}

    When we try to verify termination of busy-waiting programs, multiple challenges arise.
    Throughout this section, we describe these challenges and our approach to overcome them.
    In \S~\ref{gs:sec:Nutshell:ThreadSafePhysicalSignals} we start by discussing the core ideas of our logic.
    In order to simplify the presentation we initially consider a simple language with built-in thread-safe \emph{signals} and a corresponding minimal example where one thread busy-waits for such a signal.
    Signals are heap cells containing Boolean values that are specially marked as being solely used for busy waiting.
    Throughout this section, we generalize our setting as well as our example towards one that allows to verify programs with busy waiting for arbitrary conditions over arbitrary shared data structures.
    In \S~\ref{gs:sec:Nutshell:NonThreadSafePhysicalSignals} we present the concepts necessary to verify data-race-, deadlock-freedom and termination in the presence of  built-in signals that are not thread safe.
%    In \S~\ref{gs:sec:Nutshell:AbitraryDataStructures} we introduce general heap cells and show that busy-waiting for a shared heap cell to be set is equivalent to busy-waiting for a non-thread-safe signal.
    In \S~\ref{gs:sec:Nutshell:AbitraryDataStructures} we explain how to use these non-thread-safe signals to verify programs that wait for arbitrary conditions over shared data structures.
    We illustrate this by an example waiting for a shared heap cell to be set.
    In \S~\ref{gs:sec:Nutshell:SignalErasure} we erase the signals from our program and lift them to the verification level in the form of a concept we call \emph{ghost signals}.

    \subsection{Simplest Setting: Thread-Safe Physical Signals}\label{gs:sec:Nutshell:ThreadSafePhysicalSignals}
    We want to verify programs that busy-wait for arbitrary conditions over arbitrary shared data structures. 
%    We want to verify programs that busy-wait for arbitrary .
    As a first step towards achieving this, we first consider programs that busy-wait for simple Boolean flags, specially marked as being used for the purpose of busy waiting. 
    We call these flags \emph{signals}. 
    For now, we assume that read and write operations on signals are thread-safe.
    Consider a simple programming language with built-in signals and with the following commands:
    (i)~\cmdNewSignal for creating a new unset signal, 
    (ii)~\cmdSetSignal{x} for setting~$x$ and 
    (iii)~\cmdAwaitNoMut{\cmdIsSignalSet{\varVar}} for busy-waiting until $x$ is set.
%    Fig.~\ref{gs:fig:MinEx:RealThreadSafeSignals} presents a minimal example where two threads communicate via a shared signal~\sigProgVar.
    The main thread creates the signal \sigProgVar and forks a new thread that busy-waits for \sigProgVar to be set.
    Then, the main thread sets the signal.
    As we assume signal operations to be thread-safe in this example, we do not have to care about potential data races.
    Notice that like all busy-waiting programs, this program is guaranteed to terminate only under fair thread scheduling: 
    Indeed, it does not terminate if the main thread is never scheduled after it forks the new thread. 
    In this chapter we verify termination under fair scheduling.

    \begin{figure}
            \vspace{-0.3cm}
            $$
            \begin{array}{l}
                    \keyword{let}\ \sigProgVar := \cmdNewSignal\ \keyword{in}\\
                    \cmdFork{
                            \cmdAwaitNoMut{\cmdIsSignalSet{\sigProgVar}}
                    };
                    \\
                    \cmdSetSignal{\sigProgVar}
            \end{array}
            $$
            \vspace{-0.3cm}
            \caption{
                    Minimal example with two threads communicating via a physical thread-safe signal.
            }
            \label{gs:fig:MinEx:RealThreadSafeSignals}
            \vspace{-0.5cm}
    \end{figure}

\newpage
\subsubsection{Augmented Semantics}
    \paragraph{Obligations}
	The only construct in our language that can lead to non-termination are busy-waiting loops of the form \cmdAwaitNoMut{\cmdIsSignalSet{\sigProgVar}}.
	In order to prove that programs terminate it is therefore sufficient to prove that all created signals are eventually set.
	We use so-called \emph{obligations}~\cite{Hamin2019TransferringOT, Hamin2018DeadlockFreeM, Leino2010DeadlockFreeCA, Kobayashi2006ANT} to ensure this.
	These are \emph{ghost resources}~\cite{Jung2016HigherorderGS}, i.e., resources that do not exist during runtime and can hence not influence a program's runtime behaviour.
	They carry, however, information relevant to the program's verification.
	Generally, holding an obligation requires a thread to discharge it by performing a certain action.
	For instance, when the main thread in our example creates signal \sigProgVar, it simultaneously creates an obligation to set it.
	The only way to discharge this obligation is to set \sigProgVar.
	
	We denote thread IDs by \tidVar and describe which obligations a thread \tidVar holds by bundling them into an obligations chunk \obsQual{\tidVar}{\obBagVar}, where \obBagVar is a multiset of signals.
	We denote multisets by double braces \multiset{\dots} and multiset union by~\msCup.
	Each occurrence of a signal \sigVar in \obBagVar corresponds to an obligation by thread \tidVar to set \sigVar.
	Consequently, \noObsQual{\tidVar} asserts that thread \tidVar does not hold any obligations.

	\paragraph{Augmented Semantics} 
	In the \emph{real} semantics of the programming language we consider here, ghost resources such as obligations do not exist during runtime.
	To prove termination, we consider an \emph{augmented} version of it that keeps track of ghost resources during runtime.
	In this semantics, we maintain the invariant that every thread holds exactly one \obsPred chunk.
	That is, for every running thread \tidVar, our heap contains a unique heap cell \obsQualPred{\tidVar} that stores the thread's bag of obligations.
	Further, we let a thread get stuck if it tries to finish while it still holds undischarged obligations.
	Note that we use the term \emph{finish} to refer to thread-local behaviour while we write \emph{termination} to refer to program-global behaviour, i.e., meaning that every thread finishes.
%	We let this semantics reflect our proof rules.
%	For instance, as formulated by \prMinSigNewSignalName, creating a new signal \sigVar also spawns an obligation to set \sigVar.
%	Further, we ensure that any execution gets stuck that violates any restriction formulated by our proof rules.
%	For instance, \prMinSigSetSignalName requires the thread setting \sigVar to hold an obligation for \sigVar.
%	A program trying to set the signal without holding the necessary obligation gets stuck.
	For every augmented execution there trivially exists a corresponding execution in the real semantics.

	Fig.~\ref{gs:fig:AugmentedReductionRules} presents some of the reduction rules we use to define the augmented semantics.
	We use \augHeapVar to refer to augmented heaps, i.e., heaps that can contain ghost resources.
	A reduction step has the form 
	\augRedStep
		{\augHeapVar}{\cmdVar}
		[\tidVar]
		{\nextAugHeapVar}{\nextCmdVar}
		[\ftsVar]
	expresses that thread \tidVar reduces heap \augHeapVar (which is shared by all threads) and command \cmdVar to heap \nextAugHeapVar and command \nextCmdVar.
	Further, \ftsVar represents the set of threads forked during this step.
	It is either empty or a singleton containing the new thread's ID and the command it is going to execute, i.e., \setOf{(\tidForkedVar, \cmdForkedVar)}.
	We omit it whenever it is clear from the context that no thread is forked.
	Further, we denote disjoint union of sets by \disjCup.

%	Every time a new thread \tidForkedVar is forked, we allocate a new heap cell \obsQualPred{\tidForkedVar} to store its respective bag of obligations.
%	As mentioned above, our semantics ensures that the heap contains exactly one obligations chunk per running thread, %provided we start the execution with a compliant heap.
%	Hence, we will never run into a heap
%	$\augHeapVar
%	\disjCup \setOf{\obsQual{\tidVar}{\obBagVar}}
%	\disjCup \setOf{\nextObsQual{\tidVar}{\nextObBagVar}}
%	$
%	that contains multiple obligations chunks belonging to the same thread \tidVar.

	Our reduction rules comply with the intuition behind obligations we outlined above.
	\augRedNewSignalName creates a new signal and simultaneously a corresponding obligation.
	The only way to discharge it is by setting the signal using \augRedSetSignalName.

\begin{figure}
    \vspace{-0.5cm}
	\begin{mathpar}
		\augRedNewSignal
		\and
		\augRedSetSignal
		\and
		\augRedFork
		\and
		\augRedAwait
	\end{mathpar}

    \vspace{-0.2cm}
	\caption{Reduction rules for augmented semantics.}
	\label{gs:fig:AugmentedReductionRules}
    \vspace{-0.8cm}
\end{figure}

    \paragraph{Forking}
    Whenever a thread forks a new thread, it can pass some of its obligations to the newly forked thread, cf.\@ \augRedForkName.
    Forking a new thread with ID \tidForkedVar also allocates a new heap cell \obsQualPred{\tidForkedVar} to store its bag of obligations.
    Since this is the only way to allocate a new \obsPred heap cell, we will never run into a heap
    $\augHeapVar
    	\disjCup \setOf{\obsQual{\tidVar}{\obBagVar}}
    	\disjCup \setOf{\nextObsQual{\tidVar}{\nextObBagVar}}
    $
    that contains multiple obligations chunks belonging to the same thread \tidVar.    
    Remember that threads cannot finish while holding obligations.
    This prevents them from dropping obligations via dummy~forks.

	\paragraph{Levels}
	In order to prove that a busy-waiting loop \cmdAwaitNoMut{\cmdIsSignalSet{\sigProgVar}} terminates, we must ensure that the waiting thread does not directly or indirectly wait for itself.
	We could just check that it does not hold an obligation for the signal it is waiting for, but that is not sufficient as the following example demonstrates:
	Consider a program with two signals $\sigProgVar_1, \sigProgVar_2$ and two threads.
	Let one thread hold the obligation for $\sigProgVar_2$ and execute 
	$
	\cmdAwaitNoMut{\cmdIsSignalSet{\sigProgVar_1}};\
	\cmdSetSignal{\sigProgVar_2}
	$.
	Likewise, let the other thread hold the obligation for $\sigProgVar_1$ and let it execute
	$
	\cmdAwaitNoMut{\cmdIsSignalSet{\sigProgVar_2}};\
	\cmdSetSignal{\sigProgVar_1}
	$.

	To prevent such \emph{wait cycles} modularly, we apply the usual approach~\cite{Leino2010DeadlockFreeCA, Boyapati2002OwnershipTF, Flanagan2002ExtendedStaticCheckingJava}.
	For every program that we want to execute in our augmented semantics, we choose a partially ordered set of levels \LevelSet.
	Further, during every reduction step in the augmented semantics that creates a signal \sigVar, we pick a level $\levVar \in \LevelSet$ and associate it with \sigVar.
	Note that much like obligations, levels do not exist during runtime in the real semantics.
	Signal chunks in the augmented semantics have the form \assSignalNoBool{(\idVar, \levVar)} where \idVar is the unique signal identifier returned by \cmdNewSignal.
	The level assigned to any signal can be chosen freely, cf.\@ \augRedNewSignalName.
	In practice, determining levels boils down to solving a set of constraints that reflect the dependencies.
	In our example, however, the choice is trivial as it only involves a single signal.
	We choose $\LevelSet = \setOf{0}$ and 0 as level for \sigProgVar and thereby get \assSignalNoBool{(\sigProgVar, 0)}.
	Generally, we denote signal tuples by $\sigVar = (\idVar, \levVar)$.
	Now we can rule out cyclic wait dependencies by only allowing a thread to busy-wait for a signal \sigVar if its level \sigGetLev{\sigVar} is smaller than the level of each held obligation, cf.\@ \augRedAwaitName~\footnote
        {
        	For simplicity, our augmented semantics assumes that the level order and the level associated with any object remains fixed for the entire execution.
        	However, following the approach presented in \cite{Leino2009MultithreadedPrograms}, it would be sound to add a step rule that allows a thread to change the level of an object it has exclusive access to (cf.\@~\S~\ref{gs:sec:Nutshell:NonThreadSafePhysicalSignals}).
        }.
	Given a bag of obligations~\obBagVar, we denote this by $\sigGetLev{\sigVar} \levObsLt \obBagVar$.

	\paragraph{Proving Termination}
	As we will explain below, the augmented semantics has no fair infinite executions.
	We can use this as follows to prove that a program \cmdVar terminates under fair scheduling:
	For every fair infinite execution of \cmdVar, show that we can construct a corresponding augmented execution.
	(This requires that each step's side conditions in the augmented semantics are satisfied.
	Note that we thereby prove certain properties for the real execution, like absence of cyclic wait dependencies.)
	As there are no fair infinite executions in the augmented semantics, we get a contradiction.
	It follows that \cmdVar has no fair infinite executions in the real semantics.

    \paragraph{Soundness}
	In order to prove soundness of our approach, we must prove that there indeed are no fair infinite executions in the augmented semantics.
	This boils down to proving that no signal can be waited for infinitely often.
	Consider any program and any fair augmented execution of it.
%	Consider any execution of any program verified using our proof rules.
	Consider the execution's \emph{program order graph}, 
	(i)~whose nodes are the execution steps and 
	(ii)~which has an edge from a step to the next step of the same thread and to the first step of the forked thread, if it is a fork step. 
	Notice that for each obligation created during the execution, the set of nodes corresponding to a step made by a thread while that thread holds the obligation constitutes a path that ends when the obligation is discharged.
	We say that this path \emph{carries} the obligation.
	
	It is not possible that a signal is waited for infinitely often. 
	Indeed, suppose some signals \InfWaitSigSet are. 
	Take $\minSig \in \InfWaitSigSet$ with minimal level. 
	Since \minSig is never set, the path in the program order graph that carries the obligation must be infinite as well. 
	Indeed, suppose it is finite. 
	The final node $N$ of the path cannot discharge the obligation without setting the signal, so it must pass the obligation on either to the next step of the same thread or to a newly forked thread. 
	By fairness of the scheduler, both of these threads will eventually be scheduled.
	This contradicts $N$ being the final node of the path.    
	
	The path carrying the obligation for \minSig waits only for signals that are waited for finitely often. 
	(Remember that \augRedAwaitName requires the signal waited for to be of a lower level than all held obligations, i.e.,  a lower level than that of \minSig.)
	It is therefore a finite path.
	A contradiction. 
	%    This contradicts the fact that it carries an obligation that is never discharged.
	
	Notice that the above argument relies on the property that every non-empty set of levels has a minimal element. 
	For this reason, for termination verification we require that \LevelSet is not just partially ordered, but also well-founded.

\subsubsection{Program Logic}

%	\paragraph{}\!\!\!\!\!\!\!
	Directly using the augmented semantics to prove that our example program terminates is cumbersome.
	In the following, we present a separation logic that simplifies this task.
 
	\paragraph{Safety}
	We call a program \cmdVar \emph{safe} under a (partial) heap \augHeapVar if it provides all the resources necessary such that both \cmdVar and any threads it forks can execute without getting stuck in the augmented semantics. (This depends on the angelic choices.)
	We denote this by \safe{\augHeapVar}{\cmdVar}~\cite{VafeiadisCSLSound}~\footnote{
        For a formal definition see appendix \S~\ref{gs:appendix:sec:Soundness:ModelRelation} and the technical report~\cite{Reinhard2020GhostSignalsTR} (appendix~\S~\ref{gsTR:appendix:sec:HoareTripleModelRelation}).
    }.

	Consider a program \cmdVar that is safe under an augmented heap \augHeapVar.
	Let \pheapVar be the real heap that matches \augHeapVar apart from the ghost resources.
	Then, for every real execution that starts with \pheapVar we can construct a corresponding augmented execution.

    \paragraph{Specifications}
	We use Hoare triples \hoareTriple{\assPreVar}{\cmdVar}[\resVar]{\assPostVar(\resVar)}~\cite{Hoare1968HoareLogic} to specify the behaviour of a program \cmdVar.
%	Let \evalCtxt[\cdot] be any context that when instantiated that when instantiated to a program by plugging 
%	
	Such a triple expresses the following:
	Consider any evaluation context \evalCtxt, such that for every return value \valVar, running \evalCtxt[\valVar] from a state that satisfies $\assPostVar(\valVar)$ is safe.
	Then, running \evalCtxt[\cmdVar] from a state that satisfies~\assPreVar is safe.
% TODO: Remove
\myComment
{
	Let \assModels{\augHeapVar}{\assVar} express that assertion \assVar holds under heap \augHeapVar.
	Then we can formalize this interpretation as follows:
	$$
	\begin{array}{c}
		\hoareTriple{\assPreVar}{\cmdVar}[\resVar]{\assPostVar(\resVar)}\\
		\Longleftrightarrow\\
		\begin{array}{l}
			\forall \augHeapFrameVar.\
			\forall \evalCtxt.\
			(\forall \valVar.\
			\forall \augHeapPostVar.\
			\assModelsOR
			{\augHeapPostVar}
			{\assPostVar(\valVar)}
			\ \rightarrow\
			\safe
			{\augHeapPostVar \disjCup \augHeapFrameVar}
			{\,\evalCtxt[\valVar]}
			)
			\\
			%                            \quad\quad\quad\quad\quad \rightarrow\
			\phantom{
				\forall \augHeapFrameVar.\
			}
			\rightarrow\ \
			\forall \augHeapPreVar.\
			\assModelsOR{\augHeapPreVar}{\assPreVar}
			\ \rightarrow\
			\safe
			{\augHeapPreVar \disjCup \augHeapFrameVar}
			{\,\evalCtxt[\cmdVar]}	
		\end{array}
	\end{array}
	$$
}

    \paragraph{Proof System}
    We define a proof relation \htProvesSymb which ensures that whenever we can prove 
    \htProvesShort{\assPreVar}{\cmdVar}[\resVar]{\assPostVar(\resVar)},
    then \cmdVar complies with the specification 
    \hoareTripleShort{\assPreVar}{\cmdVar}[\resVar]{\assPostVar(\resVar)}.
    Fig.~\ref{gs:fig:MinEx:RealThreadSafeSignals:ProofRules} presents some of the proof rules we use to define~\htProvesSymb.
    As we evolve our setting throughout this section, we also adapt our proof rules.
    Rules that will be changed later are marked with a prime in their name.
    The full set of rules is presented in the appendix (cf.\@ Fig.~\ref{gs:appendix:fig:ProofRulesPartOne} and~\ref{gs:appendix:fig:ProofRulesPartTwo}).
	Our proof rules~\prMinSigSetSignalName and~\prMinSigAwaitName are similar to the rules for sending and receiving on a channel presented in~\cite{Leino2010DeadlockFreeCA}.
    
    Notice how the proof rules enforce the side-conditions of the augmented semantics.
    Hence, all we have to do to prove that a program \cmdVar terminates is to prove that every thread eventually discharges all its obligations.
    That is, we have to prove \htProves{\noObs}{\cmdVar}{\noObs}.
    Fig.~\ref{gs:fig:MinEx:RealThreadSafeSignals:ProofSketch} illustrates how we can apply our rules to verify that our minimal example terminates.

% TODO: Remove
\myComment
{
	\paragraph{Proving Termination}
	As explained above, we can prove that a program \cmdVar terminates in two steps:
	(i)~Showing that for every fair real execution of \cmdVar, we can construct an augmented counterpart that does not get stuck.
	(ii)~Using the property that there are no infinite fair executions in the augmented semantics.
	As our proof rules' restrictions reflect those in the augmented semantics, a proof in our logic corresponds to such a construction.
	For instance, during a proof, we show how to choose levels and how to delegate obligations to newly forked threads.
	Hence, all that we have to do is to prove that all threads in the constructed augmented execution eventually discharge their obligations.
	Otherwise they would get stuck when trying to finish.
	That is, we can prove termination by proving \htProves{\noObs}{\cmdVar}{\noObs}.	
}

% TODO: Remove
\myComment
{
	\paragraph{Interpreting Obligations Chunks}
	The assertion \obs{\obBagVar} asserts that the current thread's bag of obligations is exactly \obBagVar. 
	We can link this to regular separation logic as follows:
	Whenever a new thread $t$ is created in our augmented semantics, a new heap cell $t.\obsPred$ is allocated as well.
	%    	Imagine that whenever a thread $t$ is created, a heap cell $t.\obsPred$ is allocated as well.
	\obs{\obBagVar} then means that heap cell $t.\obsPred$ (where $t$ is the current thread) stores the bag of obligations \obBagVar, i.e.,
	\slPointsTo{t.\fixedPredNameFont{obs}}{\obBagVar}
	in standard separation logic syntax (cf.\@ \S~\ref{gs:sec:Nutshell:AbitraryDataStructures}).
}

% TODO: Remove
\myComment
{
	We denote the standard separating conjunction by \slStar.
	Considering the augmented semantics, we see that the assertion
	$\obs{\obBagVar \msCup \nextObBagVar}$
	is not equivalent to
	$\obs{\obBagVar} \slStar \obs{\nextObBagVar}$.
	The assertion
	$\obs{\obBagVar \msCup \nextObBagVar}$ represents a single heap cell $t.\obsPred$ that stores the union $\obBagVar \msCup \nextObBagVar$, i.e., a single heap chunk \slPointsTo{t.\fixedPredNameFont{obs}}{\obBagVar \msCup \nextObBagVar}.
	In contrast, $\obs{\obBagVar} \slStar \obs{\nextObBagVar}$ represents a heap where  \obBagVar and \nextObBagVar are stored in two different heap cells, i.e.,
	\slPointsTo{t.\obsPred}{\obBagVar}
	and
	\slPointsTo{t.\obsPred^\prime}{\nextObBagVar}.
}

% TODO: Remove
\myComment
{
	Further, our proof system and the augmented semantics both maintain the invariant that every thread holds exactly one obligations chunk \obs{\obBagVar}, provided that we start with exactly one in the precondition.
	Consider an arbitrary precondition $\obs{\obBagVar} \slStar \assPreVar$ where \assPreVar does not contain any obligations chunks.
	Then, using our proof rules it is not possible to prove a specification of the form
	\htProves
	{\obs{\obBagVar} \slStar \assPreVar}
	{\cmdVar}
	{\obs{\nextObBagVar} \slStar \obs{\nextNextObBagVar} \slStar{} \dots}.
}

    \begin{figure}
            \begin{subfigure}{\textwidth}
                    $$
                    \begin{array}{l p{0.5cm} l}
                            \progProof{\noObs}
                            \\
                            \keyword{let}\ \sigProgVar := \cmdNewSignal\ \keyword{in}
                                    &&\proofRuleHint{\prMinSigNewSignalName with $\levVar = 0$}
                            \\
                            \progProof{
                                    \obsOf{(\sigProgVar, 0)}
                                    \slStar
                                    \assSignalNoBool{(\sigProgVar, 0)}
                            }
                                    &&\proofDef{\sigVar := (\sigProgVar, 0)}
                            \\
                            \keyword{fork}\ (
                                    \progProof{
                                            \noObs
                                            \slStar
                                            \assSignalNoBool{\sigVar}
                                    }
                            \\
                            \phantom{\keyword{fork}\ (}\
                                    \cmdAwaitNoMut{\cmdIsSignalSet{\sigProgVar}}
                                            &&\proofHint{\ $\sigGetLev{\sigVar} = 0  \levObsLt \msEmpty$}
                            \\
                            \phantom{\keyword{fork}\ (}
                                    \progProof{
                                            \noObs
                                            \slStar
                                            \assSignalNoBool{\sigVar}
                                    });
                            \\
                            \progProof{
                                    \obsOf{\sigVar}
                            }
                            \\
                            \cmdSetSignal{\sigProgVar}
                            \\
                            \progProof{\noObs}
                    \end{array}
                    $$
                    \caption{
                            Proof outline for program from Fig.~\ref{gs:fig:MinEx:RealThreadSafeSignals}.
                            Applied proof rule marked in \proofRuleHint{purple}.
                            Abbreviation marked in \proofDef{\text{red}}.
                            General hint marked in \ \proofHint{grey}.
                    }
                    \label{gs:fig:MinEx:RealThreadSafeSignals:ProofSketch}
            \end{subfigure}
            
            \begin{subfigure}{\textwidth}
                    \begin{mathpar}
                            \prMinSigNewSignal
                            \and
                            \prMinSigSetSignal
                            \and
                            \and
                            \prMinSigFork
                            \and
                            \prMinSigAwait
                            \and
                            \prLet
                    \end{mathpar}
                    \caption{Proof rules. Rules only used in this section marked with \ensuremath{{}^\prime}.}
                    \label{gs:fig:MinEx:RealThreadSafeSignals:ProofRules}
            \end{subfigure}
            \caption{Verifying termination of minimal example with physical thread-safe signal.}
    \end{figure}

    \subsection{Non-Thread-Safe Physical Signals}\label{gs:sec:Nutshell:NonThreadSafePhysicalSignals}
%    As a step towards supporting waiting for arbitrary conditions on arbitrary shared data structures, including non-thread-safe ones, we now move to non-thread-safe signals. 
    As a step towards supporting waiting for arbitrary conditions over shared data structures, including non-thread-safe ones, we now move to non-thread-safe signals. 
    For simplicity, in this chapter we consider programs that use mutexes to synchronize concurrent accesses to shared data structures. 
    (Our ideas apply equally to programs that use other constructs, such as atomic machine instructions.) 
    Fig.~\ref{gs:fig:MinEx:RealNonThreadSafeSignals:CodeAndSyntacticSugar} presents our updated example.
    
    \begin{figure}
            \begin{subfigure}[b]{\textwidth}
            $$
            \begin{array}{l}
                    \keyword{let}\ \sigProgVar := \cmdNewSignal\ \keyword{in}\\
                    \keyword{let}\ \mutProgVar := \cmdNewMut\ \keyword{in}\\
                    \cmdFork{
                            \cmdAwait{\mutProgVar}{\cmdIsSignalSet{\sigProgVar}}
                    };
                    \\
                    \cmdAcquireMut{\mutProgVar};
                    \\
                    \cmdSetSignal{\sigProgVar};
                    \\
                    \cmdReleaseMut{\mutProgVar}
            \end{array}
            $$
            \caption{Code.}
            \end{subfigure}
			\\
			\\
            \begin{subfigure}[b]{\textwidth}
                    $$
                    \begin{array}{l c l}
                            \cmdAwait{\mutProgVar}{\cmdVar} &\ :=\
                                    &(\keyword{while}\
                                            \cmdAcquireMut{\mutProgVar};
                            \\
                                    &&\phantom{\keyword{while}\ }\
                                            \keyword{let}\ \resVar := \cmdVar\ \keyword{in}
                            \\
                                    &&\phantom{\keyword{while}\ }\
                                            \cmdReleaseMut{\mutProgVar};
                            \\
                                    &&\phantom{\keyword{while}\ }\
                                            \neg \resVar
                            \\
                                    &&\phantom{(}\keyword{do}\ \keyword{skip})
                    \end{array}
                    $$
                    \caption{Syntactic sugar. \resVar not free in \mutProgVar.}
%                    \label{fig:MinEx:RealNonThreadSafeSignals:Code}
            \end{subfigure}
            
            \caption{Minimal example with two threads communicating via a physical non-thread-safe signal protected by a mutex.}
            \label{gs:fig:MinEx:RealNonThreadSafeSignals:CodeAndSyntacticSugar}
    \end{figure}

    As signal \sigProgVar is no longer thread-safe, the two threads can no longer use it directly to communicate.
    Instead, we have to synchronize accesses to avoid data races.
    Hence, we protect the signal by a mutex \mutProgVar created by the main thread.
    In each iteration, the forked thread acquires the mutex, checks whether \sigProgVar has been set and releases it again.
    After forking, the main thread acquires the mutex, sets the signal and releases it again.

    \paragraph{Exposing Signal Values}
    Signals are specially marked heap cells storing Boolean values.
    We make this explicit by extending our signal chunks from \assSignalNoBool{\sigVar} to \assSignal{\sigVar}{\slBoolVar} where \slBoolVar is the current value of \sigVar and by updating our proof rules accordingly.
    Upon creation, signals are unset.
    Hence, creating a signal \sigProgVar now spawns an \emph{unset} signal chunk \assSignal{(\sigProgVar, \levVar)}{\slFalse} for some freely chosen level~\levVar and an obligation for $(\sigProgVar, \levVar)$, cf.~\prRealNonSafeSigsNewSignalName.
    We present our new proof rules in Fig.~\ref{gs:fig:MinEx:RealNonThreadSafeSignals:ProofRules} and demonstrate their application in Fig.~\ref{gs:fig:MinEx:RealNonThreadSafeSignals:ProofSketch}.

    \begin{figure}
%    \vspace{-0.7cm}
%            \begin{subfigure}{\textwidth}
                    $$
%					\hspace{-1.1cm}
\begin{array}{l l}
                            \progProof{\noObs}
                            \\
                            \keyword{let}\ \sigProgVar := \cmdNewSignal\ \keyword{in}
                                    &\proofRuleHint{\prRealNonSafeSigsNewSignalName with $\levVar = \meLevS$}
                            \\
                            \progProof{
                                    \obsOf{(\sigProgVar, \meLevS)}
                                    \slStar
                                    \assSignal{(\sigProgVar, \meLevS)}{\slFalse}
                            }
                                    &\proofRuleHint{\prViewShiftName \& \vsSemImpName}
                            \\
                            \progProof{
                                    \obsOf{(\sigProgVar, \meLevS)}
                                    \slStar
                                    \exists \slBoolVar.\ \assSignal{(\sigProgVar, \meLevS)}{\slBoolVar}
                            }
                                    &\proofDef{\sigVar := (\sigProgVar, \meLevS),\
                                            \invM := \exists \slBoolVar.\ \assSignal{\sigVar}{\slBoolVar}
                                    }
                            \\
                            \keyword{let}\ \mutProgVar := \cmdNewMut\ \keyword{in}
                                    &\proofRuleHint{\prRealNonSafeSigsNewMutexName with $\levVar = \levM$}
                            \\
                            \progProof{
                                            \obsOf{\sigVar}
                                            \slStar
                                            \assMutex{\mutVar}{\invM}
                                    }
                                            &\proofRuleHint{\prViewShiftName}
                            \\
\end{array}
                    $$
                    \vspace{-0.4cm}
                    \caption{
                            Proof outline for program \ref{gs:fig:MinEx:RealNonThreadSafeSignals:CodeAndSyntacticSugar},
                            verifying termination with mutexes \& non-thread safe signals.
                            Applied proof and view shift rules marked in \proofRuleHint{purple}.
                            Abbreviations marked in \proofDef{\text{red}}.
                            General hints marked in \ \proofHint{grey}.
                    }
%            \end{subfigure}
            
            \label{gs:fig:MinEx:RealNonThreadSafeSignals:ProofSketch}
%            \vspace{-0.8cm}
    \end{figure}

    \begin{figure}
            \begin{subfigure}{\textwidth}
            	\vspace{-0.8cm}
                    \begin{mathpar}
                            \prRealNonSafeSigsNewSignal
                            \and
                            \prRealNonSafeSigsSet
                            \and
                            \prRealNonSafeSigsIsSet
                            \and
                            {
                            	\hspace{-0.2cm}
	                            \prRealNonSafeSigsAwait
	                        }
                    \end{mathpar}
                    \caption{Signals \& busy waiting.}
            \end{subfigure}

            \begin{subfigure}{\textwidth}
                    \begin{mathpar}
                            \prRealNonSafeSigsNewMutex
                            \and
                            {
                            	\hspace{-0.2cm}
	                            \prRealNonSafeSigsAcquire
	                            \quad
	                            \prRealNonSafeSigsRelease
	                        }
                    \end{mathpar}
                    \caption{Mutexes.}
            \end{subfigure}

            \begin{subfigure}{\textwidth}
                    \begin{mathpar}
                            \prFrame
                            \and
                            \prExists
                            \and
                            {
                            	\hspace{-0.4cm}
	                            \prForkPaper
	                            \quad
	                            \prSimpleViewShiftPaper
	                        }
                    \end{mathpar}
                    \caption{Standard rules.}
            \end{subfigure}

            \begin{subfigure}{\textwidth}
                    \begin{mathpar}
                            \vsSemImp
                            \and
                            \vsTrans
                            \and
                            \vsRealNonSafeSigsCloneMut
                    \end{mathpar}
                    \caption{View shifts.}
            \end{subfigure}
            
            \caption{
                    Proof rules \& view shift rules for mutexes \& non-thread safe signals. 
                    Rules only used in this section marked with \ensuremath{{}^\prime{}^\prime}.
            }
            \label{gs:fig:MinEx:RealNonThreadSafeSignals:ProofRules}
    \end{figure}

    \paragraph{Data Races}
    As read and write operations on signals are no longer thread-safe, our logic has to ensure that two threads never try to access \sigProgVar at the same time.
    Hence, in our logic possession of a signal chunk \assSignal{\sigVar}{\slBoolVar} expresses (temporary) \emph{exclusive ownership} of \sigVar.
    Further, our logic requires threads to own any signal they are trying to access.
%    Hence, we use signal chunks to express that the threat holding the chunk \assSignal{\sigVar}{\slBoolVar} currently \emph{owns} the signals and therefore has the sole right to access the signal.
%    Following this interpretation, we require holding a chunk of the form \assSignal{(\sigProgVar, \levVar)}{\slBoolVar} to set \sigProgVar and to read its Boolean value \slBoolVar, cf.~\prRealNonSafeSigsSetName and~\prRealNonSafeSigsIsSetName.
    Specifically, when a thread wants to set \sigProgVar, it must hold a chunk of the form \assSignal{(\sigProgVar, \levVar)}{\slBoolVar}, cf.~\prRealNonSafeSigsSetName.
    The same holds for reading a signal's value, cf.~\prRealNonSafeSigsIsSetName.
    Note that signal chunks are not duplicable and only created upon creation of the signal they refer to.
    Therefore, holding a signal chunk for \sigProgVar indeed guarantees that the holding thread has the exclusive right to access \sigProgVar (while holding the signal chunk).
 
    \paragraph{Synchronization \& Lock Invariants}
    After the main thread creates \sigProgVar, it exclusively owns the signal.
    The main thread can transfer ownership of this resource during forking, cf.~\prMinSigForkName, and thereby allow the forked thread to busy-wait for \sigProgVar.
    This would, however, leave the main thread without any permission to set the signal and thereby discharge its obligation.

    We use mutexes to let multiple threads share ownership of a common set of resources in a synchronized fashion.
    Every mutex is associated with a \emph{lock invariant} \assLockInvVar, an assertion chosen by the proof author that specifies which resources the mutex protects.
    In our example, we want both threads to share \sigProgVar.
    To reflect the fact that the signal's value changes over time, we choose a lock invariant that abstracts over its concrete value.
    We choose
    $\assLockInvVar := \exists \slBoolVar.\ \assSignal{(\sigProgVar, \levVar)}{\slBoolVar}$.
    Let us ignore the chosen signal level \levVar for now.
    Creating the mutex \mutProgVar consumes this lock invariant and binds it to \mutProgVar by creating a mutex chunk \assMutex{(\mutProgVar, \dots)} {\invM}, cf.~\prRealNonSafeSigsNewMutexName.
    Thereby, the main thread loses access to \sigProgVar.
    The only way to regain access is by acquiring~\mutProgVar, cf.~\prRealNonSafeSigsAcquireName.
    Once the thread releases \mutProgVar, it again loses access to all resources protected by the mutex, cf.~\prRealNonSafeSigsReleaseName.
    
    \paragraph{Deadlocks}
    We have to ensure that any acquired mutex is eventually released, again.
    Hence, acquiring a mutex spawns a release obligation for this mutex and the only way to discharge this obligation is indeed by releasing it, cf.~\prRealNonSafeSigsAcquireName and \prRealNonSafeSigsReleaseName.
    
    Any attempt to acquire a mutex will block until the mutex becomes available.
    In order to prove that our program terminates, we have to prove that it does not get stuck during an acquisition attempt.
    To prevent wait cycles involving mutexes, we require the proof author to associate every mutex as well (just like signals) with a level \levVar.
    This level can be freely chosen during the mutex' creation, cf.~\prRealNonSafeSigsNewMutexName.
    Mutex chunks therefore have the form \assMutex{(\hlocVar, \levVar)}{\assLockInvVar} where~\hlocVar is the heap location the mutex is stored at.
    Their only purpose is to record the level and lock invariant a mutex is associated with.
    Hence, these chunks can be freely duplicated as we will see later.
    Generally, we denote mutex tuples by $\mutVar = (\hlocVar, \levVar)$.
    We only allow to acquire a mutex if its level is lower than the level of each held obligation, cf.~\prRealNonSafeSigsAcquireName.
    This also prevents any thread from attempting to acquire mutexes twice, e.g., 
    $\cmdAcquireMut{\mutProgVar}; \cmdAcquireMut{\mutProgVar}$ or \cmdAwait{\mutProgVar}{\cmdAcquireMut{\mutProgVar}}.

    \paragraph{View Shifts}
    When verifying a program, it can be necessary to reformulate the proof state and to draw semantic conclusions.
    To allow this we introduce a so-called \emph{view shift} relation~\viewShiftSymb~\cite{Jung2018IrisGroundUp}.
    By applying proof rule~\prViewShiftName and \vsSemImpName we can strengthen the precondition and weaken the postcondition.
    In our example, we use this to convert the unset signal chunk into the lock invariant which abstracts over the signal's value, i.e.,
    \viewShift
            {\assSignal{\sigVar}{\slFalse}}
            {\exists \slBoolVar.\ \assSignal{\sigVar}{\slBoolVar}}.

    The logic we present in this work is an intuitionistic separation logic that allows us to drop chunks.
    \footnote{
            This allows a thread to drop its obligations chunk \obs{\obBagVar}.
            Note, however, that by dropping this chunk the thread does not drop its obligations, but only its ability to show what its obligations are.
            In particular the thread would be unable to present an empty obligations chunk upon termination.
    }
    This allows us to simplify the postcondition of our fork proof rule's premise from $\noObs \slStar \assPostVar$ to \noObs, cf.~\prForkName, and drop all unneeded chunks via a semantic implication 
    \viewShift{\noObs \slStar \assPostVar}{\noObs}.
    
    We also allow to clone mutex chunks via view shifts, cf.~\vsRealNonSafeSigsCloneMutName.
    In our example, this is necessary to inform both threads which level and lock invariant mutex \mutProgVar is associated with.
    That is, the main thread clones the mutex chunk \assMutex{\mutVar}{\assLockInvVar} and passes one chunk on when it forks the busy-waiting thread.

    In \S~\ref{gs:sec:Nutshell:SignalErasure} we extend our view shift relation and revisit our interpretation of what a view shift expresses. 
    The full set of rules we use to define \viewShiftSymb is presented in the appendix (cf.\@~Fig.~\ref{gs:appendix:fig:ViewShiftRules}).

    \paragraph{Busy Waiting}
    In the approach presented in this chapter, for simplicity we only support busy-waiting loops of the form \cmdAwait{\mutProgVar}{\cmdVar}, which is syntactic sugar for
    \cmdWhileSkip
            {
                    \cmdAcquireMut{\mutProgVar};
                    \cmdLet{\resVar}{\cmdVar}
                            {
                                    \cmdReleaseMut{\mutProgVar};
                                    \neg \resVar
                            }
            }
    where \resVar denotes a fresh variable.~\footnote{
            As we discuss in \S~\ref{gs:sec:ToolSupport}, in the technical report accompanying this chapter (appendix~\ref{ch:appendix-ghostSignals-TR}) we present a more general logic that imposes no such syntactic restrictions.
     }
    In each iteration, the loop tries to acquire \mutProgVar, executes \cmdVar, releases \mutProgVar again and lets the result returned by \cmdVar determine whether the loop continues.
    Such loops can fail to terminate for two reasons:
    (i)~Acquiring \mutProgVar can get stuck and
    (ii)~the loop could diverge.
    
    We prevent the loop from getting stuck by requiring \mutProgVar's level to be lower than the level of each held obligation, cf.~\prRealNonSafeSigsAwaitName.
    Further, we enforce termination by requiring the loop to wait for a signal.
    That is, when verifying a busy-waiting loop using our approach, the proof author must choose a fixed signal and prove that this signal remains unset at the end of every non-finishing iteration.
    This way, we can prove that the loop terminates by proving that every signal is eventually set, just as in \S~\ref{gs:sec:Nutshell:ThreadSafePhysicalSignals}.
    And just as before, our logic requires the level of the waited-for signal to be lower than the level of each held obligation.
    
    Acquiring the mutex in every iteration makes the lock invariant available during the verification of the loop body \cmdVar.
    This lock invariant has to be restored at the end of the iteration such that it can be consumed during the mutex's release.
    \prRealNonSafeSigsAwaitName allows for an additional view shift to restore the invariant.
    In our example, we end our busy-waiting loop's non-finishing iterations with the assertion \assSignal{\sigVar}{\slFalse}.
    We use a semantic implication view shift to convert the signal chunk into the mutex invariant 
    $\exists \slBoolVar.\ \assSignal{\sigVar}{\slBoolVar}$.

    \paragraph{Choosing Levels}
    In our example, we have to assign levels to the mutex \mutProgVar and to the signal \sigProgVar.
    Our proof rules for mutex acquisition and busy waiting impose some restrictions on the levels of the involved mutexes and signals.
    By analysing the corresponding rule applications that occur in our proof, we can derive which constraints our level choice must comply with.
    Our example's verification involves one application of \prRealNonSafeSigsAcquireName and one application of \prRealNonSafeSigsAwaitName:
    (i)~Our main thread tries to acquire \mutProgVar while holding an obligation to set \sigProgVar.
    (ii)~The forked thread busy-waits for \sigProgVar while not holding any obligations.    
    Our assignment of levels must therefore satisfy the single constraint $\mutGetLev{\mutVar} \levLt \sigGetLev{\sigVar}$. 
    So, we choose $\LevelSet = \setOf{0, 1}$,
    $\mutGetLev{\mutVar} = \levM$ and
    $\sigGetLev{\sigVar} = \meLevS$.

%    \subsection{Arbitrary Data Structures}
    \subsection{Arbitrary Data Structures}\label{gs:sec:Nutshell:AbitraryDataStructures}
    
    The proof rules we introduced in \S~\ref{gs:sec:Nutshell:NonThreadSafePhysicalSignals} allow us to verify programs busy-waiting for arbitrary conditions over arbitrary shared data structures as follows:
%    Let $C_1, \dots, C_N$ be the conditions that $P$ waits for.
%    (i)~We introduce corresponding signals $\sigVar_1, \dots, \sigVar_N$ and insert them in the program.
%    (ii)~Whenever some thread performs an action establishing some condition $C_i$ it also sets $\sigVar_i$.
%    Further, we prove invariants stating that for every $i$, signal $\sigVar_i$ is set if and only if condition $C_i$ holds.
%    (iii)~The thread waiting for $C_i$ can use $\sigVar_i$.
    For every condition $C$ the program waits for, the proof author inserts a signal \sigVar into the program.
    They ensure that \sigVar is set at the same time the program establishes~$C$ and prove an invariant stating that the signal's value expresses whether $C$ holds.
    Then, the waiting thread can use \sigVar to wait for $C$.
    We illustrate this here for the simplest case of setting a single heap cell in Fig.~\ref{gs:fig:MinEx:Heap:Code}.

    \begin{figure}
            \vspace{-0.7cm}
            \begin{subfigure}[t]{0.45\textwidth}
                    $$
\begin{array}[t]{l}
                            \keyword{let}\ \x := \cmdAlloc{0}\ \keyword{in}\\
                            \keyword{let}\ \mutProgVar := \cmdNewMut\ \keyword{in}\\
                            \cmdFork{
                                    \cmdAwait{\mutProgVar}{\cmdHeapLocReadEq{\x}{1}}
                            };
                            \\
                            \cmdAcquireMut{\mutProgVar};
                            \\
        %                    \hline
        %                    \\
                            \cmdAssignToHeap{\x}{1};
                            \\
                            \cmdReleaseMut{\mutProgVar}
                    \end{array}
                    $$
                    \vspace{-0.3cm}
                    \caption{Example program with busy waiting for heap cell \x to be set.}
                    \label{gs:fig:MinEx:Heap:Code}
            \end{subfigure}
            \hspace{0.7cm}
            \begin{subfigure}[t]{0.45\textwidth}
                    $$
                    \begin{array}[t]{l}
                            \keyword{let}\ \x := \cmdAlloc{0}\ \keyword{in}\\
                            \!\!\progProofNew{
                                    \keyword{let}\ \sigProgVar := \cmdNewSignal\ \keyword{in}
                            }\\
                            \keyword{let}\ \mutProgVar := \cmdNewMut\ \keyword{in}
        \vspace{-0.072cm}
                            \\
                            \cmdFork{
                                    \cmdAwait{\mutProgVar}{\cmdHeapLocReadEq{\x}{1}}
                            };
        \vspace{-0.072cm}
                            \\
                            \cmdAcquireMut{\mutProgVar};
                            \\
        %                    \hline
        %                    \\
                            \cmdAssignToHeap{\x}{1};
                            \\
                            \!\!\progProofNew{
                                    \cmdSetSignal{\sigProgVar};
                            }
                            \\
                            \cmdReleaseMut{\mutProgVar}
                    \end{array}
                    $$
                    \vspace{-0.3cm}
                    \caption{
                            Example program~\ref{gs:fig:MinEx:Heap:Code} with additional signal \sigProgVar inserted, marked in \progProofNew{\text{green}}.
                            \sigProgVar and \x are kept in sync.
                    }
                    \label{gs:fig:MinEx:HeapSignal:Code}
            \end{subfigure}
            
            \begin{subfigure}{\textwidth}
                    $$
                    \begin{array}{l c l}
                            \cmdHeapLocReadEq{\expVar}{\nextExpVar} &\ := \
                                    &(\cmdLet{\resVar}
                                            {\cmdReadHeapLoc{\expVar}}
                                            {\resVar = \nextExpVar})
                    \end{array}
                    $$
                    \vspace{-0.4cm}
                    \caption{Syntactic sugar. \resVar free in \nextExpVar.}
            \end{subfigure}

            \caption{Minimal example illustrating busy waiting for condition over heap cell.}
            \vspace{-0.3cm}
    \end{figure}

    The program involves three new non-thread-safe commands:
    (i)~\cmdAlloc{\valVar} for allocating a new heap cell and initializing it with value~\valVar,
    (ii)~\cmdAssignToHeap{\hlocVar}{\valVar} for assigning value~\valVar to heap location~\hlocVar,
    (iii)~\cmdReadHeapLoc{\hlocVar} for reading the value stored in heap location~\hlocVar.
    We use \cmdHeapLocReadEq{\hlocVar}{\valVar} as syntactic sugar for 
    \cmdLet{\resVar}
            {\cmdReadHeapLoc{\expVar}}
            {\resVar = \nextExpVar}.
    
    In our example, the main thread allocates \x, initializes it with the value 0 and protects it using mutex \mutProgVar.
    It forks a new thread busy-waiting for \x to be set.
    Afterwards, the main thread sets \x.
    As explained above, we verify the program by inserting a signal \sigProgVar that reflects whether \x has been set, yet.
    Fig.~\ref{gs:fig:MinEx:HeapSignal:Code}~presents the resulting code.
    The main thread creates the signal and sets it when it sets \x.

    \begin{figure}
            \vspace{-0.6cm}
            \begin{subfigure}{\textwidth}
                    $$
                    \hspace{-0.2cm}
\begin{array}{l p{0.0cm} l}
                            \progProof{\noObs}
                            \\
                            \keyword{let}\ \x := \cmdAlloc{0}\ \keyword{in}
                            \\
                            \progProof{
                                    \noObs
                                    \slStar
                                    \slPointsTo{\x}{0}
                            }
                            \\
                            \keyword{let}\ \sigProgVar := \cmdNewSignal\ \keyword{in}
                                    &&\proofRuleHint{\prRealNonSafeSigsNewSignalName with $\levVar = \meLevS$}
                            \\
                            \keyword{let}\ \mutProgVar := \cmdNewMut\ \keyword{in}
                                    &&\proofRuleHint{\prRealNonSafeSigsNewMutexName with $\levVar = \levM$}
                            \\
                            \proofDef{
                                    \sigVar := (\sigProgVar, \meLevS),\
                                    \mutVar := (\mutProgVar, \levM)
                            }
                            \\
                            \proofDef{
                                    \invM := \exists \valVar.\ 
                                                    \slPointsTo{\x}{\valVar}
                                                    \slStar
                                                    \assSignal{\sigVar}{\valVar = 1}
                            }
                            \\
                            \progProof{
                                    \obsOf{\sigVar}
                                    \slStar
                                    \assMutex{\mutVar}{\invM}
                                    \slStar
                                    \assMutex{\mutVar}{\invM}
                            }
                            \\
\end{array}
                    $$
                    \vspace{-0.4cm}
                    \caption{
                            Proof outline for program~\ref{gs:fig:MinEx:HeapSignal:Code}.
                            Applied proof rules marked in \proofRuleHint{purple}.
                            Abbreviations marked in \proofDef{\text{red}}.
                            General hints marked in \ \proofHint{grey}.
                    }
                    \label{gs:fig:MinEx:HeapSignal:ProofSketch}
            \end{subfigure}

            \begin{subfigure}{\textwidth}
                    \begin{mathpar}
                            \prAlloc
                            \and
                            \prAssignToHeapPaper
                            \and
                            \prHeapCellsReadHeapLoc
                            \and
                            \prExp
                    \end{mathpar}
                    \vspace{-0.4cm}
                    \caption{
                            Proof rules.
                            Evaluation function \evalExp{\cdot}.
                            Rules only used in this section marked with~${}^\prime{}^\prime{}^\prime$.
                    }
                    \label{gs:fig:MinEx:Heap:ProofRules}
            \end{subfigure}
            \caption{Verifying termination of busy waiting for condition over heap cell.}
            \vspace{-0.5cm}
    \end{figure}

    \paragraph{Heap Cells}
    Verifying this example does not conceptually differ from the example we presented in \S~\ref{gs:sec:Nutshell:NonThreadSafePhysicalSignals}.
    Fig.~\ref{gs:fig:MinEx:Heap:ProofRules} presents the new proof rules we need and Fig.~\ref{gs:fig:MinEx:HeapSignal:ProofSketch} sketches our example's verification.
    As with non-thread-safe signals, we have to prevent multiple threads from trying to access \x at the same time in order to prevent data races.
    For this we use so-called \emph{points-to} chunks~\cite{Reynolds2002SeparationLA, OHearn2001LocalRA}.
    They have the form \slPointsTo{\hlocVar}{\valVar} and express that heap location~\hlocVar stores the value~\valVar.
    When a thread holds such a chunk, it exclusively owns the right to access heap location~\hlocVar.
    
    Heap locations are unique and the only way to create a new points-to chunk is to allocate and initialize a new heap cell via \cmdAlloc{\valVar}, cf.~\prAllocName.
    Hence, there will never be two points-to chunks involving the same heap location.
    In order to read or write a heap cell via \cmdReadHeapLoc{\hlocVar} or \cmdAssignToHeap{\hlocVar}{\expVar}, the acting thread must first acquire possession of the corresponding points-to chunk, cf.~\prAssignToHeapName and~\prHeapCellsReadHeapLocName.
    
    \paragraph{Relating Signals to Conditions}
    In our example, the forked thread busy-waits for \x to be set while our proof rules require us to justify each iteration by showing an unset signal.
    That is, we must prove an invariant stating that the value of \x matches \sigProgVar.
    As this invariant must be shared between both threads, we encode it in the lock invariant:
    $\assLockInvVar := \exists \valVar.\ \slPointsTo{\x}{\valVar} \slStar \assSignal{\sigVar}{\valVar = 1}$.
    This does not only allow both threads to share the heap cell and the signal but it also automatically enforces that they maintain the invariant whenever they acquire and release the mutex.
%    Fig.~\ref{gs:fig:MinEx:HeapSignal:ProofSketch} presents the resulting proof.

\subsection{Signal Erasure}\label{gs:sec:Nutshell:SignalErasure}
    
    In the program from Fig.~\ref{gs:fig:MinEx:HeapSignal:Code} signal \sigProgVar is never read and does hence not influence the waiting thread's runtime behaviour.
    Therefore, we can verify the original program presented in Fig.~\ref{gs:fig:MinEx:Heap:Code} by erasing the physical signal and treating it as ghost code.

    \paragraph{Ghost Signals}
    Central aspects of the proof sketch we presented in Fig.~\ref{gs:fig:MinEx:HeapSignal:ProofSketch} are that
    (i)~the main thread was obliged to set \sigProgVar and that
    (ii)~the value of \sigProgVar reflected whether \x was already set.
%    Even though our program does not need the signal \sigProgVar to run, we previously relied on it to verify that the forked thread's busy-waiting terminates.
%    In particular, we relied on the assumption that the main thread was obliged to perform an action (setting \sigProgVar) which through the lock invariant was bound to the event the forked thread was waiting for (\x being set).
    \emph{Ghost signals} allow us to keep this information but at the same to remove the physical signals from the code.
%    We can keep this information but at the same time eliminate signals from the program code by treating them as ghost resources.
    Ghost signals are essentially identical to the physical non-thread-safe signals we used so far.
    However, as ghost resources they cannot influence the program's runtime behaviour.
    They merely carry information we can use during the verification process.
    
    \paragraph{View Shifts Revisited}
    We implement ghost signals by extending our view shift relation.
    In particular, we introduce two new view shift rules: \vsNewSignalName and \vsSetSignalName presented in Fig.~\ref{gs:fig:MinExSignalsErased:ViewShiftRules}.
    The former creates a new unset signal and simultaneously spawns an obligation to set it.
    The latter can be used to set a signal and thereby discharge a corresponding obligation.
    We say that these rules change the \emph{ghost state} and therefore call their application a \emph{ghost proof step}.
    With this extension, a view shift \viewShift{\assPreVar}{\assPostVar} expresses that we can reach postcondition \assPostVar from precondition \assPreVar by
    (i)~drawing semantic conclusions or by
    (ii)~manipulating the ghost state.
    In Fig.~\ref{gs:fig:MinExSignalsErased:ProofSketch} we use ghost signals to verify the program from~\ref{gs:fig:MinEx:Heap:Code}.

    Note that lifting signals to the verification level does not affect the soundness of our approach.
    The argument we presented in \S~\ref{gs:sec:Nutshell:ThreadSafePhysicalSignals} still holds.
    We formalize our logic and provide a formal soundness proof in appendix~\S~\ref{ch:appendix-paper-2-ghostSignals} and in the technical report~\cite{Reinhard2020GhostSignalsTR} (appendix~\S~\ref{ch:appendix-ghostSignals-TR}).
    The latter contains a more general version of the presented logic that 
    (i) is not restricted to busy-waiting loops of the form \cmdAwait{\mutProgVar}{\cmdVar} and that
    (ii)~is easier to integrate into existing tools like VeriFast~\cite{Jacobs2011Verifast}, as explained in \S~\ref{gs:sec:ToolSupport}.

    \begin{figure}
            \vspace{-1.3cm}
            \begin{subfigure}{\textwidth}
                    $$
                    \hspace{-0.2cm}
\begin{array}{l l}
                            \progProof{\noObs}
                            \\
                            \keyword{let}\ \x := \cmdAlloc{0}\ \keyword{in}
                            \\
                            \progProof{\noObs \slStar \slPointsTo{\x}{0}}
                            \\
                            \proofHint{\ new\_ghost\_signal};
                                    &\proofRuleHint{\vsNewSignalNoBoolName with $\levVar = \meLevS$}
                            \\
                            \progProofML{
                                    \exists \sigProgVar.\
                                    \obsOf{(\sigProgVar, \meLevS)}
                                    \slStar
                                    \slPointsTo{\x}{0}
                                    \!\!
                                    \\
                                    \slStar\
                                    \assSignal{(\sigProgVar, \meLevS)}{\slFalse}
                                    \!\!
                            }
                                    &\proofDef{\sigVar := (\sigProgVar, \meLevS)}
                            \\
\end{array}
                    $$
%                    \vspace{-0.3cm}
                    \caption{
                            Proof outline for the program presented in Fig.~\ref{gs:fig:MinEx:Heap:Code}.
                            Auxiliary commands hinting at view shifts and general hints marked in \ \proofHint{grey (italic)}.
                            Applied proof and view shift rules marked in \proofRuleHint{purple}.
                            Abbreviations marked in \proofDef{\text{red}}.
                    }
                    \label{gs:fig:MinExSignalsErased:ProofSketch}
            \end{subfigure}
            
            \begin{subfigure}{\textwidth}
                    \begin{mathpar}
                            \vsNewSignal
                            \and\hspace{-0.05cm}
                            \vsSetSignal
                    \end{mathpar}
                    \caption{
                            Proof rules.
                    }
                    \label{gs:fig:MinExSignalsErased:ViewShiftRules}
            \end{subfigure}
            \caption{Verifying termination with ghost signals.}
            \vspace{-1.7cm}
    \end{figure}

\section{A Realistic Example}\label{gs:sec:RealisticExample}
	
	\vspace{-0.3cm}
    To demonstrate the expressiveness of the presented verification approach, we verified the termination of the program presented in Fig.~\ref{gs:fig:RealisticExample}.
    It involves two threads, a consumer and a producer, communicating via a shared bounded FIFO with a maximal capacity of 10.
    The producer enqueues numbers 100, \dots, 1 into the FIFO and the consumer dequeues those.
    Whenever the queue is full, the producer busy-waits for the consumer to dequeue an element.
    Likewise, whenever the queue is empty, the consumer busy-waits for the producer to enqueue the next element.
    Each thread's finishing depends on the other thread's productivity.
    This is, however, no cyclic dependency.
    For instance, in order to prove that the producer eventually pushes number $i$ into the queue, we only need to rely on the consumer to pop $i+10$.
    A similar property holds for the consumer.

	\vspace{-0.3cm}

    \begin{figure}
    \vspace{-0.65cm}
    \begin{subfigure}{\textwidth}
%    \vspace{-0.5cm}
    {
            \NewDocumentCommand{\fifo}{}{\progVar{fifo_{10}}}
            \NewDocumentCommand{\fifoLoaded}{}{\progVar{f}}
            %\NewDocumentCommand{\m}{}{\progVar{m}}
            \RenewDocumentCommand{\m}{}{\progVar{m}}

            \NewDocumentCommand{\counter}{m}{\progVar{c_{#1}}}
            
            % producing thread
                    \NewDocumentCommand{\producerID}{}{p}
                    \NewDocumentCommand{\pc}{}{\counter{\producerID}}
                    \NewDocumentCommand{\pcLoaded}{}{\progVar{c}}

            % consumer counter
                    \NewDocumentCommand{\consumerID}{}{c}
                    \NewDocumentCommand{\cc}{}{\counter{\consumerID}}
                    \NewDocumentCommand{\ccLoaded}{}{\progVar{c}}
                    
                    % signals
                            \NewDocumentCommand{\idGen}{m m}{\metaVar{\idVar_\text{#1}^{#2}}}
                            
                            % signal set when number is pushed into FIFO
                            \NewDocumentCommand{\idPush}{m}{\idGen{push}{#1}}
                            \NewDocumentCommand{\levPush}{m}{\metaVar{\levVar_\text{push}^{#1}}}
                            
                            % signal set when number is popped from FIFO
                            \NewDocumentCommand{\idPop}{m}{\idGen{pop}{#1}}
                            \NewDocumentCommand{\levPop}{m}{\metaVar{\levVar_\text{pop}^{#1}}}
                            
                            \NewDocumentCommand{\sigPush}{m}{\metaVar{\sigVar_\text{push}^{#1}}}
                            \NewDocumentCommand{\sigPop}{m}{\metaVar{\sigVar_\text{pop}^{#1}}}
            
            $$
            \hspace{-0.3cm}
\begin{array}{p{0.68\textwidth} l}
                \proofHint{\ alloc\_ghost\_signal\_IDs(\idPop{i}, \idPush{i})
                        \ \ \text{for}\ \ 
                        $1 \leq i \leq 100$;~\hspace{-20cm}
                }
                \vspace{0.1cm}
                \\
                \proofDef{
                        \levPop{i} := 102 - i,\ \
                        \levPush{i} := 101 - i,\ \
%                        \sigPop{i} := (\idPop{i}, \levPop{i}),\ \
%                        \sigPush{i} := (\idPush{i}, \levPush{i})
                        \sigVar_x^i := (\idVar_x^i, \levVar_x^i)\ \
                        \text{for}\ \ 1 \leq i \leq 100
                        \hspace{-20cm}
                }
                \vspace{0.1cm}
                \\
                \proofHint{\ init\_ghost\_signals(\sigPop{100}, \sigPush{100});}
                \vspace{0.1cm}
                \\
                \progProof{\obsOf{\sigPop{100}, \sigPush{100}} \slStar \dots}
                \\
                \keyword{let}\ \fifo := \cmdAlloc{\lstNil}\ \keyword{in}\\
                \keyword{let}\ \mutProgVar := \cmdNewMut\ \keyword{in}
                \hspace{-5cm}\\
                \keyword{let}\ \pc := \cmdAlloc{100}\ \keyword{in}\\
                \keyword{let}\ \cc := \cmdAlloc{100}\ \keyword{in}\
                \hspace{-5cm}
                \\
\end{array}
            $$
            
            \vspace{-0.4cm}
            \caption[Producer-consumer program with bounded FIFO.]
            {
                    Example program with two threads communicating via a shared bounded FIFO with maximal size 10.
                    Auxiliary commands hinting at view shifts 
%                    (cf.~Fig.~\ref{gs:fig:ViewShiftRules--fineGrainedSignalCreation} for new rules) 
                    \& general hints~marked in \ \proofHint{grey (italic)}.
                    Abbreviations marked in \proofDef{\text{red}}.
                    Hints on proof state marked in \ghostFont{blue}.
            }
            \label{gs:fig:RealisticExample}
    }
    \end{subfigure}

%	\vspace{-0.2cm}

    \begin{subfigure}{\textwidth}
            \begin{mathpar}
                    \vsAllocSigID
                    \and
                    \vsSigInit
            \end{mathpar}
%            \vspace{-0.5cm}
            \caption{Fine-grained view shift rules for signal creation.}
            \label{gs:fig:ViewShiftRules--fineGrainedSignalCreation}
    \end{subfigure}
    
%    \vspace{-0.2cm}
    \caption{Realistic example program.}
%    \vspace{-1.5cm}
    \end{figure}

{
    %\NewDocumentCommand{\m}{}{\progVar{m}}
    \RenewDocumentCommand{\m}{}{\progVar{m}}
    \NewDocumentCommand{\idPush}{m}{\metaVar{\idVar_\text{push}^{#1}}}
    \NewDocumentCommand{\levPush}{m}{\metaVar{\levVar_\text{push}^{#1}}}
                            
    % signal set when number is popped from FIFO
    \NewDocumentCommand{\idPop}{m}{\metaVar{\idVar_\text{pop}^{#1}}}
    \NewDocumentCommand{\levPop}{m}{\metaVar{\levVar_\text{pop}^{#1}}}
                            
    \NewDocumentCommand{\sigPush}{m}{\metaVar{\sigVar_\text{push}^{#1}}}
    \NewDocumentCommand{\sigPop}{m}{\metaVar{\sigVar_\text{pop}^{#1}}}

    \paragraph{Fine-Tuning Signal Creation}
    To simplify complex proofs involving many signals we refine the process of creating a new ghost signal.
    For simplicity, we combined the allocation of a new signal ID and its association with a level and a Boolean in one step.
    For some proofs, such as the one we outline in this section, it can be helpful to fix the IDs of all signals that will be created throughout the proof already at the beginning.
    To realize this, we replace view shift rule \vsNewSignalName by the rules presented in Fig.~\ref{gs:fig:ViewShiftRules--fineGrainedSignalCreation} and adapt our signal chunks accordingly.
    With these more fine-grained view shifts, we start by allocating a signal ID, cf.\@ \vsAllocSigIDName.
    Thereby we obtain an \emph{uninitialized} signal \sigUninit{\idVar} that is not associated with any level or Boolean, yet.
    Also, allocating a signal ID does not create any obligation because threads can only wait for \emph{initialized} (and unset) signals.
    When we initialize a signal, we bind its already allocated ID to a level of our choice and associate the signal with \slFalse, cf.~\vsSigInitName.
    This creates an obligation to set the signal.
    
%    This change does not affect the soundness of our verification approach.
%    However, it drastically simplifies proofs, such as the one discussed below, that would normally require the conditional creation of new signals within loops.
%    Using the finer-grained view shifts, we can create all the signals we need centrally at the beginning of the proof.
%    Since we thereby do not create any obligations, this step does not affect the acquisition of locks later on, as long as the signals remain uninitialized.

    \paragraph{Loops \& Signals}
    In our program, both threads have a local counter initially set to 100 and run a nested loop.
    The outer loops are controlled by their thread's counter, which is decreased in each iteration until it reaches 0 and the loop stops.
    For such loops, we introduce a conventional proof rule for total correctness of loops, cf.\@ Fig.~\ref{gs:appendix:fig:ProofRulesPartOne} in the appendix.
    Verifying termination of the inner loops is a bit more tricky and requires the use of ghost signals.

    So far, we had to fix a single signal for the verification of every \keyword{await} loop.
    We can relax this restriction to considering a finite set of signals the loop may wait for, cf.~\prAwaitGenName presented in Fig.~\ref{gs:appendix:fig:ProofRulesPartOne} in the appendix.
    Apart from being a generalisation, this rule does not differ from \prRealNonSafeSigsAwaitName introduced in~\S~\ref{gs:sec:Nutshell:NonThreadSafePhysicalSignals}.

    Initially, we allocate 200 signal IDs $\idPush{100}, \dots, \idPush{1}, \idPop{100}, \dots, \idPop{1}$.
    We are going to ensure that always at most one push signal and at most one pop signal are initialized and unset.
%    and start with \sigPush{100} and \sigPop{100} as 100 is the first number produced.
    The producer and consumer are going to hold the obligation for the push and pop signal, respectively.
    The producer will hold the obligation for \sigPush{i} while $i$ is the next number to be pushed into the FIFO and it will set \sigPush{i} when it pushes the number $i$ into the FIFO.
    Meanwhile, the consumer will use \sigPush{i} to wait for the number $i$ to arrive in the queue when it is empty.
    Similarly, the consumer will hold the obligation for \sigPop{i} while number~$i$ is the next number to be popped from the FIFO and will set \sigPop{i} when it pops the number $i$.
    The producer uses \sigPop{i} to wait for the consumer to pop $i$ from the queue when it is full.
    At any time, we let the mutex \mutProgVar protect the two active signals and thereby make them accessible to both threads.
    
    \paragraph{Choosing the Levels}
    Note that we ignored the levels so far.
    The producer and the consumer both acquire the mutex while holding an obligation for a signal.
    Hence, we choose $\LevelSet = \N$, $\sigGetLev{\m} = 0$ and $\sigGetLev{\sigVar} > 0$ for every signal \sigVar.
    Both threads will justify iterations of their respective \keyword{await} loop by using an unset signal at the end of such an iteration.
    Our proof rules allow us to ignore the mutex obligation during this step.
    Hence, the mutex level does not interfere with the level of the unset signal.
%    The mutex level is irrelevant for this the mutex level does not interfere with proving termination of the \keyword{await} loops.
%    The producer waits when the queue is full and the consumer when it is empty.
    Whenever the queue is full, the producer waits for the consumer to pop an element and whenever the queue is empty, the consumer waits for the producer to push.
    That is, the producer waits for~\sigPop{i+10} while holding an obligation for~\sigPush{i} and the consumer waits for~\sigPush{i} while holding an obligation for~\sigPop{i}.
    So, we have to choose the signal levels such that 
    $\sigGetLev{\sigPop{i+10}} < \sigGetLev{\sigPush{i}}$
    and
    $\sigGetLev{\sigPush{i}} < \sigGetLev{\sigPop{i}}$
    hold.
    We solve this by choosing
    $\sigGetLev{\sigPop{i}} = 102 - i$
    and
    $\sigGetLev{\sigPush{i}} = 101 - i$.

%    \paragraph{Distributing Resources}
%    Both threads count from 100 downwards.
%    So, we use the allocated IDs \idPush{100}, \idPop{100} to initialize our first two signals $\sigPush{100} = (\idPush{100}, 1)$, $\sigPop{100} = (\idPop{100}, 2)$  and transfer their ownership to the mutex.
%    We split the mutex chunk and pass one half to each thread such that both can access the initialized signals.
%    Additionally, the producer thread receives the 100 iteration permissions, the obligation for \sigPush{100}, the uninitialized signal chunks \sigUninit{\idPush{99}}, \dots, \sigUninit{\idPush{1}} and the wait permissions \waitPerm[\idPop{100}, 0], \dots, \waitPerm[\idPop{1}, 0] upon its forking.
%    The rest remains with the consumer, i.e., 100 iteration permissions, the obligation for \sigPop{100}, uninitialized signal chunks \sigUninit{\idPop{99}}, \dots, \sigUninit{\idPop{1}} and wait permissions \waitPerm[\idPop{100}, 0], \dots, \waitPerm[\idPop{1}, 0].
    
    \paragraph{Verifying Termination}
    This setup suffices to verify the example program.
    Via the lock invariant, each thread has access to both active signals.
    Whenever the producer pushes a number $i$ into the queue, it sets \sigPush{i} which discharges the held obligation and decreases its counter.
    Afterwards, if $i > 1$, it uses the uninitialized signal chunk  \sigUninit{\idPush{i-1}} to initialize $\sigPush{i-1} = (\idPush{i-1}, 101 - (i-1))$ and replaces \sigPush{i} in the lock invariant by \sigPush{i-1} before it releases the lock.
    If $i = 1$, the counter reached 0 and the loop ends.
    In this case, the producer holds no obligation.
    The consumer behaves similarly.
    Since we proved that each thread discharged all its obligations, we proved that the program terminates.
    Fig.~\ref{gs:fig:RealisticExample} illustrates the most important proof steps.
    We present the program's verification in full detail in appendix~\S~\ref{gs:appendix:sec:VerificationOfRealisticExample} and in the technical report~\cite{Reinhard2020GhostSignalsTR} (appendix~\ref{gsTR:appendix:subsec:BoundedFifo}).
    Furthermore, we encoded~\cite{ArtifactJacobs2020VerifastGhostSignalConsumerProducerBoundedFifo} the proof in VeriFast~\cite{Jacobs2011Verifast}.
    
    The number of threads in this program is fixed.
    However, our approach also supports the verification of programs where the number of threads is not even statically bounded.
    In the appendix in \S~\ref{gs:appendix:sec:caseStudy:unbounded} we present and verify such a program.
    It involves $N$ producer and $N$ consumer threads that communicate via a shared buffer of size 1, for a random number $N > 0$ determined during runtime.
    % TODO: Remove
	\myComment
	{
		Similarly, we can generalize our proof to support an arbitrary number of elements sent through the buffer as well as an arbitrary buffer size or an unbounded buffer.
	}
}

\section{Specifying Busy-Waiting Concurrent Objects}\label{gs:sec:FineGrainedConcurrency}
	
    Our approach can be used to verify busy-waiting concurrent objects with respect to abstract specifications. 
    For example, we have verified~\cite{ArtifactJacobs2020VerifastCLHLock} the CLH lock \cite{herlihy-multiprocessor} against a specification that is very similar to our proof rules for built-in mutexes shown in Fig.~\ref{gs:fig:MinEx:RealNonThreadSafeSignals:ProofRules}. 
    The main difference is that it is slightly more abstract: when a lock is initialized, it is associated with a \emph{bounded infinite set} of levels rather than with a single particular level.
    (To make this possible, an appropriate universe of levels should be used, such as the set of lists of natural numbers, ordered lexicographically.)
    To acquire a lock, the levels of the obligations held by the thread must be above the elements of the set; the new obligation's level is an element of the set.

\section{Tool Support}\label{gs:sec:ToolSupport}
	
        We have extended the VeriFast tool~\cite{verifast2104} for separation logic-based modular verification of C and Java programs so that it supports verifying termination of busy-waiting C or Java programs.
        When verifying termination, VeriFast consumes a \emph{call permission} at each recursive call or loop iteration.
        In the technical report~\cite{Reinhard2020GhostSignalsTR} (appendix~\S~\ref{ch:appendix-ghostSignals-TR}) we define a generalised version of our logic that instead of providing a special proof rule for busy-waiting loops, provides \emph{wait permissions} and a \emph{wait view shift}. 
        A call permission of a \emph{degree} $\delta$ can be turned into a wait permission of a degree $\delta' < \delta$ for a given signal \sigVar. 
        A wait view shift for an unset signal \sigVar for which a wait permission of degree $\delta$ exists produces a call permission of degree $\delta$, which can be used to fuel a busy-waiting loop.
        When busy-waiting for some signal \sigVar, we can generate new permissions to justify each iteration as long as \sigVar remains unset.
        
        VeriFast allows threads to freely exchange permissions.
        This is useful to verify termination of non-blocking algorithms involving compare-and-swap loops~\cite{Jacobs2018ModularTerminationVerification}. 
        However, we must be careful to prevent self-fueling busy-waiting loops.
        Hence, we restrict where a permission can be consumed based on the \emph{thread phase} it was created in.
        The main thread's initial phase is $\epsilon$.
        When a thread in phase $p$ forks a new thread, its phase changes to $p.\mathsf{Forker}$ and the new thread starts in phase $p.\mathsf{Forkee}$.
        We allow a thread in phase $p$ to consume a permission only if it was produced in an \emph{ancestor thread phase} $p' \fpLeq p$.

        The only change we had to make to VeriFast's symbolic execution engine was to enforce the thread phase rule. 
        We encoded the other aspects of the logic simply as axioms in a \emph{trusted header file}. 
        We used this tool support to verify the bounded FIFO (\S~\ref{gs:sec:RealisticExample}) and the CLH lock (\S~\ref{gs:sec:FineGrainedConcurrency}).
		The bounded FIFO proof~\cite{ArtifactJacobs2020VerifastGhostSignalConsumerProducerBoundedFifo} contains 160 lines of proof annotations for 37 lines of code (an annotation overhead of 435\%) and takes 0.08s to verify.
		The CLH lock proof~\cite{ArtifactJacobs2020VerifastCLHLock} contains 343 lines of annotations for 49 lines of code (an overhead of 700\%) and takes 0.1s to verify.
        \vspace{-0.2cm}

\section{Integrating Higher-Order Features}\label{gs:sec:HigherOrder}
	
    The logic we presented in this chapter does not support higher-order features such as assertions that quantify over assertions, or storing assertions in the (logical) heap as the values of ghost cells.
	While we did not need such features to carry out our example proofs, they are generally useful to verify higher-order program modules against abstract specifications. 
	The typical way to support such features in a program logic is by applying \emph{step indexing} \cite{appel-indexed,BIRKEDAL20104102}, where the domain of logical heaps is indexed by the number of execution steps left in the (partial) program trace under consideration. 
	Assertions stored in a logical heap at index $n+1$ talk about logical heaps at index $n$; i.e., they are meaningful only \emph{later}, after at least one more execution step has been performed.
	
	It follows that such logics apply directly only to \emph{partial} correctness properties. 
	Fortunately, we can reduce a termination property to a safety property by writing our program in a programming language \emph{instrumented} with runtime checks that guarantee termination. 
	Specifically, we can write our program in a programming language that fulfils the following criteria:
	It tracks signals, obligations and permissions at runtime and
	has constructs for signal creation, waiting and setting a signal.
	The \keyword{fork} command takes as an extra operand the list of obligations to be transferred to the new thread (and the other constructs similarly take sufficient operands to eliminate any need for angelic choice).
	Threads get stuck when these constructs' preconditions are not satisfied, such as when a thread waits for a signal while holding the obligation for that signal.
	%    Specifically, we can write our program in a programming language that tracks signals, obligations, iteration permissions, and wait permissions at run time, has constructs for iteration permission weakening, signal creation, wait permission creation, waiting, and setting a signal, and where the $\mathbf{fork}$ command takes as an extra operand the list of obligations to be transferred to the new thread (and the other constructs similarly take sufficient operands to eliminate any need for angelic choice), and that gets stuck when these constructs' preconditions are not satisfied, such as when a thread waits for a signal while holding the obligation for that signal, or when it performs a loop iteration without holding an iteration permission. 
	We can then use a step-indexing-based higher-order logic such as Iris \cite{Jung2018IrisGroundUp} to verify that no thread in our program ever gets stuck. 
	Once we established this, we know none of the instrumentation has any effect and can be safely \emph{erased} from the~program.

\section{Related \& Future Work}\label{gs:sec:RelatedWork}
	
\NewDocumentCommand{\TadaLive}{}{TaDA Live\xspace}
\NewDocumentCommand{\Lili}{}{LiLi\xspace}

   	In recent work~\cite{Reinhard2020AbruptExit} we propose a separation logic to verify termination of programs where threads busy-wait to be abruptly terminated. 
   	We generalize this work to support busy waiting for arbitrary conditions.
    
    In~\cite{Jacobs2018ModularTerminationVerification} we propose an approach based on \emph{call permissions} to verify termination of single- and multithreaded programs that involve loops and recursion.
    However, that work does not consider busy-waiting loops.
    In the technical report~\cite{Reinhard2020GhostSignalsTR} (appendix~\S~\ref{ch:appendix-ghostSignals-TR}), we present a generalised logic that uses call permissions and allows busy waiting to be implemented using arbitrary looping and/or recursion.
    Furthermore, the use of call permissions allowed us to encode our case studies in our VeriFast tool which also uses call permissions for termination verification.

    Liang and Feng~\cite{Liang2016LiliAPL, Liang2017LiliProgressOC} propose \Lili, a separation logic to verify liveness of blocking constructs implemented via busy waiting.
    In contrast to our verification approach, theirs is based on the idea of contextual refinement.
    In their approach, client code involving calls of blocking methods of the concurrent object is verified by first applying the contextual refinement result to replace these calls by code involving primitive blocking operations and then verifying the resulting client code using some other approach. 
    In contrast, specifications in our approach are regular Hoare-style triples and proofs are regular Hoare-style proofs.

    In~\cite{Jacobs2020IOLiveness} we propose a Hoare logic to verify liveness properties of the I/O behaviour of programs that do not perform busy waiting. 
    By combining that approach with the one we proposed in this chapter, we expect to be able to verify I/O~liveness of realistic concurrent programs involving both I/O and busy waiting, such as a server where one thread receives requests and enqueues them into a bounded FIFO, and another one dequeues them and responds.
    To support this claim, we encoded the combined logic in VeriFast and verified a simple server application where the receiver and responder thread communicate via a shared buffer~\cite{ArtifactJacobs2020VerifastIOLivenessServer}.

\section{Conclusion}\label{gs:sec:Conclusion}
	
    We propose what is to the best of our knowledge the first separation logic for verifying termination of programs with busy waiting.
    We offer a soundness proof of the system presented in this chapter in appendix~\S~\ref{ch:appendix-paper-2-ghostSignals}, and of a more general system in the technical report~\cite{Reinhard2020GhostSignalsTR} (appendix~\S~\ref{ch:appendix-ghostSignals-TR}).
    Further, we demonstrated its usability by verifying a realistic example.
    We encoded our logic and the realistic example in VeriFast~\cite{ArtifactJacobs2020VerifastGhostSignalConsumerProducerBoundedFifo} and used this encoding also to verify the CLH lock~\cite{ArtifactJacobs2020VerifastCLHLock}.
    Moreover, we expect that our approach can be integrated into other existing concurrent separation logics such as Iris~\cite{Jung2018IrisGroundUp}.

%%%%%%%%%%%%%%%%%%%%%%%%%%%%%%%%%%%%%%%%%%%%%%%%%%
% Keep the following \cleardoublepage at the end of this file, 
% otherwise \includeonly includes empty pages.
\cleardoublepage

% vim: tw=70 nocindent expandtab foldmethod=marker foldmarker={{{}{,}{}}}
%

%
  \graphicspath{{\chaptersdir/paper-3-ct4ms/image/}}%
  %\cleardoublepage%
  % !TeX root = ../../thesis.tex
\chapter{Completeness Thresholds for Memory Safety of Array Traversing Programs}\label{ch:paper-3-ct4ms}

This chapter was previously published as:

\citationBox{
	\textbf{T. Reinhard}, J. Fasse, and B. Jacobs. 
	\emph{Completeness Thresholds for Memory Safety of Array Traversing Programs.} 
	In Proceedings of the 12th ACM SIGPLAN 
	International Workshop on the State Of the Art in Program Analysis (SOAP) (2023), 
	ACM, pp. 47–54.
	\href
		{https://doi.org/10.1145/3589250.3596143}
		{doi.org/10.1145/3589250.3596143}
}
To streamline the notation, we replaced some constraints by constraint sets.

\section{Introduction}\label{ct:sec:Intro}

\paragraph{Unbounded vs Bounded Proofs}
Many techniques have been developed to convince ourselves of the trustworthiness of software.
A fundamental pillar for any higher-level property is memory safety.
In memory-unsafe languages the burden of proof lies with the programmer.
Yet, it remains hard to prove and in general requires us to write tedious, inductive proofs.
One way to automate the verification process is to settle on bounded proofs and accept bounded \mbox{guarantees}.

Consider a program \cmdVar that searches through an array of size \sizeVar.
An unbounded memory safety proof for \cmdVar would yield that the program is safe for any possible input, in particular for any array size, i.e., 
$\forall \sizeVar.\ \texttt{memsafe}(\cmdVar(\sizeVar))$.
A bounded proof that only considers input sizes up to ten would only guarantee that the program is safe for any such bounded array, i.e., 
$\forall \sizeVar \leq 10.\ \texttt{memsafe}(\cmdVar(\sizeVar))$.

\paragraph{Completeness Thresholds}
Approximating unbounded proofs by bounded ones is a technique often used in model checking.
Hence, the relationship between bounded and unbounded proofs about finite state transition systems has been studied extensively~\cite{Biere1999SymbolicMC, Clarke2004CompletenessAC, Kroening2003EfficientCO, Bundala2012OnTM, Abdulaziz2018FormallyVA, Heljanko2005IncrementalAC, Awedh2004ProvingMP, McMillan2003InterpolationAS}.
For a finite transition system $T$ and a property of interest $\phi$, a \emph{completeness threshold} is any number $k$ such that we can prove $\phi$ by only examining path prefixes of length $k$ in $T$, i.e., $T \models_k \phi \Rightarrow T \models \phi$~\cite{Clarke2004CompletenessAC}~\footnote{
	Note that the term completeness threshold is used inconsistently in literature.
	Some papers such as~\cite{Clarke2004CompletenessAC} use the definition above, according to which completeness thresholds are not unique.
	Others like~\cite{Kroening2003EfficientCO} define them as the minimal number $k$ such that $T \models_k \phi \Rightarrow T \models \phi$, which makes them unique.
}.
Over the years, various works characterised over-approximations of least completeness thresholds for different types of properties~$\phi$.
These over-approximations are typically described in terms of key attributes of the transition system $T$, such as the \emph{recurrence diameter} (longest loop-free path)~\cite{Kroening2003EfficientCO}.
For instance, consider the class of global safety properties of the form $Gp$ for finite transition systems $T$, where $p$ is a local property.
We know that the smallest completeness threshold for this class expressible solely in terms of $T$'s diameter is exactly $\texttt{diam}(T)$~\cite{Biere1999SymbolicMC, Kroening2011LinearCTs}.
As safety property of the form $Gp$, this also applies to memory safety of finite transition systems.

In general, heap-manipulating programs' state space can be infinite.
That is because the program's input data can be arbitrarily large and because executions can be arbitrarily long.
Therefore, the key attributes described above will generally be infinite as well.
This vast structural difference between the programs we are interested in and the transition systems for which completeness thresholds have been studied prevents us from reusing any of the existing \mbox{definitions or results}.

\section{Limitations of Bounded Proofs}\label{ct:sec:LimationsOfBoundedProofs}

\paragraph{Bounded Model Checking}
Generally, if we want unbounded memory safety guarantees, we have to consider all possible input sizes and all possible executions.
This is often hard and requires us to write tedious inductive proofs.
An alternative is to give up on the idea of unbounded guarantees and to settle for bounded ones.
One approach that has proven useful during development of critical software is bounded model checking~(BMC)~\cite{Chong2020CodeLeveMC}.

The underlying idea is to approximate the original verification problem by a finite model that we can check automatically.
With this approach, we choose a size bound \sizeBoundVar and only consider inputs with sizes up to \sizeBoundVar.
Further, we also only check finite execution prefixes.
A common approach is to unwind loops and recursion up to a certain depth.

The intuition behind this approach is that if the program contains errors, they likely already occur for small input sizes and early loop iterations.
As long as BMC does not perform abstraction~\cite{Clarke2000CounterExGuidedRefinement}, all reported counterexamples are real bugs.
However, we should be careful not to forget that this way we only obtain a bounded proof yielding bounded~guarantees.

\begin{figure}
	%\raggedright
	\begin{tabular}{l l l}
		$\left.
		\begin{minipage}{4cm}
\begin{lstlisting}
for i in [L : s-R] do
	!a[i+Z]
\end{lstlisting}
		\end{minipage}
		\quad\right\}
		$
		=: 
		&\travPatternVars{a, s}
	\end{tabular}
	\\
	\begin{tabular}{l l l}
		\travPatternClass
		&:=
		&$\{ \travPattern \quad | \quad \leftRangeBound, \rightRangeBound, \zVar \in \Z \}$
	\end{tabular}
	\caption{
		Class \travPatternClass of programs \travPattern traversing an array \arrayVar of size \sizeVar, attempting to read elements.
		$\leftRangeBound, \rightRangeBound, \zVar$ \mbox{are constants}.
	}
	\label{ct:fig:ProgClass}
\end{figure}

\paragraph{Array Traversal Pattern}
Consider the class of programs \travPatternClass presented in Fig.~\ref{ct:fig:ProgClass} in a WHILE language with pointer arithmetic.
Given a pointer \arrayVar and a variable \sizeVar, such that \arrayVar points to an array of size \sizeVar,
%Given an input array \arrayVar of size \sizeVar, 
each program iterates through the array and attempts to read elements.
The class models a basic programming pattern and common off-by-n errors~\cite{cwe193}. 
We use upper case letters for constants and lower case letters for (program) variables.
A program \travPattern from this class iterates from $\indexVar = \leftRangeBound$ to $\indexVar = \sizeVar - \rightRangeBound$ (bounds incl.) and attempts to read the array at index $\indexVar + \zVar$.
We use $!x$ to express accesses to a heap location $x$.
Whether memory errors occur for a concrete instance \travPattern depends on how the constants \leftRangeBound, \rightRangeBound, \zVar are chosen.
We use it as minimal example \mbox{throughout this chapter}.

\paragraph{What Could Go Wrong with Bounded Proofs?\!\!}\
To illustrate the issue, let us use BMC to check various instances of the array traversal pattern:
(i)~traversal of the entire array: \travPattern[0][1][0],
(ii)~traversal of the array with accesses offset by two from the index: \travPattern[0][1][2] and
(iii)~an additional reduction of the index variable's upper bound by one: \travPattern[0][2][2].
It is easy to see that (i) is memory-safe while (ii) and (iii) are not.
However, before we run a model checking algorithm we have to choose appropriate bounds.
The pattern we are looking at is quite simple.
So, we choose size bound $\sizeBoundVar = 1$ and unwinding depth $\depthBoundVar = 1$ for the BMC procedure.
Note that the latter effectively means:
we do not restrict the loop depth for the input sizes we chose.

For the standard variant (i)~we cannot find any errors within the bounds.
This is fine because the program is safe.
In variant (ii)~array accesses $\arrayVar[\indexVar+2]$ are incorrectly shifted to the right.
This already leads to an out-of-bounds error for arrays of size 1.
This size falls within our chosen bounds, so BMC reports this error and we can correct it.
Finally, (iii)'s reduction of the index variable's range to $[0,\sizeVar - 2]$ means that the program only performs loop iterations for arrays of size $\sizeVar \geq 2$.
Consequently, it is trivially safe for the sizes $0$ and $1$.
These are the sizes our bounded proof explores.
Hence, BMC does not report any errors and leads us to wrongly believe that \travPattern[0][2][2] is safe.

\section{Completeness Thresholds}\label{ct:sec:CTs}

As illustrated above, bounded proofs are in general unsound approximations of unbounded proofs.
A concrete approximation is sound iff we choose the bounds large enough, such that we can be sure that we do not miss any errors.
We focus on bounding input sizes (in our examples array sizes), ignoring loop bounds that do not depend on these~parameters.

Recall from \S~\ref{ct:sec:Intro} that completeness thresholds are a concept from model checking of finite transition systems~\cite{Clarke2004CompletenessAC}.
We borrow this terminology and apply it to memory safety verification.
Hence, for a program $\cmdVar(x)$ with input parameter $x \in \domVar$, we call any subdomain $\ctVar \subseteq \domVar$ a \emph{completeness threshold}~(\emph{CT}) for $x$ in $\cmdVar$ if we can prove memory safety of \cmdVar by only considering inputs from \ctVar, i.e., 
$
\forall x \in \ctVar.\ 
  \texttt{memsafe}(\cmdVar(x))
\Rightarrow
\forall x \in \domVar.\ 
  \texttt{memsafe}(\cmdVar(x))
$.

\paragraph{Intuitive CT Extraction}
Returning to our example class of programs \travPatternClass implementing the array traversal pattern.
Some of these are memory safe, some are not.
So, let us try to compute completeness thresholds for these programs.
First, let's take a look at the errors that might occur.
This gives us an idea which sizes a sound bounded proof must cover.
Any instance \travPatternVars{\arrayVar, \sizeVar} iterates ascendingly $i = \leftRangeBound, \dots, \sizeVar-\rightRangeBound$ and accesses $\arrayVar[i+\zVar]$.
For sizes \sizeVar that cause the ascending range $\leftRangeBound, \dots, \sizeVar-\rightRangeBound$ to be empty, we do not execute the loop at all.
Any such run is trivially memory safe.
Therefore, any meaningful bounded proof of \travPatternVars{\arrayVar, \sizeVar} must include sizes~\sizeVar with 
$\{\leftRangeBound, \dots, \sizeVar-\rightRangeBound\} \neq \emptyset$, i.e.,
$\sizeVar \geq \leftRangeBound+\rightRangeBound$.

Suppose $\sizeVar \geq \leftRangeBound+\rightRangeBound$.
An error occurs if the index $\indexVar+\zVar$ violates the array bounds, i.e., if $\indexVar+\zVar < 0$ or $\indexVar+\zVar \geq \sizeVar$.
Taking the index range into account, we see that we get an error if
$\leftRangeBound+\zVar < 0$ 
or 
$\sizeVar-\rightRangeBound+\zVar \geq \sizeVar$ holds.
We can simplify the latter to $\zVar-\rightRangeBound \geq 0$.

Note that neither $\leftRangeBound+\zVar < 0$ nor $\zVar-\rightRangeBound \geq 0$ depend on the array size \sizeVar.
This means that as long as we focus on sizes above the threshold $\sizeVar \geq \leftRangeBound+\rightRangeBound$,
the concrete choice of \sizeVar does not influence whether an error occurs or not.
In other words, it suffices for our bounded proof to only check a single (arbitrarily chosen) size
$\ctElem \geq \leftRangeBound+\rightRangeBound$
and then we can extrapolate the result, i.e.,
$$
  \forall \arrayVar.\
	  \memsafe{\travPatternVars{\arrayVar, q}}
  \ \Rightarrow\
  \forall \sizeVar.
  \forall \arrayVar.\
    \memsafe{\travPatternVars{\arrayVar, \sizeVar}}
$$

Hence, any set $\{q\}$ for $q \geq \leftRangeBound+\rightRangeBound$ is a CT for the array size parameter \sizeVar in \travPattern.
We just found a uniform characterization of CTs for the entire class~\travPatternClass.
Note that $\{q\}$ is not necessarily the smallest CT.
For safe instances such as \travPattern[0][1][0], the empty set $\emptyset$ is a valid CT as well.

\paragraph{Our Approach}
We study CTs for $x$ in $\cmdVar(x)$ by studying its \emph{verification condition}~(VC). 
%(cf.\@ \S~\ref{ct:sec:CTs:ApproxCTsViaVCs}).
The latter is an automatically generated logical formula of the form
$\forall x \in \domVar.\ \vcVar(x)$
and proving it entails memory safety of $\cmdVar(x)$ for all choices of $x$.
Next, we currently simplify $\vcVar(x)$ by hand until it becomes clear how the choice of $x$ affects the validity of $\vcVar(x)$.
Knowing this allows us to partition the domain into 
%$\domVar = \ctVar_1 \cup \dots \cup \ctVar_n$
$\domVar = \bigcup \ctVar_i$.
For each subdomain we get $\vcVar_i(x) = \forall x \in \ctVar_i.\ \vcVar(x)$.
If possible, we simplify each $\vcVar_i(x)$ into $\vcVar_i'(x)$ based on the restricted subdomain~$\ctVar_i$ with the goal to eliminate occurrences of $x$.
If $\vcVar_i'$ does not mention $x$ we pick any element of $\ctVar_i$ as representative $\ctVar_i'$.
Otherwise, $\ctVar_i' = \ctVar_i$.
Hence, $\bigcup \ctVar_i'$ is a CT for $x$ in $\cmdVar(x)$.
In the following we elaborate this in more detail.

\subsection{Approximating CTs via Verification Conditions}
\label{ct:sec:CTs:ApproxCTsViaVCs}

Now that we have an intuition for the CTs of \travPatternClass, let's turn our informal argument from above into a formal one.
Formal definitions of the language and logic we consider and proofs for the presented lemmas can be found in the technical report~\cite{Reinhard2023ct4ms-atrav-TR}.

\paragraph{Hoare Triples}
We use Hoare triples~\cite{Hoare1968HoareLogic} to express program specifications.
A triple \hoareTriple{\preVar}{\cmdVar}{\postVar} expresses that the following properties hold for every execution that starts in a state which satisfies precondition~\preVar:
Firstly, the execution does not encounter any runtime errors.
Secondly, it either
(i)~does not terminate or
(ii)~it terminates in a state complying with postcondition~\postVar.

In this work, we study the memory safety of programs that do not change the shape of the data structures they process.
Hence, we choose preconditions that merely describe the memory layout of the data structures which our programs receive as input.
For the array traversal program, we choose the predicate \arrayPred as precondition, which expresses that \arrayVar points to a contiguous memory chunk of size \sizeVar.
Given that our target programs do not change the memory layout, specifications simplify to \hoareTriple{\preVar}{\cmdVar}{\preVar}.
For the array traversal we get
\hoareTriple{\arrayPred}{\travPattern}{\arrayPred}.

\begin{definition}[Completeness Thresholds for Programs]
  Let \hoareTriple{\preVar}{\cmdVar}{\postVar} be a program specification containing a free variable $x$ with domain \domVar.
  We call a subdomain $\ctVar \subseteq \domVar$ a \emph{completeness threshold} for $x$ in \hoareTriple{\preVar}{\cmdVar}{\postVar} if
  $$
    \models 
      \forall x \in \ctVar.\
      \hoareTriple{\preVar}{\cmdVar}{\postVar}
    \quad\Rightarrow\quad
    \models
      \forall x \in \domVar.\
      \hoareTriple{\preVar}{\cmdVar}{\postVar}
  $$
\end{definition}
We omit spelling out the pre- and postconditions when they are clear from the context.
Instead we say that \ctVar is a completeness threshold for $x$ in program \cmdVar.

\paragraph{Separation Logic}
We use a first-order affine/intuitionistic separation logic with recursion predicates~\cite{DBLP:journals/cacm/OHearn19,OHearn2001LocalRA,Reynolds2002SeparationLA}
to describe memory.
Since we focus on heap-manipulating programs, we use assertions to describe heaps.
Separation logic comes with a few special operators:
(i)~The points-to chunk \slPointsTo{x}{v} describes a heap containing a location $x$ which holds the value~$v$.
We write \slPointsToAny{x} to express that we do not care about the value stored in the heap cell.
(ii)~The separating conjunction $\assVar_1 \slStar \assVar_2$ expresses that $\assVar_1$ and $\assVar_2$ describe disjoint heaps.
Hence, $\slPointsToAny{x} \slStar \slPointsToAny{y}$ implies that $x \neq y$.
(iii)~The separating implication $\assVar_1 \slWand \assVar_2$ can be read as $\assVar_2$ \emph{without} $\assVar_1$.
That is, combining the heap described by $\assVar_1 \slWand \assVar_2$ with a disjoint heap described by $\assVar_1$ yields a heap compliant with $\assVar_2$.
(iii)~In our logic, the persistence modality~$\slPersistent \assVar$ means that~\assVar does not describe resources and hence holds under the empty heap (cf.\ \cite{Jung2018IrisGroundUp}).

We assume that \arrayPredName denotes a (recursively defined) predicate, such that for every fixed size \sizeVar, we can express it as iterated separating conjunction: 
$\displaystyle
  \arrayPred 
  \equiv
  \slBigStar_{0 \leq k < s} \slPointsToAny{\arrayVar[k]}
$.

\paragraph{Verification Conditions}
A common way to verify programs is via verification conditions~\cite{Flanagan2001AvoidingExpExplosionVC, Parthasarathy2021VCGenerator}.
For any specification \hoareTriple{\preVar}{\cmdVar}{\postVar}, a \emph{verification condition}~(\emph{VC}) is any logical formula \vcVar, such that we can verify \hoareTriple{\preVar}{\cmdVar}{\postVar} by proving \vcVar, i.e.,
$
\models \vcVar
\ \Rightarrow\, \
\models \hoareTriple{\preVar}{\cmdVar}{\postVar}
$.

\begin{definition}[Verification Condition]
  We call an assertion \assVar a \emph{verification condition} for \hoareTriple{\preVar}{\cmdVar}{\postVar} if 
  $$
    \models \assVar
    \quad\Rightarrow\quad
    \models \hoareTriple{\preVar}{\cmdVar}{\postVar}.
  $$
\end{definition}

\begin{definition}[Completeness Thresholds for Assertions]
  Let \assVar be an assertion with a free variable $x$ of domain $X$.
  We call a subdomain $\ctVar \subseteq X$ a \emph{completeness threshold} for $x$ in \assVar if
  $$
    \models \forall x \in \ctVar.\ \assVar
    \quad\Rightarrow\quad
    \models \forall x \in X.\ \assVar.
  $$
\end{definition}

Consider a specification 
\hoareTriple{\preVar}{\cmdVar}{\postVar}
with a free variable $x \in X$ and a corresponding VC 
$\forall x \in X.\ \vcVar$.
Suppose we get a completeness threshold \ctVar for $x$ in \vcVar.
Knowing this threshold reduces correctness of the specification to the bounded VC, i.e.,
$
\models \forall x \in \ctVar.\ \vcVar 
\Rightarrow 
\hoareTriple{\preVar}{\cmdVar}{\postVar}
$.
That is, we can derive unbounded guarantees from a bounded proof.
We usually omit quantification domains when they are clear from the~context.

\paragraph{Weakest Liberal Preconditions}
A common way to generate VCs is via weakest liberal preconditions~\cite{Flanagan2001AvoidingExpExplosionVC, Dijsktra1976DisciplineOfProgramming}.
For any program \cmdVar and postcondition \postVar, the \emph{weakest liberal precondition}
\wlp{\cmdVar}{\postVar}
is an assertion for~which
$$
  \forall \preVar.\ \
  \big(\
    \preVar \models \wlp{\cmdVar}{\postVar}
    \ \Rightarrow\ \
    \models \hoareTriple{\preVar}{\cmdVar}{\postVar}
  \ \big)
$$
holds.
That is, if the weakest liberal precondition holds for the starting state,
then \cmdVar does either not terminate or it terminates in a state complying with postcondition \postVar.
In particular, no memory error occurs during the execution.
The canonical VC for \hoareTriple{\preVar}{\cmdVar}{\postVar} is 
$\forall \overline{x}.\ \preVar \rightarrow \wlp{\cmdVar}{\postVar}$
where $\overline{x}$ is the tuple of variables occuring freely in \preVar, \cmdVar and \postVar.

\paragraph{Limitations of CTs}
In general, VCs are over-approxima\-tions.
Hence, CTs derived from a VC do not always apply to the corrresponding program.
Consider the specification
\hoareTriple
  {\arrayPred}
  {\travPattern[0][2][2]}
  {\arrayPred}
and any unsatisfiable assertion
$\vcFalse \equiv \slFalse$.
Then 
$\forall \sizeVar \in \N.\ \vcFalse$ 
is an over-ap\-prox\-i\-ma\-ting VC, since
$
  \slFalse 
  \Rightarrow\
  \models
    \hoareTriple
      {\arrayPred}
      {\travPattern[0][2][2]}
      {\arrayPred}
$.
As discussed in \S~\ref{ct:sec:LimationsOfBoundedProofs}, a bounded proof that only covers sizes $0, 1$ does not discover the errors in \travPattern[0][2][2].
Hence, the set $\{0\}$ is not a CT for $x$ in
\hoareTriple
  {\arrayPred}
  {\travPattern[0][2][2]}
  {\arrayPred}.
However, $\{0\}$ is a CT for \sizeVar in \vcFalse, since
$
  \forall \sizeVar \in \{0\}.\ \vcFalse
  \equiv
  \slFalse
$
and 
$
  \slFalse 
  \Rightarrow \
  \models \forall \sizeVar \in \N.\ \vcFalse
$.
We see that we can only transfer a CT \ctVar for some variable $x$ derived from a VC \vcVar to the corresponding program, if \vcVar does not over-approximate with regard to $x$.
(We are currently studying this connection.)
Note that this is, however, the case for the examples we discuss in this chapter.
Though, even if \vcVar does over-approximate, CT \ctVar still applies to any proof considering a property at least as strong as \vcVar.

\deleteBlock{
  \begin{definition}[Pecision]
    Let \assVar and \hoareTriple{\preVar}{\cmdVar}{\postVar} be an assertion and specification, respectively, with a free variable $x$ of domain \domVar.
    We call \assVar \emph{precise} in $x$ for \hoareTriple{\preVar}{\cmdVar}{\postVar} if the following holds for all values $v \in \domVar$:
    $$
      \models
        \subst{\assVar}{x}{v}
      \quad\Rightarrow\quad
      \models
        \subst
          {\big(\hoareTriple{\preVar}{\cmdVar}{\postVar}\big)}
          {x}
          {v}
    $$
  \end{definition}
}

As the name suggests, VCs derived from weakest preconditions yield very weak properties.
It is reasonable to assume that often a bounded proof would imply a bounded version of the \wlpName-based VC.
Therefore, it is reasonable to use them during our study of CTs.

\paragraph{Extracting CTs via VCs}
Applying the above approach to our specification\\
\hoareTriple
  {\arrayPred}
  {\travPatternVars{\arrayVar, \sizeVar}}
  {\arrayPred},
we get a VC
$\forall \arrayVar.\ \forall \sizeVar.\ \vcTrav$,
where $\vcTrav$ is as follows~\footnote{
  The weakest precondition calculus requires us to annotate loops with loop invariants.
  In the setting we study, the initial memory layout is invariant under the program's execution.
  The preconditions we consider describe exactly the initial memory layout, nothing else.
  Hence, we can reuse preconditions as loop invariants during the \wlpName computation.
}:
$$
\begin{array}{p{0.55cm} p{0.05cm} l l}
  \!\vcTrav  
    &:=
    &\arrayPred \rightarrow
    &\vcLine{1}
    \\
    &&\quad
      \arrayPred
    &\vcLine{2}
    \\
    &&\quad
      \slStar\ \slPersistent(
        \forall \indexVar.\
          (\leftRangeBound \leq \indexVar \leq \sizeVar - \rightRangeBound)
          \wedge
          \arrayPred
          \slWand
    &\vcLine{3}
    \\
    &&\quad
      \quad\quad\quad\quad
          \exists v.\
          \slPointsTo{\arrayVar[i+\zVar]}{v}
          \wedge
          \arrayPred
      )
      \!\!
    \\
    &&\quad
      \slStar\ (\arrayPred \slWand \arrayPred)
    &\vcLine{4}
\end{array}
$$
Here, 
\vcLine{1}~is the precondition describing our memory layout.
The separating conjunction \vcLine{2}-\vcLine{4} is the weakest precondition derived from our specification and \vcTrav says that it should follow from precondition~\vcLine{1}.

The weakest precondition states that the memory layout must stay invariant under the loop execution.
\vcLine{2}~says that it should hold before the loop starts.
\vcLine{3}~demands that every loop iteration preserves the layout, i.e., that the layout description is a loop invariant.
\vcLine{4}~states this invariant implies the postcondition from our specification, which, again, is the unchanged memory layout.

Remember that we use an affine separation logic.
Clearly, this VC contains many trivially obsolete parts.
We can simplify \vcTrav to $\vcVar_1$:

$$
\begin{array}{l l l l}
  \vcTrav
  &\equiv
  &\vcTransComment{Eliminate \vcLine{2}, \vcLine{4}}
  \\
  &&\arrayPred \rightarrow
  \\
  &&\quad
    \slPersistent(
      \forall \indexVar.\
        (\leftRangeBound \leq \indexVar \leq \sizeVar - \rightRangeBound)
        \wedge
        \arrayPred
        \slWand
  \\
  &&\quad
    \quad\quad\quad\quad
        \exists v.\
        \slPointsTo{\arrayVar[i+\zVar]}{v}
        \wedge
        \arrayPred
    )
  \\
  &\equiv
  &\vcTransComment{Persistency makes pre.\! obsolete}
  \\
  &&\quad
    \slPersistent(
      \forall \indexVar.\
        (\leftRangeBound \leq \indexVar \leq \sizeVar - \rightRangeBound)
        \wedge
        \arrayPred
        \slWand
  \\
  &&\quad
    \quad\quad\quad\quad
        \exists v.\
        \slPointsTo{\arrayVar[i+\zVar]}{v}
        \wedge
        \arrayPred
    )
    \\
    &\equiv
    &\vcTransComment{
      \arrayPred equiv.\! to
      $\displaystyle\slBigStar_{0\leq k<s} \slPointsToAny{\arrayVar[k]}$
    }
    \\
    &&\quad
        \forall \indexVar.\
          (\leftRangeBound \leq \indexVar \leq \sizeVar - \rightRangeBound)
          \rightarrow
          (0 \leq \indexVar + \zVar < \sizeVar)
  \\
  &=:
  &\vcVar_1
\end{array}
$$

This equivalent VC $\forall \arrayVar.\ \forall \sizeVar.\ \vcVar_1$ does reflect the intuition we developed when analysing the program informally:
For sizes $\sizeVar < \leftRangeBound + \rightRangeBound$, the program does not perform any loop iterations and hence it is trivially memory safe.
For bigger arrays, a memory error occurs iff index $\indexVar+\zVar$ violates the array bounds.

We can justify this intuition by partitioning the domain of \sizeVar into 
$
  \N = 
  \{0, \dots, \leftRangeBound + (\rightRangeBound - 1)\} 
  \cup 
  \{\leftRangeBound + \rightRangeBound, \dots\}
$.
Let's analyse $\vcVar_1$ for both subdomains separately.
For a size $\sizeLow < \leftRangeBound + \rightRangeBound$, we get
$$
  \vcVar_1(\sizeLow)
  \quad \equiv \quad
  \forall \indexVar.\
    \slFalse \rightarrow (0 \leq \indexVar + \zVar < \sizeLow)
  \quad \equiv \quad
    \slTrue
$$
So, we do not have to bother checking sizes $\sizeVar < \leftRangeBound + \rightRangeBound$. For bigger sizes $\sizeHigh \geq \leftRangeBound + \rightRangeBound$, we get
$$
\begin{array}{l}
  \vcVar_1(\sizeHigh)
  \ \ \equiv\ \
  \forall \indexVar.\
      (\leftRangeBound \leq \indexVar \leq \sizeHigh - \rightRangeBound)
      \rightarrow
      (0 \leq \indexVar + \zVar < \sizeHigh)
  \\
  \equiv\ \
  \forall \indexVar.\
    (
      \leftRangeBound \leq \indexVar 
      \rightarrow 
      0 \leq \indexVar + \zVar
    )
    \ \wedge \
    (
      \indexVar \leq \sizeHigh - \rightRangeBound
      \rightarrow
      \indexVar + \zVar < \sizeHigh
    )
  \\
  \equiv\ \
  \forall \indexVar.\
    (
      \leftRangeBound \leq \indexVar 
      \rightarrow 
      0 \leq \indexVar + \zVar
    )
    \ \wedge \
    (
      \indexVar \leq  -\rightRangeBound
      \rightarrow
      \indexVar + \zVar < 0
    )
  \\
  =:\ \
  \vcVar_2.
\end{array}
$$

Since $\sizeHigh$ does not occur freely in $\vcVar_2$, the truth of $\vcVar_1(\sizeHigh)$ does not depend on the choice of $\sizeHigh$.
Remember that we have,
$\vcTrav(\sizeLow) \equiv \slTrue$
and
$\vcTrav(\sizeHigh) \equiv \vcVar_2$.
Hence,
$$
  \models 
    \forall \arrayVar.\ \forall \sizeVar.\ 
    \vcTrav
  \quad\Leftrightarrow\quad
  \models
    \forall \arrayVar.\
      \vcTrav(\sizeHigh)
$$
We see that it suffices to check the original VC $\vcTrav$ for any size $\sizeHigh \geq \leftRangeBound - \rightRangeBound$ to prove memory safety of our array traversing program \travPattern.
That is, $\{\sizeHigh\}$ is a CT.

\paragraph{Characterizing CTs via Constraints}
Note that the constraint
$\sizeVar \geq \leftRangeBound - \rightRangeBound$
we just derived
is a uniform representation for the CTs of the entire class \travPatternClass.
We often use constraint sets \setOf{k_1, \dots, k_n} to concisely characterize CTs.
Each $k_i$ describes a property that some element in our CT must cover to reach a specific potential error.
If $k_i$ is unsatisfiable, it means that the error is unreachable.
Hence, a set \ctVar models the constraint set if it contains models 
$k_1, \dots, k_n$ for each satisfiable constraint, i.e.,
$
	\not\models \neg k_i 
	\Rightarrow\,
	\models k_i[\sizeVar \mapsto k_i]
$.
For singleton constraint~sets~\setOf{k}, we drop the set notation and just write $k$.

Overall, a constraint set formulates a property that is sufficiently strong so that every subdomain covering it is a~CT.
For every program $\travPattern \in \travPatternClass$, every subdomain $\ctVar' \subseteq \N$ is a~CT for the array size \sizeVar if
$
	\ctVar' \cap 
	\{
		\sizeVar \in \N \ | \ 
		\sizeVar \geq \leftRangeBound - \rightRangeBound  
	\}
	\neq \emptyset
$.

\subsection{Modularity of Completeness Thresholds}
\label{ct:sec:CTs:Modularity}

\paragraph{Unrelated Data Structures}
Consider the program\\
\sumPatternVars{\arrayVar, \sizeVar, \nVar} that attempts to sum up all elements of array~\arrayVar and writes the result to heap location~\nVar:

\begin{tabular}{lll}
  $\left.
    \begin{minipage}{4cm}
\begin{lstlisting}
for i in [L : s-R] do 
  !n := !n + !a[i+Z]
\end{lstlisting}
    \end{minipage}
  \right\}$
  &=:
  &\sumPatternVars{\arrayVar, \sizeVar, \nVar}
\end{tabular}
\\
Analogously to \travPattern it iterates through the array and attempts to read array elements.
Additionally, it uses the read value to update the sum stored at heap location \nVar.
Let us assume that the array \arrayVar and the result variable \nVar do not alias.
We get the specification
\hoareTriple{\preVar}{\sumPattern}{\preVar}
for 
$\preVar := \arrayPred \slStar \slPointsToAny{\nVar}$.
We assume structured memory.
Therefore, it is not possible to access \nVar via an array access~$\arrayVar[...]$.

Intuitively, it is clear that the array size does not affect the memory accesses to heap location~\nVar.
Hence, the CTs for~\sizeVar in~\sumPattern should be the same as the ones for~\travPattern.
In fact, analysing the \wlpName-based VC
for \sumPattern confirms this intuition.
It has the form 
$\forall \arrayVar. \forall \sizeVar. \forall \resVar.\ \vcSum$
and we can rewrite \vcSum into
$$
\begin{array}{l l l}
  \vcSum
  &\equiv
  &\vcTrav
    \slStar
    (\slPointsToAny{\nVar} \rightarrow \subAssVar)
\end{array}
$$
where \vcTrav is the VC from \S~\ref{ct:sec:CTs:ApproxCTsViaVCs} for the array traversing program \travPattern.
Moreover, $\freeVars{\subAssVar} = \{\nVar\}$.
Since \sizeVar does not occur freely in
$\slPointsToAny{\nVar} \rightarrow \subAssVar$,
we can ignore it while searching for a CT for \sizeVar in \vcSum.

\begin{lemma}[VC Slicing]\label{ct:lem:vcSlicing}
  Let $\assVar, \assVar_x, \assVar_y$ be assertions with 
  $x \in \freeVars{\assVar_x}$ 
  and 
  $x \not\in \freeVars{\assVar_y}$ 
  and 
  $\assVar \equiv \assVar_x \slStar \assVar_y$.
  Let $\ctVar \subseteq \domVar$ be a CT for $x$ in $\assVar_x$.
  Then, \ctVar is also a CT for $x$ in $\assVar$, i.e.,
  $$
    \models
      \forall x \in \ctVar.\ 
      \forall \overline{y} \in \overline{Y}.\
        \assVar
    \quad\Rightarrow\quad
      \models
      \forall x \in \domVar.\ 
      \forall \overline{y} \in \overline{Y}.\
        \assVar
  $$
\end{lemma}

\begin{figure}
  \raggedright
  \begin{tabular}{l l l}
    $\left.
    \begin{minipage}{4cm}
\begin{lstlisting}
for i in [L : s-R] do (
  n := a[i+Z];
  complex_fct(n, y, k)
)
\end{lstlisting}
      \end{minipage}
    \quad\right\}$
    &=:
    &\compProgVars{\arrayVar, \sizeVar, \unrelStructVar, \unrelSizeVar} 
  \end{tabular}
  
  \begin{tabular}{l l l}
    $\memAssVar(\arrayVar, \sizeVar)$
    &:=
    &\arrayPred \slStar\ \unrelDataPred
    \\
    \compProgClass
    &:=
    &$\{ 
      \hoareTriple{\memAssVar}{\compProg}{\memAssVar}
      \quad | \quad 
      \leftRangeBound, \rightRangeBound, \zVar \in \Z 
    \}$
  \end{tabular}

  \caption{
    Class of programs involving a complex data structure and computation that do not depend on the array size~\sizeVar.
  }
  \label{ct:fig:ComplexProgClass}
\end{figure}

We can extrapolate what we saw in the \sumPattern example to more complex classes of programs.
Consider the class~\compProgClass presented in Fig.~\ref{ct:fig:ComplexProgClass}.
A program $\compProg \in \compProgClass$ receives two non-aliasing data structures: an array \arrayVar of size \sizeVar and a complex data structure \unrelStructVar of size \unrelSizeVar, described by the predicate \unrelDataPred.
\compProg reads elements from array \arrayVar, stores the result in a local variable \nVar and then calls a complex function
\verb|complex_fct(n,y,k)|
which does neither depend on \arrayVar nor \sizeVar.
The VC \vcComp will reflect this.
That is, analogous to the example above, it should be expressible as
$
  \vcComp 
  \equiv 
    \vcTrav
    \slStar 
    (\unrelDataPred \rightarrow \dots)
$
where the right conjunct does not depend on \sizeVar.

VC Slicing lemma~\ref{ct:lem:vcSlicing} tells us that whenever we want to characterize a CT for a specific parameter, 
we can ignore all separated VC conjuncts that do not involve this parameter.
Effectively, this means that we can ignore all the complex parts of \compProg that are not related to the array size while searching for a CT for \sizeVar.
This allows us to reduce the search to the CTs of~\travPatternClass.

\paragraph{Compositionality}
We can describe the CTs of complex programs in terms of the CTs of their building blocks.
Consider the program $\cmdVar_1;\cmdVar_2$ and suppose that $\cmdVar_1$ and $\cmdVar_2$ are instances of patterns we studied before.
So we know that each $\cmdVar_i$ corresponds to a VC 
$\forall x.\ \vcVar_i$ 
with a CT $\ctVar_i$ for $x$.
Let $\vcVar_{1;2}$ be the VC for $\cmdVar_1; \cmdVar_2$ that we want to prove.
Suppose it can be rewritten into 
$\forall x.\ \vcVar_1 \wedge \vcVar_2$.
Then, we know that $\ctVar_1 \cup \ctVar_2$ is a CT for $\vcVar_{1;2}$.
Therefore, our approach to studying CTs is to study patterns and combinators.

\paragraph{Basic Patterns}
We view basic patterns such as the array traversal pattern discussed above as the basic building blocks.
They tend to occur frequently in programs and they are sufficiently concise to extract CTs by studying their VCs.
In particular, we focus on traversal and access patterns that preserve the memory layout.
For now, we focus on arrays, but we are going to generalize it to arbitrary inductive data structures.

\paragraph{Managing Complexity}
One of our main goals is to describe CTs for interesting classes of programs.
VCs tend to become very complex very fast as a program gets more complex.
Hence, we need a way to deal with this complexity and to break the CT analysis down into simpler problems.
Following the structure of the program we want to reason about is a natural approach.

\paragraph{Combinator Patterns}
In order to exploit the program structure while analysing CTs, we need to study how control structures affect CTs.
For instance, as described above, we can characterize the CT of a sequence $\cmdVar_1; \cmdVar_2$ as the union of the CTs derived from $\cmdVar_1$ and $\cmdVar_2$.
Further, consider the command 
$\texttt{if}\ e\ \texttt{then}\ \cmdVar_1\ \texttt{else}\ \cmdVar_2$
and suppose that we can describe CTs for a size \sizeVar in each~$\cmdVar_i$ via a constraint set~$\constraintSetVar_i$.
Then, we can describe the CT for the entire command via the constraint set
$
	(e \wedge K_1)
	\ \cup\
	(\neg e \wedge K_2)
$.

\section{Conclusion}\label{ct:sec:Conclusion}

Past approaches to program verification either targeted unbounded guarantees and relied on unbounded, often inductive, proofs or they targeted bounded guarantees and tried to approximate the program behaviour using techniques like bounded model checking.
We have, however, seen little interaction between the two communities.

In this work we propose a new perspective on memory safety proofs that connects unbounded and bounded proofs.
We show that we can reduce unbounded memory safety proofs to bounded ones for certain programs that traverse arrays and preserve the memory layout.
For any such program considering a few select array sizes yields the same guarantees as considering arrays of all possible sizes.
We call this concept \emph{completeness thresholds} in reference to a similar concept from model checking of finite transition systems.
Moreover, we show that studying verification conditions are an adequate way to study completeness thresholds.

\section{Related Work}\label{ct:sec:RelatedWork}

Completeness thresholds were first introduced by \citeauthor{Kroening2003EfficientCO}~\cite{Kroening2003EfficientCO}.
So far, the study of CTs has been limited to finite state systems.
Indeed, the well-known CTs for classes of LTL properties are defined with respect to the (recurrence) diameter of the finite state system in question (e.g.\ \cite{Biere1999SymbolicMC,Kroening2011LinearCTs}).
Determining the worst-case execution time of a program and discovering upper bounds on loops by iterative unrolling can also be used to determine CTs~\cite{DSilva2008SurveyAutomatedVerification,Clarke2004CBMC}.
For a possibly infinite state system those CTs can naturally be infinite as well.
By specializing in just one property, memory safety, we are able to characterize and possibly find useful CTs for these systems as well.
%  Practically speaking, a classical CT for arbitrarily large transitions systems (even if finite!) will be infeasibly large.
Model checking for parameterized network topologies of identical (e.g.\ bisimilar~\cite{Browne1989NetworksIdenticalProcesses} or isomorphic~\cite{Emerson1995ReasoningAboutRings}) processes features a related concept to completeness thresholds called \emph{cutoff}.
That is, model-checking up to the cutoff implies correctness of scaling the topology up to infinitely many processes.
Positive results exist for properties of such token rings~\cite{Browne1989NetworksIdenticalProcesses,Emerson1995ReasoningAboutRings} but also other topologies~\cite{Clarke2004NetworkDecomposition,Aminof2014ParamMCTokenPassing}.

The model checking literature (cf.~\cite{Clarke2018ModelCheckingBook}) boasts a wealth of alternative approaches to obtain unbounded guarantees on finite state systems, e.g.,
k-induction~\cite{Bjesse2000SatVerificationWithoutSpaceTrav,Sheeran2000k_induction}, 
Craig interpolation~\cite{McMillan2003InterpolationAS} and 
property-directed reachability~\cite{Bradley2011SatMCWithoutUnrolling,Een2011ImplPDR}
with adaptions to the software verification setting (e.g.~\cite{Donaldson2011k_induction,Beyer2022SatMCRevisited,Beyer2020PDR}).
% Abstraction techniques~\cite{Cousot1977AbstractInterpretation} may be applied to obtain feasibly large finite state systems from large or infinite ones so that these approaches apply.
% This process necessarily introduces false-positives although partial mitigations exist~\cite{Clarke2000CounterExGuidedRefinement}.

Array-manipulating programs are well-studied across different domains~\cite{Bozga2009AutoVerificationIntArray, Bradley2006DecidableArrays, JhalaM2007ArrayAbstraction,
Chakraborty2020ArrayFullInduction}.
However, we consider our main contribution to be a novel approach to connect unbounded and bounded proofs about memory safety.
Ultimately, as discussed in the outlook, we aim to generalize and automate our approach to tree-like data structures.
In that regard, \citet{Mathur2020DecidingMemorySafetySinglePass} consider the special case of proving memory safety of heap-manipulating programs as well.
They prove that memory safety is decidable if the initial heap is forest-like and the program only performs a single-pass over the data-structure (see also \S~\ref{ct:sec:Outlook}).
%
%While we support arrays and reasoning about buffer overflows, \cite{Mathur2020DecidingMemorySafetySinglePass} allows memory (de-)allocation.
%
They do not cover arrays and buffer overflows, but they support (de-)allocation.

\section{Outlook}\label{ct:sec:Outlook}

\paragraph{CTs for Programs}
Consider any program \cmdVar with a free variable $x$ and a corresponding VC \vcVar.
Suppose we derived a CT \ctVar for $x$ in \vcVar.
In general, this does not allow us to conclude that \ctVar is also a CT for $x$ in \cmdVar.
Intuitively, this is only true if \vcVar does not over-approximate with regard to $x$.
This holds for the programs and VCs studied in this chapter.
While CTs for VCs still tell us something about proofs targeting these VCs, our ultimate goal is to derive thresholds for programs.
Hence, we are currently studying this connection.

\paragraph{Scalability}
In this work, we focus on a restricted array traversal pattern to illustrate CTs.
Our goal is to scale this approach to complex programs.
Therefore, we are currently studying more array traversal and access patterns and combinators.
This will allow us to better understand how the structure of programs affects the relation between bounded and unbounded proofs.
Knowing this will allow us to characterise CTs for complex classes of programs that cover errors besides off-by-n errors.

Afterwards, we are going to extend our approach to include arbitrary inductive data types.
In particular, we plan to describe CTs for a class of programs $\mathit{Sort}$ that includes (safe and unsafe) implementations of in-place sorting algorithms involving nested loops.
Once we managed that, we are going to investigate how allocation and deallocation affect CTs.

\paragraph{Decidability}
The memory accesses we observe in sorting functions typically mainly depend on the size of the sorted data structure, the traversal strategy and a comparison relation $<$.
Suppose we managed to derive finite CTs for the class $\mathit{Sort}$.
We conjecture that this will be sufficient to conclude decidability of memory safety for $\mathit{Sort}$.
Note that this targeted result would escape the scope of the related work by 
\citeauthor{Mathur2020DecidingMemorySafetySinglePass}~\cite{Mathur2020DecidingMemorySafetySinglePass}.
The latter showed decidability of memory safety for a certain form of single-pass programs, i.e., programs that traverse a datastructure exactly once.

\paragraph{Improving BMC Guarantees}
Bounded model checking suffers from the state space explosion problem~\cite{Clarke2008BmcStateExplosion, Park2000JavaMC, Clarke2000CounterExGuidedRefinement}
with respect to the chosen bounds.
It is often only practical to check very small bounds on each parameter to keep the verification time practical.
We plan to automate the approach introduced in this chapter and to leverage existing static analysis techniques (e.g.~\cite{Weiser84ProgramSlicing, MastroeniDataDependencyProgramSlicing, Asavoae2018ChiselProgramSlicing, Moser90DataDependencyGraphs}) to simplify the generated VCs. 
Once we are able to automatically compute a small CT for one parameter, say the traversed data structure's size, we can adjust the corresponding bound.
We can be sure that we didn't miss to check any size that could lead to memory errors.
Hence, it would strengthen the guarantees we get from our bounded proof.
Ultimately, we would like to integrate with the industrial-strength BMC tool CBMC~\cite{DBLP:conf/dac/ClarkeKY03,Clarke2004CBMC}.

Further, it would speed up the verification time and free resources that can be spend on exploring other parts of the program.
That is, the lowered size bound might allow us to increase other bounds that seem more important.
We would be able to do this while keeping the verification time stable and without sacrificing guarantees.

%%%%%%%%%%%%%%%%%%%%%%%%%%%%%%%%%%%%%%%%%%%%%%%%%%
% Keep the following \cleardoublepage at the end of this file, 
% otherwise \includeonly includes empty pages.
\cleardoublepage

% vim: tw=70 nocindent expandtab foldmethod=marker foldmarker={{{}{,}{}}}
%

%
  \graphicspath{{\chaptersdir/conclusion/image/}}%
  %\cleardoublepage%
  %

%%%%%%%%%%%%%%%%%%%%%%%%%%%%%%%%%%%%%%%%%%%%%%%%%%%%%%%%%%%%%%%%%%%%%%

\appendix

\includepublications{publications}

%\includeappendix{appendix-paper-1-abruptExit}
% !TeX root = ../../thesis.tex
\chapter{Ghost Signals: Formalization \& Case Studies}\label{ch:appendix-paper-2-ghostSignals}

\NewDocumentCommand{\appGsSecDir}{}{chapters/appendix-paper-2-ghostSignals/CAV_paper_appendix}

This chapter formalizes the work described in chapter~\ref{ch:paper-2-ghostSignals} and presents additional case studies.
The contents of this chapter were previously published on arXiv in the form of an appendix of the extended version of the ghost signal paper:
\citationBox
{
	\textbf{T. Reinhard} and Bart Jacobs.
	\emph{Ghost Signals: Verifying Termination of Busy Waiting (Extended Version)}. 
	arXiv, 
	CoRR, 
	abs/2010.11762, 2021. 
	https://doi.org/10.48550/arXiv.2010.11762
}

In this appendix, we formalize our approach and prove its soundness.
We start in \S~\ref{gs:appendix:sec:General} by defining the notations we use.
In \S~\ref{gs:appendix:sec:Language} we define the simple programming language we consider in this work.
In \S~\ref{gs:appendix:sec:Logic} we define our logic and in particular provide a full overview of all the proof and view shift rules we use and state a soundness theorem.
In \S~\ref{gs:appendix:sec:Soundness} we prove our approach sound.
Our proof utilizes an annotated semantics that keeps track of ghost resources and thereby connects the runtime and the verification level.
We define it in \S~\ref{gs:appendix:sec:Soundness:AnnotatedSemantics}.
Afterwards, in \S~\ref{gs:appendix:sec:Soundness:ModelRelation} we define a model for Hoare triples and prove that every triple that we can derive with our proof rules also holds in our model.
In \S~\ref{gs:appendix:sec:Soundness:Proof} we use this model and the annotated semantics to prove our soundness theorem.
At last, in \S~\ref{gs:appendix:sec:Case Studies} we present two verification case studies.
In \S~\ref{gs:appendix:sec:VerificationOfRealisticExample} we provide a detailed proof outline for the realistic example presented in \S~\ref{gs:sec:RealisticExample}.
In \S~\ref{gs:appendix:sec:caseStudy:unbounded} we present and verify a similar program but with a statically unbounded number of producer and consumer threads.

\section{General}\label{gs:appendix:sec:General}
    \begin{definition}[Projections]
            For any Cartesian product $C = \prod_{i \in I} A_i$ and any index $k \in I$, we denote the $k^\text{th}$ projection by
            $\tupleProj[C]{k}: \prod_{i \in I} A_i \rightarrow A_k$.
            We define
            $$
                    \tupleProj[C]{k}[(a_i)_{i\in I}] 
                    \  := \ 
                    a_k.
            $$
            In case the domain $C$ is clear from the context, we write \tupleProj{k} instead of \tupleProj[C]{k}.
    \end{definition}

    \begin{definition}[Disjoint Union]
            Let $A, B$ be sets.
            We define their disjoint union as
            $$
                    A \disjCup B \ := \ A \cup B
            $$
            if $A \cap B = \emptyset$
            and leave it undefined otherwise.
    \end{definition}

    \begin{definition}[Bags]\label{gs:def:Bags}
            For any set $X$ we define the set of bags \BagsOf{X} and the set of finite bags \FinBagsOf{X} over $X$ as
            $$
\begin{array}{l l l}
                    \BagsOf{X} \ &:= \
                            &X  \rightarrow  \N,\\
                    \FinBagsOf{X} \ &:= \
                            &\setOf{
                                    B \in \BagsOf{X} 
                                    \ \ | \ \
                                    \setOf{x \in B \ | \ B(x) > 0}\
                                    \text{finite}
                            }.
            \end{array}
            $$
            We define union and subtraction of bags as
            $$
            \begin{array}{l c l}
                    (B_1 \msCup B_2)(x)  \ &:= \
                            &B_1(x) + B_2(x),\\
                    (B_1 \setminus B_2)(x) \ &:= \
                            &\max(0,\ B_1(x) - B_2(x)).
            \end{array}
            $$
            For finite bags where the domain is clear from the context, we define the following set-like notation:
            $$
            %
            $$
            Further, we define for any $\natVar \in \N$ the notation:
            $$
                    n \cdot \multiset{x}
                    \ := \
                    \multiset{\underbrace{x, \dots, x}_\text{\natVar times}}.
            $$
            We define the following set-like notations for element and subset relationship:
            $$
            \begin{array}{l l l}
                    x \in B \ &\Leftrightarrow \ & B(x) > 0,
                    \\
                    B_1 \subseteq B_2 \ &\Leftrightarrow  \
                            &\forall x \in B_1.\ B_1(x) \leq B_2(x) ,
                    \\
                    B_1 \subset B_2 \ \, &\Leftrightarrow \ \,
                            &\exists C \subseteq B_1.\ 
                                    C \neq \msEmpty
                                    \ \wedge\
                                    B_1 = B_2 \setminus C.
            \end{array}
            $$
            For any bag $B \in \BagsOf{X}$ and predicate $P \subseteq X$ we define the following refinement:
            $$
                    \multiset{x \in B \ | \ P(x)}
                    \ := \ 
                    \left\{
                            \begin{array}{l c l l}
                                    x &\mapsto &B(x)  \ &\text{if}\ P(x),\\
                                    x &\mapsto &0 &\text{otherwise}.
                            \end{array}
                    \right.
            $$
    \end{definition}

    \begin{definition}[Disjoint Union]
            Let $A, B$ be sets.
            We define their disjoint union as
            $$
                    A \disjCup B \ := \ A \cup B
            $$
            if $A \cap B = \emptyset$
            and leave it undefined otherwise.
    \end{definition}

\section{Language}\label{gs:appendix:sec:Language}
        In this section and the next, we present our approach formally. 
        In this section, we define the programming language; 
        in \S~\ref{gs:appendix:sec:Logic} we define the proof system. 
%        We state the soundness theorem and prove it in \S~\ref{gs:appendix:sec:Soundness}.

        We consider a simple imperative programming language with support for multi-threading, shared memory and synchronization via mutexes.
        For its definition we assume 
        (i)~an infinite set of program variables $\varVar \in \VarSet$,
        (ii)~an infinite set of heap locations $\hlocVar \in \HeapLocSet$,
        (iii)~a set of values $\valVar \in \ValueSet$ which includes heap locations, Booleans $\B = \setOf{\slTrue, \slFalse}$ and the unit value \valUnit,
        (iv)~a set of operations $\opVar \in \OpSet$
        and
        (v)~an infinite, totally ordered and well-founded set of thread IDs $\tidVar \in \ThreadIDSet$.
        
        \begin{definition}[Syntax]\label{gs:def:Syntax}
%                Let \VarSet be an infinite set of variables, \LogHeapSet an infinite set of heap locations and \OpSet a set of operations.
%                Further let \ValueSet be a set of values with 
%                $\setOf{\valUnit} \cup \B \cup \HeapLocSet \subseteq \ValueSet$.
                We define the sets of commands \CmdSet and expressions \ExpSet according to the syntax presented in Fig.~\ref{gs:appendix:gs:fig:def:Syntax}.
        \end{definition}

    \begin{figure}
            \begin{subfigure}{\textwidth}
                    $$
\begin{array}{c c c c c}
                            \varVar \in \VarSet
                            : \text{Program variables}
                            \quad
                            &\hlocVar \in \HeapLocSet
                            : \text{Heap locations}
                            \quad
                            \\
                            \valVar \in \ValueSet
                                    \supseteq \setOf{\valUnit} \cup \B \cup \HeapLocSet
                            : \text{Values}
                            \quad
                            &\opVar \in \OpSet
                            : \text{Operations}
                    \end{array}
                    $$
%                    \vspace{-0.2cm}
                    \caption{
                            Assumed sets and variables. 
                            \VarSet and \HeapLocSet infinite.
                    }
                    \label{gs:appendix:fig:AssumedSetsForLanguage}
            \end{subfigure}
            
            \begin{subfigure}{\textwidth}
                    $$
                    %
                    $$
                    \vspace{-0.3cm}
                    \caption{Expressions and commands.}
                    \label{gs:appendix:fig:def:ExpCmdSyntax}
            \end{subfigure}
            
            \begin{subfigure}{\textwidth}
                    $$
                    \begin{array}{l c l }
                            \cmdAwait{\expVar}{\cmdVar}\ &:= \ \
                                    &(\cmdWhile
                                            {
                                                    \cmdAcquireMut{\expVar};
                                                    \cmdLet
                                                            {\resVar}
                                                            {\cmdVar}
                                                            {\cmdReleaseMut{\expVar}; \neg\resVar}
                                            }
                                            {\cmdSkip})
                            \\
                            \cmdVar\ ;\, \nextCmdVar &:=\ \
                                    &(\cmdLet{\resVar}{\cmdVar}{\nextCmdVar})
                            \quad\quad\quad\quad
                            \\
                            \expVar \neq \nextExpVar &:=\ \
                                    &\neg(\expVar = \expVar)
                            \\
                            \nextNextExpVar(\cmdReadHeapLoc{\expVar}) &:=\ \
                                    &(\cmdLet{\resVar}{\cmdReadHeapLoc{\expVar}}{\nextNextExpVar(\resVar)})
                    \end{array}
                    $$
                    \vspace{-0.3cm}
                    \caption{Syntactic sugar. $\nextNextExpVar(y)$ expression with free variable $y$. $\resVar \in \VarSet$ not free in \expVar, \nextNextExpVar, \nextCmdVar.}
                    \label{gs:appendix:fig:def:SyntacticSugar}
            \end{subfigure}
            
            \caption{Syntax.}
            \label{gs:appendix:gs:fig:def:Syntax}
    \end{figure}
        
        The language contains standard sets of pure expressions \ExpSet and (potentially) side-effectful commands \CmdSet.
        The latter includes commands for heap allocation and manipulation, forking and loops.
        We define \emph{physical heaps}~\cite{Jacobs2018ModularTerminationVerification} (as opposed to \emph{logical heaps}~\cite{Jacobs2018ModularTerminationVerification} presented in the next section) as a finite set of \emph{physical resource chunks}.
        A points-to chunk \slPointsTo{\hlocVar}{\valVar} expresses that heap location~\hlocVar points to value \valVar~\cite{Reynolds2002SeparationLA, Jacobs2018ModularTerminationVerification}.
        Moreover, we have chunks to represent unlocked and locked mutexes.

        \begin{definition}[Physical Resources]
                We define the set of physical resources \PhysResSet syntactically as follows:
                $$
                \begin{array}{l c l}
                        \presVar \in \PhysResSet &\ ::=\
                                &\slPointsTo{\hlocVar}{\valVar} \alt
                                    \presUnlocked{\hlocVar}\alt
                                    \presLocked{\hlocVar}
                \end{array}
                $$
                $$
                \begin{array}{c p{0.5cm} c}
                        \hlocVar \in \HeapLocSet
                        &&\valVar \in \ValueSet
                \end{array}
                $$
        \end{definition}

        \begin{definition}[Physical Heaps]
                We define the set of physical heaps as 
                $$\PhysHeapSet\ :=\ \finPowerSetOf{\PhysResSet}$$
                and the function 
                $\phGetLocs: \PhysHeapSet \rightarrow \finPowerSetOf{\HeapLocSet}$
                mapping physical heaps to the sets of allocated heap locations as
                $$
                \begin{array}{l c l l}
                        \phGetLocs(\pheapVar) &\ :=\
                                &\setOf{
                                        \hlocVar \in \HeapLocSet \ \ |\
                                        &\presUnlocked{\hlocVar} \in \pheapVar\ \vee\
                                                \presLocked{\hlocVar} \in \pheapVar\ \vee\\
                                        &&&\exists \valVar \in \ValueSet.\
                                                \slPointsTo{\hlocVar}{\valVar} \in \pheapVar
                                }.
                \end{array}
                $$
                We denote physical heaps by \pheapVar.
        \end{definition}

        We represent a program state by a physical heap and a \emph{thread pool}, which we define as a partial function mapping a finite number of thread IDs to threads.
        Thread IDs are unique and never reused.
        Hence, we represent running threads by commands and terminated ones by \thTerminated instead of removing threads from the pool.
       For the following definition, remember that we assume the set of thread IDs \ThreadIDSet to be infinite and well-founded.

        \begin{definition}[Thread Pools]\label{gs:appendix:def:ThreadPools}
                We define the set of thread pools \ThreadPoolSet as the set of finite partial functions mapping thread IDs to threads:
                $$
                        \ThreadPoolSet \ := \
                                \finPartialFunSet
                                        {\ThreadIDSet}
                                        {(\CmdSet \cup \setOf{\thTerminated})}.
                $$
                The symbol \thTerminated represents a terminated thread.
                We denote thread pools by \tpVar, thread IDs by \tidVar and the empty thread pool by \tpEmpty, i.e.,
                $$
                \begin{array}{l}
                        \tpEmpty \ : \ \ThreadIDSet \finPartialFunArrow  (\CmdSet \cup \setOf{\thTerminated}),
                        \\
                        \defDom{\tpEmpty} \, = \, \emptyset.
                \end{array}
                $$
                We define the operation
                $\tpAddSymb : \ThreadPoolSet \times \setOf{C \subset \CmdSet \ |\ \cardinalityOf{C} \leq 1} \rightarrow \ThreadPoolSet$
%                $\tpAddSymb : \ThreadPoolSet \times \finPowerSetOf{\CmdSet} \partialFunArrow \ThreadPoolSet$
                as follows:
                $$
                \begin{array}{l c l}
                        \tpAdd{\tpVar}{\emptyset} &\ := \
                                &\tpVar,\\
                        \tpAdd{\tpVar}{\setOf{\cmdVar}} &\ :=\
                                &\tpUpdate{\tpVar}{\newTidVar}{\cmdVar}
                                \quad\text{for}\quad
                                \newTidVar := 
                                        \min(\ThreadIDSet \setminus \defDom{\tpVar}).
                \end{array}
                $$
        \end{definition}

        We define the operational semantics of our language in terms of two small-step reduction relations: \stRedStepSymb for single threads and \tpRedStepSymb for thread pools.
        Since expressions are pure and their evaluation is deterministic we identify closed expressions with their ascribed value.
        (i)~\stRedStep
                        {\cmdVar}
                        [\nextPheapVar]
                        {\nextCmdVar}
                        [\ftsVar]
                expresses that heap \pheapVar and command \cmdVar are reduced in a single step to \nextPheapVar and \nextCmdVar and that this thread forks a set of threads~\ftsVar.
                This set is either empty or a singleton as no step forks more than one thread.
        (ii)~\tpRedStep
                        {\tpVar}
                        {\nextTpVar}
                expresses that heap \pheapVar and thread pool \tpVar are reduced in a single step to \nextPheapVar and \nextTpVar.
                ID \tidVar identifies the thread reduced in this step.

        \begin{definition}[Evaluation of Closed Expressions]\label{gs:def:EvaluationOfClosedExpressions}
                We define a partial evaluation function $\evalExp{\cdot}: \ExpSet \partialFunArrow \ValueSet$ on expressions by recursion on the structure of expressions as follows:
                $$
\begin{array}{l l l l}
                        \evalExp{\valVar} &:= &\valVar 
                                &\text{if}\quad \valVar \in \ValueSet,\\
                        \evalExp{\expEq{\expVar}{\nextExpVar}} &:= &\valTrue
                                &\text{if}\quad
                                        \evalExp{\expVar} = \evalExp{\nextExpVar} \neq \bot,\\
                        \evalExp{\expEq{\expVar}{\nextExpVar}} \ &:= \ &\valFalse\ \
                                &\text{if}\quad
                                        \evalExp{\expVar} \neq \evalExp{\nextExpVar}\ \wedge\ 
                                        \evalExp{\expVar} \neq \bot\ \wedge\
                                        \evalExp{\nextExpVar} \neq \bot,\\
                        \evalExp{\expNot{\expVar}} &:= &\valFalse
                                &\text{if}\quad
                                        \evalExp{\expVar} = \valTrue,\\
                        \evalExp{\expNot{\expVar}} &:= &\valTrue
                                &\text{if}\quad
                                        \evalExp{\expVar} = \valFalse,\\
                        \evalExp{\expVar} &:= &\bot &\text{otherwise}.
                \end{array}
                $$
                We identify closed expressions \expVar with their ascribed value \evalExp{\expVar}.
        \end{definition}
        
        \begin{definition}[Evaluation Context]
                We define the set of evaluation contexts \EvalCtxtSet syntactically as follows:
                $$
                \begin{array}{l c l}
                        \evalCtxt \in \EvalCtxtSet &\ ::=\
                                &\cmdIf{\evalCtxtHole}{\cmdVar} \alt
                                        \cmdLet{\varVar}{\evalCtxtHole}{\cmdVar}
                \end{array}
                $$
                $$
                \begin{array}{c p{0.5cm} c}
                        \cmdVar \in \CmdSet
                        &&\varVar \in \VarSet
                \end{array}
                $$
                For any $\cmdVar \in \CmdSet$ and $\evalCtxt \in \EvalCtxtSet$, we define 
                $\evalCtxt[\cmdVar] := \subst{\evalCtxt}{\evalCtxtHole}{\cmdVar}$.
        \end{definition}
        
        Note that for every $\cmdVar \in \CmdSet, \evalCtxt \in \EvalCtxtSet$, we have $\evalCtxt[\cmdVar] \in \CmdSet$.

        \begin{definition}[Single Thread Reduction Relation]
                We define a reduction relation \stRedStepSymb for single threads according to the rules presented in Figure~\ref{gs:appendix:fig:SingleThreadReductionRules}.
                A reduction step has the form
                $$\stRedStep
                        {\cmdVar}
                        [\nextPheapVar]
                        {\nextCmdVar}
                        [\ftsVar]
                $$
                for a set of forked threads 
                $\ftsVar \subset \CmdSet$ with 
                $\cardinalityOf{\ftsVar} \leq 1$.
                
                For simplicity of notation, we omit \ftsVar if it is clear from the context that no thread is forked and $\ftsVar = \emptyset$.
        \end{definition}
        
        \begin{definition}[Thread Pool Reduction Relation]
                We define a thread pool reduction relation \tpRedStepSymb according to the rules presented in Figure~\ref{gs:appendix:fig:ThreadPoolReductionRules}.
                A reduction step has the form
                $$\tpRedStep{\tpVar}{\nextTpVar}.$$
        \end{definition}

        \begin{figure}
                \begin{mathpar}
                        \stRedEvalCtxt
                        \and
                        \stRedFork
                        \and
                        \stRedIfTrue
                        \and
                        \stRedIfFalse
                        \and
                        \stRedLet
                        \and
                        \stRedWhile
                        \and
                        \stRedAlloc
                        \and
                        \stRedReadHeapLoc
                        \and
                        \stRedAssign
                        \and
                        \stRedNewMutex
                        \and
                        \stRedAcquire
                        \and
                        \stRedRelease
                \end{mathpar}

        \caption{Single thread reduction rules.}
        \label{gs:appendix:fig:SingleThreadReductionRules}
        \end{figure}
        
        \begin{figure}
                \begin{mathpar}
                        \tpRedLift
                        \and
                        \tpRedTerm
                \end{mathpar}
                \caption{Thread pool reduction rules}
                \label{gs:appendix:fig:ThreadPoolReductionRules}
        \end{figure}

        As thread scheduling is non-deterministic, so is our thread pool reduction relation \tpRedStepSymb.
        Consider the minimal example we presented in Fig.~\ref{gs:fig:MinEx:Heap:Code} in \S~\ref{gs:sec:Nutshell:AbitraryDataStructures} to illustrate busy waiting for a shared memory flag to be set.
        It does not terminate if the main thread is never scheduled after the new thread was forked.
        Hence, our verification approach relies on the assumption of fair scheduling.
        That is, we assume that every thread is always eventually scheduled while it remains running.
        Further, we represent program executions by sequences of reduction steps.
        As we primarily consider infinite sequences in this work, we define reduction sequences to be infinite to simplify our terminology.

        \begin{definition}[Reduction Sequence]
                Let $(\pheapVar_i)_{i\in \N}$ and $(\tpVar_i)_{i \in \N}$ be infinite sequences of physical heaps and thread pools, respectively.
                We call \phtpRSeq a \emph{reduction sequence} if
                there exists a sequence of thread IDs $(\tidVar_i)_{i \in \N}$ such that 
                $
                \tpRedStep
                        [\pheapVar[i]]
                        {\tpVar[i]}
                        [\tidVar_i]
                        [\pheapVar[i+1]]
                        {\tpVar[i+1]}
                $
                holds for every $i \in \N$.
        \end{definition}

        \begin{definition}[Fairness]\label{gs:def:Fairness}
                We call a reduction sequence \phtpRSeq \emph{fair} iff for all $i \in \N$ and $\tidVar \in \defDom{\tpVar[i]}$ with $\tpVar[i](\tidVar) \neq \thTerminated$ there exists some $k \geq i$ with
                $$
                \tpRedStep
                        [\pheapVar[k]]{\tpVar[k]}
                        [\tidVar]
                        [\pheapVar[k+1]]{\tpVar[k+1]}.
                $$
        \end{definition}

\section{Logic}\label{gs:appendix:sec:Logic}
    In this section we formalize the logic we sketched in \S~\ref{gs:sec:ApproachInANutshell}.
    For the definition we assume 
    (i)~an infinite set of ghost signal IDs $\idVar \in \IDSet$
    and
    (ii)~an infinite, partially ordered and well-founded set of levels $\levVar \in \LevelSet$.
    We denote the level order relation by \levLt.

    \begin{definition}[Fractions]
            We define the set of fractions as
            $$\FracSet \ := \
                        \setOf{ \fracVar \in \Q \ \ | \ \ 0 < \fracVar \leq 1}
            $$
            and denote fractions by \fracVar.
    \end{definition}
    
    \begin{definition}[Obligations, Signals \& Mutexes]
            We define the set of obligations \ObSet and signals \SigSetVar as
            $$
            \begin{array}{l c l}
                    \ObSet &\ :=\  &(\HeapLocSet \cup \IDSet) \times \LevelSet,
                    \\
                    \SignalSet &\ :=\ &\IDSet \times \LevelSet.
            \end{array}
            $$
            We denote bags of obligations by $\obBagVar \in \BagsOf{\ObSet}$, signals by $\sigVar \in \SignalSet$ and mutexes by $\mutVar \in \HeapLocSet \times \LevelSet$.
            For convenience of notation, we define the following selector functions:
            $$
            \begin{array}{l c l p{0.5cm} l}
                    \sigGetID{(\idVar, \slWildcard)} &\ :=\ &\idVar 
                            &&\text{for signals},
                    \\
                    \mutGetLoc{(\hlocVar, \slWildcard)} &\ :=\ &\hlocVar
                            &&\text{for mutexes},
                    \\
                    \sigGetLev{(\slWildcard, \levVar)} &\ :=\ &\levVar
                            &&\text{for signals, mutexes and obligations}.
            \end{array}
            $$
    \end{definition}

        \begin{definition}[Assertions]
                We define the set of \emph{assertions} \AssertionSet according to the syntax presented in Figure~\ref{gs:appendix:fig:def:Assertions}.\footnote{
                        That is, we define \AssertionSet as the least fixpoint of $F$ where 
                        $F(\assSetVar) = 
                                \setOf{\slTrue, \slFalse}
                                \cup
                                \setOf{\neg \assVar \ | \
                                                \assVar \in \assSetVar}
                                \cup
                                \setOf{\assVar_1 \wedge \assVar_2 \ | \ 
                                                \assVar_1, \assVar_2 \in \assSetVar} 
                                \cup
                                \dots
                                \cup
                                \setOf{\bigvee \oAssSetVar \ | \
                                                \oAssSetVar \subseteq \assSetVar}
                                \cup
                                \dots
                        $.
                        Since $F$ is a monotonic function over a complete lattice, it has a least fixpoint according to the Knaster-Tarski theorem~\cite{Tarski1955KnasterTarskiTheorem}.
                }
                We omit the index set $I$ in quantifications when its choice becomes clear from the context and write 
                $\exists \assIndexVar.\, \assFam[\assIndexVar]$
                and
                $\forall \assIndexVar.\, \assFam[\assIndexVar]$
                instead of                
                $\exists \assIndexVar \in \assIndexSetVar.\, \assFam[\assIndexVar]$
                and
                $\forall \assIndexVar \in \assIndexSetVar.\, \assFam[\assIndexVar]$,
                respectively.
        \end{definition}

    \begin{figure}
            \begin{subfigure}{\textwidth}
                    $$
\begin{array}{c p{0.5cm} c}
                            \assSetVar \subseteq \AssertionSet
                            &
                            &\text{Index set}\ \assIndexSetVar
                    \end{array}%
                    $$
                    \caption{
                            Assertion syntax.
                            We omit quantification domain \assIndexSetVar when it is clear from the context.
                    }
                    \label{gs:appendix:fig:def:Assertionsyntax}
            \end{subfigure}
            
            \begin{subfigure}{\textwidth}
                    $$
                    \begin{array}{l c l}
                            \assVar_1 \rightarrow \assVar_2
                                    \ \ &:= \ \
                                    &\neg\assVar_1 \vee \assVar_2
                            \\
                            \assVar_1 \leftrightarrow \assVar_2
                                    \ \ &:= \ \
                                    &(\assVar_1 \rightarrow \assVar_2) \wedge
                                        (\assVar_2 \rightarrow \assVar_1)
                            \\
                            \exists \assIndexVar \in \assIndexSetVar.\ \assFam[\assIndexVar]
                                    \ \ &:= \ \
                                    &\bigvee \setOf{\assFam[\assIndexVar] \ | \ \assIndexVar \in \assIndexSetVar}
                            \\
                            \forall \assIndexVar \in \assIndexSetVar.\ \assFam[\assIndexVar] 
                                    \ \ &:=\ \
                                    &\neg\exists \assIndexVar \in \assIndexSetVar.\, \neg \assFam[\assIndexVar]
                    \end{array}
                    $$
                    \caption{Syntactic sugar.}
            \end{subfigure}
            
            \caption{Assertions.}
            \label{gs:appendix:fig:def:Assertions}
    \end{figure}

    In \S~\ref{gs:appendix:sec:Language}, we used physical resources and heaps to model a program's state.
    We use assertions to capture which fraction of a physical resource and which ghost resources a thread owns.
    Therefore, we have to interpret assertions in an extended notion of state.
    We define \emph{logical heaps} and \emph{logical resources}~\cite{Jacobs2018ModularTerminationVerification} which correspond to physical ones but additionally encompass ghost resources and ownership.
    For instance, logical resources include signal chunks and initialized mutex chunks are associated with a lock invariant.
    Rather than being a set of resources, logical heaps map logical resources to fractions.
    This allows us to express which portion of a resource a thread owns.

    \begin{definition}[Logical Resources]
            We define the set of logical resources \LogResSet syntactically as follows:
            $$
\begin{array}{l c l}
                    \lresVar \in \LogResSet &\ ::=\
                            &\slPointsTo{\hlocVar}{\valVar} \alt
                                    \lresSignal{(\idVar, \levVar)}{\slBoolVar} \alt
                                    \lresUninit{\hlocVar} \alt
                    \\
                            &&\!\!
                            		\lresMutex{(\hlocVar, \levVar)}{\assVar} \alt
                                    \lresLocked{(\hlocVar, \levVar)}{\assVar}{\fracVar} \alt
                                    \lresObs{\obBagVar}
            \end{array}
            $$
            Further, we define the function
            $\lresGetHLocs: \LogResSet \rightarrow \HeapLocSet$
            mapping logical resources to their respective (either empty or singleton) set of involved heap locations as
            $$
            \begin{array}{l c c l}
                    \lresGetHLocs[\slPointsTo{\hlocVar}{\valVar}] 
                            \ \ &:= \ \
                            &\setOf{\hlocVar},
                    \\
                    \lresGetHLocs[\lresUninit{\hlocVar}]
                            \ \ &:= \ \
                            &\setOf{\hlocVar},
                    \\
                    \lresGetHLocs[\lresMutex{(\hlocVar, \levVar)}{\assVar}]
                            \ \ &:= \ \
                            &\setOf{\hlocVar},
                    \\
                    \lresGetHLocs[\lresLocked{(\hlocVar, \levVar)}{\assVar}{\fracVar}]
                            \ \ &:= \ \
                            &\setOf{\hlocVar},
                    \\
                    \lresGetHLocs[\slWildcard]
                            \ \ &:= \ \
                            &\emptyset
                            &\text{otherwise}.
            \end{array}
            $$
    \end{definition}

    \begin{definition}[Logical Heaps]
            We define the set of logical heaps as
            $$\lheapVar \in \LogHeapSet 
                    \ := \
                    \LogResSet \rightarrow \setOf{q \in \Q~|~ q \geq 0}.
            $$
            We define the empty logical heap \lhEmpty as the constant zero function
            $$
                    \lhEmpty : \lresVar \mapsto 0.
            $$
            We denote logical heaps by \lheapVar, point-wise addition by $+$ and multiplication with non-negative rationals by $\cdot$, i.e.,
            $$
            \begin{array}{l c l l}
                    (\lheapVar_1 + \lheapVar_2)(\lresVar) &\ :=\
                            &\lheapVar_1(\lresVar) + \lheapVar_2(\lresVar),\\
                    (q \cdot \lheapVar)(\lresVar) &\ :=\
                            &q \cdot (\lheapVar(\lresVar))
            \end{array}
            $$
            for $q\in \Q$ with $q\geq 0$.
            We give $\cdot$ a higher precedence than $+$.
            For convenience of notation we represent logical heaps containing finitely many resources by sets of resources as follows
            $$
            %
                            \right.
            \end{array}
            $$
    \end{definition}
    
    \begin{definition}[Logical Heap Predicates]
            Let \lheapVar be a logical heap.
            We call \lheapVar \emph{complete} and write \lhComplete{\lheapVar} if it contains exactly one obligations chunk, i.e.,
            if there exists a bag of obligations \obBagVar with
            $\lheapVar(\lresObs{\obBagVar} ) = 1$
            and if there does not exist any bag of obligations $\obBagVar'$ with 
            $\obBagVar \neq \obBagVar'$ and
            $\lheapVar(\lresObs{\obBagVar'}) > 0$.
            
            We call \lheapVar \emph{finite} and write \lhFinite{\lheapVar} if 
            it contains only finitely many resources, i.e.,
            if the set 
            \setOf{\lresVar \in \LogResSet \ \ | \ \ \lheapVar(\lresVar) > 0}
            is finite.
            
            We call \lheapVar \emph{consistent} and write \lhConsistent{\lheapVar} if 
            (i)~it contains only full obligations chunks, i.e.,
            if
            $$
            \begin{array}{l l l}
                    \lheapVar(\lresObs{\obBagVar}) 
                            &\in &\N
            \end{array}
            $$
            holds for all $\obBagVar \in \BagsOf{\ObSet}$
            and if
            (ii)~heap locations are unique in \lheapVar, i.e.,
            if there are no $\lresVar[1], \lresVar[2] \in \LogResSet$ with 
            $\lresVar[1] \neq \lresVar[2]$,
            $\lheapVar(\lresVar[1]) > 0$,
            $\lheapVar(\lresVar[2]) > 0$ 
            and with
            $\lresGetHLocs[\lresVar[1]] \cap \lresGetHLocs[\lresVar[2]] \neq \emptyset$.
    \end{definition}

    We interpret assertions in terms of a model relation.
    \assModels{\lheapVar}{\assVar} expresses that assertion \assVar holds with respect to logical heap \lheapVar.
    Further, we define the view shift relation~\viewShiftSymb and the proof relation~\htProvesSymb we sketched in \S~\ref{gs:sec:ApproachInANutshell}.

    \begin{definition}[Assertion Model Relation]
            We define a model relation 
            $\assModelsSymb \,\subset \, \LogHeapSet \times \AssertionSet$
            for assertions by recursion on the structure of assertions according to the rules presented in Figure~\ref{gs:appendix:fig:def:AssertionModelRelation}.
            We write \assModels{\lheapVar}{\assVar} to express that logical heap \lheapVar models assertion \assVar and
            \assNotModels{\lheapVar}{\assVar} to express that \assModels{\lheapVar}{\assVar} does not hold.
    \end{definition}

    \begin{figure}
                    $$
\begin{array}{l c l}
                            \assModels{\lheapVar}{\slTrue}\\
                            \assNotModels{\lheapVar}{\slFalse}\\
                            \assModels{\lheapVar}{\neg \assVar}
                                    &\text{if}
                                    &\assNotModels{\lheapVar}{\assVar}\\
                            \assModels{\lheapVar}{\assVar_1 \wedge \assVar_2}
                                    &\text{if}
                                    &\assModels{\lheapVar}{\assVar_1}
                                            \wedge
                                            \assModels{\lheapVar}{\assVar_2}\\
                            \assModels{\lheapVar}{\assVar_1 \vee \assVar_2}
                                    &\text{if}
                                    &\assModels{\lheapVar}{\assVar_1}
                                            \vee
                                            \assModels{\lheapVar}{\assVar_2}\\
                            \assModels{\lheapVar}{\assVar_1 \slStar \assVar_2}
                                    &\text{if}
                                    &\exists \lheapVar[1], \lheapVar[2] \in \LogHeapSet.\
                                            \lheapVar = \lheapVar[1] + \lheapVar[2]
                                            \ \wedge
                           	\\
                           			&&\phantom{\exists \lheapVar[1], \lheapVar[2] \in \LogHeapSet}\ \
                                            \assModels{\lheapVar[1]}{\assVar_1}
                                            \wedge
                                            \assModels{\lheapVar[2]}{\assVar_2}\\
                            \assModels
                                {\lheapVar}
                                {\slPointsTo[\fracVar]{\hlocVar}{\valVar}}
                                    &\text{if}
                                    & \lheapVar(\slPointsTo{\hlocVar}{\valVar}) \geq \fracVar\\
\end{array}
                    $$
                
                    \caption{
                            Assertion model relation.
                            \assNotModels{\lheapVar}{\assVar} expresses that
                            \assModels{\lheapVar}{\assVar} does not hold.
                    }
                    \label{gs:appendix:fig:def:AssertionModelRelation}
    \end{figure}

    \begin{definition}
            Let $\levVar \in \LevelSet$, $\obBagVar \in \BagsOf{\ObSet}$.
            We define $\levObsLt\, \subset \LevelSet \times \ObSet$ as
            $$
                    \levVar \levObsLt \obBagVar
                    \ \Longleftrightarrow \
                    \forall \obVar \in \obBagVar.\
                            \levVar \levLt \sigGetLev{\obVar}.
            $$
    \end{definition}

    \begin{definition}[View Shift]
            We define a view shift relation 
            $\viewShiftSymb \ \subset\ \AssertionSet \times \AssertionSet$
            according to the rules presented in Fig.~\ref{gs:appendix:fig:ViewShiftRules}.
    \end{definition}

    \begin{definition}[Proof Relation]
            We define a proof relation
            $\htProvesSymb \ \subset \ \AssertionSet \times \CmdSet \times (\ValueSet \rightarrow \AssertionSet)$
            according to the rules presented in Fig.~\ref{gs:appendix:fig:ProofRulesPartOne} and \ref{gs:appendix:fig:ProofRulesPartTwo}.
            We state the provability of a Hoare triple in the form of\, \htProves{\assPreVar}{\cmdVar}[\resVar]{\assPostVar(\resVar)} where \resVar captures the value returned by \cmdVar.
            To simplify the notation, we omit the result value if it is clear from the context or irrelevant.
    \end{definition}

    \begin{figure}
            \begin{mathpar}
                    \vsSemImp
                    \and
                    \vsTrans
                    \and
                    \vsOr
                    \and
                    \vsNewSignal
                    \and
                    \vsSetSignal
                    \and
                    \vsMutInit
                    \and
                    \vsGhostLoop
            \end{mathpar}
            
            \caption{View shift rules.}
            \label{gs:appendix:fig:ViewShiftRules}
    \end{figure}

    \begin{figure}
\vspace{-0.9cm}
            \begin{subfigure}{\textwidth}
                    \begin{mathpar}
                            \prFrame
                            \and
                            \prSimpleViewShiftPaper
                            \and
                            \prExp
                            \and
                            \prExists
                            \and
                            \prForkPaperWithResult
                    \end{mathpar}
                    \subcaption{Basic Proof Rules.}
            \end{subfigure}

            \begin{subfigure}{\textwidth}
                    \begin{mathpar}
                            \prIf
                            \and
                            \prAwaitGen
                            \and
                            \prWhileDecStrict
                            \and
                            \prLet
                    \end{mathpar}
                    \subcaption{Control Structures.}
            \end{subfigure}

            \caption{Proof rules (part 1).}
            \label{gs:appendix:fig:ProofRulesPartOne}
    \end{figure}

    \begin{figure}
            \begin{subfigure}{\textwidth}
                    \begin{mathpar}
                            \prAcquire
                            \and
                            \prReleaseSimplifiedPaper
                            \and
                            \prNewMutex
                    \end{mathpar}
                    \subcaption{Mutexes.}
            \end{subfigure}

            \begin{subfigure}{\textwidth}
                    \begin{mathpar}
                            \prAlloc
                            \and
                            \prReadHeapLoc
                            \and
                            \prAssignToHeap
                    \end{mathpar}
                    \subcaption{Heap Access.}
            \end{subfigure}

            \caption{Proof rules (part 2).}
            \label{gs:appendix:fig:ProofRulesPartTwo}
    \end{figure}

    Following the intuition provided in \S~\ref{gs:sec:ApproachInANutshell}, we can prove that a program terminates by proving that it discharges all obligations.
    The following theorem states that this approach is sound.

    \begin{restatable}[Soundness]{theorem}{theoSoundness}\label{gs:theo:Soundness}
            Let \htProves{\noObs}{\cmdVar}{\noObs} hold.
             There exists no fair, infinite reduction sequence \phtpRSeq with
            $\pheapVar[0] = \emptyset$ and 
            $\tpVar[0] = 
                    \setOf{(\tidStartVar, \cmdVar)}
            $
            for any choice of \tidStartVar.
    \end{restatable}

\section{Soundness}\label{gs:appendix:sec:Soundness}

    We already sketched the high-level intuition behind our soundness argument in \S~\ref{gs:sec:Nutshell:ThreadSafePhysicalSignals}.
    In this section, we prove it formally.
    In \S~\ref{gs:appendix:sec:Soundness:AnnotatedSemantics} we define an annotated reduction semantics that tracks which resources threads own (including ghost resources).
    In \S~\ref{gs:appendix:sec:Soundness:ModelRelation} we define a model relation for Hoare triples and prove that every specification we can derive with our proof rules also holds in our model.
    In \S~\ref{gs:appendix:sec:Soundness:Proof} we use the annotated semantics and the model relation to prove the soundness theorem we stated above.

\subsection{Annotated Semantics}\label{gs:appendix:sec:Soundness:AnnotatedSemantics}

    During runtime, all threads share one physical heap where every thread is free to access every resource.
    This does not reflect the notions of ownership and lock invariants which we maintain on the verification level.
    It also does not allow us to restrict actions based on levels, e.g., only allowing the acquisition of a mutex if its level is lower than the level of each held obligation.
    Hence:
    (i)~We annotate every thread by a logical heap to express which resources it owns (including ghost resources) and thereby obtain an \emph{annotated thread pool}.
    (ii)~We represent the program state by an \emph{annotated heap} that keeps track of lock invariants and levels.
    In particular, we associate unlocked mutexes with logical heaps to represent the resources they protect.
    Since annotated heaps keep track of levels, they also keep track of signals.

        \begin{definition}[Intermediate Representation]
                We define an extended set of commands \CmdSetExt according to the syntax presented in Figure~\ref{gs:appendix:fig:ExtendedSyntax}.
        \end{definition}
        
        For the rest of this appendix, commands \cmdVar refer to the extended set of commands, i.e., $\cmdVar \in \CmdSetExt$.
        
        \begin{figure}
                    $$
\begin{array}{l c l l}        
                            \cmdVar \in \CmdSetExt  &\ ::= \ \
                                    &\lowlightBack{
                                            \expVar \alt
                                                    \cmdWhile{\cmdVar}{\cmdSkip} \alt
                                                    \cmdFork{\cmdVar}\alt
                                            }\\
                                    &&\lowlightBack{
                                                    \cmdLet{\varVar}{\cmdVar}{\cmdVar} \alt
                                                    \cmdIf{\cmdVar}{\cmdVar} \alt
                                                    \cmdAlloc{\expVar} \alt
                                            }\\
                                    &&\lowlightBack{
                                                    \cmdReadHeapLoc{\expVar} \alt
													\cmdAssignToHeap{\expVar}{\expVar} \alt
                                    		}\\
                                    &&\lowlightBack{
                                                    \cmdNewMut \alt
                                                    \cmdAcquireMut{\expVar} \alt
                                                    \cmdReleaseMut{\expVar} \alt
                                            }\\
                                    &&\icmdAwaitStarted{\SigSetVar}{\hlocVar}{\cmdVar} \alt
                                            \icmdWhileDecStarted{\natVar}{\cmdVar}
%                                            \alt \icmdWait{\SigSetVar}
                        \\
                        \natVar \in \N
                        \\
                        \SigSetVar \finSubset \SignalSet
                                &&\lowlightText{\text{finite set of signals}}
                    \end{array}
                    $$
                    \caption{Extended set of commands for intermediate representation.}
                    \label{gs:appendix:fig:ExtendedSyntax}
        \end{figure}

     \begin{definition}[Annotated Resources]
                We define the set of annotated resources \AnnoResSet as
                $$
                \begin{array}{l l l}
                        \aresVar \in \AnnoResSet \ &::= \
                                &\slPointsTo{\hlocVar}{\valVar} \alt
                                        \aresUninit{\hlocVar} \alt\\
                                &&\aresUnlocked{(\hlocVar, \levVar)}{\assVar}{\lheapVar}
                                        \alt
                                        \aresLocked{(\hlocVar, \levVar)}{\assVar}{\fracVar} \alt\\
                                &&\aresSignal{(\idVar, \levVar)}{\slBoolVar}
                \end{array}
                $$
                where \lheapVar does not contain any obligations chunks.
        \end{definition}

        \begin{definition}[Annotated Heaps]
                We define the set of annotated heaps as
                $$\AnnoHeapSet \ := \ \finPowerSetOf{\AnnoResSet},$$
                the function 
                $\ahGetLocs: \AnnoHeapSet \rightarrow \finPowerSetOf{\HeapLocSet}$
                mapping annotated heaps to the sets of allocated heap locations as
                $$
                \begin{array}{l c l l}
                        \ahGetLocs(\aheapVar) &\ := \
                                &\setOf{ 
                                        \hlocVar \in \HeapLocSet \ \ |\
                                        & \exists \valVar \in \ValueSet.\
                                                \exists \levVar \in \LevelSet.\
                                                \exists \assVar \in \AssertionSet.\\
                                        &&&\exists \lheapVar \in \LogHeapSet.\
                                                \exists \fracVar \in \FracSet.\\
                                        &&&\quad
                                                \slPointsTo{\hlocVar}{\valVar} \in \aheapVar\ \vee\
                                                \aresUninit{\hlocVar} \in \aheapVar\ \vee\\
                                        &&&\quad
                                                \aresUnlocked
                                                        {(\hlocVar, \levVar)}
                                                        {\assVar}
                                                        {\lheapVar}
                                                    \in \aheapVar\ \vee\\
                                        &&&\quad
                                                \aresLocked{(\hlocVar, \levVar)}{\assVar}{\fracVar} \in \aheapVar
                                }
                \end{array}
                $$
                and the function
                $\ahGetIDs : \AnnoHeapSet \rightarrow \finPowerSetOf{\IDSet}$
                mapping annotated heaps to sets of allocated signal IDs as
                $$
                \begin{array}{l c l }
                        \ahGetIDs(\aheapVar) &\ :=\
                                &\setOf{
                                         \idVar \in \IDSet  \ \ | \ \
                                        \exists \levVar \in \LevelSet.\
                                        \exists \slBoolVar \in \B.\
                                                \aresSignal{(\idVar, \levVar)}{\slBoolVar}
                                                \in \aheapVar
                                }.
                \end{array}
                $$
                We denote annotated heaps by \aheapVar.
                
                We call an annotated heap \aheapVar \emph{finite} and write \ahFinite{\aheapVar} if there exists no chunk
                $\aresUnlocked{(\hlocVar, \levVar)}{\assVar}{\lheapVar} \in \aheapVar$
                for which \lhFinite{\lheapVar} does not hold.
        \end{definition}

    Physical, annotated and logical heaps represent program states on different abstraction levels.
    While each level focuses on different aspects of the program state, they also share information.
    For instance, all three kinds of heaps use points-to chunks.
    Also, annotated and logical heaps both have signal chunks.
    Therefore, when referring to multiple kinds of heaps to talk about the same program state, we have to ensure that their contents match and do not contradict each other.
    We call such heaps \emph{compatible}.
    The following definitions make this precise.
        
    \begin{definition}[Compatibility of Annotated and Physical Heaps]
            We inductively define a relation
            $\aphCorrespSymb\ \subset \AnnoHeapSet \times \PhysResSet$
            between annotated and physical heaps such that the following holds:
            $$
\begin{array}{l c l}
                    \emptyset &\aphCorrespSymb &\emptyset,\\
                    \slPointsTo{\hlocVar}{\valVar}\ \cup\ \aheapVar
                            &\aphCorrespSymb
                            &\slPointsTo{\hlocVar}{\valVar}\ \cup\ \pheapVar,\\
                    \aresUninit{\hlocVar}\ \cup\ \aheapVar
                            &\aphCorrespSymb
                            &\presUnlocked{\hlocVar}\ \cup\ \pheapVar,\\
                    \aresUnlocked
                            {(\hlocVar, \levVar)}
                            {\assLockInvVar}
                            {\lheapLockInvVar}
                        \ \cup\ \aheapVar
                            &\aphCorrespSymb
                            &\presUnlocked{\hlocVar}\ \cup\ \pheapVar,\\
                    \aresLocked{(\hlocVar, \levVar)}{\assLockInvVar}{\fracVar}\ \cup\ \aheapVar
                            &\aphCorrespSymb
                            &\presLocked{\hlocVar}\ \cup\ \pheapVar,\\
                    \aresSignal{\sigVar}{\slBoolVar}\ \cup\ \aheapVar
                            &\aphCorrespSymb
                            &\pheapVar,
            \end{array}
            $$
            where 
            $\aheapVar \in \AnnoHeapSet$ 
            and
            $\pheapVar \in \PhysHeapSet$ 
            are annotated and physical heaps with
            \aphCorresp{\aheapVar}{\pheapVar}.
    \end{definition}

        \begin{definition}[Compatibility of Annotated and Logical Heaps]
                We inductively define a relation
                $\alhCorrespSymb\, \subset\, \AnnoHeapSet \times \LogHeapSet$
                between annotated and logical heaps such that the following holds:
                $$
                \begin{array}{l c l}
                        \emptyset &\alhCorrespSymb &\lhEmpty,
                        \\
                        \ahAddRes
                                {\aheapVar}
                                {\slPointsTo{\hlocVar}{\valVar}}
                        &\alhCorrespSymb
                        &\lhAddRes
                                {\lheapVar}
                                {\slPointsTo{\hlocVar}{\valVar}},
                        \\
                        \ahAddRes
                                {\aheapVar}
                                {\aresUninit{\hlocVar}}
                        &\alhCorrespSymb
                        &\lhAddRes
                                {\lheapVar}
                                {\lresUninit{\hlocVar}},
                        \\
                        \ahAddRes
                                {\aheapVar}
                                {\aresUnlocked
                                        {\mutVar}
                                        {\assLockInvVar}
                                        {\lheapLockInvVar}
                                }
                        &\alhCorrespSymb
                        &\lhAddRes
                                {\lheapVar}
                                {\lresMutex{\mutVar}{\assLockInvVar}}
                            + \lheapLockInvVar,
                        \\
                        \ahAddRes
                                {\aheapVar}
                                {\aresLocked{\mutVar}{\assLockInvVar}{\fracVar}}
                        &\alhCorrespSymb
                        &\lheapVar
                                + \lheapOf{\lresLocked{\mutVar}{\assLockInvVar}{\fracVar}}
                        \\
                                &&\phantom{\lheapVar}
                                        + (1-\fracVar) \cdot \lheapOf{\lresMutex{\mutVar}{\assLockInvVar}},
                        \\
\end{array}
                $$
                where $\aheapVar \in \AnnoHeapSet$ and $\lheapVar \in \LogHeapSet$ are annotated and logical heaps with
                $\hlocVar, \mutGetLoc{\mutVar} \not\in \ahGetLocs[\aheapVar]$,
                $\sigGetID{\sigVar} \not\in \ahGetIDs[\aheapVar]$
                and \alhCorresp.
        \end{definition}

    We define annotated versions \atpRedStepSymb and \astRedStepSymb of the relations \tpRedStepSymb and \stRedStepSymb, respectively.
    The annotated reduction semantics we thereby obtain needs to reflect ghost proof steps implemented by view shifts.
    Hence, we define \atpRedStepSymb in terms of two relations: 
    (i)~\gtpRedStepSymb for ghost steps and 
    (i)~\rtpRedStepSymb for actual program execution steps.
    The annotated semantics ensure that a reduction gets stuck if a thread violates any of the restrictions formulated by our proof rules.

        \begin{definition}[Annotated Single Thread Reduction Relation]
                We define a reduction relation \astRedStepSymb for annotated threads according to the rules presented in Fig.~\ref{gs:appendix:fig:AnnotatedSingleThreadReductionRulesPartOne} and~\ref{gs:appendix:fig:AnnotatedSingleThreadReductionRulesPartTwo}.
                A reduction step has the form
                $$
                        \astRedStep
                                {\cmdVar}
                                [\tidVar]
                                [\nextAheapVar]
                                [\nextLheapVar]
                                {\nextCmdVar}
                                [\aftsVar]
                $$
                for a set of annotated forked threads
                $\aftsVar \subset \LogHeapSet \times \CmdSet$
                with $\cardinalityOf{\aftsVar} \leq 1$.

                It indicates that given annotated heap \aheapVar and a logical heap \lheapVar, 
                command \cmdVar can be reduced to annotated heap \nextAheapVar, logical heap \nextLheapVar and command \nextCmdVar.
                The either empty or singleton set \aftsVar represents whether a new thread is forked in this step.
                
                For simplicity of notation we omit \aftsVar if it is clear from the context that no thread is forked and $\aftsVar = \emptyset$.
        \end{definition}

        \begin{figure}
	        	\vspace{-1cm}
                \begin{subfigure}{\textwidth}
                        \begin{mathpar}
                                \astRedEvalCtxt
                                \\
                                \astRedForkPaper
                        \end{mathpar}
                        \caption{Basic constructs.}
                \end{subfigure}

                \begin{subfigure}{\textwidth}
                        \begin{mathpar}
%                                \astRedWhile
%                                \and
                                \astRedIfTrue
                                \and
                                \astRedIfFalse
                                \and
%                                \astRedConsumeItPermPaper
%                                \and
                                \astRedLet
                        \end{mathpar}
                        \caption{Control structures.}
                \end{subfigure}
                
                \begin{subfigure}{\textwidth}
                        \begin{mathpar}
                                \astRedAlloc
                                \and
                                \astRedReadHeapLoc
                                \and
                                \astRedAssign
                        \end{mathpar}
                        \caption{Heap access.}
                \end{subfigure}
                
                \begin{subfigure}{\textwidth}
                        \begin{mathpar}
                                \astRedNewMutex
                                \and
                                {
                                	\hspace{-0.65cm}
	                                \astRedAcquire
                                }
                                \and
                                {
                                	\hspace{-0.2cm}
	                                \astRedReleaseNoPerm
	                            }
	                            \vspace{-0.2cm}
                        \end{mathpar}
                        \caption{Mutexes.}
                \end{subfigure}
                
                \caption{
                        Annotated single thread reduction rules (part 1).
                }
                \label{gs:appendix:fig:AnnotatedSingleThreadReductionRulesPartOne}
        \end{figure}

        \begin{figure}
                \begin{subfigure}{\textwidth}
                        \begin{mathpar}
                                \astRedWhileDecInit
                                \and
                                \astRedWhileDec
                                \and
                                \astRedAwaitInit
                                \and
                                \astRedAwait
                        \end{mathpar}
                        \caption{Loops and auxiliary commands.}
                \end{subfigure}
                
                \caption{
                        Annotated single thread reduction rules (part 2).
                }
                \label{gs:appendix:fig:AnnotatedSingleThreadReductionRulesPartTwo}
        \end{figure}

        \begin{definition}[Annotated Thread Pools]
                We define the set of annotated thread pools \AnnoThreadPoolSet as the set of finite partial functions mapping thread IDs to annotated threads:
                $$
                        \AnnoThreadPoolSet \ := \ 
                                \finPartialFunSet
                                        {\ThreadIDSet}
                                        {
                                                \LogHeapSet \times
                                                (\CmdSetExt \cup \setOf{\thTerminated})
                                        }.
                $$
                We denote annotated thread pools by \atpVar and the empty thread pool by \atpEmpty, i.e.,
                $$
                \begin{array}{l}
                        \atpEmpty \ : \
                                \ThreadIDSet \finPartialFunArrow 
                                \LogHeapSet \times (\CmdSetExt \cup \setOf{\thTerminated}),
                        \\
                        \defDom{\atpEmpty} \, = \, \emptyset.
                \end{array}
                $$
                We define the extension operation \atpAddSymb analogously to \tpAddSymb, cf. Definition~\ref{gs:appendix:def:ThreadPools}.
                
                For convenience of notation we define selector functions for annotated threads as
                $$
                \begin{array}{l c l}
                        \atGetHeap{(\lheapVar, \cmdVar)} &:=
                                &\lheapVar,
                        \\
                        \atGetCmd{(\lheapVar, \cmdVar)} &:=
                                &\cmdVar.
                \end{array}
                $$
        \end{definition}

        \begin{definition}[Ghost Reduction Relation]
                We define a thread pool reduction relation \gtpRedStepSymb according to the rules presented in Fig.~\ref{gs:appendix:fig:GhostThreadPoolReductionRulesPart} to express ghost steps.
                A ghost reduction step has the form
                $$
                        \gtpRedStep
                                {\atpVar}
                                {\nextAtpVar}.
                $$
                We denote its reflexive transitive closure by \gtpRedStepMultSymb.
        \end{definition}

        \begin{figure}
                \begin{mathpar}
                        \gtpRedNewSignal
                        \and
                        \gtpRedSetSignal
                        \and
                        \gtpRedMutInitNoPerm
                \end{mathpar}

                \caption{Ghost thread pool reduction rules.}
                \label{gs:appendix:fig:GhostThreadPoolReductionRulesPart}
        \end{figure}

        \begin{definition}[Non-ghost Thread Pool Reduction Relation]
                We define a thread pool reduction relation \rtpRedStepSymb according to the rules presented in Fig.~\ref{gs:appendix:fig:RealThreadPoolReductionRules} to express real (i.e. non-ghost) reduction steps.
                A reduction step has the form
                $$\rtpRedStep
                        {\atpVar}
                        {\nextAtpVar}.
                $$
        \end{definition}

        \begin{figure}
                \begin{mathpar}
                        \rtpRedLift
                        \and
                        \rtpRedTerm
                \end{mathpar}
                
                \caption{Non-ghost thread pool reduction rules.}
                \label{gs:appendix:fig:RealThreadPoolReductionRules}
        \end{figure}

        \begin{definition}[Annotated Thread Pool Reduction Relation]
                We define the annotated thread pool reduction relation \atpRedStepSymb as
                $$
                \atpRedStepSymb \ \ := \ \ \gtpRedStepSymb \cup \rtpRedStepSymb.
                $$
        \end{definition}

        Note that the  reduction relation \atpRedStepSymb indeed reflects the restrictions our proof rules impose.
        For instance, proof rule \prAwaitGenName only allows a thread to wait for a signal if the signal's level is below the level of all held obligations.
        \astRedAwaitName ensures that any thread that does not comply with this restriction gets stuck.

        \begin{definition}[Annotated Reduction Sequence]\label{gs:appendix:def:AnnotatedRedSeq}
                Let $(\aheapVar_i)_{i\in \N}$ and $(\atpVar_i)_{i \in \N}$ be infinite sequences of annotated heaps and annotated thread pools, respectively.
                Let
                $\sigAnno : \N \partialFunArrow \SignalSet$ be a partial function mapping indices to signals.
                
                We call \ahtpRSeqFull an \emph{annotated reduction sequence} if
                there exists a sequence of thread IDs $(\tidVar_i)_{i \in \N}$ such that the following holds for every $i \in \N$:
                \begin{itemize}
                        \item
                \atpRedStep
                        [\aheapVar[i]]
                        {\atpVar[i]}
                        [\tidVar_i]
                        [\aheapVar[i+1]]
                        {\atpVar[i+1]}
                        \item If this reduction step does not result from an application of \rtpRedLiftName in combination with \astRedAwaitName, then $\sigAnno[i] = \funUndefVal$.
                                If \astRedAwaitName is applied to some signal \sigVar, then $\sigAnno[i] = \sigVar$.
                \end{itemize}
                In case the signal annotation \sigAnno is  clear from the context or not relevant, we omit it and write \ahtpRSeq instead of \ahtpRSeqFull.
                
                We call $(\aheapVar[i], \atpVar[i])$ an \emph{annotated machine configuration}.
        \end{definition}

        \begin{definition}[Fairness of Annotated Reduction Sequences]
                We call an annotated reduction sequence \ahtpRSeq \emph{fair} iff for all $i \in \N$ and $\tidVar \in \defDom{\atpVar[i]}$ with
                $\atGetCmd{\atpVar[i](\tidVar)} \neq \thTerminated$
                there exists some $k \geq i$ with
                $$
                \rtpRedStep
                        [\aheapVar[k]]{\atpVar[k]}
                        [\tidVar]
                        [\aheapVar[k+1]]{\atpVar[k+1]}.
                $$
        \end{definition}

        \begin{lemma}[Preservation of Finiteness]\label{gs:lem:PreservationOfFiniteness}
                Let \ahtpRSeq be an annotated reduction sequence with 
                \ahFinite{\aheapVar[0]}
                and
                \lhFinite{\atGetHeap{\atpVar[0](\tidVar)}}
                for all $\tidVar \in \defDom{\atpVar[0]}$.
                
                Then,
                \lhFinite{\atGetHeap{\atpVar[i](\tidVar)}}
                holds for all $i \in \N$ and all $\tidVar \in \defDom{\atpVar[i]}$.
        \end{lemma}
        \begin{proof}
                Proof by induction on $i$.
        \end{proof}

        \begin{lemma}[Preservation of Completeness]\label{gs:lem:PreservationOfCompleteness}
                Let \ahtpRSeq be an annotated reduction sequence with
                \lhComplete{\atGetHeap{\atpVar[0](\tidVar)}}
                for all $\tidVar \in \defDom{\atpVar[0]}$.
                Then, 
               \lhComplete{\atGetHeap{\atpVar[i](\tidVar)}}
                holds for every $i \in \N$ and every $\tidVar \in \defDom{\atpVar[i]}$.
        \end{lemma}
        \begin{proof}
                Proof by induction on $i$.
        \end{proof}

        Every thread of an annotated thread pool is annotated by a thread-local logical heap that expresses which resources are owned by this thread.
        In the following we define a function to extract the logical heap expressing which resources are owned by threads of a thread pool (i.e. the sum of all thread-local logical heaps).
        
        \begin{definition}
                We define the function 
                $\atpOwnedResHeap : \AnnoThreadPoolSet \rightarrow \LogHeapSet$
                mapping annotated thread pools to logical heaps as follows:
                $$
                \begin{array}{l c l}
                        \atpVar &\ \mapsto
                                &\quadBack
                                        \bigSum[\quad\tidVar\, \in\, \defDom{\atpVar}]
                                        \quadBack\!\!
                                        \atGetHeap{\atpVar(\tidVar)}
                \end{array}
                $$
        \end{definition}

        Annotated resources representing unlocked locks, i.e.,
        \aresUnlocked{\mutVar}{\assVar}{\lheapVar[\assVar]},
        contain a logical heap \lheapVar[\assVar] that expresses which resources are protected by this lock.
        In the following, we define a function that extracts a logical heap from an annotated heap \aheapVar expressing which resources are protected by unlocked locks in \aheapVar.

        \begin{definition}
                We define the function 
                $\ahProtectedResHeap : \AnnoHeapSet \rightarrow$ \phantom{.} $\LogHeapSet$
                mapping  annotated heaps to logical heaps as follows:
                
                For any annotated heap \aheapVar let 
                $$
                \begin{array}{l c l l}
                        \multiLetterSetFont{LockInvs}(\aheapVar) \ &:= \
                                &\leftMsBrace \lheapLockInvVar \in \LogHeapSet \ \ | 
                                        &\exists \mutVar \in \HeapLocSet \times \LevelSet.\
                                            \exists \assLockInvVar \in \AssertionSet. \\
                                        &&&\quad
                                            \aresUnlocked{\mutVar}{\assLockInvVar}{\lheapLockInvVar} \in \aheapVar
                                \rightMsBrace
                \end{array}
                $$
                be the auxiliary bag aggregating all logical heaps corresponding to lock invariants of unlocked locks stored in \aheapVar.
                We define \ahProtectedResHeap as
                $$
                        \aheapVar \ \ \mapsto 
                                \bigSum[\lheapLockInvVar
                                                    \,\in\, \multiLetterSetFont{LockInvs}(\aheapVar)]
                                        \quadBack\quadBack\lheapLockInvVar.
                $$
                
        \end{definition}

        We consider a machine configuration $(\aheapVar, \atpVar)$ to be \emph{consistent} if it fulfils three criteria:
        (i)~Every thread-local logical heap is consistent, i.e.,
%        \lhConsistent{\atGetHeap{\atpVar(\tidVar)}}
%        holds for all $\tidVar \in \defDom{\atpVar}$.
        for all used thread IDs \tidVar, \atGetHeap{\atpVar(\tidVar)} only stores full obligations chunks.
        (ii)~Every logical heap protected by an unlocked lock in \aheapVar is consistent.
        (iii)~\aheapVar is compatible with the logical heap that represents
        (a)~the resources owned by threads in \atpVar and
        (b)~the resources protected by unlocked locks stored in \aheapVar.

        \begin{definition}[Consistency of Annotated Machine Configurations]
                We call an annotated machine configuration $(\aheapVar, \atpVar)$ \emph{consistent} and write \\
                \amConfConsistent{\aheapVar}{\atpVar}
                if all of the following hold:
                \begin{itemize}
                        \item \lhConsistent{\atGetHeap{\atpVar(\tidVar)}}
                                    for all $\tidVar \in \defDom{\atpVar}$,
                        \item
                                $
                                        \forall \mutVar.\
                                        \forall \assLockInvVar.\
                                        \forall \lheapLockInvVar.\ 
                                                \aresUnlocked{\mutVar}{\assLockInvVar}{\lheapLockInvVar} \in \aheapVar
                                                \ \rightarrow\
                                                \lhConsistent{\lheapLockInvVar},
                                $
                        \item
                                $
                                        \aheapVar
                                        \alhCorrespSymb\
                                        \atpOwnedResHeap[\atpVar] + 
                                        \ahProtectedResHeap[\aheapVar].
                                $
                \end{itemize}
        \end{definition}

        \begin{lemma}[Preservation of Consistency]
                Let \ahtpRSeq be an annotated reduction sequence with \amConfConsistent{\aheapVar[0]}{\atpVar[0]}.
                Then, \amConfConsistent{\aheapVar[i]}{\atpVar[i]} holds for every $i \in \N$.
        \end{lemma}
        \begin{proof}
                Proof by induction on $i$.
        \end{proof}

\subsection{Hoare Triple Model Relation}\label{gs:appendix:sec:Soundness:ModelRelation}
    
    We interpret program specifications \hoareTriple{\assPreVar}{\cmdVar}{\assPostVar} in terms of a model relation \htModels{\assPreVar}{\cmdVar}{\assPostVar} and an auxiliary safety relation \safe{\lheapVar}{\cmdVar}.
    
    In the annotated semantics we annotate threads by local logical heaps that express which resources they own (including ghost resources) and use an extended intermediate representation for commands.
    We say that an annotated thread pool \atpVar is an \emph{annotation} of a plain thread pool \tpVar if they represent matching information.
    That is, they must store matching threads under the same thread IDs.

    \begin{definition}[Command Annotation]
            We define the predicate
            $\cmdAnnotName \subset \CmdSetExt \times \CmdSet$
            such that \cmdAnnot{\cmdVar^+}{\cmdVar} holds iff \cmdVar results from $\cmdVar^+$ by replacing all subcommands of the form
            (i)~\icmdWhileDecStarted{\natVar}{\nextCmdVar} by
                    \cmdWhileSkip{\nextCmdVar}
            and
            (ii)~\icmdAwaitStarted{\SigSetVar}{\expVar}{\nextCmdVar} by
                    \cmdAwait{\expVar}{\nextCmdVar}.
    \end{definition}

    \begin{definition}[Thread Pool Annotation]
            We define a predicate
            $\tpAnnotName \subset \AnnoThreadPoolSet \times \ThreadPoolSet$
            such that:
            $$
            \begin{array}{c}
                    \tpAnnot{\atpVar}{\tpVar}\\
                    \Longleftrightarrow\\
                    \dom{\atpVar} = \dom{\tpVar}\ \wedge\
                    \forall \tidVar \in \dom{\tpVar}.\
%                    \exists \lheapVar \in \LogHeapSet.\
%                            \atpVar(\tidVar) = (\lheapVar, \tpVar(\tidVar)).
                    \cmdAnnot
                            {\atGetCmd{\atpVar(\tidVar)}}
                            {\tpVar(\tidVar)}
            \end{array}
            $$
    \end{definition}

    Intuitively, a command \cmdVar is \emph{safe} under a logical heap \lheapVar if \lheapVar provides the necessary resources so that for every execution of \cmdVar, there is a corresponding annotated execution of \cmdVar that does not get stuck.
    That is, if we consider a reduction step from \cmdVar to \nextCmdVar in the plain operational semantics, then the resources that thread \cmdVar owns according to \lheapVar, allow us to perform a corresponding sequence of annotated reduction steps leading to \nextCmdVar.
    Specifically, we can perform a finite number of ghost steps to manipulate ghost resources followed by one real step to reduce \cmdVar to \nextCmdVar.
    Furthermore, safety requires \nextCmdVar to be safe under its respective local logical heap.
    The same must hold for any potentially forked thread.

    \begin{definition}[Safety]
            We define the safety predicate
            $\safeName \subseteq \LogHeapSet \times \CmdSet$
            coinductively as the greatest solution (with respect to $\subseteq$) of the following equation:
            $$

            \end{array}
            $$
    \end{definition}

	We interpret Hoare triples \hoareTriple{\assPreVar}{\cmdVar}{\assPostVar} in terms of safety as follows:
	Let \evalCtxt be an evaluation context that (when instantiated) is safe under any heap \lheapPostVar which fulfills postcondition \assPostVar, i.e., let \evalCtxt be a context for which \phantom{\!\!}
	$
	\forall \valVar.\ \forall \lheapPostVar.\
	\assModelsOR{\lheapPostVar}{\assPostVar(\valVar)}
	\rightarrow
	\safe{\lheapPostVar}{\evalCtxt[\valVar]}
	$
	holds.
	Then, any heap \lheapPreVar that satisfies precondition \assPreVar provides all resources necessary to safely run \evalCtxt[\cmdVar], i.e.,
	$\forall \lheapPreVar.\
		\assModels{\lheapPreVar}{\assPreVar}
		\ \rightarrow\
		\safe
			{\lheapPreVar}
			{\,\evalCtxt[\cmdVar]}
	$.
	Thereby, if the reduction of \cmdVar under \lheapPreVar finishes and returns a value \valVar, then postcondition $\assPostVar(\valVar)$ holds in the final state.
	In the following definition, we also allow \lheapPreVar and \lheapPostVar to be embedded into arbitrary heap frames \lheapFrameVar.

    \begin{definition}[Hoare Triple Model Relation]
            We define the model relation for Hoare triples
            $\htModelsSymb\ \subset \AssertionSet \times \CmdSet \times (\ValueSet \rightarrow \AssertionSet)$
            such that:
            $$
            \begin{array}{c}
                    \htModels{\assPreVar}{\cmdVar}[\resVar]{\assPostVar(\resVar)}\\
                    \Longleftrightarrow\\
                    \begin{array}{l}
                            \forall \lheapFrameVar.\
                            \forall \evalCtxt.\
                            (\forall \valVar.\
                                \forall \lheapPostVar.\
                                \assModelsOR
                                        {\lheapPostVar}
                                        {\assPostVar(\valVar)}
                                \ \rightarrow\
                                \safe
                                        {\lheapPostVar + \lheapFrameVar}
                                        {\,\evalCtxt[\valVar]}
                            )
                            \\
%                            \quad\quad\quad\quad\quad \rightarrow\
                            \phantom{
                                    \forall \lheapFrameVar.\
                            }
                            \rightarrow\ \
                                \forall \lheapPreVar.\
                                \assModelsOR{\lheapPreVar}{\assPreVar}
                                \ \rightarrow\
                                \safe
                                        {\lheapPreVar + \lheapFrameVar}
                                        {\,\evalCtxt[\cmdVar]}
                                                        
                    \end{array}
            \end{array}
            $$
    \end{definition}
    
    We can instantiate context \evalCtxt in above definition to \cmdLet{\varVar}{\evalCtxtHole}{\valUnit}, 
    which yields the consequent
    \safe
            {\lheapPreVar + \lheapFrameVar}
            {\cmdLet{\varVar}{\cmdVar}{\valUnit}}.
    Note that this implies
    \safe
            {\lheapPreVar + \lheapFrameVar}
            {\cmdVar}.
	Also note that compliance with the frame rule directly follows from above definition, i.e., 
	\htModels{\assPreVar}{\cmdVar}{\assPostVar}
	implies
	\htModels
		{\assPreVar \slStar \assFrameVar}
		{\cmdVar}
		{\assPostVar \slStar \assFrameVar}
	for any frame \assFrameVar.
    Further, every specification we can derive in our proof system also holds in our model.
    
    \begin{lemma}[Hoare Triple Soundness]\label{gs:lem:HoareTripleSoundness}
            Let 
            \htProves{\assPreVar}{\cmdVar}{\assPostVar} hold,
            then also
            \htModels{\assPreVar}{\cmdVar}{\assPostVar}
            holds.
    \end{lemma}
    \begin{proof}
            Proof by induction on the derivation of \htProves{\assPreVar}{\cmdVar}{\assPostVar}.
    \end{proof}

\subsection{Soundness Proof}\label{gs:appendix:sec:Soundness:Proof}
    In this section we prove our soundness theorem.

    \paragraph{Constructing Annotated Executions}
    Given a command \cmdVar which provably discharges all obligations and a fair reduction sequence for \cmdVar,
    we can construct a corresponding annotated reduction sequence \ahtpRSeq. 
    This is a useful tool to analyse program executions as \ahtpRSeq carries much more information than the original sequence, e.g., which obligations a thread holds.
    % in which every machine configuration $(\aheapVar[i], \atpVar[i])$ is consistent.
    Note that our definition of fairness forbids \ahtpRSeq to perform ghost steps forever and that we use \alhCorresp[\aheapVar][\lheapVar] to express that \aheapVar is compatible with \lheapVar.

    \begin{restatable}[Construction of Annotated Reduction Sequences]{lemma}{lemConstructionOfAnnotatedRedSeq}\label{gs:lem:ConstructionOfAnnotatedRedSeq}
%    \begin{lemma}[Construction of Annotated Reduction Sequences]
%    \label{gs:lem:ConstructionOfAnnotatedRedSeq}
           Suppose we can prove
           \htModels{\assPreVar}{\cmdVar}{\noObs}.
           Let \lheapPreVar be a logical heap with
           \assModels{\lheapPreVar}{\assPreVar} and \lhComplete{\lheapPreVar}
           and \aheapVar[0] an annotated heap with
           \alhCorresp[\aheapVar[0]][\lheapPreVar].
           Let \phtpRSeq be a fair plain reduction sequence with
           \aphCorresp{\aheapVar[0]}{\pheapVar[0]} and
           $\tpVar[0] = \setOf{(\tidVar_0, \cmdVar)}$
           for some thread ID $\tidVar_0$ and command \cmdVar.

           Then, there exists a fair annotated reduction sequence \ahtpRSeq with
           $\atpVar = \setOf{(\tidVar_0,\, (\lheapPreVar, \cmdVar))}$
           and \amConfConsistent{\aheapVar[i]}{\atpVar[i]} for all $i \in \N$.
%    \end{lemma}
    \end{restatable}
    \begin{proof}
            We can construct the annotated reduction sequence inductively from the plain reduction sequence.
    \end{proof}

    \paragraph{Program Order Graph}
    In the remainder of this section, we prove that programs where every thread discharges all obligations terminate.
    For this, we need to analyse each thread's control flow, i.e., the subsequence of execution steps belonging to the thread.
    We do this by taking a sequence \phtpRSeq representing a program execution, constructing an annotation \ahtpRSeq carrying additional information and then analysing its \emph{program order graph} \progOrdGraph[\ahtpRSeq] which represents the execution's control flow.

    \begin{definition}[Program Order Graph]\label{gs:appendix:def:ProgramOrderGraph}
    { % local scope
            \NewDocumentCommand{\RuleNameSet}{}{\metaVar{N^r}}
            \NewDocumentCommand{\AnnoRuleNameSet}{}{\metaVar{N^\contextIdFont{a}}}
            
            \NewDocumentCommand{\ruleNameVar}{o}
                    {\metaVar{
                            n^\contextIdFont{a}
                            \IfValueT{#1}{_{#1}}
                    }}
            \NewDocumentCommand{\stRuleNameVar}{o}
                    {\metaVar{
                            n^\contextIdFont{st}
                            \IfValueT{#1}{_{#1}}
                    }}

            \NewDocumentCommand{\baseNode}{}{\metaVar{a}} % i
            \NewDocumentCommand{\succNode}{}{\metaVar{b}} % j
            \NewDocumentCommand{\auxNode}{}{\metaVar{k}} % k
            \NewDocumentCommand{\genNode}{}{\metaVar{\ell}} % l
        
            Let \ahtpRSeqFull be an annotated reduction sequence.
            Let \RuleNameSet be the set of names referring to reduction rules defining the relations \rtpRedStepSymb, \gtpRedStepSymb and \astRedStepSymb.
            We define the set of annotated reduction rule names \AnnoRuleNameSet where \astRedAwaitName is annotated by signals as
            $$
            \begin{array}{l c l}
                    \AnnoRuleNameSet \ &:= \
                            &(\RuleNameSet \setminus \setOf{\astRedAwaitName})\\
                            &&\!\phantom{\RuleNameSet}
                            \ \cup\
                            (\setOf{\astRedAwaitName}  \times \SignalSet)
            \end{array}
            $$
            We define the program order graph
            $\progOrdGraph[\ahtpRSeqFull] = (\N, \EdgeSetVar)$
            with root 0 where 
            $\EdgeSetVar \subset
                    \N \times \ThreadIDSet \times \AnnoRuleNameSet \times \N
            $.

            A node $\baseNode \in \N$ corresponds to the sequence's $\baseNode^\text{th}$ reduction step, i.e., to the step
            \atpRedStep
                    [\aheapVar[\baseNode]]{\atpVar[\baseNode]}
                    [\tidVar]
                    [\aheapVar[\baseNode+1]]{\atpVar[\baseNode+1]}
            for some $\tidVar \in \defDom{\atpVar[\baseNode]}$.
            An edge from node \baseNode to node \succNode expresses that the $\succNode^\text{th}$ reduction step continues the control flow of step $\baseNode$.
            For any $\genNode \in \N$, let $\tidVar_\genNode$ denote the ID of the thread reduced in step \genNode.
            Furthermore, let \ruleNameVar[\genNode] denote the name of the reduction rule applied in the $\genNode^\text{th}$ step, in the following sense:
            \begin{itemize}
                    \item If 
                            \atpRedStep
                                    [\aheapVar[\genNode]]{\atpVar[\genNode]}
                                    [\tidVar]
                                    [\aheapVar[\genNode+1]]{\atpVar[\genNode+1]}
                            results from an application of \rtpRedLiftName in combination with single-thread reduction rule \stRuleNameVar other than \astRedAwaitName,
                            then $\ruleNameVar[\genNode] = \stRuleNameVar$.
                    \item If 
                            \atpRedStep
                                    [\aheapVar[\genNode]]{\atpVar[\genNode]}
                                    [\tidVar]
                                    [\aheapVar[\genNode+1]]{\atpVar[\genNode+1]}
                            results from an application of \rtpRedLiftName in combination with \astRedAwaitName,
                            then $\ruleNameVar[\genNode] = (\astRedAwaitName, \sigAnno[\genNode])$.
                    \item Otherwise, \ruleNameVar denotes the applied (real or ghost) thread pool reduction rule.
            \end{itemize}
            
            An edge 
            $(\baseNode, \tidVar, \ruleNameVar, \succNode) \in 
                    \N \times \ThreadIDSet \times \AnnoRuleNameSet \times \N$
            is contained in \EdgeSetVar if
            $\ruleNameVar = \ruleNameVar[\baseNode]$
            and one of the following conditions applies:
            \begin{itemize}
                    \item $\tidVar = \tidVar_\baseNode = \tidVar_\succNode$ and 
                            $\succNode = \min(\setOf{\auxNode > \baseNode \ | \
                                                    \atpRedStep
                                                            [\aheapVar[\auxNode]]{\atpVar[\auxNode]}
                                                            [\tidVar_\baseNode]
                                                            [\aheapVar[\auxNode+1]]{\atpVar[\auxNode+1]}
                                                })$.
                            \\
                            In this case, the edge expresses that step \succNode marks the first time that thread $\tidVar_\baseNode$ is rescheduled for reduction (after step \baseNode).
%                            In this case the edge expresses that the $j^\text{th}$ reduction step continues the execution of thread \tidVar reduced in step $i$.
                    \item $\defDom{\atpVar[\baseNode+1]} \setminus \defDom{\atpVar[\baseNode]}
                                    = \setOf{\tidVar}$
                            and\\
                            $\succNode = \min{\setOf{ \auxNode \in \N \ | \
                                                    \atpRedStep
                                                            [\aheapVar[\auxNode]]{\atpVar[\auxNode]}
                                                            [\tidVar]
                                                            [\aheapVar[\auxNode+1]]{\atpVar[\auxNode+1]}
                                            }}$.
                            \\
                            In this case, \tidVar identifies the thread forked in step \baseNode.
                            The edge expresses that step \succNode marks the first reduction of the forked thread.
            \end{itemize}
            
            In case the choice of reduction sequence \ahtpRSeqFull is clear from the context, we write \progOrdGraph instead of \progOrdGraph[\ahtpRSeqFull].
    } % end of local scope
    \end{definition}

    \begin{observation}\label{gs:obs:ProgramOrderGraphBinaryTree}
            Let \ahtpRSeq be an annotated reduction sequence with\\
            $\cardinalityOf{\defDom{\atpVar[0]}} = 1$.
            The sequence's program order graph \progOrdGraph[\ahtpRSeq] is a binary tree.
    \end{observation}

    For any reduction sequence \ahtpRSeq, the paths in its program order graph \progOrdGraph[\ahtpRSeq] represent the sequence's control flow paths.
    Hence, we are going to use program order graphs to analyse reduction sequences' control flows.
    We refer to a program order graph's edges by the kind of reduction step they represent.
    For instance, we call any edge $(a, \tidVar, R, b)$ a \emph{loop edge} where $R$ refers to one of the rules related to the execution loops,
    i.e., \astRedWhileDecInitName, 
    \astRedWhileDecName,
    \astRedAwaitInitName or
    \astRedAwaitName.
    In the following, we prove that any path in a program order graph that does not involve a loop edge is finite.
    This follows from the fact that the size of the command reduced along this path decreases with each non-ghost non-loop step.

    \begin{lemma}\label{gs:lem:InfinitePathsCotainInfinitelyManyLoopEdges}
            Let \ahtpRSeq be a fair annotated reduction sequence.
            Let $\pathVar = (\NodeSetVar, \EdgeSetVar)$ be a path in \progOrdGraph[\ahtpRSeq].
            Let 
            $$
            \begin{array}{l c l}
                    R &\ =\ &\{\,
                            \astRedWhileDecInitName,\
                            \astRedWhileDecName,\
                            \\
                            &&\phantom{\{}\
                            \astRedAwaitInitName\
                    \}
                    \ \cup \
                    (\setOf{\astRedAwaitName} \times \SignalSet)
            \end{array}
            $$
            be the set of names of single-thread reduction rules related to loops and let
            $L = \setOf{\edgeVar \in \EdgeSetVar \ | \ \tupleProj{3}[\edgeVar] \in R }$
            be the set of loop edges contained in \pathVar.
            Then, \pathVar is infinite if and only if $L$ is infinite.
    \end{lemma}
    \begin{proof}
            If $L$ is infinite, \pathVar is obviously infinite as well.
            So, suppose $L$ is finite.
            
            For any command, we consider its size to be the number of nodes contained in its abstract syntax tree.
            By structural induction over the set of commands, it follows that the size of a command $\cmdVar = \atGetCmd{\atpVar(\tidVar)}$ decreases in every non-ghost reduction step
            \atpRedStep
                    [\aheapVar]
                    {\atpVar}
                    [\tidVar]
                    [\nextAheapVar]
                    {\nextAtpVar}            
            that is not an application of \rtpRedLiftName in combination with some $r \in R$.
            
%            For any node $i \in \NodeSetVar$, let $\cmdVar_i$ denote the command reduced in the reduction step that $i$ represents.
%            Since \ahtpRSeq is fair, \pathVar contains infinitely many edges corresponding to non-ghost reduction steps.
            Since $L$ is finite, there exists a node $x$ such that the suffix $\pathVar_{\geq x}$ starting at node $x$ does not contain any loop edges.
%            It contains, however, infinitely many edges corresponding to non-ghost reduction steps.
            By fairness of \ahtpRSeq, every non-empty suffix of $\pathVar_{\geq x}$ contains an edge corresponding to a non-ghost reduction step.
            For any edge $\edgeVar = (i, \tidVar, n, j)$ consider the command 
            $\cmdVar_\edgeVar = \atGetCmd{\atpVar[i](\tidVar)}$
            reduced in this edge.
            The size of these commands decreases along $\pathVar_{\geq x}$.
            So, $\pathVar_{\geq x}$ must be finite and thus \pathVar must be finite as well.
    \end{proof}

    \begin{definition}
            Let
            $\progOrdGraph = (\NodeSetVar, \EdgeSetVar)$
            be a subgraph of some program order graph.
            We define the function
            $\awaitEdges[\progOrdGraph] : \SignalSet \rightarrow \powerSetOf{\EdgeSetVar}$
            mapping any signal \sigVar to the set of await edges in \progOrdGraph concerning \sigVar as:
            $$
            \begin{array}{c c l}
                    \awaitEdges[\progOrdGraph][\sigVar] &\ := \
                            &\setOf{
                                    (a, \tidVar, (\astRedAwaitName, \nextSigVar), b) \in \EdgeSetVar 
                                    \ |\
                                    \nextSigVar = \sigVar
                                }.
            \end{array}
            $$
            Furthermore, we define the set
            $\WaitSignalSet[\progOrdGraph] \subset \SignalSet$
            of signals being waited for in \progOrdGraph and its subset
            $\InfWaitSigSet[\progOrdGraph] \subseteq \WaitSignalSet[\progOrdGraph]$
            of signals waited-for infinitely often in \progOrdGraph
            as follows:
            $$
            \begin{array}{c c l}
                    \WaitSignalSet[\progOrdGraph] &\ := \
                            &\setOf{
                                    \sigVar \in \SignalSet
                                    \ | \
                                    \awaitEdges[\progOrdGraph](\sigVar) \neq \emptyset
                                },
                    \\
                    \InfWaitSigSet[\progOrdGraph] &\ := \
                            &\setOf{
                                            \infSig \in \WaitSignalSet[\progOrdGraph] \ | \
                                            \awaitEdges[\progOrdGraph][\infSig]\ \text{infinite}
                                    }.
            \end{array}
            $$
    \end{definition}

    \begin{definition}[Partial Order on Finite Bags]
    {
            \NewDocumentCommand{\baseSet}{}{\metaVar{X}}
            \NewDocumentCommand{\baseLt}{}{\ensuremath{<_\baseSet}}
            
            \NewDocumentCommand{\smallerBag}{}{\metaVar{A}}
            \NewDocumentCommand{\biggerBag}{}{\metaVar{B}}
            
            \NewDocumentCommand{\removedBag}{}{\metaVar{C}}
            \NewDocumentCommand{\addedBag}{}{\metaVar{D}}
            
            \NewDocumentCommand{\elemRemovedBag}{}{\metaVar{c}}
            \NewDocumentCommand{\elemAddedBag}{}{d}
            
            Let \baseSet be a set and $\baseLt \, \subset \baseSet \times \baseSet$ a partial order on \baseSet.
            We define the partial order 
            $\bagLt[\baseSet]\, \subset \FinBagsOf{\baseSet} \times \FinBagsOf{\baseSet}$
            on finite bags over \baseSet
            as the 
            Dershowitz-Manna ordering~\cite{Dershowitz1979MultisetOrder} induced by \baseLt:
            $$
            \begin{array}{c}
                    \smallerBag \bagLt[X] \biggerBag
                    \ \ \Longleftrightarrow
                    \begin{array}[t]{l l}
                    \exists \removedBag, \addedBag \in \FinBagsOf{X}.
                            &\msEmpty \neq \removedBag \subseteq \biggerBag\\
                            &\wedge\ 
                            \smallerBag = 
                                    (\biggerBag \setminus \removedBag) \msCup \addedBag
                            \\
                            &\wedge\
                            \forall \elemAddedBag \in \addedBag.\
                            \exists \elemRemovedBag \in \removedBag.\
                            \elemAddedBag \baseLt \elemRemovedBag.
                    \end{array}
            \end{array}
            $$
            We define $\bagLeq[X]\, \subset \FinBagsOf{X} \times \FinBagsOf{X}$ such that
            $$
                    \smallerBag \bagLeq[\baseSet] \biggerBag
                    \ \ \Longleftrightarrow \ \
                    \smallerBag = \biggerBag \ \vee\  \smallerBag \bagLt[\baseSet] \biggerBag
            $$
            holds.
    }
    \end{definition}

    \begin{corollary}\label{gs:appendix:cor:natBagLtWellFounded}
            The partial order 
            $\bagLt[\N] \ \subset\ \FinBagsOf{\N} \times \FinBagsOf{\N}$
            is well-founded.
    \end{corollary}
    \begin{proof}
            Follows from~\cite{Dershowitz1979MultisetOrder} and and well-foundedness of $<_\N$.
    \end{proof}

    Below we define a metric on commands in a graph that allows us to prove that every control flow path in which no signal is waited-for infinitely often is finite.
    We construct the metric in two steps.
    (i)~We consider the maximal degree for which the command contains a nested loop.
    Here, we consider an uninitialized loop \cmdWhileSkip{\cmdVar} as a doubly nested one since its reduction in the annotated semantics involves its conversion into either a decrease loop or into an await loop, cf.~\astRedWhileDecInitName and~\astRedAwaitInitName.
    (ii)~We use the extracted degree to construct a rank (in form of a finite bag of degrees) that intuitively spoken measures the number of loop iterations the command's execution causes in the graph we consider.

        \begin{definition}[Degree Extraction]\label{appendix:def:DegreeExtraction}
                We define the function
                $\extractDegree{\cdot} : \CmdSet \rightarrow \N$
                mapping augmented commands to degrees by recursion on the structure of commands as follows:
                $$
\begin{array}{l c l l}
                        \extractDegree{\cmdWhileSkip{\cmdVar}} &\ := \
                                &\extractDegree{\cmdVar} + 2,\\
                        \extractDegree{\icmdWhileDecStarted{\natVar}{\cmdVar}} &:=
                                &\extractDegree{\cmdVar} + 1,\\
                        \extractDegree{\icmdAwaitStarted{\SigSetVar}{\mutVar}{\cmdVar}} &:=
                                &\extractDegree{\cmdVar} + 1,\\
                        \extractDegree{\cmdFork{\cmdVar}} &:=
                                &\extractDegree{\cmdVar},\\
                        \extractDegree{\cmdLet{\varVar}{\cmdVar}{\nextCmdVar}} &:=
                                &\max(\extractDegree{\cmdVar}, \extractDegree{\nextCmdVar}),\\
                        \extractDegree{\cmdIf{\cmdVar}{\nextCmdVar}} &:=
                                &\max(\extractDegree{\cmdVar}, \extractDegree{\nextCmdVar}),\\
                        \extractDegree{\cmdVar} &:= 
                                &0
                                &\ \text{otherwise}.
                \end{array}
                $$
        \end{definition}
        
        \begin{definition}[Rank Extraction]\label{gs:appendix:def:RankExtraction}
                For any subgraph \progOrdGraph of a program order graph with $\InfWaitSigSet[\progOrdGraph] = \msEmpty$, we define the function
                $\extractDegBag{\cdot} : \CmdSet \rightarrow \FinBagsOf{\N}$ mapping commands to finite bags of degrees by recursion on the structure of commands as follows:
                $$
                \begin{array}{l c l}
                        \extractDegBag{\cmdWhileSkip{\cmdVar}} &\ :=\
                                &\multiset{\extractDegree{\cmdWhileSkip{\cmdVar}}},
                        \\
                        \extractDegBag{\icmdWhileDecStarted{\natVar}{\cmdVar}} &\ :=\
                                &\displaystyle
                                        \natVar \cdot \multiset{\extractDegree{\icmdWhileDecStarted{\natVar}{\cmdVar}}},
                        \\
                        \extractDegBag{\icmdAwaitStarted{\SigSetVar}{\mutVar}{\cmdVar}} &\ :=\
                                &\displaystyle
                                        \cardinalityOf{
                                                \bigcup_{\sigVar \in \SigSetVar} \awaitEdges[\progOrdGraph](\sigVar)
                                        }
                                \cdot\ \multiset{\extractDegree{\icmdAwaitStarted{\SigSetVar}{\mutVar}{\cmdVar}}},
                        \\
                        \extractDegBag{\cmdFork{\cmdVar}} &\ :=\
                                &\extractDegBag{\cmdVar},
                        \\
                        \extractDegBag{\cmdLet{\varVar}{\cmdVar}{\nextCmdVar}} &\ :=\
                                &\extractDegBag{\cmdVar} \msCup \extractDegBag{\nextCmdVar},
                        \\
                        \extractDegBag{\cmdIf{\cmdVar}{\nextCmdVar}} &\ :=\
                                &\extractDegBag{\cmdVar} \msCup \extractDegBag{\nextCmdVar},
    %                    \\
    %                    \extractDegBag{\icmdWait{\SigSetVar}} &\ :=\  
    %                            &\multiset{\extractDegree{\icmdWait{\SigSetVar}}},
    %                            &\msEmpty
                        \\
\end{array}
                $$
        \end{definition}

    We view paths in a program order graph as single-branched subgraphs.
    This allows us to apply above definition on graphs to paths.
    In the proof of the following lemma, we see that in any control flow path where every signal is waited-for only finitely often, the rank of the reduced command decreases step by step.

    \begin{restatable}{lemma}{lemNoInfSigOnPathImpliesPathFinite}\label{gs:lem:NoInfSigOnPathImpliesPathFinite}
            Let 
            $\progOrdGraph[\ahtpRSeq]$
            be a program order graph and let
            $\pathVar = (\NodeSetVar, \EdgeSetVar)$
            be a path in \progOrdGraph with 
            $\InfWaitSigSet[\pathVar] = \emptyset$.
            For every 
            $\tidVar \in \defDom{\atpVar[0]}$
            let $\atGetHeap{\atpVar[0](\tidVar)}$ be finite and complete.
            Then, \pathVar is finite.
    \end{restatable}
    { %local scope for proof
            \NewDocumentCommand{\rootNode}{}{\metaVar{r}}
            \NewDocumentCommand{\nodeToDegBag}{}{\metaVar{f}}
            \NewDocumentCommand{\pSuff}{m}{\metaVar{\pathVar_{\geq #1}}}
            
            Assume \pathVar is infinite.
            We prove a contradiction by assigning a decreasing metric to every node along the path.
            For every $i \in \NodeSetVar$, let \pSuff{i} be the suffix of \pathVar starting at node $i$.
            Every node $i \in \NodeSetVar$ corresponds to a reduction step
            \atpRedStep[\aheapVar[i]]{\atpVar[i]}[\tidVar_i][\aheapVar[j]]{\atpVar[j]}.
            In the following, we let $\cmdVar_i := \atpVar[i](\tidVar_i)$ denote the command reduced in this step.
            Consider the function
            $\nodeToDegBag : \NodeSetVar \rightarrow \FinBagsOf{\N}$,
            $i \mapsto \extractDegBag{\cmdVar_i}[\pSuff{i}]$
            mapping every node on the path to the rank of the command whose reduction the node represents.
            
            Consider the sequence
            $(\nodeToDegBag(i))_{i \in \NodeSetVar}$.
            Since every element is a finite bag of natural numbers, we can order it by \bagLt[\N].
            We are going to prove a contradiction by proving that the sequence
            is an infinitely descending chain.
            
            Consider any edge
            $\edgeVar = (i, \tidVar, r, j) \in \EdgeSetVar$.
            There are only 4 cases in which 
            $\nodeToDegBag(i) \neq \nodeToDegBag(j)$
            holds.
            \begin{itemize}
                    \item $r = \astRedWhileDecInitName$:\\
                            In this case we have 
                            $\cmdVar_i = \cmdWhileSkip{\cmdVar}$ and 
                            $\cmdVar_j = \icmdWhileDecStarted{\natVar}{\cmdVar}$
                            for some $\cmdVar, \natVar$.
                            We get
                            $$\nodeToDegBag(j) 
                                    \ = \
                                    \natVar \cdot \multiset{\extractDegree{\cmdVar}+1}
                                    \ \bagLt[\N]\
                                    \multiset{\extractDegree{\cmdVar}+2}
                                    \ =\
                                    \nodeToDegBag(i).
                            $$
                    \item $r = \astRedWhileDecName$:\\
                            In this case we have
                            $\cmdVar_i = \icmdWhileDecStarted{\natVar}{\cmdVar}$
                            and 
                            $\cmdVar_j = \cmdIf{\cmdVar}{\icmdWhileDecStarted{\natVar-1}{\cmdVar}}$
                            for some $\natVar, \cmdVar$.
                            We get
                            $$\nodeToDegBag(j)
                                    \ = \
                                    \multiset{\extractDegree{\cmdVar}} 
                                            \msCup
                                            (\natVar-1) \cdot \multiset{\extractDegree{\cmdVar}+1}
                                    \ \bagLt[\N]\
                                    \natVar \cdot \multiset{\extractDegree{\cmdVar}+1}
                                    \ = \
                                    \nodeToDegBag(i).
                            $$
                    \item $r = \astRedAwaitInitName$:\\
                            In this case we have
                            $\cmdVar_i = \cmdAwait{\mutVar}{\cmdVar}$
                            and\\
                            $\cmdVar_j = 
                                    \cmdIf
                                            {
                                                    \cmdAcquireMut{\mutVar};
                                                    \cmdLet{\varVar}{\cmdVar}
                                                            {
                                                                    \cmdReleaseMut{\mutVar};
                                                                    \neg\varVar
                                                            }
                                            }
                                            {\icmdAwaitStarted{\SigSetVar}{\mutVar}{\cmdVar}}
                            $
                            for some $\mutVar, \varVar, \cmdVar, \SigSetVar$.
                            We get
                            $$\nodeToDegBag(j)
                                    \ = \
                                    \multiset{\extractDegree{\cmdVar}}
                                    \msCup
                                    \cardinalityOf{\dots} \cdot \multiset{\extractDegree{\cmdVar} + 1}
                                    \ \bagLt[\N] \
                                    \multiset{\extractDegree{\cmdVar} + 2}
                                    \ = \
                                    \nodeToDegBag(i).
                            $$
                    \item $r = (\astRedAwaitName, \nextSigVar)$ for some \nextSigVar:\\
                            In this case we have 
                            $\cmdVar_i = \icmdAwaitStarted{\SigSetVar}{\mutVar}{\cmdVar}$
                            and\\
                            $\cmdVar_j = 
                                    \cmdIf
                                            {
                                                    \cmdAcquireMut{\mutVar};
                                                    \cmdLet{\varVar}{\cmdVar}
                                                            {
                                                                    \cmdReleaseMut{\mutVar};
                                                                    \varVar
                                                            }
                                            }
                                            {\icmdAwaitStarted{\SigSetVar}{\mutVar}{\cmdVar}}
                            $
                            for some $\mutVar, \varVar, \cmdVar, \SigSetVar$.
                            We get
                            $$\begin{array}{c c c c c}
                                    \displaystyle
                                    \nodeToDegBag(j)
                                    &\ =\
                                    &\multiset{\extractDegree{\cmdVar}}
                                            \msCup
                                            \cardinalityOf{
                                                    \bigcup_{\sigVar \in \SigSetVar} \awaitEdges[\pSuff{j}](\sigVar)
                                            }
                                                    \cdot \multiset{\extractDegree{\cmdVar}+1}
                                    \\
                                    \displaystyle
                                    &\ \bagLt[\N]\
                                            &\left(
                                                    1 + 
                                                    \cardinalityOf{
                                                            \bigcup_{\sigVar \in \SigSetVar} \awaitEdges[\pSuff{j}](\sigVar)
                                                    }
                                            \right)
                                                    \cdot \multiset{\extractDegree{\cmdVar}+1}
                                    \\
                                    \\
                                    \displaystyle
                                    &\ = \
                                            &\cardinalityOf{
                                                    \bigcup_{\sigVar \in \SigSetVar} \awaitEdges[\pSuff{i}](\sigVar)
                                            }
                                                    \cdot \multiset{\extractDegree{\cmdVar}+1}
                                    \ = \
                                    \nodeToDegBag(i)
                            \end{array}
                            $$
            \end{itemize}
            
            By application of Lemma~\ref{gs:lem:InfinitePathsCotainInfinitelyManyLoopEdges} we get that \pathVar contains infinitely many of the loop edges listed above.
            Hence, $(\nodeToDegBag(i))_{i \in \NodeSetVar}$ is an infinitely decreasing chain.
            By Corollary~\ref{gs:appendix:cor:natBagLtWellFounded}, \bagLt[\N] is well-founded.
            A contradiction.
            So, \pathVar is finite.
    } %end of local scope

    We proceed to prove that no signals are waited for infinitely often.

    \begin{restatable}{lemma}{lemNoSignalWaitedForInfinitelyOften}\label{gs:lem:NoSignalWaitedForInfinitelyOften}
            Let \ahtpRSeq be a fair annotated reduction sequence with
            $\atpVar[0] = \setOf{(\tidStartVar, (\lheapVar[0], \cmdVar))}$,
            \ahFinite{\aheapVar[0]},
            \lhComplete{\lheapVar[0]},
            \lhFinite{\lheapVar[0]}
            and also
            \amConfConsistent{\aheapVar[0]}{\atpVar[0]}.
            Let \lheapVar[0] contain no signal chunks.
            Further, let \aheapVar[0] contain no chunks of the form
            \aresUnlocked{\mutVar}{\assLockInvVar}{\lheapLockInvVar}
            where \lheapLockInvVar contains any signal chunks.
            Let \progOrdGraph be the program order graph of \ahtpRSeq.
            Then, $\InfWaitSigSet[\progOrdGraph] = \emptyset$.
    \end{restatable}
    \begin{proof}
    { % local scope
            \NewDocumentCommand{\obTid}{}{\metaVar{\tidVar_\text{ob}}}
            \NewDocumentCommand{\lowerSig}{}{\metaVar{z}}
            \NewDocumentCommand{\obPath}{}{\metaVar{p_\text{ob}}}
            \NewDocumentCommand{\obExtPath}{}{\metaVar{\bar{\obPath}}}
            
            \NewDocumentCommand{\obPathNodes}{}{\metaVar{V_\text{ob}}}
            
            \NewDocumentCommand{\tmpCapacity}{o}
                    {\ensuremath{
                            \fixedFuncNameFont{cap}
                            \IfValueT{#1}{(#1)}
                    }\xspace}

            Suppose $\InfWaitSigSet[\progOrdGraph] \neq \emptyset$.
            Since \LevelSet is well-founded, the same holds for the set \setOf{\sigGetLev{\sigVar} \ | \ \sigVar \in \InfWaitSigSet}.
            Hence, there is some $\minSig \in \InfWaitSigSet$ for which no $\lowerSig \in \InfWaitSigSet$ with
            $\sigGetLev{\lowerSig} \levLt \sigGetLev{\minSig}$ exists.
            
            Since neither the initial logical heap \lheapVar[0] nor any unlocked lock invariant stored in \aheapVar[0] does contain any signals, \minSig must be created during the reduction sequence.
            The reduction step creating signal \minSig is an application of \gtpRedNewSignalName, which simultaneously creates an obligation to set \minSig.
            By preservation of completeness, Lemma~\ref{gs:lem:PreservationOfCompleteness}, every thread-local logical heap \atGetHeap{\atpVar[i](\tidVar)} annotating some thread \tidVar in some step $i$ is complete.
            According to reduction rule \astRedAwaitName, every await edge
            $(a, \tidVar, (\astRedAwaitName, \minSig), b)$
            implies together with completeness that in step $a$ (i)~thread \tidVar does not hold any obligation for \minSig 
            (i.e.\@ 
                $\atGetHeap{\atpVar[a](\tidVar)}(\lresObs{\obBagVar}) = 1$
                for some bag of obligations \obBagVar with
                $\minSig \not\in \obBagVar$
            )
            and (ii)~\minSig has not been set, yet
            (i.e.\@
                $\aresSignal{\minSig}{\slFalse} \in \aheapVar[a]$).
            Hence, in step $a$ another thread $\obTid \neq \tidVar$ must hold the obligation for \minSig
            (i.e.\@
                $\atGetHeap{\atpVar[a](\obTid)}(\lresObs{\obBagVar}) = 1$
                for some bag of obligations \obBagVar with
                $\minSig \in \obBagVar$).
            Since there are infinitely many await edges concerning \minSig in \progOrdGraph, the signal is never set.
            
            By fairness, for every await edge as above, there must be a non-ghost reduction step 
            \atpRedStep
                    [\aheapVar[k]]{\atpVar[k]}
                    [\obTid]
                    [\aheapVar[k+1]]{\atpVar[k+1]}
            of the thread \obTid holding the obligation for \minSig with $k \geq a$.
            Hence, there exists an infinite path \obPath in \progOrdGraph where each edge $(e, \obTid, n, f) \in \edges{\obPath}$ concerns some thread \obTid holding the obligation for \minSig.
            (Note that this thread ID does not have to be constant along the path, since the obligation can be passed on during fork steps.)
            
            The path \obPath does not contain await edges
            $(e, \obTid, (\astRedAwaitName, \infSig), f)$
            for any $\infSig \in \InfWaitSigSet$,
            since reduction rule \astRedAwaitName would (together with completeness of \atGetHeap{\atpVar[e](\obTid)}) require
            \infSig to be of a lower level than all held obligations.
            This restriction implies
            $\sigGetLev{\infSig} \levLt \sigGetLev{\minSig}$
            and would hence contradict the minimality of \minSig.
            That is, $\InfWaitSigSet[\obPath] = \emptyset$.

%            We can conclude that \lheapVar[0] is finite, from the consistency of the initial annotated machine configuration (\aheapVar[0], \atpVar[0]) and the fact that \lheapVar[0] does not contain any signal or registration chunks and only finitely many permission chunks.
            By preservation of finiteness, Lemma~\ref{gs:lem:PreservationOfFiniteness}, we get that every logical heap associated with the root of \obPath is finite.
            This allows us to apply Lemma~\ref{gs:lem:NoInfSigOnPathImpliesPathFinite}, by which we get that \obPath is finite.
            A contradiction.
    } % end of local scope
    \end{proof}

    Finally, we got everything we need to prove that any program that discharges all its obligations terminates.

    \begin{lemma}\label{gs:lem:NoFairInfiniteAnnotatedRedSeq}
            Let \htModels{\noObs}{\cmdVar}{\noObs} hold.
            There exists no fair, infinite annotated reduction sequence \ahtpRSeq with
            $\atpVar[0] = \setOf{(\tidStartVar, (\lheapVar[0], \cmdVar))}$,
            $\aheapVar[0] = \emptyset$
            and
            $\lheapVar[0]  =  \lheapOf{\lresObs{\msEmpty}}$.
    \end{lemma}
    \begin{proof}
    { % local scope
            \NewDocumentCommand{\infPath}{}{\metaVar{\pathVar_\infty}}
            Suppose a reduction sequence as described above exists.
            We are going to prove a contradiction by considering its infinite program order graph \progOrdGraph.

            Since \atpVar[0] contains only a single thread, \progOrdGraph is a binary tree with an infinite set of vertices.
            By the Weak König's Lemma~\cite{Simpson1999WKL} \progOrdGraph has an infinite branch, i.e. an infinite path \pathVar starting at root 0.

            The initial logical heap \lheapVar[0] is complete and finite and the initial annotated machine configuration $(\aheapVar[0], \atpVar[0])$ is consistent.
            By Lemma~\ref{gs:lem:NoSignalWaitedForInfinitelyOften} we know that $\InfWaitSigSet[\progOrdGraph] = \emptyset$.
            Since
            $\InfWaitSigSet[\pathVar] \subseteq \InfWaitSigSet[\progOrdGraph]$,
            we get
            $\InfWaitSigSet[\pathVar] = \emptyset$.            
            This allows us to apply Lemma~\ref{gs:lem:NoInfSigOnPathImpliesPathFinite}, by which we get that \pathVar is finite, which is a contradiction.
    } % end of local scope
    \end{proof}

    \theoSoundness*
    \begin{proof}
    { % local scope
            Assume that such a reduction sequence exists.
            By Hoare triple soundness, Lemma~\ref{gs:lem:HoareTripleSoundness},
            we get
            \htModels{\noObs}{\cmdVar}{\noObs}
            from
            \htProves{\noObs}{\cmdVar}{\noObs}.
            Consider the logical heap 
            $\lheapVar[0] = \lheapOf{\lresObs{\msEmpty}}$
            and the annotated heap
            $\aheapVar[0] = \emptyset$.
            It holds
            \assModels
                    {\lheapVar[0]}
                    {\noObs},
            \alhCorresp[\aheapVar[0]][\lheapVar[0]] (since \lheapVar[0] does not contain any logical resources with an annotated counterpart)
            and
            \aphCorresp{\aheapVar[0]}{\pheapVar[0]} (since both heaps are empty).
            This allows us to apply Lemma~\ref{gs:lem:ConstructionOfAnnotatedRedSeq}, by which
            we can construct a corresponding fair annotated reduction sequence \ahtpRSeq that starts with
            $\aheapVar[0] = \emptyset$ and 
            $\atpVar[0] = 
                    \setOf{(\tidStartVar, (\lheapVar[0], \cmdVar))}
            $.
            By Lemma~\ref{gs:lem:NoFairInfiniteAnnotatedRedSeq} \ahtpRSeq does not exist.
            A contradiction.
    } % end of local scope
    \end{proof}

\section{Case Studies}\label{gs:appendix:sec:Case Studies}

In this section, we verify two example programs in detail.
In \S~\ref{gs:appendix:sec:VerificationOfRealisticExample} we verify the realistic example program presented in \S~\ref{gs:sec:RealisticExample}.
It involves two threads communicating via a shared bounded FIFO.
In \S~\ref{gs:appendix:sec:caseStudy:unbounded} we present and verify a program similar to the one from \S~\ref{gs:sec:RealisticExample} but with an unbounded number of producer and consumer threads.
In order to lower the visual burden on the reader, we use the following colour coding:

As in the paper body, we present the proof state in \ghostFont{blue}, 
applied proof and view shift rules in \proofRuleHint{\text{purple}},
and abbreviations and definition in \proofDef{red}.
Since the verification outlines we present in this section span multiple figures, we include hints and explanations concerning other figures (e.g. a hint pointing to the figure where an invariant was defined).
To set these remarks and in general any hints off from the essential parts of the proof, we present them in a \ \proofHint{grey and italic font}.
We occasionally remind the reader of earlier proof steps performed in a previous figure by repeating them in \ \proofHint{the same grey font} at the beginning of the current figure.

Further, we highlight how our proof steps effect the proof state as follows:
Consider a proof state of the form \progProof{A \slStar \slPointsTo{\hlocVar}{\valVar}}.
(i)~When a proof step adds a new chunk~$C$, we highlight it in \progProofNew{\text{green}}, i.e.
\progProof{A \slStar \slPointsTo{\hlocVar}{\valVar} \slStar \progProofNew{C}}.
(ii)~When a proof step removes the chunk $A$, we highlight this change by underlying the removed part of the assertion with a \progProofCancel{\text{dark grey}} background, i.e., 
\progProof{\progProofCancel{A\ \slStar}\ \slPointsTo{\hlocVar}{\valVar}}.
Note that in this case, the greyed out \progProofCancel{A\ \slStar} is not a part of the proof state anymore.
(iii)~When a proof step changes only part of a chunk, we highlight this change in \progProofChanged{\text{yellow}}.
For instance,  if the step changes the value of heap location \hlocVar from \valVar to \nextValVar, we highlight it in the resulting state as
\progProof{A \slStar \slPointsTo{\hlocVar}{\progProofChanged{\nextValVar}}}.

\subsection{Verification of Realistic Example}\label{gs:appendix:sec:VerificationOfRealisticExample}

In this section, we verify the realistic example program from \S~\ref{gs:sec:RealisticExample} presented in Fig.~\ref{gs:fig:RealisticExample}.
We present the full proof outline in Fig.~\ref{gs:fig:VerificationExampleBoundedFifoInitSL} -- \ref{gs:fig:VerificationExampleBoundedFifoConsumptionStepSL}.
We scaled the proof outlines down to fit them into the page layout. 
You can find the original proof outlines in the extended version of the ghost signal paper~\cite{Reinhard2021GhostSignalsExtended} and in the technical report~\cite{Reinhard2020GhostSignalsTR}.

    {% local scope of verification figures
            % Variables copied from program figure.
            % ====================================================
                    \NewDocumentCommand{\fifo}{}{\progVar{fifo_{10}}}
                    \NewDocumentCommand{\fifoLoaded}{}{\progVar{f}}
                    %\NewDocumentCommand{\m}{}{\mutProgVar}
                    \RenewDocumentCommand{\m}{}{\mutProgVar}

                    \NewDocumentCommand{\counter}{m}{\progVar{c_{#1}}}
                    
                    % producing thread
                            \NewDocumentCommand{\producerID}{}{p}
                            \NewDocumentCommand{\pc}{}{\counter{\producerID}}
                            \NewDocumentCommand{\pcLoaded}{}{\progVar{c}}

                    % consumer counter
                            \NewDocumentCommand{\consumerID}{}{c}
                            \NewDocumentCommand{\cc}{}{\counter{\consumerID}}
                            \NewDocumentCommand{\ccLoaded}{}{\progVar{c}}

            % Variables used solely for verification figure
            % ====================================================
                    
                    % fifo
                            \NewDocumentCommand{\locFifo}{}{\metaVar{\hlocVar_\fifo}}
                            \NewDocumentCommand{\valFifoInv}{}{\metaVar{\valVar_\fifo^\m}}

                    % consumer
                            \NewDocumentCommand{\locCC}{}{\metaVar{\hlocVar_{\cc}}}
                            \NewDocumentCommand{\valCC}{}{\metaVar{\valVar_\cc}}
                            \NewDocumentCommand{\nextValCC}{}{\metaVar{\valVar_\cc'}}
                            \NewDocumentCommand{\valCCInv}{}{\metaVar{\valVar_\cc^\m}}
                            \NewDocumentCommand{\cObBag}{}{\metaVar{\obBagVar_\consumerID}}
                            \NewDocumentCommand{\cNextObBag}{}{\metaVar{\obBagVar_\consumerID'}}
                            
                            \NewDocumentCommand{\cLoopInv}{m}{\metaVar{L_\consumerID(#1)}}

                            \NewDocumentCommand{\cPostIf}{}{\metaVar{\fixedPredNameFont{PostIf}_\consumerID}}

                    % producer
                            \NewDocumentCommand{\locPC}{}{\metaVar{\hlocVar_\pc}}
                            \NewDocumentCommand{\valPC}{}{\metaVar{\valVar_\pc}}
                            \NewDocumentCommand{\nextValPC}{}{\metaVar{\valVar_\pc'}}
                            \NewDocumentCommand{\valPCInv}{}{\metaVar{\valVar_\pc^\m}}
                            \NewDocumentCommand{\pObBag}{}{\metaVar{\obBagVar_\producerID}}
                            \NewDocumentCommand{\pNextObBag}{}{\metaVar{\obBagVar_\producerID'}}
                            
                            \NewDocumentCommand{\pLoopInv}{m}{\metaVar{L_\producerID(#1)}}
                            
                            \NewDocumentCommand{\pPostIf}{}{\metaVar{\fixedPredNameFont{PostIf}_\producerID}}
            
                    % mutex
                            \NewDocumentCommand{\mut}{}{\mutVar}
                            %\NewDocumentCommand{\locM}{}{\metaVar{\hlocVar_\m}}
                            \RenewDocumentCommand{\locM}{}{\metaVar{\hlocVar_\m}}
                            %\NewDocumentCommand{\invM}{}{\metaVar{P_\m}}
                            \RenewDocumentCommand{\invM}{}{\metaVar{P_\m}}
                            %\NewDocumentCommand{\levM}{}{\metaVar{0}}
                            \RenewDocumentCommand{\levM}{}{\metaVar{0}}
                            \NewDocumentCommand{\invMNoExFifo}{m}{\metaVar{P_\m'(#1)}}
                            \NewDocumentCommand{\invMNoSigPush}{}
                                    {
                                            \invM^{\substack{\text{no:\sigPush{}}}}
                                    }
                            \NewDocumentCommand{\invMNoSigPop}{}
                                    {
                                            \invM^{\substack{\text{no:\sigPop{}}}}
                                    }
            
                    % signals
                            \NewDocumentCommand{\idGen}{m m}{\metaVar{\idVar_\text{#1}^{#2}}}
                            
                            % signal set when number is pushed into FIFO
                            \NewDocumentCommand{\idPush}{m}{\idGen{push}{#1}}
                            \NewDocumentCommand{\levPush}{m}{\metaVar{\levVar_\text{push}^{#1}}}
                            
                            % signal set when number is popped from FIFO
                            \NewDocumentCommand{\idPop}{m}{\idGen{pop}{#1}}
                            \NewDocumentCommand{\levPop}{m}{\metaVar{\levVar_\text{pop}^{#1}}}
                            
                            \NewDocumentCommand{\sigPush}{m}{\metaVar{\sigVar_\text{push}^{#1}}}
                            \NewDocumentCommand{\sigPop}{m}{\metaVar{\sigVar_\text{pop}^{#1}}}

%$$
%\begin{array}{p{\textwidth}}
%\hline
%\\commands\\
%\hline
%\end{array}$$
%\setlength{\arraycolsep}{0.1cm}

    \begin{figure}
    	\caseStudyFont
            % line with approx. width of array
            % helps visualizing whether space to left and to right of array are equal            
%            \noindent\makebox[\linewidth]{\rule{1.35\textwidth}{1pt}} 

            \makebox[\textwidth][c]
            {$
            % box treats separator center between left and right column as "center of array"
%            \hspace{0.025\textwidth}  % adjusting center
            \hspace{0.105\textwidth}  % adjusting center
%            \hspace{0.22\textwidth}  % adjusting center
            \arxivShift			% adjusting center for arxiv submission
            \begin{array}{l | p{0.266\pdfpagewidth}}
                    \progProof{
                            \noObs 
                    }
                    \\
                    \keyword{let}\ \fifo \mathop{:=} \cmdAlloc{\lstNil}\ \keyword{in}
                    \
                    \keyword{let}\ \m \mathop{:=} \cmdNewMut\ \keyword{in}
                            &\ \ \proofRuleHint{\prLetName (2x) \& \prAllocName}
                    \\
                    \ghostFont{\forall \locFifo, \locM.}
                            &\ \ \proofRuleHint{\& \prNewMutexName}
                    \\
                    \progProof{ 
                            \noObs
                            \slStar
                            \progProofNew{
                                    \slPointsTo{\locFifo}{\lstNil}
                                    \slStar
                                    \assMutUninit{\locM}
                            }
                    }
                            &\ \ \proofRuleHintML{\prViewShiftName \& \vsAllocSigIDName\\
                                    \& \prExistsName (200x)
                            }
                    \\
                    \ghostFont{\forall \idPop{1}, \dots \idPop{100}, \idPush{1}, \dots, \idPush{100}.}
                    \\
                    \progProof{
                            \progProofNew{
                                    \slBigStar_{i = 1, \dots, 100}
                                            \quadBack\quadBack\quadBack
                                            \sigUninit{\idPop{i}}
                                    \slStar
                                    \slBigStar_{i = 1, \dots, 100}
                                            \quadBack\quadBack\quadBack
                                            \sigUninit{\idPush{i}}
                            }
                            \slStar \dots
                    }
                    \\
\end{array}
            $}

            \caption{Verification of realistic example~\ref{gs:fig:RealisticExample}: Initialization.}
            \label{gs:fig:VerificationExampleBoundedFifoInitSL}
    \end{figure}

    \begin{figure}
    	\caseStudyFont
            % line with approx. width of array
            % helps visualizing whether space to left and to right of array are equal            
%            \noindent\makebox[\linewidth]{\rule{1.35\textwidth}{1pt}} 
            \makebox[\textwidth][c]
            {$
            % box treats separator center between left and right column as "center of array"
            \hspace{0.17\textwidth}  % adjusting center
            \arxivShift			% adjusting center for arxiv submission
\begin{array}{l | p{0.266\pdfpagewidth}}
                    \proofDef{
                            \invMNoExFifo{\valFifoInv} \ := \
                            \exists \valPCInv, \valCCInv.
                    }
                    	&\phantom{a}
                   	\vspace{0.2cm}
                    \\
                    \proofDef{
                            \phantom{\invM \ :=\ }
                            \slPointsTo[\frac{1}{2}]{\locPC}{\valPCInv}
                            \ \slStar\
                            \slPointsTo[\frac{1}{2}]{\locCC}{\valCCInv}
                            \ \slStar \
                            0 \leq \valPCInv \leq 100
                            \ \slStar \
                            0 \leq \valCCInv \leq 100
                    }
                            &\proofHintML{Producer \& consumer
                            			\\
                            			counters.
                            			}
                    \vspace{0.2cm}
                    \\
                    \proofDef{
                            \phantom{\invM \ :=\ }
                            \ \slStar\
                            \slPointsTo{\locFifo}{\valFifoInv}
                            \ \slStar\
                            \valCCInv = \valPCInv + \lstSize{\valFifoInv}
                            \ \slStar \
                            0 \leq \lstSize{\valFifoInv} \leq 10
                    }
                    \ \ \
                            &\proofHintML{Bounded FIFO \& its
                            			\\
                            			relationship to counters}
                    \vspace{0.1cm}
                    \\
\end{array}
            $}

            \caption
            {Verification of realistic example~\ref{gs:fig:RealisticExample}: Lock invariant.}
            \label{gs:fig:VerificationExampleBoundedFifoLockInvariantSL}
    \end{figure}

    \begin{figure}
    	\caseStudyFont
            % line with approx. width of array
            % helps visualizing whether space to left and to right of array are equal            
%            \noindent\makebox[\linewidth]{\rule{1.35\textwidth}{1pt}} 

            \makebox[\textwidth][c]
            {$
            % box treats separator center between left and right column as "center of array"
            \hspace{0.025\textwidth}  % adjusting center
            \arxivShift			% adjusting center for arxiv submission
            \begin{array}{l | p{0.266\pdfpagewidth}}
                    \proofDef{
                            \pLoopInv{n, \pObBag} \ := \ 
                            0 < n \leq 100
                            \ \slStar \
                            \displaystyle
                            \quadBack
                            \slBigStar_{i = 1, \dots, n-1}
                                    \quadBack
                                    \sigUninit{\idPush{i}}
                    }
                            &\ \ \proofHint{Loop invariant of producer.}
                    \\
                    \proofDef{
                            \phantom{\pLoopInv{n, \pObBag} \ := \ }
                            \ \slStar \
                            (n > 0 \ \leftrightarrow \ \pObBag = \multiset{\sigPush{n}})
                            \ \slStar \
                            (n = 0 \ \leftrightarrow \ \pObBag = \msEmpty)
                    }
                    \ \ \
            \end{array}
            $}

            \caption{Verification of realistic example~\ref{gs:fig:RealisticExample}: Producer's loop invariant.}
            \label{gs:fig:VerificationExampleBoundedFifoProducerLoopInvariantSL}
    \end{figure}

    \begin{figure}
    	\caseStudyFont
            % line with approx. width of array
            % helps visualizing whether space to left and to right of array are equal            
%            \noindent\makebox[\linewidth]{\rule{1.35\textwidth}{1pt}} 

            \makebox[\textwidth][c]
            {$
            % box treats separator center between left and right column as "center of array"
            \hspace{0.025\textwidth}  % adjusting center
            \arxivShift			% adjusting center for arxiv submission
            \begin{array}{l | p{0.266\pdfpagewidth}}
                    \proofDef{
                            \cLoopInv{n, \cObBag} \ := \ 
                            0 < n \leq 100
                            \ \slStar \
                            \displaystyle
                            \quadBack
                            \slBigStar_{i = 1, \dots, n-1}
                                    \quadBack
                                    \sigUninit{\idPop{i}}
                    }
                            &\ \ \proofHint{Loop invariant of consumer.}
                    \\
                    \proofDef{
                            \phantom{\cLoopInv{n, \cObBag} \ := \ }
                            \ \slStar \
                            (n > 0 \ \leftrightarrow \ \cObBag = \multiset{\sigPop{n}})
                            \ \slStar \
                            (n = 0 \ \leftrightarrow \ \cObBag = \msEmpty)
                    }
                    \ \ \
            \end{array}
            $}

            \caption{Verification of realistic example~\ref{gs:fig:RealisticExample}: Consumer's loop invariant.}
            \label{gs:fig:VerificationExampleBoundedFifoConsumerLoopInvariantSL}
    \end{figure}

    \begin{figure}
    		\caseStudyFont
            % line with approx. width of array
            % helps visualizing whether space to left and to right of array are equal            
%            \noindent\makebox[\linewidth]{\rule{1.35\textwidth}{1pt}} 
%\vspace{-0.7cm}

            \makebox[\textwidth][c]
            {$
            % box treats separator center between left and right column as "center of array"
            \hspace{0.05\textwidth}  % adjusting center
            \arxivShift			% adjusting center for arxiv submission
\begin{array}{l | p{0.22\pdfpagewidth}}
                    \dots\ref{gs:fig:VerificationExampleBoundedFifoInitSL}\dots
                            &\ \ \referenceHintML{Continuation of Fig.~\ref{gs:fig:VerificationExampleBoundedFifoInitSL}.}
                    \\
                    \progProofML{
                            \obsOf{\sigPush{100}, \sigPop{100}}
                            \slStar
                            \slPointsTo[\frac{1}{2}]{\locPC}{100}
                            \slStar
                            \slPointsTo[\frac{1}{2}]{\locCC}{100}
                            \\
                            \slStar
                            \slBigStar_{i = 1, \dots, 99}
                                    \quadBack\quadBack\quadBack\!
                                    (\sigUninit{\idPush{i}} \slStar \sigUninit{\idPop{i}})
                            \\
                            \slStar\,
                            \assMutex[\frac{1}{2}]{\mut}{\invM}
                            \slStar
                            \assMutex[\frac{1}{2}]{\mut}{\invM}
                    }
                    \\
                    \keyword{fork}\ (
                            &\ \ \proofRuleHint{\prForkName}
                    \\
\end{array}
            $}

            \caption{Verification of realistic example~\ref{gs:fig:RealisticExample}: Main thread (consumer) forks producer thread.}
            \label{gs:fig:VerificationExampleBoundedFifoForkingSL}
    \end{figure}

    \begin{figure}
    	\caseStudyFont
            % line with approx. width of array
            % helps visualizing whether space to left and to right of array are equal            
%            \noindent\makebox[\linewidth]{\rule{1.35\textwidth}{1pt}} 

			\hspace{0.25cm}
            \makebox[\textwidth][c]
            {$
            % box treats separator center between left and right column as "center of array"
%            \hspace{0.025\textwidth}  % adjusting center
            \hspace{0.07\textwidth}  % adjusting center
            \arxivShift			% adjusting center for arxiv submission
\begin{array}{l | p{0.21\pdfpagewidth}}
                    \lowlightText{\forall \locFifo, \locM, \locPC, \locCC.}
                    \\
                    \dots &\ \ \referenceHintML{Continuation of\\ Fig.~\ref{gs:fig:VerificationExampleBoundedFifoForkingSL}.}
                    \\
                    \progProof{
                            \obsOf{\sigPush{100}}
                            \slStar
                            \slPointsTo[\frac{1}{2}]{\locPC}{100}
                            \slStar
                            \pLoopInv{100, \multiset{\sigPush{100}}}
                            \slStar
                            \assMutex[\frac{1}{2}]{\mut}{\invM}
                    }
                    \\
                    \keyword{while}\ (
                            &\ \ \proofRuleHint{\prWhileDecStrictName}
                    \\\quad\quad
                            \ghostFont{\forall \valPC.}
% TODO
%                    \\\vspace{-0.6cm}
					\\\vspace{-0.22cm}
                    \\\quad\quad
                            \progProof{
                                    \obsOf{\sigPush{\progProofChanged{\valPC}}}
                                    \slStar
                                    \slPointsTo[\frac{1}{2}]{\locPC}{\progProofChanged{\valPC}}
                                    \slStar
                                    \pLoopInv{\progProofChanged{\valPC}, \multiset{\sigPush{\progProofChanged{\valPC}}}}
                                    \slStar
                                    \assMutex[\frac{1}{2}]{\mut}{\invM}
                            }
                    \\
\end{array}
            $}

            \caption{Verification of realistic example~\ref{gs:fig:RealisticExample}: Producer loops.}
            \label{gs:fig:VerificationExampleBoundedFifoProducerLoopSL}
    \end{figure}

    \begin{figure}
    	\caseStudyFont
            % line with approx. width of array
            % helps visualizing whether space to left and to right of array are equal            
%            \noindent\makebox[\linewidth]{\rule{1.35\textwidth}{1pt}} 

            \makebox[\textwidth][c]
            {$
            % box treats separator center between left and right column as "center of array"
            \hspace{0.08\textwidth}  % adjusting center
            \arxivShift			% adjusting center for arxiv submission
\begin{array}{l | p{0.22\pdfpagewidth}}
                    \lowlightText{\forall \locFifo, \locM, \locPC, \locCC, \valPC, \pObBag, \valFifoInv.}
                    \\
                    \dots &\ \ \referenceHint{Continuation of Fig.~\ref{gs:fig:VerificationExampleBoundedFifoProducerLoopSL}.}
                    \\
                    \referenceHintML{For definition of \invM, \pLoopInv{n, \obBagVar} and variations  cf.\@ 
                            Fig.~\ref{gs:fig:VerificationExampleBoundedFifoLockInvariantSL}
                            and~\ref{gs:fig:VerificationExampleBoundedFifoProducerLoopInvariantSL}.
                    }
                    \\
                    \progProof{
                            \lstSize{\valFifoInv} < 10
                            \slStar 
                            \pLoopInv{\valPC, \pObBag}
                            \slStar
                            \obs{\pObBag \msCup \multiset{\mut}}
                            \slStar
                            \slPointsTo[\frac{1}{2}]{\locPC}{\valPC}
                            \slStar
                            \invMNoExFifo{\valFifoInv}
                    }
                            &\ \ \proofRuleHint{\prViewShiftName \& \vsSemImpName}
                  	\vspace{0.2cm}
                    \\
\end{array}
            $}

            \caption{Verification of realistic example~\ref{gs:fig:RealisticExample}: Producer thread's production step.}
            \label{gs:fig:VerificationExampleBoundedFifoProductionStepSL}
    \end{figure}

    \begin{figure}
    	\caseStudyFont
            % line with approx. width of array
            % helps visualizing whether space to left and to right of array are equal            
%            \noindent\makebox[\linewidth]{\rule{1.35\textwidth}{1pt}} 

			\hspace{-0.4cm}
            \makebox[\textwidth][c]
            {$
            % box treats separator center between left and right column as "center of array"
%            \hspace{0.025\textwidth}  % adjusting center
            \hspace{0.2\textwidth}  % adjusting center
            \arxivShift			% adjusting center for arxiv submission
\begin{array}{l | p{0.22\pdfpagewidth}}
                    \lowlightText{\forall \locFifo, \locM, \locPC, \locCC.}
                    \\
                    \dots &\ \ \referenceHintML{Continuation of
                    												\\
                    												Fig.~\ref{gs:fig:VerificationExampleBoundedFifoForkingSL}.}
                    \\
                    \progProof{
                            \obsOf{\sigPop{100}}
                            \slStar
                            \slPointsTo[\frac{1}{2}]{\locCC}{100}
                            \slStar
                            \cLoopInv{100, \multiset{\sigPop{100}}}
                            \slStar
                            \assMutex[\frac{1}{2}]{\mut}{\invM}
                    }
                    \\
                    \keyword{while}\ (
                            &\ \ \proofRuleHint{\prWhileDecStrictName}
                    \\\quad\quad
                            \ghostFont{\forall \valCC.}
% TODO
%                    \\\vspace{-0.6cm}
					\\\vspace{-0.3cm}
                    \\\quad\quad
                            \progProof{
                                    \obsOf{\sigPop{\progProofChanged{\valCC}}}
                                    \slStar
                                    \slPointsTo[\frac{1}{2}]{\locCC}{\progProofChanged{\valCC}}
                                    \slStar
                                    \cLoopInv{\progProofChanged{\valCC}, \multiset{\sigPop{\progProofChanged{\valCC}}}}
                                    \slStar
                                    \assMutex[\frac{1}{2}]{\mut}{\invM}
                            }
                    \\
\end{array}
            $}

            \caption{Verification of realistic example~\ref{gs:fig:RealisticExample}: Consumer loops.}
            \label{gs:fig:VerificationExampleBoundedFifoConsumerLoopSL}
    \end{figure}

    \begin{figure}
    	\caseStudyFont
            % line with approx. width of array
            % helps visualizing whether space to left and to right of array are equal            
%            \noindent\makebox[\linewidth]{\rule{1.35\textwidth}{1pt}} 

            \makebox[\textwidth][c]
            {$
            % box treats separator center between left and right column as "center of array"
%            \hspace{0.025\textwidth}  % adjusting center
            \hspace{0.12\textwidth}  % adjusting center
            \arxivShift			% adjusting center for arxiv submission
\begin{array}{l | p{0.266\pdfpagewidth}}
                    \lowlightText{\forall \locFifo, \locM, \locPC, \locCC, \valCC, \cObBag, \valFifoInv.}
                    \\
                    \dots &\ \ \referenceHintML{Continuation of Fig.~\ref{gs:fig:VerificationExampleBoundedFifoConsumerLoopSL}.}
                    \\
                    \referenceHintML{\ \ For definition of \invM, \cLoopInv{n, \obBagVar}  and variations cf.\@ 
                            Figures~\ref{gs:fig:VerificationExampleBoundedFifoLockInvariantSL}
                            and~\ref{gs:fig:VerificationExampleBoundedFifoConsumerLoopInvariantSL}.
                    }
                    \\
                    \progProof{
                            \lstSize{\valFifoInv} > 0
                            \slStar 
                            \cLoopInv{\valCC, \cObBag}
                            \slStar
                            \obs{\cObBag \msCup \multiset{\mut}}
                            \slStar
                            \slPointsTo[\frac{1}{2}]{\locCC}{\valCC}
                            \slStar
                            \invMNoExFifo{\valFifoInv}
                    }
                            &\ \ \proofRuleHint{\prViewShiftName \& \vsSemImpName}
                  	\vspace{0.2cm}
                    \\
\end{array}
            $}

            \caption{Verification of realistic example~\ref{gs:fig:RealisticExample}: Consumer thread's consumption step.}
            \label{gs:fig:VerificationExampleBoundedFifoConsumptionStepSL}
    \end{figure}

    }% end of local scope for verification figures

\clearpage    
\subsection{Case Study: Statically Unbounded Number of Communicating Parties}\label{gs:appendix:sec:caseStudy:unbounded}

{   % local scope for case study
    % program variables
    \NewDocumentCommand{\numThs}{}{\progVar{N}}
    
%    \NewDocumentCommand{\numProds}{}{\progVar{np}}
%    \NewDocumentCommand{\numConss}{}{\progVar{nc}}
    
    \NewDocumentCommand{\buffer}{}{\progVar{buf}}
    \NewDocumentCommand{\bufferLoaded}{}{\progVar{b}}
    \NewDocumentCommand{\locBuffer}{}{\metaVar{\hlocVar_\buffer}}
    \NewDocumentCommand{\valBuffer}{}{\metaVar{\valVar_\buffer}}
    \NewDocumentCommand{\nextValBuffer}{}{\metaVar{\nextValVar_\buffer}}
    
    \NewDocumentCommand{\locMut}{}{\metaVar{\hlocVar_\mutProgVar}}
    \NewDocumentCommand{\levMut}{}{0}
    
    % decreasing counter for number of elements that are left to be pushed
%    \NewDocumentCommand{\inputSize}{}{\progVar{inputSize}}
    
    % fork loop counter
    \NewDocumentCommand{\flCnt}{}{\progVar{c}}
    \NewDocumentCommand{\locFlCnt}{}{\metaVar{\hlocVar_\flCnt}}
    \NewDocumentCommand{\valFlCnt}{}{\metaVar{\valVar_\flCnt}}
    
    % signals
    \NewDocumentCommand{\sigPush}{m m}{\metaVar{\sigVar_{\text{push}, #1}^{#2}}}
    \NewDocumentCommand{\sigPop}{m m}{\metaVar{\sigVar_{\text{pop}, #1}^{#2}}}
    \NewDocumentCommand{\idPush}{m m}{\metaVar{\idVar_{\text{push}, #1}^{#2}}}
    \NewDocumentCommand{\idPop}{m m}{\metaVar{\idVar_{\text{pop}, #1}^{#2}}}
    \NewDocumentCommand{\levPush}{m}{\metaVar{\levVar_\text{push}^{#1}}}
    \NewDocumentCommand{\levPop}{m}{\metaVar{\levVar_\text{pop}^{#1}}}
    
    % variables used by lock invariant
    \NewDocumentCommand{\numWritten}{}{\metaVar{W}}
    \NewDocumentCommand{\numRead}{}{\metaVar{R}}
    
    % ghost heap cells
    \NewDocumentCommand{\prodFinished}{m}{\metaVar{w_{#1}}}
    \NewDocumentCommand{\consFinished}{m}{\metaVar{r_{#1}}}
    \NewDocumentCommand{\glocProdFin}{m}{\metaVar{\glocVar_{#1}^w}}
    \NewDocumentCommand{\glocConsFin}{m}{\metaVar{\glocVar_{#1}^r}}

    % loop invariants
    \NewDocumentCommand{\flInv}{m}{\metaVar{L_f(#1)}}
    \NewDocumentCommand{\pInv}{m m}{\metaVar{L_p^{#1}(#2)}}
    \NewDocumentCommand{\cInv}{m m}{\metaVar{L_c^{#1}(#2)}}

    \NewDocumentCommand{\nextPush}{}{\metaVar{n_\text{push}}}
    \NewDocumentCommand{\nextPop}{}{\metaVar{n_\text{pop}}}

    % lock invariant
    \NewDocumentCommand{\lockInvNoQuant}{}{\metaVar{\widehat{\assLockInvVar}}}
    \NewDocumentCommand{\pLockInvRest}{}{\metaVar{\assLockInvVar_p^\text{rest}}}
    \NewDocumentCommand{\cLockInvRest}{}{\metaVar{\assLockInvVar_c^\text{rest}}}

    % state reached after if statement in producer and consumer loop
    \NewDocumentCommand{\pPostIf}{}{\metaVar{\fixedPredNameFont{PostIf}_p}}
    \NewDocumentCommand{\cPostIf}{}{\metaVar{\fixedPredNameFont{PostIf}_c}}

    In this section we present and verify a program similar to the one from the previous section, but where the number of producer and consumer threads is not statically bounded.
    Fig.~\ref{gs:appendix:caseStudy:undbounded:code} presents this program.
    It involves three parties: the main thread, producer threads and consumer threads.
    The main thread creates a shared buffer of size 1, generates a random number $\numThs > 0$ and spawns \numThs producer and \numThs consumer threads., which communicate via the shared buffer.
    Each producer tries to push a single random number into the buffer.
    In case the buffer is full, the producer busy-waits for it to become empty, i.e., it busy-waits for a consumer to pop the number currently stored in the buffer.
    After it pushed, the producer terminates.
    Each consumer tries to pop a single number from the buffer.
    In case the buffer is empty, it waits for some producer to push a number into the buffer.
    After it popped, the consumer terminates.

    \begin{figure}
            $$
\begin{array}{l p{0cm} l}
                    \keyword{let}\ \numThs := \cmdRandomNat + 1\ \keyword{in}
                            &&\lowlightText{\text{Random number of threads.}}
                    \\
                    \keyword{let}\ \buffer := \cmdAlloc{-1}\ \keyword{in}
                            &&\lowlightText{\text{Shared buffer. }}
                    \\
                    		&&\lowlightText{\text{Negative value $\leftrightarrow$ empty.}}
                    \\
                    \keyword{let}\ \mutProgVar := \cmdNewMut\ \keyword{in}
                    \\
                    \keyword{let}\ \flCnt := \cmdAlloc{\numThs}\ \keyword{in}
                    \\
                    \keyword{while}\ (
                            &&\lowlightText{\text{Main thread forks \numThs producer}}
                    \\&&\lowlightText{\text{and \numThs consumer threads.}}
                    \\\quad\quad
                            \keyword{fork}\ (
                                    &&\lowlightText{\text{Producer}}
                    \\
\end{array}
            $$
            \caption{Program with statically unbounded number of threads communicating via a shared buffer of size 1.}
            \label{gs:appendix:caseStudy:undbounded:code}
    \end{figure}

    \paragraph{Threads Racing for Buffer}
    Since the program is conceptionally similar to the one from the previous section, the same holds for its termination proof.
    There is, however, one fundamental difference.
    The previous program involved one producer and one consumer.
    So, it was clear which thread would push the $k^\text{th}$ element, set the $k^\text{th}$ push signal and discharge the corresponding obligation.
    The same was true for the consumer and the pop signals.
    Now, \numThs producers and \numThs consumers race for the buffer.
    We cannot statically determine which ones will win.
    Therefore, we cannot statically decide which threads we should delegate the push and pop obligations to.
    
    \paragraph{Signals}
    To solve this, we create \numThs variations of every push and pop signal.
    That is, we create signals  \sigPush{i}{k} and \sigPop{i}{k} for $1 \leq i,k \leq \numThs$ where $i$ refers to the $i$'s producer or consumer thread and $k$ to the $k^\text{th}$ element pushed to or popped from the buffer.
    We delegate the obligations for $\sigPush{i}{1}, \dots, \sigPush{i}{\numThs}$ to the $i^\text{th}$ producer and the obligations for $\sigPop{i}{1}, \dots, \sigPop{i}{\numThs}$ to the $i^\text{th}$ consumer.

    When the $i^\text{th}$ producer waits because the buffer is full and contains the $k^\text{th}$ number, it sets all its remaining unset signals up to the $k^\text{th}$ one.
    That is, signals $\sigPush{i}{k}, \dots, \sigPush{i}{\numThs}$ remain unset.
    (Reducing the number of obligations that a producer holds while waiting for a consumer to pop, reduces the potential for level conflicts and simplifies ordering the signals.)
    When the $i^\text{th}$ producer pushes any number to the buffer, it sets all its remaining signals, before it terminates.
    A consumer can use any unset \sigPush{j}{k} for any $j$ to wait for the $k^\text{th}$ element to arrive in the buffer.
    We proceed analogously with the consumers and their pop signals.
    Similarly to the proof from the previous section, we include our signals in the lock invariant associated with the mutex that protects the shared buffer.
    Thereby, we allow all threads to share the signals in a synchronised fashion.
    
    \paragraph{Levels}
    The $i^\text{th}$ producer uses some unset \sigPop{j}{k} to wait for the $k^\text{th}$ element to be popped from the buffer.
    Meanwhile it holds obligations $\sigPush{i}{k}, \dots, \sigPush{i}{\numThs}$.
    Analogously, the $i^\text{th}$ consumer holds obligations $\sigPop{i}{k}, \dots, \sigPop{i}{\numThs}$ while using some unset \sigPush{j}{k} to wait for the $k^\text{th}$ element to be pushed.
    We assign one common level \levPush{k} to all $\sigPush{1}{k}, \dots, \sigPush{\numThs}{k}$ and one common level \levPop{k} to all $\sigPop{1}{k}, \dots, \sigPop{\numThs}{k}$.
    It must hold
    $\levPop{k} < \levPush{j}$ for all $j > k$ and
    $\levPush{k} < \levPop{h}$ for all $h\geq k$.
    Hence, we chose
    $\levPush{k} = 2\cdot k - 1$ and
    $\levPop{k} = 2\cdot k$.

    \paragraph{Ghost Variables}
    In order to simplify the verification of this program, we introduce \emph{ghost variables}, which are standard~\cite{Owicki1976VerifyingPropiertiesGhostVariables, Jung2016HigherorderGS, Jacobs2018ModularTerminationVerification}.
    Ghost variables are ghost resources that behave like regular program variables, but since they only exist on the verification level, they cannot affect the runtime behaviour.
    We assume an infinite set of ghost locations \GhostLocSet with $\GhostLocSet \cap \HeapLocSet = \emptyset$ and a set of ghost values \GhostValueSet.
    We implement ghost variables in the form of \emph{ghost heap cells} and represent them by \emph{ghost points-to chunks} \slPointsTo{\glocVar}{\gvalVar}.
    Fig.~\ref{gs:appendix:caseStudy:undbounded:ghostVariables} presents two new view shift rules for ghost variables.
    Rule \vsNewGhostCellName allows to create a new ghost variables and \vsMutateGhostCellName allows to change the value of an existing one.
    Adding ghost variables does not affect the soundness of our verification approach.
    In fact, the generalised logic we present and prove sound in the technical report~\cite{Reinhard2020GhostSignalsTR} (appendix~\S~\ref{ch:appendix-ghostSignals-TR}) includes ghost variables.
    In our termination proof, we introduce ghost variables \prodFinished{i} and \consFinished{i}.
    Variable \prodFinished{i} tracks whether the $i^\text{th}$ producer has already pushed ("written") an element to the buffer and 
    \consFinished{i} tracks whether the $i^\text{th}$ consumer has already popped ("read") an element from the buffer.
    These variables conveniently allow us to to refer to the number of elements written to and read from the buffer in the lock invariant.
    In Fig.~\ref{gs:appendix:caseStudy:unbounded:init} -- \ref{gs:appendix:caseStudy:unbounded:consumerLoop}, we present the full proof outline for our program from Fig.~\ref{gs:appendix:caseStudy:undbounded:code}.
    We scaled the proof outlines down to fit them into the page layout. 
    You can find the original proof outlines in the extended version of the ghost signal paper~\cite{Reinhard2021GhostSignalsExtended} and in the technical report~\cite{Reinhard2020GhostSignalsTR}.

    \begin{figure}
    		\ifArxiv{\vspace{-0.6cm}}
            \begin{mathpar}
                    \vsNewGhostCell
                    \and
                    \vsMutateGhostCell
            \end{mathpar}
            \ifArxiv{\vspace{-0.6cm}}
            \caption{View shift rules for ghost variables.}
            \label{gs:appendix:caseStudy:undbounded:ghostVariables}
    \end{figure}

    \begin{figure}
    	\caseStudyFont
            % line with approx. width of array
            % helps visualizing whether space to left and to right of array are equal            
%            \noindent\makebox[\linewidth]{\rule{1.35\textwidth}{1pt}} 

            \hspace{1.05cm}
            \makebox[\textwidth][c]
            {$
            % box treats separator center between left and right column as "center of array"
%            \hspace{0.025\textwidth}  % adjusting center
%            \hspace{0.2\textwidth}  % adjusting center
            \arxivShift			% adjusting center for arxiv submission
\begin{array}{l  |p{0.224\pdfpagewidth}}
%                    b b b b b b b b b b b b b b b b b b b b b b b b b b b b b b b b b b b b b b b b b b b b b b b b b b b b b b b b b b b b b b b b b b b b b b b 
%                    & a a a a a a a a a a a a a a a a a a a a a a
%                    \\
                    \progProof{\noObs}&
                    \\
                    \keyword{let}\ \numThs := \cmdRandomNat+1\ \keyword{in}\
                    \keyword{let}\ \buffer := \cmdAlloc{-1}\ \keyword{in}\
                            &\proofRuleHintML{\prLetName (2x) 
                            									\\
                            									\& \prAllocName}
                    \vspace{0.2cm}
                    \\
                    \keyword{let}\ \mutProgVar := \cmdNewMut\ \keyword{in}\
                    \keyword{let}\ \flCnt := \cmdAlloc{\numThs}\ \keyword{in}\
                            &\proofRuleHintML{\prLetName (2x) 
                            									\\
                            									\& \prNewMutexName 
                            									\\
                            									\& \prAllocName}
                            \hspace{-1cm}
                    \\
                    \ghostFont{\forall \locBuffer, \locMut, \locFlCnt.}&
                    \\
\end{array}
            $}
            
            \caption{Verification of program~\ref{gs:appendix:caseStudy:undbounded:code}: Initialisation.}
            \label{gs:appendix:caseStudy:unbounded:init}
    \end{figure}

    \begin{figure}
    	\caseStudyFont
            % line with approx. width of array
            % helps visualizing whether space to left and to right of array are equal            
%            \noindent\makebox[\linewidth]{\rule{1.35\textwidth}{1pt}} 

            \hspace{0.4cm}
            \makebox[\textwidth][c]
            {$
            % box treats separator center between left and right column as "center of array"
%            \hspace{0.025\textwidth}  % adjusting center
%            \hspace{0.04\textwidth}  % adjusting center
            \arxivShift			% adjusting center for arxiv submission
\begin{array}{l c l p{0.25cm} | p{0.25cm} p{0.2\pdfpagewidth}}
                    \proofDef{
                            \nextAssLockInvVar(
                                    \valBuffer, \numWritten, \numRead, 
                                    \prodFinished{1}, \dots, \prodFinished{\numThs},
                                    \consFinished{1}, \dots, \consFinished{\numThs}
                            )
                    	\hspace{-20.5cm}
                    }
                    \\
                    &\proofDef{\ := \ }
                    &\proofDef{
                            \slPointsTo{\locBuffer}{\valBuffer}
                            \ \wedge\
                            \valBuffer \in \Z
                            \ \wedge\ 
                            \numThs > 0
                    }
                            &&&\proofHintML{Shared buffer.
                            								\\
                            								Negative $\leftrightarrow$ empty.}
                  	\vspace{0.2cm}
                    \\
\end{array}
            $}
            
            \caption{Verification of program~\ref{gs:appendix:caseStudy:undbounded:code}: Lock invariant.}
            \label{gs:appendix:caseStudy:unbounded:lockInv}
    \end{figure}

    \begin{figure}
    	\caseStudyFont
            % line with approx. width of array
            % helps visualizing whether space to left and to right of array are equal            
%            \noindent\makebox[\linewidth]{\rule{1.35\textwidth}{1pt}} 

            \makebox[\textwidth][c]
            {$
            % box treats separator center between left and right column as "center of array"
%            \hspace{0.025\textwidth}  % adjusting center
            \hspace{0.1\textwidth}  % adjusting center
            \arxivShift			% adjusting center for arxiv submission
            \ifArxiv{\hspace{0.5cm}}
\begin{array}{l c l p{0.25cm} | p{0.25cm} p{0.3\pdfpagewidth}}
                    \proofDef{\flInv{\flCnt}} &\ \proofDef{:=} \
                            &\proofDef{
                                            \slPointsTo{\locFlCnt}{\flCnt}
                                            \ \wedge\
                                            0 \leq \flCnt \leq \numThs
                                            \ \wedge\
                                            \numThs > 0
                            }
                            &&&\proofHintML{
                                    Decreasing counter.\\
                                    Prods.\@ \& cons.\@ 
                                    \\
                                    $\flCnt+1, \dots, \numThs$ already forked.
                            }
               	\vspace{0.2cm}
                \\
                            &&\proofDef{
                                    \slStar\
                                    \obs{
                                            \multiset{\sigPush{i}{k}, \sigPop{i}{k} \ | \
%                                                    \numThs - \flCnt + 1 \leq i \leq \numThs
                                                    1 \leq i \leq \flCnt
                                                    \ \wedge \
                                                    1 \leq k \leq \numThs
                                            }
                                    }
                            }
                            &&&\proofHintML{
                                    Remaining obligations for threads\\
                                    that have not been forked, yet.
                            }
                   	\vspace{0.1cm}
                    \\
\end{array}
            $}
            
            \caption{Verification of program~\ref{gs:appendix:caseStudy:undbounded:code}: Fork loop invariant.}
            \label{gs:appendix:caseStudy:unbounded:forkLoopInv}
    \end{figure}

    \begin{figure}
        \caseStudyFont
        % line with approx. width of array
        % helps visualizing whether space to left and to right of array are equal            
%            \noindent\makebox[\linewidth]{\rule{1.35\textwidth}{1pt}} 

        \makebox[\textwidth][c]
        {$
        % box treats separator center between left and right column as "center of array"
%            \hspace{0.025\textwidth}  % adjusting center
        \hspace{0.04\textwidth}  % adjusting center
        \arxivShift			% adjusting center for arxiv submission
        \begin{array}{l c l p{0.3cm} | p{0.3cm} p{0.3\pdfpagewidth}}
                \proofDef{\pInv{i}{\obBagVar}} &\ \proofDef{:=} \
                        &\proofDef{
                                        \exists \nextPush.\
                                        1 \leq \nextPush \leq \numThs
                        }
                        &&&\proofHintML{
                                Lower bound for index of next element\\
                                that will be pushed by any thread.
                        }
            \\
                        &&\proofDef{
                                \slStar\
                                \obBagVar = \multiset{\sigPush{i}{\nextPush}, \dots, \sigPush{i}{\numThs}}
                        }
                        &&&\proofHintML{}
            \\
                        &&\proofDef{
                                \slStar\
                                \slPointsTo[\frac{1}{2}]{\glocProdFin{i}}{\slFalse}
                        }
                        &&&\proofHintML{}
        \end{array}
        $}
        
        \caption{Verification of program~\ref{gs:appendix:caseStudy:undbounded:code}: Loop invariant for producer $i$.}
        \label{gs:appendix:caseStudy:unbounded:producerLoopInv}
    \end{figure}

    \begin{figure}
    	\caseStudyFont
            % line with approx. width of array
            % helps visualizing whether space to left and to right of array are equal            
%            \noindent\makebox[\linewidth]{\rule{1.35\textwidth}{1pt}} 

            \makebox[\textwidth][c]
            {$
            % box treats separator center between left and right column as "center of array"
%            \hspace{0.025\textwidth}  % adjusting center
            \hspace{0.04\textwidth}  % adjusting center
            \arxivShift			% adjusting center for arxiv submission
            \begin{array}{l c l p{0.3cm} | p{0.3cm} p{0.3\pdfpagewidth}}
                    \proofDef{\cInv{i}{\obBagVar}} &\ \proofDef{:=} \
                            &\proofDef{
                                            \exists \nextPop.\
                                            1 \leq \nextPop \leq \numThs
                            }
                            &&&\proofHintML{
                                    Lower bound for index of next element\\
                                    that will be popped by any thread.
                            }
                \\
                            &&\proofDef{
                                    \slStar\
                                    \obBagVar = \multiset{\sigPop{i}{\nextPop}, \dots, \sigPush{i}{\numThs}}
                            }
                            &&&\proofHintML{}
                \\
                            &&\proofDef{
                                    \slStar\
                                    \slPointsTo[\frac{1}{2}]{\glocConsFin{i}}{\slFalse}
                            }
                            &&&\proofHintML{}
            \end{array}
            $}
            
            \caption{Verification of program~\ref{gs:appendix:caseStudy:undbounded:code}: Loop invariant for consumer $i$.}
            \label{gs:appendix:caseStudy:unbounded:consumerLoopInv}
    \end{figure}

    \begin{figure}
    	\caseStudyFont
            % line with approx. width of array
            % helps visualizing whether space to left and to right of array are equal            
%            \noindent\makebox[\linewidth]{\rule{1.35\textwidth}{1pt}} 

			\hspace{0.5cm}
            \makebox[\textwidth][c]
            {$
            % box treats separator center between left and right column as "center of array"
%            \hspace{0.025\textwidth}  % adjusting center
            \hspace{0.04\textwidth}  % adjusting center
            \arxivShift			% adjusting center for arxiv submission
\right)
                            }&
                    \\\quad\quad
                            \keyword{fork}\ (       % consumer fork
                                    &\proofRuleHint{\prForkName}
                    \\\quad\quad\quad\quad
                                    \progProofML{
                                            \progProofNew{
                                                    \obsOf{\sigPop{\valFlCnt}{1}, \dots, \sigPop{\valFlCnt}{\numThs}}
                                                    \slStar
                                                    \assMutex[\frac{1}{\numThs}]{\mutVar}{\assLockInvVar}
                                            }
                                            \\
                                            \progProofNew{
                                                    \slStar\,
                                                    \slPointsTo[\frac{1}{2}]{\glocConsFin{\valFlCnt}}{\slFalse}
                                            }
                                    }&
                    \\\quad\quad\quad\quad
                                    \dots
                                    &\referenceHintML{Consumer loop on
                                    									\\
                                    									Fig.~\ref{gs:appendix:caseStudy:unbounded:consumerLoop}.}
                    \\\quad\quad\quad\quad
                                    \progProof{
                                            \obs{\progProofChanged{\msEmpty}}
                                            \,
                                            \progProofCancel{
                                                    \slStar
                                                    \assMutex[\frac{1}{\numThs}]{\mutVar}{\assLockInvVar}
                                            }\
                                    }&
                    \\\quad\quad
                            ); & % end of consumer fork
                    \\\quad\quad
                            \progProofML{
                                    \slPointsTo{\locFlCnt}{\valFlCnt}
                                    \slStar
                                    \quadBack\!\!\!\!\!
                                    \displaystyle
                                    \slBigStar_{ i = 1, \dots, \progProofChanged[\scriptsize]{\valFlCnt-1}}
                                            \!\!\!\quadBack
                                            \slPointsTo[\textstyle\frac{1}{2}]{\glocProdFin{i}}{\slFalse}
                                            \slStar
                                            \slPointsTo[\textstyle\frac{1}{2}]{\glocConsFin{i}}{\slFalse}
                                  	\\
                                    \slStar\,
                                    \assMutex[\progProofChanged{\frac{2\cdot\valFlCnt-2}{2\cdot\numThs}}]{\mutVar}{\assLockInvVar}
                                    \\
                                    \slStar\,
                                    \fixedPredNameFont{obs}
                                    \!\left(\begin{array}{l}
                                            \progProofCancel{\multiset{\sigPop{\valFlCnt}{1}, \dots, \sigPop{\valFlCnt}{\numThs}}\, \msCup}
                                            \\
                                            \multiset{
                                                    \sigPush{i}{k}, \sigPop{i}{k} \ | \
                                                    1 \leq i \leq \valFlCnt-1
                                                    \wedge
                                                    1 \leq k \leq \numThs
                                            }
                                    \end{array}\right)
                            }&
                    \\\quad\quad
                    \cmdAssignToHeap{\flCnt}{\cmdReadHeapLoc{\flCnt}-1};
                            &\proofHintML{
                                    \proofRuleHintML{
	                                    \prLetName 
	                                    \\
	                                    \& \prReadHeapLocName
	                                    \\
	                                    \& \prAssignToHeapName
                                    }
                                    \\
                                    Remember that this
                                    \\
                                    command is syntactic
                                    \\
                                    sugar.
                            }
                    \\\quad\quad
                            \progProofML{
                                    \slPointsTo{\locFlCnt}{\progProofChanged{\valFlCnt-1}}
                                    \slStar
                                    \quadBack\!
                                    \displaystyle
                                    \slBigStar_{i = 1, \dots, \valFlCnt-1}
                                            \!\!\!\!\!\!
                                            \slPointsTo[\textstyle\frac{1}{2}]{\glocProdFin{i}}{\slFalse}
                                            \slStar
                                            \slPointsTo[\textstyle\frac{1}{2}]{\glocConsFin{i}}{\slFalse}
                                   	\\
                                    \slStar\,
                                    \assMutex[\textstyle\frac{2\cdot\valFlCnt-2}{2\cdot\numThs}]{\mutVar}{\assLockInvVar}
                                    \vspace{0.1cm}
                                    \\
                                    \slStar\,
                                    \obs{\multiset{
                                            \sigPush{i}{k}, \sigPop{i}{k} \ | \
                                            1 \leq i \leq \valFlCnt-1
                                            \wedge
                                            1 \leq k \leq \numThs
                                    }}
                            }&
                    \\\quad\quad
                            \progProof{\flInv{\valFlCnt-1}}
                            \phantom{A^{^A}}	% increase distance to line above
                    \\\quad\quad
                            \flCnt > 0
                                    &\proofRuleHint{\prExpName}
                  	\vspace{0.1cm}
                    \\\quad\quad
                            \progProof{\flInv{\valFlCnt-1}}
                                    &\proofRuleHintML{\prViewShiftName 
                                    									\\
                                    									\& \vsSemImpName}
                    \\\quad\quad
                            \progProof{
                                    \slIfElse{\valFlCnt-1 > 0}
                                            {\flInv{\valFlCnt-1}}
                                            {\noObs}
                            }&
                    \\
                    )\ \keyword{do}\ \cmdSkip & % end of fork loop
                    \\
                    \progProof{
                            \progProofCancel{\flInv{\numThs}}
                            \,
                            \progProofNew{\noObs}
                    }&
            \end{array}
            $}
            
            \caption{Verification of program~\ref{gs:appendix:caseStudy:undbounded:code}: Fork loop.}
            \label{gs:appendix:caseStudy:unbounded:forkLoop}
    \end{figure}

    \begin{figure}
    	\caseStudyFont
            % line with approx. width of array
            % helps visualizing whether space to left and to right of array are equal            
%            \noindent\makebox[\linewidth]{\rule{1.35\textwidth}{1pt}} 

            \hspace{0.22cm}
            \makebox[\textwidth][c]
            {$
            % box treats separator center between left and right column as "center of array"
%            \hspace{0.025\textwidth}  % adjusting center
            \hspace{0.04\textwidth}  % adjusting center
            \arxivShift			% adjusting center for arxiv submission
\begin{array}{l  |  p{0.2\pdfpagewidth}}
                    \dots&\referenceHintML{Continuation of 
                    											\\
                    											Fig.~\ref{gs:appendix:caseStudy:unbounded:forkLoop}}
                    \\
                    \lowlightText{\forall \locBuffer, \locMut, \locFlCnt,
                            \idPush{1}{1}, \idPop{1}{1}, \dots, \idPush{\numThs}{\numThs}, \idPop{\numThs}{\numThs},\
                            \glocProdFin{1},\glocConsFin{1}, \dots, \glocProdFin{\numThs},\glocConsFin{\numThs}.
                            \valFlCnt.
                    }&
                    \\
                    \progProof{
                            \obsOf{\sigPush{\valFlCnt}{1}, \dots, \sigPush{\valFlCnt}{\numThs}}
                            \slStar
                            \assMutex[\frac{1}{\numThs}]{\mutVar}{\assLockInvVar}
                            \slStar
                            \slPointsTo[\frac{1}{2}]{\glocProdFin{\valFlCnt}}{\slFalse}
                    }
                            &\proofRuleHintML{\prViewShiftName 
                            									\\
                            									\& \vsSemImpName}
                   	\vspace{0.1cm}
                    \\
\end{array}
            $}
            
            \caption{Verification of program~\ref{gs:appendix:caseStudy:undbounded:code}: Producer loop.}
            \label{gs:appendix:caseStudy:unbounded:producerLoop}
    \end{figure}

    \begin{figure}
    	\caseStudyFont
            % line with approx. width of array
            % helps visualizing whether space to left and to right of array are equal            
%            \noindent\makebox[\linewidth]{\rule{1.35\textwidth}{1pt}} 

            \makebox[\textwidth][c]
            {$
            % box treats separator center between left and right column as "center of array"
%            \hspace{0.025\textwidth}  % adjusting center
            \hspace{0.25\textwidth}  % adjusting center
            \arxivShift			% adjusting center for arxiv submission
\begin{array}{l  |  p{0.3\pdfpagewidth}}
                    \dots&\referenceHintML{Continuation of
                    											\\
                    									 		Fig.~\ref{gs:appendix:caseStudy:unbounded:forkLoop}}.
                    \\
                    \lowlightText{\forall \locBuffer, \locMut, \locFlCnt,
                            \idPush{1}{1}, \idPop{1}{1}, \dots, \idPush{\numThs}{\numThs}, \idPop{\numThs}{\numThs},\
                            \glocProdFin{1},\glocConsFin{1}, \dots, \glocProdFin{\numThs},\glocConsFin{\numThs}.
                            \valFlCnt.
                    }
                    	&\phantom{a}\vspace{0.2cm}
                    \\
                    \progProof{
                            \obsOf{\sigPop{\valFlCnt}{1}, \dots, \sigPop{\valFlCnt}{\numThs}}
                            \slStar
                            \assMutex[\frac{1}{\numThs}]{\mutVar}{\assLockInvVar}
                            \slStar
                            \slPointsTo[\frac{1}{2}]{\glocConsFin{\valFlCnt}}{\slFalse}
                    }
                            &\proofRuleHintML{\prViewShiftName 
                            									\\
                            									\& \vsSemImpName}
                    \\
\end{array}
            $}
            
            \caption{Verification of program~\ref{gs:appendix:caseStudy:undbounded:code}: Consumer loop.}
            \label{gs:appendix:caseStudy:unbounded:consumerLoop}
    \end{figure}

}   % end of local scope for case study
\end{document}

%%%%%%%%%%%%%%%%%%%%%%%%%%%%%%%%%%%%%%%%%%%%%%%%%%
% Keep the following \cleardoublepage at the end of this file, 
% otherwise \includeonly includes empty pages.
\cleardoublepage

% vim: tw=70 nocindent expandtab foldmethod=marker foldmarker={{{}{,}{}}}

% !TeX root = ../../thesis.tex
% !TeX root = ../../thesis.tex
\chapter{Ghost~Signals: Generalised Logic}\label{ch:appendix-ghostSignals-TR}

% TODO: update path
\NewDocumentCommand{\appGsTRSecDir}{}{chapters/appendix-ghostSignals-TR/TR}

This chapter formalizes a generalized version of the logic presented in chapter~\ref{ch:paper-2-ghostSignals} and presents additional case studies.
The contents of this chapter were previously published as stand-alone technical report on Zenodo:
\citationBox
{
	\textbf{T. Reinhard} and Bart Jacobs. 
	\emph{Ghost Signals: Verifying Termination of Busy Waiting (Technical Report)}.
	Zenodo, 2021. 
	https://doi.org/10.5281/zenodo.4775100
}

    \NewDocumentCommand{\exProg}{}{\metaVar{\cmdVar_\text{ex}}}

\section{Universe}
    Throughout this work we assume the existence of the following sets:
    \begin{itemize}
            \item \VarSet: An infinite set of program variables.
            \item \HeapLocSet: An infinite set of heap locations.
            \item \GhostLocSet: An infinite set of ghost locations.
            \item \LevelSet, \levLt: An infinite, well-founded partially ordered set of levels.
            \item \DegreeSet, \degLt: An infinite, well-founded partially ordered set of degrees.
            \item \IDSet: An infinite set of IDs.
            \item \ThreadIDSet: An infinite, totally ordered and well-founded set of thread IDs.
            \item \ValueSet: A set of values which includes:
                    \begin{itemize}
                            \item A unit value $\valUnit \in \ValueSet$
                            \item Booleans $\BoolSet = \setOf{\valTrue, \valFalse} \subset \ValueSet$
                            \item Heap locations $\HeapLocSet \subset \ValueSet$
                    \end{itemize}
            \item \GhostValueSet: A set of ghost values.
            \item \OpSet: A set of operations (i.e. partial functions) on values.
    \end{itemize}
    
    We denote program variables by \varVar, heap locations by \hlocVar, ghost locations by \glocVar, levels by \levVar, degrees by \degVar, IDs by \idVar, thread IDs by \tidVar, values by \valVar, ghost values by \gvalVar, boolean by \slBoolVar and operations by \opVar.

\section{General}
    \begin{definition}[Projections]
            For any Cartesian product $C = \prod_{i \in I} A_i$ and any index $k \in I$, we denote the $k^\text{th}$ projection by
            $\tupleProj[C]{k}: \prod_{i \in I} A_i \rightarrow A_k$.
            We define
            $$
                    \tupleProj[C]{k}[(a_i)_{i\in I}] 
                    \ \ := \ \
                    a_k.
            $$
            In case the domain $C$ is clear from the context, we write \tupleProj{k} instead of \tupleProj[C]{k}.
    \end{definition}

    In the following we define our notion of \emph{bags}, in the literature also referred to as \emph{multisets}.

    \begin{definition}[Bags]\label{gsTR:appendix:def:Bags}
            For any set $X$ we define the set of bags \BagsOf{X} and the set of finite bags \FinBagsOf{X} over $X$ as
            $$
\begin{array}{l c l}
                    \BagsOf{X} &:= 
                            &X  \rightarrow  \N,\\
                    \FinBagsOf{X} &:= 
                            &\setOf{
                                    B \in \BagsOf{X} 
                                    \ \ | \ \
                                    \setOf{x \in B \ | \ B(x) > 0}\
                                    \text{finite}
                            }.
            \end{array}
            $$
            We define union and subtraction of bags as
            $$
            \begin{array}{l c l}
                    (B_1 \msCup B_2)(x)  &:=
                            &B_1(x) + B_2(x),\\
                    (B_1 \setminus B_2)(x) &:=
                            &\max(0,\ B_1(x) - B_2(x)).
            \end{array}
            $$
            For finite bags where the domain is clear from the context, we define the following set-like notation:
            $$
            %
            $$
            We define the following set-like notations for element and subset relationship:
            $$
            \begin{array}{l c l}
                    x \in B &\Leftrightarrow & B(x) > 0,
                    \\
                    B_1 \subseteq B_2 &\Leftrightarrow 
                            &\forall x \in B_1.\ B_1(x) \leq B_2(x) ,
                    \\
                    B_1 \subset B_2 &\Leftrightarrow
                            &\exists C \subseteq B_1.\ 
                                    C \neq \msEmpty
                                    \ \wedge\
                                    B_1 = B_2 \setminus C.
            \end{array}
            $$
            For any bag $B \in \BagsOf{X}$ and predicate $P \subseteq X$ we define the following refinement:
            $$
                    \multiset{x \in B \ | \ P(x)}
                    \ \ := \ \
                    \left\{
                            \begin{array}{l c l l}
                                    x &\mapsto &B(x)  &\text{if}\ P(x),\\
                                    x &\mapsto &0 &\text{otherwise}.
                            \end{array}
                    \right.
            $$
    \end{definition}

    \begin{definition}[Disjoint Union]
            Let $A, B$ be sets.
            We define their disjoint union as
            $$
                    A \disjCup B := A \cup B
            $$
            if $A \cap B = \emptyset$
            and leave it undefined otherwise.
    \end{definition}

\section{Syntax}
    \begin{definition}
            We define the sets of 
            commands~\CmdSet and
            expressions~\ExpSet
            according to the syntax presented in Figure~\ref{gsTR:appendix:fig:Syntax}.

            We define 
            $\cmdVar\ ;\, \nextCmdVar$
            as shorthand for
            \cmdLet{\varVar}{\cmdVar}{\nextCmdVar}
            where \varVar does not occur free in \nextCmdVar but let
            $\cdot\,;\cdot$ bind stronger.
            Further, we define 
            $\expNeq{\expVar}{\nextExpVar}$
            as abbreviation for 
            $\expNot{(\expEq{\expVar}{\nextExpVar})}$.
    \end{definition}

    \begin{figure}
            $$
            \begin{array}{c c c}
                    \valVar \in \ValueSet
                    & \varVar \in \VarSet
                    & \opVar \in \OpSet
            \end{array}
            $$
            $$
            \begin{array}{l c l l}        
                    \expVar \in \ExpSet &::=
                            & \varVar \alt
                                    \valVar \alt
                                    \expEq{\expVar}{\expVar} \alt
                                    \expNot{\expVar} \alt
                                    \opVar(\expListVar)
                    \\
                    \cmdVar \in \CmdSet  &::=
                            &\cmdWhile{\cmdVar}{\cmdSkip} \alt
                                    \cmdFork{\cmdVar}\alt\\
%                            &&\cmdIf{\cmdVar}{\cmdVar}{\cmdVar} \alt\\
                            &&\cmdLet{\varVar}{\cmdVar}{\cmdVar} \alt
                                    \cmdIf{\cmdVar}{\cmdVar} \alt\\
                            &&\cmdAlloc{\expVar} \alt
                                    \cmdReadHeapLoc{\expVar} \alt
                                    \cmdAssignToHeap{\expVar}{\expVar} \alt\\
                            &&\cmdNewMut \alt
                                    \cmdAcquireMut{\expVar} \alt
                                    \cmdReleaseMut{\expVar} \alt\hspace{-0.3cm}{}\\
                            &&\expVar \alt\\
                            &&\lowlightBack{\icmdConsumeItPerm}
                                    &\lowlightBack{intermediate}
                	\\
                			&&&\lowlightBack{representation}
            \end{array}
            $$
            \caption{Syntax.}
            \label{gsTR:appendix:fig:Syntax}
    \end{figure}

\section{Example}
    Figure~\ref{gsTR:appendix:fig:MinimalExample} presents the example program we plan to verify.
    For this example we let \ValueSet include natural numbers.

    \begin{figure}
    {% local scope            
            $$
            \cmdLetML{\x}{\cmdAlloc{0}}
            {
                    \cmdLetML{\m}{\cmdNewMut}
                    {
                            \cmdFork
                            {(
                                    \cmdWhileML
                                    {
                                            (\cmdAcquireMut \m;\\
                                            \phantom{(}
                                            \cmdLetML{\y}{\cmdReadHeapLoc{\x}}
                                            {
                                                    \cmdReleaseMut{\m};\\
                                                    \expEq{\y}{0})
                                            }
                                    }
                                    {
                                            \cmdSkip);
                                    }
                            }\\
                            \cmdAcquireMut{\m};\\
                            \cmdAssignToHeap{\x}{1};\\
                            \cmdReleaseMut{\m}
                    }
            }
            $$

            \caption{Example Program.}
            \label{gsTR:appendix:fig:MinimalExample}
    }% end of local scope
    \end{figure}

\section{Resources}

    In this section we define physical resources.
    We will use the physical resources to define the semantics of our programming language.

    \begin{definition}[Physical Resources]
            We define the set of physical resources \PhysResSet syntactically as follows:
            $$

            $$
    \end{definition}

    \begin{definition}[Physical Heaps]
            We define the set of physical heaps as 
            $$\PhysHeapSet\ :=\ \finPowerSetOf{\PhysResSet}$$
            and the function 
            $\phGetLocs: \PhysHeapSet \rightarrow \finPowerSetOf{\HeapLocSet}$
            mapping physical heaps to the sets of allocated heap locations as
            $$
            %
            $$
            We denote physical heaps by \pheapVar.
    \end{definition}

\section{Semantics}
        \begin{definition}[Evaluation of Closed Expressions]
                We define a partial evaluation function $\evalExp{\cdot}: \ExpSet \partialFunArrow \ValueSet$ on expressions by recursion on the structure of expressions as follows:
                $$
                %
                $$
                
                We identify closed expressions \expVar with their ascribed value \evalExp{\expVar}.
        \end{definition}

        \begin{definition}[Evaluation Context]
                We define the set of evaluation contexts \EvalCtxtSet as follows:
                $$
                %
                $$
                For any $\cmdVar \in \CmdSet$ and $\evalCtxt \in \EvalCtxtSet$, we define 
                $\evalCtxt[\cmdVar] := \subst{\evalCtxt}{\evalCtxtHole}{\cmdVar}$.
        \end{definition}
        
        Note that for every $\cmdVar \in \CmdSet, \evalCtxt \in \EvalCtxtSet$, we have $\evalCtxt[\cmdVar] \in \CmdSet$.

        \begin{definition}[Single Thread Reduction Relation]
                We define a reduction relation \stRedStepSymb for single threads according to the rules presented in Figure~\ref{gsTR:appendix:fig:SingleThreadReductionRules}.
                A reduction step has the form
                $$\stRedStep
                        {\cmdVar}
                        [\nextPheapVar]
                        {\nextCmdVar}
                        [\ftsVar]
                $$
                for a set of forked threads 
                $\ftsVar \subset \CmdSet$ with 
                $\cardinalityOf{\ftsVar} \leq 1$.
                
                For simplicity of notation, we omit \ftsVar if it is clear from the context that no thread is forked and $\ftsVar = \emptyset$.
        \end{definition}
        
        Note that we do not provide a reduction rule for \icmdConsumeItPerm, since we only use it as an intermediate representation for the annotated reduction relation presented in Section~\ref{gsTR:appendix:sec:AnnotatedSemantics}.

        \begin{figure}
                \begin{subfigure}{\textwidth}
                        \begin{mathpar}
                                \stRedEvalCtxt
                                \and
                                \stRedFork
                        \end{mathpar}
                        \caption{Basic Constructs.}
                \end{subfigure}

                \begin{subfigure}{\textwidth}
                        \begin{mathpar}
                                \stRedWhile
                                \and
                                \stRedIfTrue
                                \and
                                \stRedIfFalse
                                \and
                                \stRedLet
                        \end{mathpar}
                        \caption{Control Structures.}
                \end{subfigure}

                \begin{subfigure}{\textwidth}
                        \begin{mathpar}
                                \stRedAlloc
                                \and
                                \stRedReadHeapLoc
                                \and
                                \stRedAssign
                        \end{mathpar}
                        \caption{Heap Access.}
                \end{subfigure}

                \begin{subfigure}{\textwidth}
                        \begin{mathpar}
                                \stRedNewMutex
                                \and
                                \stRedAcquire
                                \and
                                \stRedRelease
                        \end{mathpar}
                        \caption{Mutexes.}
                \end{subfigure}
                
                \caption{Single thread reduction rules.}
                \label{gsTR:appendix:fig:SingleThreadReductionRules}
        \end{figure}

        \begin{definition}[Thread Pools]\label{gsTR:appendix:def:ThreadPools}
                We define the set of thread pools \ThreadPoolSet as the set of finite partial functions mapping thread IDs to threads:
                $$
                        \ThreadPoolSet \ \ := \ \
                                \finPartialFunSet
                                        {\ThreadIDSet}
                                        {(\CmdSet \cup \setOf{\thTerminated})}.
                $$
                The symbol \thTerminated represents a terminated thread.
                We denote thread pools by \tpVar, thread IDs by \tidVar and the empty thread pool by \tpEmpty, i.e.,
                $$
                \begin{array}{l}
                        \tpEmpty \ : \ \ThreadIDSet \finPartialFunArrow  (\CmdSet \cup \setOf{\thTerminated}),
                        \\
                        \defDom{\tpEmpty} \, = \, \emptyset.
                \end{array}
                $$
                We define the operation
                $\tpAddSymb : \ThreadPoolSet \times \setOf{C \subset \CmdSet \ |\ \cardinalityOf{C} \leq 1} \rightarrow \ThreadPoolSet$
%                $\tpAddSymb : \ThreadPoolSet \times \finPowerSetOf{\CmdSet} \partialFunArrow \ThreadPoolSet$
                as follows:
                $$
                \begin{array}{l c l}
                        \tpAdd{\tpVar}{\emptyset} &:=
                                &\tpVar,\\
                        \tpAdd{\tpVar}{\setOf{\cmdVar}} &:=
                                &\tpUpdate{\tpVar}{\newTidVar}{\cmdVar}
                                \quad\text{for}\quad
                                \newTidVar := 
                                        \min(\ThreadIDSet \setminus \defDom{\tpVar}).
                \end{array}
                $$
        \end{definition}

        \begin{definition}[Thread Pool Reduction Relation]
                We define a thread pool reduction relation \tpRedStepSymb according to the rules presented in Figure~\ref{gsTR:appendix:fig:ThreadPoolReductionRules}.
                A reduction step has the form
                $$\tpRedStep{\tpVar}{\nextTpVar}.$$
        \end{definition}

        \begin{figure}
                \begin{mathpar}
                        \tpRedLift
                        \and
                        \tpRedTerm
                \end{mathpar}
                
                \caption{Thread pool reduction rules.}
                \label{gsTR:appendix:fig:ThreadPoolReductionRules}
        \end{figure}

        \begin{definition}[Reduction Sequence]
                Let $(\pheapVar_i)_{i\in \N}$ and $(\tpVar_i)_{i \in \N}$ be infinite sequences of physical heaps and thread pools, respectively.
                
                We call \phtpRSeq a \emph{reduction sequence} if
                there exists a sequence of thread IDs $(\tidVar_i)_{i \in \N}$ such that 
                $$
                \tpRedStep
                        [\pheapVar[i]]
                        {\tpVar[i]}
                        [\tidVar_i]
                        [\pheapVar[i+1]]
                        {\tpVar[i+1]}
                $$
                holds for every $i \in \N$.
        \end{definition}

        \begin{definition}[Fairness]
                We call a reduction sequence \phtpRSeq \emph{fair} iff for all $i \in \N$ and $\tidVar \in \defDom{\tpVar[i]}$ with $\tpVar[i](\tidVar) \neq \thTerminated$ there exists some $k \geq i$ with
                $$
                \tpRedStep
                        [\pheapVar[k]]{\tpVar[k]}
                        [\tidVar]
                        [\pheapVar[k+1]]{\tpVar[k+1]}.
                $$
        \end{definition}

\section{Assertions}
    \begin{definition}[Fractions]
            We define the set of fractions as
            $$
            \begin{array}{l c l}
                    \FracSet &:= &\setOf{ \fracVar \in \Q \ \ | \ \ 0 < \fracVar \leq 1}.
            \end{array}
            $$
    \end{definition}

    \begin{definition}[Thread Phase IDs]
            We define the set of thread phase literals as
            $$
            \begin{array}{l c l}
                    \ThreadPhaseLiteralSet &:= &\setOf{\fplForker, \fplForkee}.
            \end{array}
            $$
            We call a finite sequence of thread phase literals a \emph{phase ID} and denote it by $\fpidVar \in \ThreadPhaseIDSet$.
            We write $\fpidVar_1 \fpLeq \fpidVar_2$ to express that $\fpidVar_1$ is a (non-strict) prefix of $\fpidVar_2$.
    \end{definition}
    
    \begin{definition}
            We define the sets of \emph{ghost signals} \SignalSet, \emph{obligations} \ObSet, \emph{wait permission} \WaitPermSet and \emph{iteration permissions} \ItPermSet as follows:
            $$
            \begin{array}{l c l}
                    \SignalSet &:=
                            &\IDSet \times \LevelSet,\\
                    \ObSet  &:=
                            &(\HeapLocSet \cup \IDSet) \times \LevelSet,\\
                    \WaitPermSet &:=
                            &\ThreadPhaseIDSet \times \IDSet \times \DegreeSet,\\
                    \ItPermSet &:=
                            &\ThreadPhaseIDSet \times \DegreeSet.
            \end{array}
            $$
            We denote ghost signals by \sigVar, obligations by \obVar, and bags of obligations by \obBagVar.
            For convenience of notation we define the selector function:
            $$
            \begin{array}{l c l}
                    \sigGetID{(\idVar, \levVar)} &:= &\levVar.
            \end{array}
            $$
    \end{definition}
    
        \begin{definition}[Assertions]
                We define the set of \emph{assertions} \AssertionSet according to the syntax presented in Figure~\ref{gsTR:appendix:fig:AssertionSyntax}.\footnote{
                        That is, we define \AssertionSet as the least fixpoint of $F$ where 
                        $F(\assSetVar) = 
                                \setOf{\slTrue, \slFalse}
                                \cup
                                \setOf{\neg \assVar \ | \
                                                \assVar \in \assSetVar}
                                \cup
                                \setOf{\assVar_1 \wedge \assVar_2 \ | \ 
                                                \assVar_1, \assVar_2 \in \assSetVar} 
                                \cup
                                \dots
                                \cup
                                \setOf{\bigvee \oAssSetVar \ | \
                                                \oAssSetVar \subseteq \assSetVar}
                                \cup
                                \dots
                        $.
                        Since $F$ is a monotonic function over a complete lattice, it has a least fixpoint according to the Knaster-Tarski theorem~\cite{Tarski1955KnasterTarskiTheorem}.
                }
%                We denote the set of variables occurring \emph{free} in an assertion $\assVar \in \AssertionSet$ by $\assFreeVarSet{\assVar}$.
                Further, we define implication and equivalence as the usual abbreviations:
                $$

                $$
                Let \assSeq  be a family of assertions indexed by some set \assIndexSetVar.
                We define quantification over \assIndexSetVar as the following abbreviations:
                $$
                %
                $$
                We omit the index set $I$ when its choice becomes clear from the context and write 
                $\exists \assIndexVar.\, \assFam[\assIndexVar]$
                and
                $\forall \assIndexVar.\, \assFam[\assIndexVar]$
                instead of                
                $\exists \assIndexVar \in \assIndexSetVar.\, \assFam[\assIndexVar]$
                and
                $\forall \assIndexVar \in \assIndexSetVar.\, \assFam[\assIndexVar]$,
                respectively.
        \end{definition}

        \begin{figure}
                $$
                %
                $$
                
                \caption{Assertion syntax.}
                \label{gsTR:appendix:fig:AssertionSyntax}
        \end{figure}

    \begin{definition}[Logical Resources]
            We define the set of logical resources \LogResSet syntactically as follows:
            $$
            %
            $$
            Further, we define the functions
            $\lresGetHLocs: \LogResSet \rightarrow \HeapLocSet$
            and
            $\lresGetGLocs: \LogResSet \rightarrow \finPowerSetOf{\GhostLocSet}$
            mapping logical resources to their respective (either empty or singleton) set of involved heap locations and ghost locations, respectively, as
            $$
            %
            $$
    \end{definition}
    
    \begin{definition}[Mutexes]
            We define the set of mutexes as $\MutexSet := \HeapLocSet \times \LevelSet$
            and denote mutexes by \mutVar.
            For convenience of notation we define the selector function
            $$
            \begin{array}{l c l}
                    \mutGetLoc{(\hlocVar, \levVar)} &:= &\hlocVar.
            \end{array}
            $$
    \end{definition}

    \begin{definition}[Logical Heaps]
            We define the set of logical heaps as
            $$
            %
            $$
            We define the empty logical heap \lhEmpty as the constant zero function
            $$
                    \lhEmpty : \lresVar \mapsto 0.
            $$
            We denote logical heaps by \lheapVar, point-wise addition by $+$ and multiplication with non-negative rationals by $\cdot$, i.e.,
            $$
            %
            $$
            for $q\in \Q$ with $q\geq 0$.
            For convenience of notation we represent logical heaps containing finitely many resources by sets of resources and define left-associative functions
            $\lhPlusRes,\, \lhMinusRes: \LogHeapSet \rightarrow \LogResSet \rightarrow \LogHeapSet$
            as follows
            $$
            %
                            \right.
                    \\
                    \lheapVar \lhPlusRes \lresVar &:=
                            &\lhUpdate
                                    {\lheapVar}
                                    {\lresVar}
                                    {\lheapVar(\lresVar) + 1},
                    \\
                    \lheapVar \lhMinusRes \lresVar &:=
                            &\lhUpdate
                                    {\lheapVar}
                                    {\lresVar}
                                    {\max(0,\, \lheapVar(\lresVar) - 1)}.
            \end{array}
            $$
            We give $\cdot$ a higher precedence than $+$, \lhPlusRes and \lhMinusRes.
            
            Further, we define the function
            $\lhGetGLocs: \LogHeapSet \rightarrow \powerSetOf{\GhostLocSet}$
            mapping logical heaps to their respective set of allocated ghost locations as
            $$
                    \lhGetGLocs[\lheapVar] 
                    \ \ := \ \
                    \bigcup_{\substack{
                                    \lresVar \in \LogResSet\\
                                    \lheapVar(\lresVar) > 0
                            }}
                            \lresGetGLocs[\lresVar].
            $$
            
            We call a logical heap \lheapVar \emph{complete} and write \lhComplete{\lheapVar} if it contains exactly one obligations chunk and exactly one phase chunk, i.e.,
            if there exist a bag of obligations \obBagVar and a phase ID \fpidVar with
            $\lheapVar(\lresObs{\obBagVar} ) = 1$
            and
            $\lheapVar(\lresThreadPhase{\fpidVar}) = 1$
            and if there do not exist any bag of obligations $\obBagVar'$ nor any phase ID \nextFpidVar with 
            (i)~$\obBagVar \neq \obBagVar'$ and
            $\lheapVar(\lresObs{\obBagVar'}) > 0$
            or with
            (ii)~$\fpidVar \neq \nextFpidVar$ and
            $\lheapVar(\lresThreadPhase{\nextFpidVar}) > 0$.
            
            We call a logical heap \lheapVar \emph{finite} and write \lhFinite{\lheapVar} if 
            it contains only finitely many resources, i.e.,
            if the set 
            \setOf{\lresVar \in \LogResSet \ \ | \ \ \lheapVar(\lresVar) > 0}
            is finite.

            We call a logical heap \lheapVar \emph{consistent} and write \lhConsistent{\lheapVar} if 
            (i)~it contains only full phase, obligations, wait and iteration permission chunks, i.e.,
            if
            $$
            \begin{array}{l l l}
                    \lheapVar(\lresThreadPhase{\fpidVar})
                            &\in &\N,\\
                    \lheapVar(\lresObs{\obBagVar}) 
                            &\in &\N,\\
                    \lheapVar(\lresWaitPerm{\fpidVar}{\idVar}{\degVar}) 
                            &\in &\N,\\
                    \lheapVar(\lresItPerm{\fpidVar}{\degVar}) 
                            &\in &\N
            \end{array}
            $$
            holds for all $\fpidVar \in \ThreadPhaseIDSet$, $\obBagVar \in \BagsOf{\ObSet}$, $\idVar \in \IDSet$ and $\degVar \in \DegreeSet$
            and if
            (ii)~heap locations and ghost locations are unique in \lheapVar, i.e.,
            if there are no $\lresVar[1], \lresVar[2] \in \LogResSet$ with 
            $\lresVar[1] \neq \lresVar[2]$,
            $\lheapVar(\lresVar[1]) > 0$,
            $\lheapVar(\lresVar[2]) > 0$ 
            and with
            $\lresGetHLocs[\lresVar[1]] \cap \lresGetHLocs[\lresVar[2]] \neq \emptyset$
            or
            $\lresGetGLocs[\lresVar[1]] \cap \lresGetGLocs[\lresVar[2]] \neq \emptyset$.
            
            To simplify the specification of logical heaps containing only a single obligations chunk with certain properties, we introduce the abbreviation
            $$
                    (\lhHasObs)
                    \ \ := \ \
                    (
                            \lhComplete{\lheapVar} 
                            \ \wedge \
                            \lheapVar(\lresObs{\obBagVar}) = 1
                    ).
            $$
    \end{definition}

    \begin{definition}[Assertion Model Relation]
            We define a model relation 
            $\assModelsSymb \,\subset \, \LogHeapSet \times \AssertionSet$
            for assertions by recursion on the structure of assertions according to the rules presented in Figure~\ref{gsTR:appendix:fig:AssertionModelRelation}.
            We write \assModels{\lheapVar}{\assVar} to express that logical heap \lheapVar models assertion \assVar and
            \assNotModels{\lheapVar}{\assVar} to express that \assModels{\lheapVar}{\assVar} does not hold.
    \end{definition}

        \begin{figure}
                $$

                $$
                
                \caption{Assertion model relation.}
                \label{gsTR:appendix:fig:AssertionModelRelation}
        \end{figure}

\section{Proof Rules}
    \begin{definition}[Level Ascriptions]
            We define a function 
            $\levName : 
                (\IDSet \cup \HeapLocSet) \times \LevelSet \rightarrow \LevelSet$ as
            $$
            %
            $$
    \end{definition}

    \begin{definition}[View Shift]
            We define a view shift relation 
            $\viewShiftSymb \ \subset\ \AssertionSet \times \AssertionSet$ 
            according to the rules presented in Figure~\ref{gsTR:appendix:fig:ViewShiftRules}.
    \end{definition}

    \begin{figure}
            \begin{mathpar}
                    \vsSemImp
                    \and
                    \vsTrans
                    \and
                    \vsOr
                    \and
                    \vsNewSignal
                    \and
                    \vsSetSignal
                    \and
%                    \vsSigIdUnique
%                    \and
                    \vsWaitPerm
                    \and
                    \vsWait
                    \and
                    \vsSpecItPerm
                    \and
                    \vsSpecWaitPerm
                    \and
                    \vsWeakItPerm
                    \and
                    \vsMutInit
                    \and
                    \vsNewGhostCell
                    \and
                    \vsMutateGhostCell
            \end{mathpar}
            
            \caption{View shift rules.}
            \label{gsTR:appendix:fig:ViewShiftRules}
    \end{figure}

    \begin{definition}[Proof Relation]
            We define a proof relation 
            $\htProvesSymb \ \subset \ \AssertionSet \times \CmdSet \times (\ValueSet \rightarrow \AssertionSet)$
            according to the rules presented in Figures~\ref{gsTR:appendix:fig:ProofRulesPartOne} and~\ref{gsTR:appendix:fig:ProofRulesPartTwo}.
    \end{definition}
    
    Note that our proof rules do not allow us to reason about the command \icmdConsumeItPerm, since we only use it as an intermediate representation during reduction.

    \begin{figure}
            \begin{subfigure}{\textwidth}
                    \begin{mathpar}
                            \prFrame
                            \and
                            \prViewShift
                            \and
                            \prSimpleViewShift
%                            \and
%                            \prSeq
%                            \and
%                            \prSkip
                            \and
                            \prExp
                            \and
                            \prExists
                            \and
%                            \prOr
%                            \and
                            {
                            	\hspace{-0.3cm}
                            	\prFork
                            }
                    \end{mathpar}
                    \subcaption{Basic proof rules.}
            \end{subfigure}

            \begin{subfigure}{\textwidth}
                    \begin{mathpar}
                            \prIf
                            \and
                            {
                            	\hspace{-0.3cm}
	                            \prWhile
	                        }
                            \and
                            \prLet
                    \end{mathpar}
                    \subcaption{Control structures.}
            \end{subfigure}

            \caption{Proof rules (part 1).}
            \label{gsTR:appendix:fig:ProofRulesPartOne}
    \end{figure}

    \begin{figure}
            \begin{subfigure}{\textwidth}
                    \begin{mathpar}
                            \prAcquire
                            \and
                            \prRelease
                            \and
                            \prNewMutex
                    \end{mathpar}
                    \subcaption{Mutexes.}
            \end{subfigure}

            \begin{subfigure}{\textwidth}
                    \begin{mathpar}
                            \prAlloc
                            \and
                            \prReadHeapLoc
                            \and
                            \prAssignToHeap
                    \end{mathpar}
                    \subcaption{Heap access.}
            \end{subfigure}

            \caption{Proof rules (part 2).}
            \label{gsTR:appendix:fig:ProofRulesPartTwo}
    \end{figure}

    \begin{lemma}
            We can derive the proof rule presented in Figure~\ref{gsTR:appendix:fig:DerivedProofRules}.
    \end{lemma}
    \begin{proof}
            Trivial.
    \end{proof}
    
    \begin{figure}
            \begin{mathpar}
                    \prSimpleWhile
            \end{mathpar}
            
            \caption{Derived proof rule.}
            \label{gsTR:appendix:fig:DerivedProofRules}
    \end{figure}

\section{Annotated Semantics}\label{gsTR:appendix:sec:AnnotatedSemantics}
     \begin{definition}[Annotated Resources]
                We define the set of annotated resources \AnnoResSet as follows:
                $$

                $$
        \end{definition}

        \begin{definition}[Annotated Heaps]
                We define the set of annotated heaps as
                $$
                %
                $$
                the function 
                $\ahGetLocs: \AnnoHeapSet \rightarrow \finPowerSetOf{\HeapLocSet}$
                mapping annotated heaps to the sets of allocated heap locations as
                $$
                %
                $$
                and the function
                $\ahGetIDs : \AnnoHeapSet \rightarrow \finPowerSetOf{\IDSet}$
                mapping annotated heaps to sets of allocated signal IDs as
                $$
                %
                $$
                We denote annotated heaps by \aheapVar.
                
                We call an annotated heap \aheapVar \emph{finite} and write \ahFinite{\aheapVar} if there exists no chunk
                $\aresUnlocked{(\hlocVar, \levVar)}{\assVar}{\lheapVar} \in \aheapVar$
                for which \lhFinite{\lheapVar} does not hold.
        \end{definition}

        \begin{definition}[Annotated Single Thread Reduction Relation]
                We define a reduction relation \astRedStepSymb for annotated threads according to the rules presented in Figures~\ref{gsTR:appendix:fig:AnnotatedSingleThreadReductionRulesPartOne} and~\ref{gsTR:appendix:fig:AnnotatedSingleThreadReductionRulesPartTwo}.
                A reduction step has the form
                $$
                        \astRedStep
                                {\cmdVar}
                                [\tidVar]
                                [\nextAheapVar]
                                [\nextLheapVar]
                                {\nextCmdVar}
                                [\aftsVar]
                $$
                for a set of annotated forked threads
                $\aftsVar \subset \LogHeapSet \times \CmdSet$
                with $\cardinalityOf{\aftsVar} \leq 1$.

                It indicates that given annotated heap \aheapVar and a logical heap \lheapVar, 
                command \cmdVar can be reduced to annotated heap \nextAheapVar, logical heap \nextLheapVar and command \nextCmdVar.
                The either empty or singleton set \aftsVar represents whether a new thread is forked in this step.
                
                For simplicity of notation we omit \aftsVar if it is clear from the context that no thread is forked and $\aftsVar = \emptyset$.
        \end{definition}

        \begin{figure}
                \begin{subfigure}{\textwidth}
                        \begin{mathpar}
                                \astRedEvalCtxt
                                \and
                                {\hspace{-1.3cm}\astRedFork}
                        \end{mathpar}
                        \caption{Basic constructs.}
                \end{subfigure}

                \begin{subfigure}{\textwidth}
                        \begin{mathpar}
                        		{
                        			\hspace{-0.3cm}
	                                \astRedWhile
	                            }
                                \and
                                \astRedIfTrue
                                \and
                                \astRedIfFalse
                                \and
                                \astRedLet
                        \end{mathpar}
                        \caption{Control structures.}
                \end{subfigure}
                
                \begin{subfigure}{\textwidth}
                        \begin{mathpar}
                                \astRedConsumeItPerm
                        \end{mathpar}
                        \caption{Intermediate representation.}
                \end{subfigure}

                \begin{subfigure}{\textwidth}
                        \begin{mathpar}
                                \astRedAlloc
                                \and
                                \astRedReadHeapLoc
                                \and
                                \astRedAssign
                        \end{mathpar}
                        \caption{Heap access.}
                \end{subfigure}

                \caption{Annotated single thread reduction rules (part 1).}
                \label{gsTR:appendix:fig:AnnotatedSingleThreadReductionRulesPartOne}
        \end{figure}

        \begin{figure}
                \begin{subfigure}{\textwidth}
                        \begin{mathpar}
                                \astRedNewMutex
                                \and
                                \astRedAcquire
                                \and
                                \astRedRelease
                        \end{mathpar}
                        \caption{Mutexes.}
                \end{subfigure}
                
                \caption{
                        Annotated single thread reduction rules (part 2).
                }
                \label{gsTR:appendix:fig:AnnotatedSingleThreadReductionRulesPartTwo}
        \end{figure}

        \begin{definition}[Annotated Thread Pools]
                We define the set of annotated thread pools \AnnoThreadPoolSet as the set of finite partial functions mapping thread IDs to annotated threads:
                $$
                \begin{array}{l c l}
                        \AnnoThreadPoolSet &:=
                                &\finPartialFunSet
                                        {\ThreadIDSet}
                                        {
                                                \LogHeapSet \times
                                                (\CmdSet \cup \setOf{\thTerminated})
                                        }.
                \end{array}
                $$
                We denote annotated thread pools by \atpVar and the empty thread pool by \atpEmpty, i.e.,
                $$
                \begin{array}{l}
                        \atpEmpty \ : \
                                \ThreadIDSet \finPartialFunArrow 
                                \LogHeapSet \times (\CmdSet \cup \setOf{\thTerminated}),
                        \\
                        \defDom{\atpEmpty} \, = \, \emptyset.
                \end{array}
                $$
                We define the modification operations \atpAddSymb and \atpTermSymb analogously to \tpAddSymb and \tpTermSymb, respectively, cf. Definition~\ref{gsTR:appendix:def:ThreadPools}.
                
                For convenience of notation we define selector functions for annotated threads as
                $$
                \begin{array}{l c l}
                        \atGetHeap{(\lheapVar, \cmdVar)} &:=
                                &\lheapVar,
                        \\
                        \atGetCmd{(\lheapVar, \cmdVar)} &:=
                                &\cmdVar.
                \end{array}
                $$
        \end{definition}

        \begin{definition}[Ghost Reduction Relation]
                We define a thread pool reduction relation \gtpRedStepSymb according to the rules presented in Figures~\ref{gsTR:appendix:fig:GhostThreadPoolReductionRulesPartOne} and~\ref{gsTR:appendix:fig:GhostThreadPoolReductionRulesPartTwo} to express ghost steps.
                A ghost reduction step has the form
                $$
                        \gtpRedStep
                                {\atpVar}
                                {\nextAtpVar}.
                $$
                We denote its reflexive transitive closure by \gtpRedStepMultSymb.
        \end{definition}

        \begin{figure}
                \begin{mathpar}
                        \gtpRedNewSignal
                        \and
                        \gtpRedSetSignal
                        \and
                        \gtpRedWaitPerm
                        \and
                        \gtpRedWait
                        \and
                        \gtpRedSpecItPerm
                        \and
                        \gtpRedSpecWaitPerm
                        \and
                        \gtpRedWeakItPerm
                        \and
                        \gtpRedMutInit
                \end{mathpar}

                \caption{Ghost thread pool reduction rules (part 1).}
                \label{gsTR:appendix:fig:GhostThreadPoolReductionRulesPartOne}
        \end{figure}

        \begin{figure}
                \begin{mathpar}
                        \gtpRedNewGhostCell
                        \and
                        \gtpRedMutateGhostCell
                \end{mathpar}

                \caption{Ghost thread pool reduction rules (part 2)}
                \label{gsTR:appendix:fig:GhostThreadPoolReductionRulesPartTwo}
        \end{figure}

        \begin{definition}[Non-ghost Thread Pool Reduction Relation]
                We define a thread pool reduction relation \rtpRedStepSymb according to the rules presented in Figure~\ref{gsTR:appendix:fig:RealThreadPoolReductionRules} to express real reduction steps.
                A reduction step has the form
                $$\rtpRedStep
                        {\atpVar}
                        {\nextAtpVar}.
                $$
        \end{definition}

        \begin{figure}
                \begin{mathpar}
                        \rtpRedLift
                        \and
                        \rtpRedTerm
                \end{mathpar}
                
                \caption{Non-ghost thread pool reduction rules.}
                \label{gsTR:appendix:fig:RealThreadPoolReductionRules}
        \end{figure}

        \begin{definition}[Annotated Thread Pool Reduction Relation]
                We define the annotated thread pool reduction relation \atpRedStepSymb as
                $$
                \begin{array}{l c l}
                        \atpRedStepSymb &:= &\gtpRedStepSymb \cup \rtpRedStepSymb.
                \end{array}
                $$
        \end{definition}

        \begin{definition}[Annotated Reduction Sequence]\label{gsTR:appendix:def:AnnotatedRedSeq}
                Let $(\aheapVar_i)_{i\in \N}$ and $(\atpVar_i)_{i \in \N}$ be infinite sequences of annotated heaps and annotated thread pools, respectively.
                Let
                $\sigAnno : \N \partialFunArrow \SignalSet$ be a partial function mapping indices to signals.
                
                We call \ahtpRSeqFull an \emph{annotated reduction sequence} if
                there exists a sequence of thread IDs $(\tidVar_i)_{i \in \N}$ such that the following holds for every $i \in \N$:
                \begin{itemize}
                        \item
                \atpRedStep
                        [\aheapVar[i]]
                        {\atpVar[i]}
                        [\tidVar_i]
                        [\aheapVar[i+1]]
                        {\atpVar[i+1]}
                        \item If this reduction step results from an application of \gtpRedWaitName to some signal \sigVar, then
                                $\sigAnno[i] = \sigVar$ 
                                holds and otherwise 
                                $\sigAnno[i] = \funUndefVal$.
                \end{itemize}
                In case the signal annotation \sigAnno is  clear from the context or not relevant, we omit it and write \ahtpRSeq instead of \ahtpRSeqFull.
                
                We call $(\aheapVar[i], \atpVar[i])$ an \emph{annotated machine configuration}.
        \end{definition}

        \begin{lemma}[Preservation of Finiteness]\label{gsTR:appendix:lem:PreservationOfFiniteness}
                Let \ahtpRSeq be an annotated reduction sequence with 
                \ahFinite{\aheapVar[0]}
                and
                \lhFinite{\atGetHeap{\atpVar[0](\tidVar)}}
                for all $\tidVar \in \defDom{\atpVar[0]}$.
                
                Then,
                \lhFinite{\atGetHeap{\atpVar[i](\tidVar)}}
                holds for all $i \in \N$ and all $\tidVar \in \defDom{\atpVar[i]}$.
        \end{lemma}
        \begin{proof}
                Proof by induction on $i$.
        \end{proof}

        \begin{lemma}[Preservation of Completeness]\label{gsTR:appendix:lem:PreservationOfCompleteness}
                Let \ahtpRSeq be an annotated reduction sequence with
                \lhComplete{\atGetHeap{\atpVar[0](\tidVar)}}
                for all $\tidVar \in \defDom{\atpVar[0]}$. 
                Furthermore, let there be no chunk
                $\aresUnlocked{\mutVar}{\assLockInvVar}{\lheapLockInvVar} \in \aheapVar[0]$
                such that 
                \phantom{aaa}
                $\lheapLockInvVar(\lresThreadPhase{\fpidVar}) > 0$
                or
                $\lheapLockInvVar(\lresObs{\obBagVar}) > 0$
                holds for any \fpidVar, \obBagVar.

                Then, 
               \lhComplete{\atGetHeap{\atpVar[i](\tidVar)}}
                holds for every $i \in \N$ and every $\tidVar \in \defDom{\atpVar[i]}$.
        \end{lemma}
        \begin{proof}
                Proof by induction on $i$.
        \end{proof}

        \begin{definition}[Fairness of Annotated Reduction Sequences]
                We call an annotated reduction sequence \ahtpRSeq \emph{fair} iff for all $i \in \N$ and $\tidVar \in \defDom{\atpVar[i]}$ with
                $\atGetCmd{\atpVar[i](\tidVar)} \neq \thTerminated$
                there exists some $k \geq i$ with
                $$
                \rtpRedStep
                        [\aheapVar[k]]{\atpVar[k]}
                        [\tidVar]
                        [\aheapVar[k+1]]{\atpVar[k+1]}.
                $$
        \end{definition}

        Every thread of an annotated thread pool is annotated by a thread-local logical heap that expresses which resources are owned by this thread.
        In the following we define a function to extract the logical heap expressing which resources are owned by threads of a thread pool (i.e. the sum of all thread-local logical heaps).
        
        \begin{definition}
                We define the function 
                $\atpOwnedResHeap : \AnnoThreadPoolSet \rightarrow \LogHeapSet$
                mapping annotated thread pools to logical heaps as follows:
                $$
                \begin{array}{l c l}
                        \atpVar &\mapsto
                                &\quadBack\quadBack
                                        \bigSum[\quad\tidVar\, \in\, \defDom{\atpVar}]
                                        \quadBack\!\!
                                        \atGetHeap{\atpVar(\tidVar)}
                \end{array}
                $$
        \end{definition}

        Annotated resources representing unlocked locks, i.e.,
        \aresUnlocked{\mutVar}{\assVar}{\lheapVar[\assVar]},
        contain a logical heap \lheapVar[\assVar] that expresses which resources are protected by this lock.
        In the following, we define a function that extracts a logical heap from an annotated heap \aheapVar expressing which resources are protected by unlocked locks in \aheapVar.
        
        \begin{definition}
                We define the function 
                $\ahProtectedResHeap : \AnnoHeapSet \rightarrow \LogHeapSet$
                mapping  annotated heaps to logical heaps as follows:
                
                For any annotated heap \aheapVar let 
                $$

                $$
                be the auxiliary set aggregating all logical heaps corresponding to lock invariants of unlocked locks stored in \aheapVar.
                We define \ahProtectedResHeap as
                $$
                        \aheapVar \ \ \mapsto 
                                \bigSum[\lheapLockInvVar
                                                    \,\in\, \multiLetterSetFont{LockInvs}(\aheapVar)]
                                        \quadBack\quadBack\lheapLockInvVar.
                $$
                
        \end{definition}

        \begin{definition}[Compatibility of Annotated and Logical Heaps]
                We inductively define a relation
                $\alhCorrespSymb\, \subset\, \AnnoHeapSet \times \LogHeapSet$
                between annotated and logical heaps such that the following holds
                $$
                %
                $$
                where $\aheapVar \in \AnnoHeapSet$ and $\lheapVar \in \LogHeapSet$ are annotated and logical heaps with
                $\hlocVar, \mutGetLoc{\mutVar} \not\in \ahGetLocs[\aheapVar]$,
                $\sigGetID{\sigVar} \not\in \ahGetIDs[\aheapVar]$
                and \alhCorresp.
        \end{definition}

        We consider a machine configuration $(\aheapVar, \atpVar)$ to be \emph{consistent} if it fulfils three criteria:
        (i)~Every thread-local logical heap is consistent, i.e.,
%        \lhConsistent{\atGetHeap{\atpVar(\tidVar)}}
%        holds for all $\tidVar \in \defDom{\atpVar}$.
        for all used thread IDs \tidVar, \atGetHeap{\atpVar(\tidVar)} only stores full phase, obligations, wait permission and iteration permission chunks.
        (ii)~Every logical heap protected by an unlocked lock in \aheapVar is consistent.
        (iii)~\aheapVar is compatible with the logical heap that represents
        (a)~the resources owned by threads in \atpVar and
        (b)~the resources protected by unlocked locks stored in \aheapVar.
        
        \begin{definition}[Consistency of Annotated Machine Configurations]
                We call an annotated machine configuration $(\aheapVar, \atpVar)$ \emph{consistent} and write\\
                \amConfConsistent{\aheapVar}{\atpVar}
                if all of the following hold:
                \begin{itemize}
                        \item \lhConsistent{\atGetHeap{\atpVar(\tidVar)}}
                                    for all $\tidVar \in \defDom{\atpVar}$,
                        \item
                                $
                                        \forall \mutVar.\
                                        \forall \assLockInvVar.\
                                        \forall \lheapLockInvVar.\ 
                                                \aresUnlocked{\mutVar}{\assLockInvVar}{\lheapLockInvVar} \in \aheapVar
                                                \ \rightarrow\
                                                \lhConsistent{\lheapLockInvVar},
                                $
                        \item
                                $
                                        \aheapVar
                                        \alhCorrespSymb\
                                        \atpOwnedResHeap[\atpVar] + 
                                        \ahProtectedResHeap[\aheapVar].
                                $
                \end{itemize}
        \end{definition}

        \begin{lemma}[Preservation of Consistency]
                Let \ahtpRSeq be an annotated reduction sequence with \amConfConsistent{\aheapVar[0]}{\atpVar[0]}.
                Then, \amConfConsistent{\aheapVar[i]}{\atpVar[i]} holds for every $i \in \N$.
        \end{lemma}
        \begin{proof}
                Proof by induction on $i$.
        \end{proof}

\section{Hoare Triple Model Relation}\label{gsTR:appendix:sec:HoareTripleModelRelation}
    \begin{definition}[Command Annotation]
            We define the predicate
            $\cmdAnnotName \subset \CmdSet \times \CmdSet$
            such that \cmdAnnot{\nextCmdVar}{\cmdVar} holds iff \nextCmdVar results from \cmdVar by removing all occurrences of \icmdConsumeItPerm.
    \end{definition}

    \begin{definition}[Thread Pool Annotation]
            We define a predicate
            $\tpAnnotName \subset \AnnoThreadPoolSet \times \ThreadPoolSet$
            such that:
            $$

            $$
    \end{definition}

    \begin{definition}[Compatibility of Annotated and Physical Heaps]
            We inductively define a relation
            $\aphCorrespSymb\ \subset \AnnoHeapSet \times \PhysResSet$
            between annotated and physical heaps such that the following holds:
            $$
            %
            $$
            where 
            $\aheapVar \in \AnnoHeapSet$ 
            and
            $\pheapVar \in \PhysHeapSet$ 
            are annotated and physical heaps with
            \aphCorresp{\aheapVar}{\pheapVar}.
    \end{definition}

    \begin{definition}[Safety]
            We define the safety predicate
            $\safeName \subseteq \LogHeapSet \times \CmdSet$
            coinductively as the greatest solution (with respect to $\subseteq$) of the following equation:
            $$
            %
            $$
    \end{definition}

    \begin{definition}[Hoare Triple Model Relation]
            We define the model relation for Hoare triples
            $\htModelsSymb\ \subset \AssertionSet \times \CmdSet \times (\ValueSet \rightarrow \AssertionSet)$
            such that:
            $$
            %
            \end{array}
            $$
    \end{definition}
    
    We can instantiate context \evalCtxt in above definition to \cmdLet{\varVar}{\evalCtxtHole}{\valUnit}, 
    which yields the consequent
    \safe
            {\lheapPreVar + \lheapFrameVar}
            {\cmdLet{\varVar}{\cmdVar}{\valUnit}}.
    Note that this implies
    \safe
            {\lheapPreVar + \lheapFrameVar}
            {\cmdVar}.

    \begin{lemma}[Hoare Triple Soundness]\label{gsTR:appendix:lem:HoareTripleSoundness}
            Let 
            \htProves{\assPreVar}{\cmdVar}{\assPostVar} hold, 
            then also
            \htModels{\assPreVar}{\cmdVar}{\assPostVar}
            holds.
    \end{lemma}
    \begin{proof}
            Proof by induction on the derivation of \htProves{\assPreVar}{\cmdVar}{\assPostVar}.
    \end{proof}

    \begin{restatable}[Soundness]{theorem}{theoSoundness}\label{gsTR:appendix:theo:Soundness}
            Let
            $$
                    \htProves
                            {
                                    \assPhase
                                    \slStar
                                    \noObs \slStar 
                                    \!\!\!\!
                                    {\displaystyle
                                    \slBigStar^{i \,=\, 1, \dots, N}}
                                    \quadBack\
                                    \itPerm[\fpidVar, \degVar_i]
                            }
                            {\cmdVar}
                            {\noObs}
            $$
            hold.
             There exists no fair, infinite reduction sequence \phtpRSeq with
            $\pheapVar[0] = \emptyset$ and 
            $\tpVar[0] = 
                    \setOf{(\tidStartVar, \cmdVar)}
            $
            for any choice of \tidStartVar.
    \end{restatable}

\section{Soundness}\label{gsTR:appendix:sec:Soundness}
    In this section, we prove the soundness theorem~\ref{gsTR:appendix:theo:Soundness}.

    \begin{restatable}[Construction of Annotated Reduction Sequences]{lemma}{lemConstructionOfAnnotatedRedSeq}\label{gsTR:appendix:lem:ConstructionOfAnnotatedRedSeq}
%    \begin{lemma}[Construction of Annotated Reduction Sequences]
%    \label{gsTR:appendix:lem:ConstructionOfAnnotatedRedSeq}
           Suppose we can prove
           \htModels{\assPreVar}{\cmdVar}{\noObs}.
           Let \lheapPreVar be a logical heap with
           \assModels{\lheapPreVar}{\assPreVar} and \lhComplete{\lheapPreVar}
           and \aheapVar[0] an annotated heap with
           \alhCorresp[\aheapVar[0]][\lheapPreVar].
           Let \phtpRSeq be a fair plain reduction sequence with
           \aphCorresp{\aheapVar[0]}{\pheapVar[0]} and
           $\tpVar[0] = \setOf{(\tidVar_0, \cmdVar)}$
           for some thread ID $\tidVar_0$ and command \cmdVar.

           Then, there exists a fair annotated reduction sequence \ahtpRSeq with
           $\atpVar = \setOf{(\tidVar_0,\, (\lheapPreVar, \cmdVar))}$
           and \amConfConsistent{\aheapVar[i]}{\atpVar[i]} for all $i \in \N$.
%    \end{lemma}
    \end{restatable}
    \begin{proof}
            We can construct the annotated reduction sequence inductively from the plain reduction sequence.
    \end{proof}

    \begin{definition}[Program Order Graph]\label{gsTR:appendix:def:ProgramOrderGraph}
    { % local scope
            \NewDocumentCommand{\RuleNameSet}{}{\metaVar{N^r}}
            \NewDocumentCommand{\AnnoRuleNameSet}{}{\metaVar{N^\contextIdFont{a}}}
            
            \NewDocumentCommand{\ruleNameVar}{o}
                    {\metaVar{
                            n^\contextIdFont{a}
                            \IfValueT{#1}{_{#1}}
                    }}
            \NewDocumentCommand{\stRuleNameVar}{o}
                    {\metaVar{
                            n^\contextIdFont{st}
                            \IfValueT{#1}{_{#1}}
                    }}

            \NewDocumentCommand{\baseNode}{}{\metaVar{a}} % i
            \NewDocumentCommand{\succNode}{}{\metaVar{b}} % j
            \NewDocumentCommand{\auxNode}{}{\metaVar{k}} % k
            \NewDocumentCommand{\genNode}{}{\metaVar{\ell}} % l
        
            Let \ahtpRSeqFull be an annotated reduction sequence.
            Let \RuleNameSet be the set of names referring to reduction rules defining the relations \rtpRedStepSymb, \gtpRedStepSymb and \astRedStepSymb.
            We define the set of annotated reduction rule names \AnnoRuleNameSet where \gtpRedWaitName is annotated by signals as
            $$
            \begin{array}{l c l}
                    \AnnoRuleNameSet \!\!&:=\!
                            &(\RuleNameSet \setminus \setOf{\gtpRedWaitName})\\
                            &&\!\phantom{\RuleNameSet}\
                            \cup\
                            (\setOf{\gtpRedWaitName} 
                                    \times \SignalSet).
            \end{array}
            $$
            We define the program order graph
            $\progOrdGraph[\ahtpRSeqFull] = (\N, \EdgeSetVar)$
            with root 0 where 
            $\EdgeSetVar \subset
                    \N \times \ThreadIDSet \times \AnnoRuleNameSet \times \N
            $.

            A node $\baseNode \in \N$ corresponds to the sequence's $\baseNode^\text{th}$ reduction step, i.e.,
            \atpRedStep
                    [\aheapVar[\baseNode]]{\atpVar[\baseNode]}
                    [\tidVar]
                    [\aheapVar[\baseNode+1]]{\atpVar[\baseNode+1]}
            for some $\tidVar \in \defDom{\atpVar[\baseNode]}$.
            An edge from node \baseNode to node \succNode expresses that the $\succNode^\text{th}$ reduction step continues the control flow of step $\baseNode$.
            For any $\genNode \in \N$, let $\tidVar_\genNode$ denote the ID of the thread reduced in step \genNode.
            Furthermore, let \ruleNameVar[\genNode] denote the name of the reduction rule applied in the $\genNode^\text{th}$ step, in the following sense:
            \begin{itemize}
                    \item If 
                            \atpRedStep
                                    [\aheapVar[\genNode]]{\atpVar[\genNode]}
                                    [\tidVar]
                                    [\aheapVar[\genNode+1]]{\atpVar[\genNode+1]}
                            results from an application of \rtpRedLiftName in combination with single-thread reduction rule \stRuleNameVar,
                            then $\ruleNameVar[\genNode] = \stRuleNameVar$.
                    \item If 
                            \atpRedStep
                                    [\aheapVar[\genNode]]{\atpVar[\genNode]}
                                    [\tidVar]
                                    [\aheapVar[\genNode+1]]{\atpVar[\genNode+1]}
                            results from an application of \gtpRedWaitName,
                            then $\ruleNameVar[\genNode] = (\gtpRedWaitName, \sigAnno[\genNode])$.
                    \item Otherwise, \ruleNameVar denotes the applied (real or ghost) thread pool reduction rule.
            \end{itemize}
            
            An edge 
            $(\baseNode, \tidVar, \ruleNameVar, \succNode) \in 
                    \N \times \ThreadIDSet \times \AnnoRuleNameSet \times \N$
            is contained in \EdgeSetVar if
            $\ruleNameVar = \ruleNameVar[\baseNode]$
            and one of the following conditions applies:
            \begin{itemize}
                    \item $\tidVar = \tidVar_\baseNode = \tidVar_\succNode$ and 
                            $\succNode = \min(\setOf{\auxNode > \baseNode \ \ | \ \
                                                    \atpRedStep
                                                            [\aheapVar[\auxNode]]{\atpVar[\auxNode]}
                                                            [\tidVar_\baseNode]
                                                            [\aheapVar[\auxNode+1]]{\atpVar[\auxNode+1]}
                                                })$.
                            \\
                            In this case, the edge expresses that step \succNode marks the first time that thread $\tidVar_\baseNode$ is rescheduled for reduction (after step \baseNode).
%                            In this case the edge expresses that the $j^\text{th}$ reduction step continues the execution of thread \tidVar reduced in step $i$.
                    \item $\defDom{\atpVar[\baseNode+1]} \setminus \defDom{\atpVar[\baseNode]}
                                    = \setOf{\tidVar}$
                            and\\
                            $\succNode = \min{\setOf{ \auxNode \in \N \ \ | \ \
                                                    \atpRedStep
                                                            [\aheapVar[\auxNode]]{\atpVar[\auxNode]}
                                                            [\tidVar]
                                                            [\aheapVar[\auxNode+1]]{\atpVar[\auxNode+1]}
                                            }}$.
                            \\
                            In this case, \tidVar identifies the thread forked in step \baseNode.
                            The edge expresses that step \succNode marks the first reduction of the forked thread.
            \end{itemize}
            
            In case the choice of reduction sequence \ahtpRSeqFull is clear from the context, we write \progOrdGraph instead of \progOrdGraph[\ahtpRSeqFull].
    } % end of local scope
    \end{definition}
    
    \begin{observation}\label{gsTR:appendix:obs:ProgramOrderGraphBinaryTree}
            Let \ahtpRSeq be an annotated reduction sequence with\\
            $\cardinalityOf{\defDom{\atpVar[0]}} = 1$.
            The sequence's program order graph \progOrdGraph[\ahtpRSeq] is a binary tree.
    \end{observation}

    For any reduction sequence \ahtpRSeq, the paths in its program order graph \progOrdGraph[\ahtpRSeq] represent the sequence's control flow paths.
    Hence, we are going to use program order graphs to analyse reduction sequences' control flows.

    We refer to a program order graph's edges by the kind of reduction step they represent.
    For instance, we call edges of the form
    $(a, \tidVar, \stRedWhileName, b)$ \emph{loop edges} because they represent a loop backjump
    and edges of the form
    $(a, \tidVar, (\gtpRedWaitName, \sigVar), b)$ \emph{wait edges}.
    Wait edges of this form represent applications of \gtpRedWaitName to signal \sigVar.

    In the following, we prove that any path in a program order graph that does not involve a loop edge is finite.
    This follows from the fact that the size of the command reduced along this path decreases with each non-ghost non-loop step.
    
    \begin{lemma}\label{gsTR:appendix:lem:InfinitePathsCotainInfinitelyManyLoopEdges}
            Let \ahtpRSeq be a fair annotated reduction sequence.
            Let $\pathVar = (\NodeSetVar, \EdgeSetVar)$ be a path in \progOrdGraph[\ahtpRSeq].
            Let 
            $L = \setOf{ \edgeVar \in \EdgeSetVar \ | \ \tupleProj{3}[\edgeVar] = \astRedWhileName }$
            be the set of loop edges contained in \pathVar.
            Then, \pathVar is infinite if and only if $L$ is infinite.
    \end{lemma}
    \begin{proof}
            If $L$ is infinite, \pathVar is obviously infinite as well.
            So, suppose $L$ is finite.
            
            For any command, we consider its size to be the number of nodes contained in its abstract syntax tree.
            By structural induction over the set of commands, it follows that the size of a command $\cmdVar = \atGetCmd{\atpVar(\tidVar)}$ decreases in every non-ghost reduction step
            \atpRedStep
                    [\aheapVar]
                    {\atpVar}
                    [\tidVar]
                    [\nextAheapVar]
                    {\nextAtpVar}            
            that is not an application of \rtpRedLiftName in combination with \astRedWhileName.
            
%            For any node $i \in \NodeSetVar$, let $\cmdVar_i$ denote the command reduced in the reduction step that $i$ represents.
%            Since \ahtpRSeq is fair, \pathVar contains infinitely many edges corresponding to non-ghost reduction steps.
            Since $L$ is finite, there exists a node $x$ such that the suffix $\pathVar_{\geq x}$ starting at node $x$ does not contain any loop edges.
%            It contains, however, infinitely many edges corresponding to non-ghost reduction steps.
            By fairness of \ahtpRSeq, every non-empty suffix of $\pathVar_{\geq x}$ contains an edge corresponding to a non-ghost reduction step.
            For any edge $\edgeVar = (i, \tidVar, n, j)$ consider the command 
            $\cmdVar_\edgeVar = \atGetCmd{\atpVar[i](\tidVar)}$
            reduced in this edge.
            The size of these commands decreases along $\pathVar_{\geq x}$.
            So, $\pathVar_{\geq x}$ must be finite and thus \pathVar must be finite as well.
    \end{proof}

    \begin{corollary}\label{gsTR:appendix:cor:InfinitePathsCotainInfinitelyManyIfTrueEdges}
            Let \ahtpRSeq be a fair annotated reduction sequence.
            Let $\pathVar = (\NodeSetVar, \EdgeSetVar)$ be a path in \progOrdGraph[\ahtpRSeq].
            Let 
            $$
            \begin{array}{l c l}
                    C &= &\setOf{ \edgeVar \in \EdgeSetVar \ | \ \tupleProj{3}[\edgeVar] = \astRedConsumeItPermName }
            \end{array}
            $$
            be the set of consume edges contained in \pathVar.
            Then, \pathVar is infinite if and only if $C$ is infinite.
    \end{corollary}
    \begin{proof}
            Follows from Lemma~\ref{gsTR:appendix:lem:InfinitePathsCotainInfinitelyManyLoopEdges} by the fact that the set
            \setOf{ \edgeVar \in \EdgeSetVar \ | \ \tupleProj{3}[\edgeVar] = \astRedWhileName }
            is infinite if and only if $C$ is infinite.
    \end{proof}

    \begin{definition}
            Let
            $\progOrdGraph = (\NodeSetVar, \EdgeSetVar)$
            be a subgraph of some program order graph.
            We define the function
            $\waitEdges[\progOrdGraph] : \SignalSet \rightarrow \powerSetOf{\EdgeSetVar}$
            mapping any signal \sigVar to the set of wait edges in \progOrdGraph concerning \sigVar as:
            $$
            \begin{array}{c c l}
                    \waitEdges[\progOrdGraph][\sigVar] &:=
                            &\setOf{
                                    (a, \tidVar, (\gtpRedWaitName, \nextSigVar), b) \in \EdgeSetVar 
                                    \ |\
                                    \nextSigVar = \sigVar
                                }.
            \end{array}
            $$
            Furthermore, we define the set
            $\WaitSignalSet[\progOrdGraph] \subset \SignalSet$
            of signals being waited for in \progOrdGraph and its subset
            $\InfWaitSigSet[\progOrdGraph] \subseteq \WaitSignalSet[\progOrdGraph]$
            of signals waited-for infinitely often in \progOrdGraph
            as follows:
            $$
            \begin{array}{c c l}
                    \WaitSignalSet[\progOrdGraph] &:=
                            &\setOf{
                                    \sigVar \in \SignalSet 
                                    \ \ | \ \
                                    \waitEdges[\progOrdGraph](\sigVar) \neq \emptyset
                                },
                    \\
                    \InfWaitSigSet[\progOrdGraph] &:=
                            &\setOf{
                                            \infSig \in \WaitSignalSet[\progOrdGraph] \ \ | \ \
                                            \waitEdges[\progOrdGraph][\infSig]\ \text{infinite}
                                    }.
            \end{array}
            $$
    \end{definition}

        \begin{definition}
            Let \ahtpRSeq be a fair annotated reduction sequence and let
            $\progOrdGraph = (\NodeSetVar, \EdgeSetVar)$
            be a subgraph of the sequence's program order graph.
            We define the function
            $\itpermsOfEdge{\progOrdGraph} : \EdgeSetVar \rightarrow \FinBagsOf{\ItPermSet}$
            mapping any edge \edgeVar to the (potentially empty) finite bag of iteration permissions derived in the reduction step corresponding to \edgeVar as follows:

            Let $(i, \tidVar, n, j) \in \EdgeSetVar$ be an edge.
            \begin{itemize}
                    \item If $n = (\gtpRedWaitName, \sigVar)$ for some signal $\sigVar \in \SignalSet$,
                                then the $i^\text{th}$ reduction step spawns a single iteration permission $(\fpidVar, \degVar)$, i.e.,\\
                                $\atpVar[i+1] =
                                        \tpUpdate
                                                {\atpVar[i]}
                                                {\tidVar}
                                                {(
                                                        \lhAddRes
                                                        {\atGetHeap{\atpVar[i](\tidVar)}}
                                                        {\lresItPerm{\fpidVar}{\degVar}}
                                                        ,\,
                                                        \atGetCmd{\atpVar[i](\tidVar)}
                                                )}
                                $.\\
                                In this case, we define
                                $$
                                \begin{array}{l c l}
                                        \itpermsOfEdge{\progOrdGraph}[(i, \tidVar, (\gtpRedWaitName, \sigVar), j)]
                                                &:=
                                                &\multiset{(\fpidVar, \degVar)}.
                                \end{array}
                                $$
                    \item If $n = \gtpRedWeakItPermName$,
                                then the $i^\text{th}$ reduction step consumes an iteration permission $(\nextFpidVar, \degVar)$ and produces $N$ permissions $(\nextFpidVar, \lowerDegVar)$ of lower degree, i.e.,
                                $\atGetHeap{\atpVar[i](\tidVar)} = 
                                        \lheapVar + \lheapOf{\itPerm[\nextFpidVar, \degVar]}
                                $
                                for some heap \lheapVar and
                                $\atpVar[i+1] = 
                                        \tpUpdate
                                                {\atpVar[i]}
                                                {\tidVar}
                                                {(\nextLheapVar,\, \atGetCmd{\atpVar[i](\tidVar)})}$
                                for
                                $$
                                \nextLheapVar \ =  \
                                        \lheapVar
                                        +
                                        N \cdot \lheapOf{\lresItPerm{\nextFpidVar}{\lowerDegVar}}.
                                $$
                                In this case, we define
                                $$
                                \hspace{-0.2cm}
                                \begin{array}{l c l}
                                \itpermsOfEdge{\progOrdGraph}[(i, \tidVar, \gtpRedWeakItPermName, j)]
                                        &:=
                                        &\multiset{
                                                \underbrace{(\nextFpidVar, \lowerDegVar), \dots, (\nextFpidVar, \lowerDegVar)}_{\text{$N$ times}}
                                        }.
                                \end{array}
                                $$
                    \item Otherwise, we define
                                $$
                                \begin{array}{l c l}
                                \itpermsOfEdge{\progOrdGraph}[(i, \tidVar, n, j)]
                                        &:=
                                        &\msEmpty.
                                \end{array}
                                $$
            \end{itemize}
        \end{definition}

    \begin{definition}[Signal Capacity]
            Let \ahtpRSeq be a fair annotated reduction sequence and 
            $\progOrdGraph = (\NodeSetVar, \EdgeSetVar)$
            be a subgraph of the sequence's program order graph.
            We define the function
            $\signalCapacity{\progOrdGraph}:
                    (\SignalSet \setminus \InfWaitSigSet[\progOrdGraph]) % signal
                    \times
                    \N % sequence index
                    \rightarrow
                    \FinBagsOf{\ItPermSet}
            $
            mapping signals and indices to bags of iteration permissions as follows:
            $$
            \signalCapacity{\progOrdGraph}(\sigVar, i) \ \ := \ \
                    \msBigCup_{\substack{
                            (a, \tidVar, n, b)\, \in\, \waitEdges[\progOrdGraph][\sigVar]
                            \\
                            a\, \geq\, i
                    }}
                    \quadBack\quadBack\quadBack
                    \itpermsOfEdge{\progOrdGraph}[(a, \tidVar, n, b)].
            $$
           We call $\signalCapacity{\progOrdGraph}(\sigVar, i)$ the \emph{capacity} of signal \sigVar at index $i$.
    \end{definition}
    
    Note that the signal capacity above is indeed finite.
    For every \progOrdGraph and every signal 
    $\sigVar \in \SignalSet \setminus \InfWaitSigSet[\progOrdGraph]$
    the set of wait edges \waitEdges[\progOrdGraph][\sigVar] is finite.
    Hence, the big union above is a finite union over finite iteration permission bags.

    \begin{definition}[Partial Order on Permissions]
            We define the partial order on iteration permissions $\itPermLt\ \subset\ \ItPermSet \times \ItPermSet$ induced by \degLt such that
            $$
                    (\fpidVar_1, \degVar_1) \itPermLt (\fpidVar_2, \degVar_2)
                    \ \ \Longleftrightarrow \ \
                    \degVar_1 \degLt \degVar_2.
            $$
    \end{definition}
    
    \begin{lemma}\label{gsTR:appendix:lem:PermissionOrderWellFounded}
            The partial order \itPermLt is well-founded.
    \end{lemma}
    \begin{proof}
            Follows directly from well-foundedness of \itPermLt.
    \end{proof}

    \begin{definition}[Partial Order on Finite Bags]
    {
            \NewDocumentCommand{\baseSet}{}{\metaVar{X}}
            \NewDocumentCommand{\baseLt}{}{\ensuremath{<_\baseSet}}
            
            \NewDocumentCommand{\smallerBag}{}{\metaVar{A}}
            \NewDocumentCommand{\biggerBag}{}{\metaVar{B}}
            
            \NewDocumentCommand{\removedBag}{}{\metaVar{C}}
            \NewDocumentCommand{\addedBag}{}{\metaVar{D}}
            
            \NewDocumentCommand{\elemRemovedBag}{}{\metaVar{c}}
            \NewDocumentCommand{\elemAddedBag}{}{d}
            
            Let \baseSet be a set and let $\baseLt \ \subset \ \baseSet \times \baseSet$ a partial order on \baseSet.
            We define the partial order 
            $\bagLt[\baseSet]\ \subset\ \FinBagsOf{\baseSet} \times \FinBagsOf{\baseSet}$
            on finite bags over \baseSet
            as the 
            Dershowitz-Manna ordering~\cite{Dershowitz1979MultisetOrder} induced by \baseLt:
            $$
            \begin{array}{c}
                    \smallerBag \bagLt[X] \biggerBag
                    \ \ \Longleftrightarrow
                    \begin{array}[t]{l l}
                    \exists \removedBag, \addedBag \in \FinBagsOf{X}.\!\!\!
                            &\msEmpty \neq \removedBag \subseteq \biggerBag\\
                            &\wedge\ 
                            \smallerBag = 
                                    (\biggerBag \setminus \removedBag) \msCup \addedBag
                            \\
                            &\wedge\
                            \forall \elemAddedBag \in \addedBag.\
                            \exists \elemRemovedBag \in \removedBag.\
                            \elemAddedBag \baseLt \elemRemovedBag.
                    \end{array}
            \end{array}
            $$
            We define $\bagLeq[X]\ \subset\ \FinBagsOf{X} \times \FinBagsOf{X}$ such that
            $$
                    \smallerBag \bagLeq[\baseSet] \biggerBag
                    \ \ \Longleftrightarrow \ \
                    \smallerBag = \biggerBag \ \vee\  \smallerBag \bagLt[\baseSet] \biggerBag
            $$
            holds.
    }
    \end{definition}

    \begin{corollary}\label{gsTR:appendix:cor:permBagLtWellFounded}
            The partial order 
            $\itPermBagLt \ \subset\ \FinBagsOf{\ItPermSet} \times \FinBagsOf{\ItPermSet}$
            is well-founded.
    \end{corollary}
    \begin{proof}
            Follows from~\cite{Dershowitz1979MultisetOrder} and Lemma~\ref{gsTR:appendix:lem:PermissionOrderWellFounded}.
    \end{proof}

    In the following, we view paths in a program order graph as single-branched subgraphs.
    This allows us to apply above definitions on graphs to paths.
    In particular, this allows us to refer to the capacity of a signal \sigVar on a path \pathVar by referring to \signalCapacity{\pathVar}.

    For the following definition, remember that a bag $B \in \BagsOf{X}$ is a function $B : X \rightarrow \N$
    while a logical heap $\lheapVar \in \LogHeapSet$ is a function $\lheapVar: \LogResSet \rightarrow \Q_{\geq 0}$.
    Also remember the signatures
    $\atpOwnedResHeap : \AnnoThreadPoolSet \rightarrow \LogHeapSet$
    and
    $\ahProtectedResHeap: \AnnoHeapSet \rightarrow \LogHeapSet$.

    \begin{definition}
            We define the functions
            $\amConfGetIterPerms: 
                    \AnnoHeapSet \times \AnnoThreadPoolSet 
                    \rightarrow
                    \BagsOf{\ItPermSet}
            $
            and
            $\amConfGetWaitPerms:
                    \AnnoHeapSet \times \AnnoThreadPoolSet
                    \rightarrow
                    \BagsOf{\WaitPermSet}
            $
            mapping annotated machine configurations to bags of iteration and wait permissions, respectively, as follows:
            $$
            \hspace{-0.14cm}
            \begin{array}{l}
                    \amConfGetIterPerms[\aheapVar, \atpVar](\fpidVar, \degVar)
                    \\\ \ :=\ \,
                            \left\lfloor
                            \big(
                                    \atpOwnedResHeap[\atpVar] + 
                                    \ahProtectedResHeap[\aheapVar]
                            \big)
                            (\lresItPerm{\fpidVar}{\degVar})
                            \right\rfloor\!,
                    \\
                    \\
                    \amConfGetWaitPerms[\aheapVar, \atpVar](\fpidVar, \idVar, \degVar)
                    \\\ \ :=\ \,
                            \left\lfloor
                            \big(
                                    \atpOwnedResHeap[\atpVar] + 
                                    \ahProtectedResHeap[\aheapVar]
                            \big)
                            (\lresWaitPerm{\fpidVar}{\idVar}{\degVar})
                            \right\rfloor\!.
            \end{array}
            $$
    \end{definition}
    
    Note that for consistent annotated machine configurations $(\aheapVar, \atpVar)$ the above flooring is without any affect.

    \begin{corollary}\label{gsTR:appendix:cor:ExtractedPermissionBagsFinite}
            Let \ahtpRSeq be an annotated reduction sequence such that
            \ahFinite{\aheapVar[0]}
            and
            \lhFinite{\atGetHeap{\atpVar[0](\tidVar)}}
            hold for every $\tidVar \in \defDom{\atpVar[0]}$.
            
            Then,
            \amConfGetIterPerms[\aheapVar[i], \atpVar[i]]
            and
            \amConfGetWaitPerms[\aheapVar[i], \atpVar[i]]
            are finite for every choice of $i \in \N$.
    \end{corollary}
    \begin{proof}
            Follows by preservation of finiteness, Lemma~\ref{gsTR:appendix:lem:PreservationOfFiniteness}.
    \end{proof}

    { % local scope
            \NewDocumentCommand{\tmpCapacity}{o}
                    {\ensuremath{
                            \fixedFuncNameFont{nodeCap}
                            \IfValueT{#1}{(#1)}
                    }\xspace}
            
            \NewDocumentCommand{\rootTid}{}{\metaVar{\tidVar_\rootVar}}
            \NewDocumentCommand{\rootWaitsBag}{}{\metaVar{\waitsBagVar_\rootVar}}
            \NewDocumentCommand{\rootPermBag}{}{\metaVar{\itPermBagVar_\rootVar}}
    \begin{restatable}{lemma}{lemNoInfSigOnPathImpliesPathFinite}\label{gsTR:appendix:lem:NoInfSigOnPathImpliesPathFinite}
            Let 
            $\progOrdGraph[\ahtpRSeq]$
            be a program order graph and let
            $\pathVar = (\NodeSetVar, \EdgeSetVar)$
            be a path in \progOrdGraph with 
            $\InfWaitSigSet[\pathVar] = \emptyset$.
            For every 
            $\tidVar \in \defDom{\atpVar[0]}$
            let $\atGetHeap{\atpVar[0](\tidVar)}$ be finite and complete.
            Further, let \aheapVar[0] be finite and contain no chunks \aresUnlocked{\mutVar}{\assLockInvVar}{\lheapLockInvVar} where \lheapLockInvVar contains any phase or obligations chunk.
            
            Then, \pathVar is finite.
    \end{restatable}
    \begin{proof}
            Assume \pathVar is infinite.
            We prove a contradiction by assigning a finite capacity to every node along the path.
            Let $\tidVar_i$ be the ID of the thread reduced in step $i$.
            For every $\tidVar \in \defDom{\atpVar[\rootVar]}$ the logical heap \atGetHeap{\atpVar[0](\tidVar)} is complete and \aheapVar[0] contains no chunks \aresUnlocked{\mutVar}{\assLockInvVar}{\lheapLockInvVar} where \lheapLockInvVar contains any phase or obligations chunk.
            By preservation of completeness, Lemma~\ref{gsTR:appendix:lem:PreservationOfCompleteness}, \atGetHeap{\atpVar[i](\tidVar_i)} is also complete and hence it contains exactly one phase chunk \lresThreadPhase{\fpidVar_i}.
            That is, for every step $i$, the phase ID $\fpidVar_i$ of the thread reduced in step $i$ is uniquely defined.
            
            Consider the function
            $\tmpCapacity : \NodeSetVar \rightarrow \FinBagsOf{\ItPermSet}$ defined as
            $$
            \begin{array}{l c l}
            \tmpCapacity[i]  &:=
                        &\multiset{
                                (\fpidAncVar, \degVar) \in \amConfGetIterPerms[\aheapVar[i], \atpVar[i]]
                                \ \ | \ \
                                \fpidAncVar \fpLeq \fpidVar_i
                        }
                        \\
                        &&\displaystyle
                        \msCup
                        \msBigCup_{\substack{
                                \idVar \,\in \, \fixedFuncNameFont{waitIDs}(\fpidVar_i)
                                \\
                                (\fpidAncVar, \idVar, \degVar) \,\in\, \amConfGetWaitPerms[\aheapVar[i], \atpVar[i]]
                                \\
                                \levVar \,\in\, \LevelSet
                        }}
                                \quadBack\quadBack\quadBack
                                \signalCapacity{\pathVar}((\idVar, \levVar), i).
            \end{array}
            $$
            where 
            $
                    \fixedFuncNameFont{waitIDs}(\fpidVar_i)
                    :=
                    \setOf{
                            \idVar
                            \ | \
                            \exists \fpidAncVar.\,
                            (\fpidAncVar, \idVar, \slWildcard) \in \amConfGetWaitPerms[\aheapVar[i], \atpVar[i]]
                            \,\wedge\,
                            \fpidAncVar \fpLeq \fpidVar_i
                    }
            $.
            
            For every $i \in \NodeSetVar$, the capacity of node $i$, i.e., \tmpCapacity[i], is the union of two finite iteration permission bags:
            (i)~Above
            \multiset{
                    (\fpidAncVar, \degVar) \in \amConfGetIterPerms[\aheapVar[i], \atpVar[i]]
                    \ | \
                    \fpidAncVar \fpLeq \fpidVar_i
            }
            captures all iteration permissions contained in \aheapVar and \atpVar[i] that are qualified by an ancestor $\fpidAncVar$ of phase ID $\fpidVar_i$ and are hence usable by the thread reduced in node $i$.
            This includes the permissions $(\fpidAncVar, \degVar)$ held by thread $\tidVar_i$ as well as such (temporarily) transferred to another thread via a lock invariant.
            (ii)~Below
            $\msBigCup \signalCapacity{\pathVar}((\idVar, \levVar), i)$
            captures all iteration permissions that will be created along the suffix of \pathVar that starts at node $i$ by waiting for signals for which thread $\tidVar_i$ already holds a wait permission $(\fpidAncVar, \idVar, \degVar)$ in step $i$.
            
            Note that for every $i \in \NodeSetVar$, the bag of iteration permissions returned by \tmpCapacity[i] is indeed finite.
            The initial annotated heap and all initial thread-local logical heaps are finite.
            This allows us to apply Corollary~\ref{gsTR:appendix:cor:ExtractedPermissionBagsFinite}, by which we get that 
            \amConfGetIterPerms[\aheapVar[i], \atpVar[i]]
            and
            \amConfGetWaitPerms[\aheapVar[i], \atpVar[i]]
            are finite.
            
            Since signal IDs are unique, for every fixed choice of $i$ and \idVar, there is at most one level \levVar, for which
            $\signalCapacity{\pathVar}((\idVar, \levVar), i) \neq \emptyset$.
            By assumption, along \pathVar all signals are waited for only finitely often, i.e., $\InfWaitSigSet[\pathVar] = \emptyset$.
            Hence, also the big union
            $
                    \msBigCup
                    \signalCapacity{\pathVar}((\idVar, \levVar), i)
            $
            is defined and finite.

            Consider the sequence
            $(\tmpCapacity[i])_{i \in \NodeSetVar}$.
            Since every element is a finite bag of permissions, we can order it by \itPermBagLt.
            We are going to prove a contradiction by proving that the sequence
            is an infinitely descending chain.
            
            Consider any edge
            $(i, \tidVar, n, j) \in \EdgeSetVar$.
            There are only three cases in which 
            $\tmpCapacity[i] \neq \tmpCapacity[j]$
            holds.
            \begin{itemize}
                    \item $n = \gtpRedWaitPermName$:\\
                            In this case, there are degrees $\degVar, \lowerDegVar$ with $\lowerDegVar \degLt \degVar$, a signal \sigVar and $N \in \N$ for which we get
                            $$
                            \tmpCapacity[j] = 
                            (\tmpCapacity[i] 
                                    \setminus 
                                    \multiset{(\nextFpidVar, \degVar)})
                            \msCup
                            \multiset{\underbrace{(\nextFpidVar, \lowerDegVar)}_{\text{$N$ times}}}.
                            $$
                            That is,
                            $\tmpCapacity[j] \itPermBagLt \tmpCapacity[i]$.
                    \item $n = \gtpRedWeakItPermName$:
                            Same as above.
                    \item $n = \astRedConsumeItPermName$:\\
                            In this case, we know that
                            $\tmpCapacity[j] = 
                                    \tmpCapacity[i] \setminus \multiset{(\fpidAncVar, \degVar)}
                                    \itPermBagLt
                                    \tmpCapacity[i]
                            $
                            holds for some \fpidAncVar and \degVar.
            \end{itemize}
            (Note that in case of $n = \gtpRedWaitName$, we have $\tmpCapacity[i] = \tmpCapacity[j]$ since
                $$
                \begin{array}{c}
                        \multiset{
                                (\fpidAncVar, \degVar) \in \amConfGetIterPerms[\aheapVar[j], \atpVar[j]]
                                \ | \
                                \fpidAncVar \fpLeq \fpidVar_j
                        }
                        \\
                        =
                        \\
                        \multiset{
                                (\fpidAncVar, \degVar) \in \amConfGetIterPerms[\aheapVar[i], \atpVar[i]]
                                \ | \
                                \fpidAncVar \fpLeq \fpidVar_i
                        }
                        \ \msCup\
                        \multiset{(\fpidVar, \degVar)}
                        
                \end{array}
                $$
                and
                $$
                        \msBigCup
                        \signalCapacity{\pathVar}((\idVar, \levVar), j)
                        \ \ = \ \
                        \left(
                                \msBigCup
                                \signalCapacity{\pathVar}((\idVar, \levVar), i)
                        \right)
                        \setminus \multiset{(\fpidVar, \degVar)}
                $$
                for some \degVar.)
            So, \tmpCapacity is monotonically decreasing.
            
            By assumption \pathVar is infinite.
            According to Corollary~\ref{gsTR:appendix:cor:InfinitePathsCotainInfinitelyManyIfTrueEdges} this implies that the path contains infinitely many consume edges,
            i.e., edges with a labelling $n = \astRedConsumeItPermName$.
            Hence, the sequence
            $(\tmpCapacity[i])_{i\in \NodeSetVar}$
            forms an infinitely descending chain.
            However, according to Corollary~\ref{gsTR:appendix:cor:permBagLtWellFounded}, \itPermBagLt is well-founded.
            A contradiction.
    \end{proof}
    } % end of local scope

    \begin{restatable}{lemma}{lemNoSignalWaitedForInfinitelyOften}\label{gsTR:appendix:lem:NoSignalWaitedForInfinitelyOften}
            Let \ahtpRSeq be a fair annotated reduction sequence with
            \amConfConsistent{\aheapVar[0]}{\atpVar[0]},
            $\atpVar[0] = \setOf{(\tidStartVar, (\lheapVar[0], \cmdVar))}$,
            \lhComplete{\lheapVar[0]},
            \lhFinite{\lheapVar[0]}
            and with \ahFinite{\aheapVar[0]}.
            Let \lheapVar[0] contain no signal or wait permission chunks.
            Further, let \aheapVar[0] contain no chunks 
            \aresUnlocked{\mutVar}{\assLockInvVar}{\lheapLockInvVar}
            where \lheapLockInvVar contains any obligations, phase or signal chunks.
            Let \progOrdGraph be the program order graph of \ahtpRSeq.
            Then, $\InfWaitSigSet[\progOrdGraph] = \emptyset$.
    \end{restatable}
    \begin{proof}
    { % local scope
            \NewDocumentCommand{\obTid}{}{\metaVar{\tidVar_\text{ob}}}
            \NewDocumentCommand{\lowerSig}{}{\metaVar{z}}
            \NewDocumentCommand{\obPath}{}{\metaVar{p_\text{ob}}}
            \NewDocumentCommand{\obExtPath}{}{\metaVar{\bar{\obPath}}}
            
            \NewDocumentCommand{\obPathNodes}{}{\metaVar{V_\text{ob}}}
            
            \NewDocumentCommand{\tmpCapacity}{o}
                    {\ensuremath{
                            \fixedFuncNameFont{cap}
                            \IfValueT{#1}{(#1)}
                    }\xspace}

            Suppose $\InfWaitSigSet[\progOrdGraph] \neq \emptyset$.
            Since \LevelSet is well-founded, the same holds for the set \setOf{\levOf{\sigVar} \ | \ \sigVar \in \InfWaitSigSet}.
            Hence, there is some $\minSig \in \InfWaitSigSet$ for which no $\lowerSig \in \InfWaitSigSet$ with
            $\levOf{\lowerSig} \levLt \levOf{\minSig}$ exists.
            
            Since neither the initial logical heap \lheapVar[0] nor any unlocked lock invariant stored in \aheapVar[0] does contain any signals, \minSig must be created during the reduction sequence.
            The reduction step creating signal \minSig is an application of \gtpRedNewSignalName, which simultaneously creates an obligation to set \minSig.
            By preservation of completeness, Lemma~\ref{gsTR:appendix:lem:PreservationOfCompleteness}, every thread-local logical heap \atGetHeap{\atpVar[i](\tidVar)} annotating some thread \tidVar in some step $i$ is complete.
            According to reduction rule \gtpRedWaitName, every wait edge
            $(a, \tidVar, (\gtpRedWaitName, \minSig), b)$
            implies together with completeness that in step $a$ (i)~thread \tidVar does not hold any obligation for \minSig 
            (i.e.\@ 
                \lhHasObs[\atGetHeap{\atpVar[a](\tidVar)}] 
                for some bag of obligations \obBagVar with
                $\minSig \not\in \obBagVar$)
            and (ii)~\minSig has not been set, yet
            (i.e.\@
                $\aresSignal{\minSig}{\slFalse} \in \aheapVar[a]$).
            Hence, in step $a$ another thread $\obTid \neq \tidVar$ must hold the obligation for \minSig
            (i.e.\@
                \lhHasObs[\atGetHeap{\atpVar[a](\obTid)}]
                for some bag of obligations \obBagVar with
                $\minSig \in \obBagVar$).
            Since there are infinitely many wait edges concerning \minSig in \progOrdGraph, the signal is never set.
            
            By fairness, for every wait edge as above, there must be a non-ghost reduction step 
            \atpRedStep
                    [\aheapVar[k]]{\atpVar[k]}
                    [\obTid]
                    [\aheapVar[k+1]]{\atpVar[k+1]}
            of the thread \obTid holding the obligation for \minSig with $k \geq a$.
            Hence, there exists an infinite path \obPath in \progOrdGraph where each edge $(e, \obTid, n, f) \in \edges{\obPath}$ concerns some thread \obTid holding the obligation for \minSig.
            (Note that this thread ID does not have to be constant along the path, since the obligation can be passed on during fork steps.)
            
            The path \obPath does not contain wait edges
            $(e, \obTid, (\gtpRedWaitName, \infSig), f)$
            for any $\infSig \in \InfWaitSigSet$,
            since reduction rule \gtpRedWaitName would (together with completeness of \atGetHeap{\atpVar[e](\obTid)}) require
            \infSig to be of a lower level than all held obligations.
            This restriction implies
            $\levOf{\infSig} \levLt \levOf{\minSig}$
            and would hence contradict the minimality of \minSig.
            That is, $\InfWaitSigSet[\obPath] = \emptyset$.

%            We can conclude that \lheapVar[0] is finite, from the consistency of the initial annotated machine configuration (\aheapVar[0], \atpVar[0]) and the fact that \lheapVar[0] does not contain any signal or registration chunks and only finitely many permission chunks.
            By preservation of finiteness, Lemma~\ref{gsTR:appendix:lem:PreservationOfFiniteness}, we get that every logical heap associated with the root of \obPath is finite.
            This allows us to apply Lemma~\ref{gsTR:appendix:lem:NoInfSigOnPathImpliesPathFinite}, by which we get that \obPath is finite.
            A contradiction.
    } % end of local scope
    \end{proof}

    \begin{restatable}{lemma}{lemNoFairInfiniteAnnotatedRedSeq}\label{gsTR:appendix:lem:NoFairInfiniteAnnotatedRedSeq}
            Let
            $$\htModels
                    {
                            \assPhase[\fpidVar_0]
                            \slStar
                            \noObs 
                            \slStar 
                            \!\!\!\!
                            {\displaystyle
                            \slBigStar^{i \,=\, 1, \dots, N}}
                            \quadBack\
                            \itPerm[\fpidVar_0, \degStartVar[i]]
                    }
                    {\cmdVar}
                    {\noObs}
            $$
            hold.
            There exists no fair, infinite annotated reduction sequence \ahtpRSeq with
            $\aheapVar[0] = \emptyset$,
            $\atpVar[0] = \setOf{(\tidStartVar, (\lheapVar[0], \cmdVar))}$
            and
            $$\lheapVar[0] \ =  \
                    \lheapOf{
                            \lresThreadPhase{\fpidVar_0},
                            \lresObs{\msEmpty},
                            \lresItPerm{\fpidVar_0}{\degStartVar[1]},
                            \dots,
                            \lresItPerm{\fpidVar_0}{\degStartVar[N]}
                    }.
            $$
    \end{restatable}
    \begin{proof}
    { % local scope
            \NewDocumentCommand{\infPath}{}{\metaVar{\pathVar_\infty}}
            Suppose a reduction sequence as described above exists.
            We are going to prove a contradiction by considering its infinite program order graph \progOrdGraph.

            According to Observation~\ref{gsTR:appendix:obs:ProgramOrderGraphBinaryTree}, \progOrdGraph is a binary tree with an infinite set of vertices.
            By the Weak König's Lemma~\cite{Simpson1999WKL} \progOrdGraph has an infinite branch, i.e. an infinite path \pathVar starting at root 0.

            The initial logical heap \lheapVar[0] is complete and finite and the initial annotated machine configuration $(\aheapVar[0], \atpVar[0])$ is consistent.
            By Lemma~\ref{gsTR:appendix:lem:NoSignalWaitedForInfinitelyOften} we know that $\InfWaitSigSet[\progOrdGraph] = \emptyset$.
            Since
            $\InfWaitSigSet[\pathVar] \subseteq \InfWaitSigSet[\progOrdGraph]$,
            we get
            $\InfWaitSigSet[\pathVar] = \emptyset$.            
            This allows us to apply Lemma~\ref{gsTR:appendix:lem:NoInfSigOnPathImpliesPathFinite}, by which we get that \pathVar is finite, which is a contradiction.
    } % end of local scope
    \end{proof}

    \theoSoundness*
    \begin{proof}
    { % local scope
            Suppose a reduction sequence as described above exists.
            Since we can prove 
            \htProves
                    {
                            \assPhase
                            \slStar
                            \noObs 
                            \slStar 
                            \slBigStar^{\!\!i \,=\, 1, \dots, N}
                            \quadBack\quadBack\quadBack\!\!
                            \itPerm[\fpidVar, \degVar_i]
                    }
                    {\cmdVar}
                    {\noObs},
            we can also conclude
            \htModels
                    {
                            \assPhase
                            \slStar
                            \noObs 
                            \slStar 
                            \slBigStar^{\!\!i \,=\, 1, \dots, N}
                            \quadBack\quadBack\quadBack\!\!
                            \itPerm[\fpidVar, \degVar_i]
                    }
                    {\cmdVar}
                    {\noObs}
            by Hoare triple soundness, Lemma~\ref{gsTR:appendix:lem:HoareTripleSoundness}.
            Consider the logical heap 
            $$\lheapVar[0] = 
                    \lheapOf{
                            \lresThreadPhase{\fpidVar},
                            \lresObs{\msEmpty},
                            \lresItPerm{\fpidVar}{\degVar_1},
                            \dots,
                            \lresItPerm{\fpidVar}{\degVar_N}
                    }
            $$
            and the annotated heap
            $\aheapVar[0] = \emptyset$.
            It holds
            \assModels
                    {\lheapVar[0]}
                    {
                            \assPhase
                            \slStar
                            \noObs 
                            \slStar 
                            \slBigStar^{\!\!i \,=\, 1, \dots, N}
                            \quadBack\quadBack\quadBack\!\!
                            \itPerm[\fpidVar, \degVar_i]
                    },
            \alhCorresp[\aheapVar[0]][\lheapVar[0]]
            and
            \aphCorresp{\aheapVar[0]}{\pheapVar[0]}.
            This allows us to apply Lemma~\ref{gsTR:appendix:lem:ConstructionOfAnnotatedRedSeq}, by which
            we can construct a corresponding fair annotated reduction sequence \ahtpRSeq that starts with
            $\aheapVar[0] = \emptyset$ and 
            $\atpVar[0] = 
                    \setOf{(\tidStartVar, (\lheapVar[0], \cmdVar))}
            $.
            By Lemma~\ref{gsTR:appendix:lem:NoFairInfiniteAnnotatedRedSeq} \ahtpRSeq does not exist.
            A contradiction.
    } % end of local scope
    \end{proof}

\section{Verification Example}	
    We scaled all proof outlines presented in this section down to fit them into the page layout. 
	You can find the original proof outlines in the official version of the technical report~\cite{Reinhard2020GhostSignalsTR}.

	As in the paper body, we present the proof state in \ghostFont{blue}, 
	applied proof and view shift rules in \proofRuleHint{\text{purple}},
	and abbreviations and definition in \proofDef{red}.
	Since the verification outlines we present in this section span multiple figures, we include hints and explanations concerning other figures (e.g. a hint pointing to the figure where an invariant was defined).
	To set these remarks and in general any hints off from the essential parts of the proof, we present them in a \ \proofHint{grey and italic font}.
	We occasionally remind the reader of earlier proof steps performed in a previous figure by repeating them in \ \proofHint{the same grey font} at the beginning of the current figure.
	
	Further, we highlight how our proof steps effect the proof state as follows:
	Consider a proof state of the form \progProof{A \slStar \slPointsTo{\hlocVar}{\valVar}}.
	(i)~When a proof step adds a new chunk~$C$, we highlight it in \progProofNew{\text{green}}, i.e.
	\progProof{A \slStar \slPointsTo{\hlocVar}{\valVar} \slStar \progProofNew{C}}.
	(ii)~When a proof step removes the chunk $A$, we highlight this change by underlying the removed part of the assertion with a \progProofCancel{\text{dark grey}} background, i.e., 
	\progProof{\progProofCancel{A\ \slStar}\ \slPointsTo{\hlocVar}{\valVar}}.
	Note that in this case, the greyed out \progProofCancel{A\ \slStar} is not a part of the proof state anymore.
	(iii)~When a proof step changes only part of a chunk, we highlight this change in \progProofChanged{\text{yellow}}.
	For instance,  if the step changes the value of heap location \hlocVar from \valVar to \nextValVar, we highlight it in the resulting state as
	\progProof{A \slStar \slPointsTo{\hlocVar}{\progProofChanged{\nextValVar}}}.
	
    \subsection{Minimal Example}
        Figures~\ref{gsTR:appendix:fig:VerificationExampleMainThread} and~\ref{gsTR:appendix:fig:VerificationExampleBusyWaitingThread} sketch the verification of the example program presented in Figure~\ref{gsTR:appendix:fig:MinimalExample}.
        For this verification we let the set of values \ValueSet include natural numbers and choose 
        $\LevelSet = \DegreeSet = \N$.

    \begin{figure}[h]
    {% local scope
            \RenewDocumentCommand{\valX}{}{\progVar{\valVar_\x}}
            
            \NewDocumentCommand{\tidMain}{}{\progVar{\tidVar_m}}
            \NewDocumentCommand{\startPermDeg}{}{1}

            \NewDocumentCommand{\s}{}{\progVar{s}}
            \NewDocumentCommand{\idS}{}{\progVar{id_\s}}
            \NewDocumentCommand{\levS}{}{1}
            
            \RenewDocumentCommand{\locM}{}{\progVar{\hlocVar_\m}}
            \RenewDocumentCommand{\levM}{}{0}
            \RenewDocumentCommand{\invM}{}{\progVar{P}}

%            \vspace{-3cm}

            % line with approx. width of array
            % helps visualizing whether space to left and to right of array are equal            
%            \noindent\makebox[\linewidth]{\rule{1.35\textwidth}{1pt}} 

            \caseStudyFont
            \makebox[\textwidth][c]
            {$
            % box treats separator center between left and right column as "center of array"
            \hspace{0.13\textwidth}  % adjusting center
            \arxivShift
            \ifArxiv{\hspace{0.3cm}}

            $}

            \caption
            [Verification example (main thread).]
            {
                    Verification sketch of main thread of example program presented in Figure~\ref{gsTR:appendix:fig:MinimalExample}.
                    For readability we omit information about a command's return value if it is not relevant to the proof.
            }
            \label{gsTR:appendix:fig:VerificationExampleMainThread}
    }% end of local scope
    \end{figure}

    \begin{figure}[h]
    { % local scope
            \RenewDocumentCommand{\valX}{}{\progVar{\valVar_\x}}
            
            \NewDocumentCommand{\valY}{}{\progVar{\valVar_\y}}
            
            \NewDocumentCommand{\tidMain}{}{\progVar{\tidVar_m}}
            \NewDocumentCommand{\startPermDeg}{}{1}
            \NewDocumentCommand{\waitPermDeg}{}{0}

            \NewDocumentCommand{\s}{}{\progVar{s}}
            \NewDocumentCommand{\idS}{}{\progVar{id_\s}}
            \NewDocumentCommand{\levS}{}{1}
            
            \RenewDocumentCommand{\locM}{}{\progVar{\hlocVar_\m}}
            \RenewDocumentCommand{\levM}{}{0}
            \RenewDocumentCommand{\invM}{}{\progVar{P}}

            \NewDocumentCommand{\degWait}{}{0}

            %\vspace{-2.5cm}

            % line with approx. width of array
            % helps visualizing whether space to left and to right of array are equal
%            \noindent\makebox[\linewidth]{\rule{1.35\textwidth}{1pt}} 

            \caseStudyFont
            \makebox[\textwidth][c]
            {$
            % box treats separator center between left and right column as "center of array"
            \hspace{0.18\textwidth}  % adjusting center
            \arxivShift

            $}
            
            \caption
            [Verification example (busy-waiting thread).]
            {
                    Verification sketch of busy-waiting thread of example program presented in Figure~\ref{gsTR:appendix:fig:MinimalExample}.
                    For readability we omit information about a command's return value if it is not relevant to the proof.
            }
            \label{gsTR:appendix:fig:VerificationExampleBusyWaitingThread}
            \vspace{1cm}
    } % end of local scope
    \end{figure}

\end{document}

    \subsection{Bounded FIFO}\label{gsTR:appendix:subsec:BoundedFifo}
        For this section, we let the set of values \ValueSet include natural numbers and finite sequences, aka lists, of natural numbers.
        Further, the set of operations \OpSet includes the canonical operations on natural numbers and lists, i.e.,
        (i)~$<$, $\leq$, $-$ and
        (ii)~list concatenation \lstAppend{l_1}{l_2}, 
        prepending an element \lstCons{e}{l}, 
        getting the head and tail of a list \lstHead{l} (defined for non-empty $l$), \lstTail{l} and
        getting the size of a list \lstSize{l}.
        We denote the empty list by \lstNil.
        We use the abbreviation 
        $a \mathrel{R_1} b \mathrel{R_2} c$
         for $R_1, R_2 \in \setOf{<, \leq}$ to denote 
        $a \mathrel{R_1} b \ \slStar\ b \mathrel{R_2} c$.
        Furthermore, we choose $\LevelSet = \DegreeSet = \N$.
        Figure~\ref{gsTR:appendix:fig:MinimalExampleBoundedFifo} presents an example program involving a bounded FIFO.

        To simplify its verification, we refine the process of creating a new ghost signal, i.e., we split it in two steps: allocating a new signal ID and initializing a signal.
        To implement this, we replace view shift rule \vsNewSignalName by the rules \vsAllocSigIDName and \vsSigInitName presented in Figure~\ref{gsTR:appendix:fig:ViewShiftRules--fineGrainedSignalCreation}.
        This way we can fix the IDs of all the signals we need throughout the proof at its beginning.
        This refinement does not affect the soundness of our verification approach.
        Figures~\ref{gsTR:appendix:fig:VerificationExampleBoundedFifoInit} -- \ref{gsTR:appendix:fig:MinimalVerificationExampleBoundedFifoConsumerWaitStep} sketch the program's verification using fine-grained signals.

    \begin{figure}
    	\vspace{-1cm}
		\begin{mathpar}
			\vsAllocSigID
			\and
			\vsSigInit
		\end{mathpar}
		\vspace{-0.3cm}
		\caption{Fine-grained view shift rules for signal creation.}
		\label{gsTR:appendix:fig:ViewShiftRules--fineGrainedSignalCreation}
%		\vspace{-1cm}
	\end{figure}

    \begin{figure}
    {
    		\vspace{-0cm}
    		
            \NewDocumentCommand{\fifo}{}{\progVar{fifo_{10}}}
            \NewDocumentCommand{\fifoLoaded}{}{\progVar{f}}

            \NewDocumentCommand{\counter}{m}{\progVar{c_{#1}}}
            
            % producing thread
                    \NewDocumentCommand{\producerID}{}{p}
                    \NewDocumentCommand{\pc}{}{\counter{\producerID}}
                    \NewDocumentCommand{\pcLoaded}{}{\progVar{c}}

            % consumer counter
                    \NewDocumentCommand{\consumerID}{}{c}
                    \NewDocumentCommand{\cc}{}{\counter{\consumerID}}
                    \NewDocumentCommand{\ccLoaded}{}{\progVar{c}}
            
            $$

            $$

            \caption[Producer-consumer program with bounded FIFO.]
            {
                    Example program with two threads communicating via a shared bounded FIFO with maximal size 10.
                    Producer thread writes numbers 100, \dots, 1 to shared FIFO and busy-waits until FIFO is not full and next element can be pushed.
                    Consumer thread pops 100 numbers from FIFO and busy-waits for next number to arrive.
            }
            \label{gsTR:appendix:fig:MinimalExampleBoundedFifo}
            \vspace{-1cm}
    }
    \end{figure}

    {% local scope of verification figures
            % Variables copied from program figure.
            % ====================================================
                    \NewDocumentCommand{\fifo}{}{\progVar{fifo_{10}}}
                    \NewDocumentCommand{\fifoLoaded}{}{\progVar{f}}

                    \NewDocumentCommand{\counter}{m}{\progVar{c_{#1}}}
                    
                    % producing thread
                            \NewDocumentCommand{\producerID}{}{p}
                            \NewDocumentCommand{\pc}{}{\counter{\producerID}}
                            \NewDocumentCommand{\pcLoaded}{}{\progVar{c}}

                    % consumer counter
                            \NewDocumentCommand{\consumerID}{}{c}
                            \NewDocumentCommand{\cc}{}{\counter{\consumerID}}
                            \NewDocumentCommand{\ccLoaded}{}{\progVar{c}}

            % Variables used solely for verification figure
            % ====================================================
                    \NewDocumentCommand{\startDeg}{}{2}
                    \NewDocumentCommand{\iterDeg}{}{1}
                    \NewDocumentCommand{\waitDeg}{}{0}
                    
                    % initial iteration permission
                    \NewDocumentCommand{\startItPerm}{}{\itPerm[\fpidRoot, \startDeg]}
                    
                    % iteration permission used to justify productive loop iterations
                    \NewDocumentCommand{\iterItPerm}{}{\itPerm[\fpidRoot, \iterDeg]}
                    
                    % iteration permission generated by producer via waiting
                    \NewDocumentCommand{\pWaitItPerm}{}{\itPerm[\fpidRootForked, \waitDeg]}

                    % iteration permission generated by consumer via waiting
                    \NewDocumentCommand{\cWaitItPerm}{}{\itPerm[\fpidRootAfterFork, \waitDeg]}
                    
                    % wait permission used to wait for sigPop{i}
                    \NewDocumentCommand{\wpermPop}{m}{\waitPerm[\fpidRoot, \idPop{#1}, \waitDeg]}

                    % wait permission used to wait for sigPush{i}
                    \NewDocumentCommand{\wpermPush}{m}{\waitPerm[\fpidRoot, \idPush{#1}, \waitDeg]}

                    % fifo
                            \NewDocumentCommand{\locFifo}{}{\metaVar{\hlocVar_\fifo}}
                            \NewDocumentCommand{\valFifoInv}{}{\metaVar{\valVar_\fifo^\m}}

                    % consumer
                            \NewDocumentCommand{\locCC}{}{\metaVar{\hlocVar_{\cc}}}
                            \NewDocumentCommand{\valCC}{}{\metaVar{\valVar_\cc}}
                            \NewDocumentCommand{\nextValCC}{}{\metaVar{\valVar_\cc'}}
                            \NewDocumentCommand{\valCCInv}{}{\metaVar{\valVar_\cc^\m}}
                            \NewDocumentCommand{\cDeg}{}{\metaVar{1}}
                            \NewDocumentCommand{\cObBag}{}{\metaVar{\obBagVar_\consumerID}}
                            \NewDocumentCommand{\cNextObBag}{}{\metaVar{\obBagVar_\consumerID'}}
                            
                            \NewDocumentCommand{\cLoopInv}{m}{\metaVar{L_\consumerID(#1)}}
                            \NewDocumentCommand{\cLoopInvLockedMut}{m}{\metaVar{L_\consumerID^{\substack{\text{locked}}}(#1)}}
                            \NewDocumentCommand{\cLoopInvNoMutNoObs}{m}
                                    {\metaVar{
                                            L_\consumerID^{\substack{\text{no:\assMutexPredName}\\\text{no:\obsPred}}}\!\!(#1)
                                    }}

                            \NewDocumentCommand{\cPostIf}{}{\metaVar{\fixedPredNameFont{PostIf}_\consumerID}}
                            \NewDocumentCommand{\cPostReleaseVS}{}{\metaVar{\fixedPredNameFont{PostReleaseVS}_\consumerID}}
                            
                            \NewDocumentCommand{\cGenDeg}{}{\metaVar{\degVar_\consumerID}}
                            \NewDocumentCommand{\cAncPid}{}{\metaVar{\fpidVar^{\text{anc}}_\consumerID}}

                    % producer
                            \NewDocumentCommand{\locPC}{}{\metaVar{\hlocVar_\pc}}
                            \NewDocumentCommand{\valPC}{}{\metaVar{\valVar_\pc}}
                            \NewDocumentCommand{\nextValPC}{}{\metaVar{\valVar_\pc'}}
                            \NewDocumentCommand{\valPCInv}{}{\metaVar{\valVar_\pc^\m}}
                            \NewDocumentCommand{\pObBag}{}{\metaVar{\obBagVar_\producerID}}
                            \NewDocumentCommand{\pNextObBag}{}{\metaVar{\obBagVar_\producerID'}}
                            
                            \NewDocumentCommand{\pLoopInv}{m}{\metaVar{L_\producerID(#1)}}
                            \NewDocumentCommand{\pLoopInvLockedMut}{m}{\metaVar{L_\producerID^{\substack{\text{locked}}}(#1)}}
                            \NewDocumentCommand{\pLoopInvNoMutNoObs}{m}
                                    {\metaVar{
                                            L_\producerID^{\substack{\text{no:\assMutexPredName}\\\text{no:\obsPred}}}\!\!(#1)
                                    }}

                            \NewDocumentCommand{\pPostIf}{}{\metaVar{\fixedPredNameFont{PostIf}_\producerID}}
                            \NewDocumentCommand{\pPostReleaseVS}{}{\metaVar{\fixedPredNameFont{PostReleaseVS}_\producerID}}
                            
                            \NewDocumentCommand{\pGenDeg}{}{\metaVar{\degVar_\producerID}}
                            \NewDocumentCommand{\pAncPid}{}{\metaVar{\fpidVar^{\text{anc}}_\producerID}}
            
                    % mutex
                            \NewDocumentCommand{\mut}{}{\metaVar{mut}}
                            \RenewDocumentCommand{\invM}{}{\metaVar{P_\m}}
                            \RenewDocumentCommand{\levM}{}{\metaVar{0}}
                            \NewDocumentCommand{\invMNoExFifo}{m}{\metaVar{P_\m'(#1)}}
            
                    % signals
                            \NewDocumentCommand{\idGen}{m m}{\metaVar{\idVar_\text{#1}^{#2}}}
                            
                            % signal set when number is pushed into FIFO
                            \NewDocumentCommand{\idPush}{m}{\idGen{push}{#1}}
                            \NewDocumentCommand{\levPush}{m}{\metaVar{\levVar_\text{push}^{#1}}}
                            
                            % signal set when number is popped from FIFO
                            \NewDocumentCommand{\idPop}{m}{\idGen{pop}{#1}}
                            \NewDocumentCommand{\levPop}{m}{\metaVar{\levVar_\text{pop}^{#1}}}
                            
                            \NewDocumentCommand{\sigPush}{m}{\metaVar{\sigVar_\text{push}^{#1}}}
                            \NewDocumentCommand{\sigPop}{m}{\metaVar{\sigVar_\text{pop}^{#1}}}

%$$
            $}

            \caption
            [Verification example bounded FIFO.]
            {
                    Verification example bounded FIFO, initialisation.
                    To lighten the notation, we do not show applications of the frame rule.
            }
            \label{gsTR:appendix:fig:VerificationExampleBoundedFifoInit}
    \end{figure}

    \begin{figure}
            % line with approx. width of array
            % helps visualizing whether space to left and to right of array are equal            
%            \noindent\makebox[\linewidth]{\rule{1.35\textwidth}{1pt}} 
%\vspace{-0.7cm}

			\caseStudyFont
            \makebox[\textwidth][c]
            {$
            % box treats separator center between left and right column as "center of array"
            \hspace{2.8cm}  % adjusting center
            \arxivShift

            $}

            \caption
            [Verification example bounded FIFO, forking \& outer loop of producer.]
            {
                    Verification example bounded FIFO, forking.
            }
            \label{gsTR:appendix:fig:VerificationExampleBoundedFifoProducerOuterLoop}
    \end{figure}

    \begin{figure}
            % line with approx. width of array
            % helps visualizing whether space to left and to right of array are equal            
%            \noindent\makebox[\linewidth]{\rule{1.35\textwidth}{1pt}} 
%\vspace{-2.1cm}
			\caseStudyFont
            \makebox[\textwidth][c]
            {$
            % box treats separator center between left and right column as "center of array"
            \hspace{0.15\textwidth}  % adjusting center
            \arxivShift

            $}

            \caption
            [Lock invariant]
            {
                    Lock invariant
            }
            \label{gsTR:appendix:fig:VerificationExampleBoundedFifoLockInvariant}
    \end{figure}

    \begin{figure}
            % line with approx. width of array
            % helps visualizing whether space to left and to right of array are equal            
%            \noindent\makebox[\linewidth]{\rule{1.35\textwidth}{1pt}} 

			\caseStudyFont
            \makebox[\textwidth][c]
            {$
            % box treats separator center between left and right column as "center of array"
            \hspace{0.025\textwidth}  % adjusting center
            \hspace{2cm}	% additional adjustment to algn margin's for producer and consumer invariant figures
            \arxivShift

            $}

            \caption{Producer's loop invariant.}
            \label{gsTR:appendix:fig:VerificationExampleBoundedFifoProducerLoopInvariant}
    \end{figure}

    \begin{figure}
        % line with approx. width of array
        % helps visualizing whether space to left and to right of array are equal            
%            \noindent\makebox[\linewidth]{\rule{1.35\textwidth}{1pt}} 

                    \caseStudyFont
        \makebox[\textwidth][c]
        {$
        % box treats separator center between left and right column as "center of array"
        \hspace{0.15\textwidth}  % adjusting center
        \arxivShift
        %
        $}

        \caption{Consumer's loop invariant.}
        \label{gsTR:appendix:fig:VerificationExampleBoundedFifoConsumerLoopInvariant}
    \end{figure}

    \begin{figure}
            % line with approx. width of array
            % helps visualizing whether space to left and to right of array are equal            
%            \noindent\makebox[\linewidth]{\rule{1.35\textwidth}{1pt}} 

			\caseStudyFont
            \makebox[\textwidth][c]
            {$
            % box treats separator center between left and right column as "center of array"
            \hspace{0.125\textwidth}  % adjusting center
            \arxivShift
            \ifArxiv{\hspace{0.3cm}}
            %
                                            \!\!\!\!\right\}
                                            \\
                                            {}
                                    }
                    \\\quad\quad
                            )\ \ghostFont{\keyword{else}\ (\ }
                    \\\quad\quad\quad\quad
                                    \progProof{
                                            \progProofNew{\lstSize{\valFifoInv} = 10}
                                            \slStar
                                            \assRootForkedPhase
                                            \slStar
                                            \pLoopInvLockedMut{\valPC, \pObBag}
                                            \slStar
                                            \invMNoExFifo{\valFifoInv}
                                            \slStar
                                            \valPC \neq 0
                                    }\
                                            &\proofRuleHintML{\prSimpleViewShiftName \& \vsSemImpName}
                    \\\quad\quad\quad\quad
                            \progProofML{
                                    \progProofCancel{
                                            \lstSize{\valFifoInv} = 10
                                    }
                                    \slStar
                                    \assRootForkedPhase
                                    \,
                                    \progProofCancel{
                                            \slStar\,
                                            \pLoopInvLockedMut{\valPC, \pObBag}
                                            \slStar
                                            \invMNoExFifo{\valFifoInv}
                                            \,\slStar
                                    }\!\!\\
                                    \progProofCancel{
                                            \valPC \neq 0
                                    }\,
                                    \progProofNew{
                                            \exists \nextValPC, \pNextObBag.\
                                            \obs{\pNextObBag \msCup \multiset{\mut}}
                                            \slStar
                                            \assMutLocked{\mut}{\invM}{\frac{1}{2}}
                                            \slStar
                                            \pPostIf
                                    }
                            }
                    \\\quad\quad
                            \ghostFont{)};
                    \\\quad\quad
                    \progProofML{
                            \assRootForkedPhase
                            \slStar
                            \progProofCancel{
                                    \pLoopInvLockedMut{\valPC, \pObBag}
                                    \slStar
                                    \invMNoExFifo{\valFifoInv}
                                    \slStar
                                    \valPC \neq 0
                            }
                            \\
                            \progProofNew{
                                    \exists \nextValPC, \pNextObBag.\
                                    \obs{\pNextObBag \msCup \multiset{\mut}}
                                    \slStar
                                    \assMutLocked{\mut}{\invM}{\frac{1}{2}}
                                    \slStar
                                    \pPostIf
                            }
                    }
                            &\proofRuleHint{\prExistsName (2x)}
                    \\\quad\quad
                    \ghostFont{\forall \nextValPC, \pNextObBag.}
%                    \\\vspace{-0.6cm}
					\\\vspace{-0.2cm}
                    \\\quad\quad
                    \cmdReleaseMut{\m};\ \
                            \quad\quad\quad\!\!\!
                            \referenceHintML{Wait step presented on Figure~\ref{gsTR:appendix:fig:VerificationExampleBoundedFifoProducerWaitStep},
                                    i.e., view shift performed
                                    \\
                                    after releasing \mut but before consuming \invM.
                            }
                            &\proofRuleHint{\prReleaseName \& \prExistsName (2x)}
%                    \\\vspace{-0.6cm}
					\\\vspace{-0.1cm}
                    \\\quad\quad
                            \ghostFont{\forall \pAncPid, \pGenDeg.}
%                    \\\vspace{-0.6cm}
					\\\vspace{-0.3cm}
                    \\\quad\quad
                            \keyword{let}\ \pcLoaded \mathop{:=} \cmdReadHeapLoc{\pc}\ \keyword{in}\ \expNeq{\pcLoaded}{0}
                                    &\proofRuleHintML{\prLetName \& \prReadHeapLocName\\\& \prExpName}
%                    \\\vspace{-0.5cm}
					\\\vspace{-0.3cm}
                    \\\quad\quad
                            \progProofML{
                                    \pAncPid \fpLeq \fpidRootForked
                                    \slStar
                                    \assRootForkedPhase
                                    \\
                                    \slStar\,
                                    (\nextValPC \neq 0
                                            \ \rightarrow\
                                            \exists \valPC, \pObBag.\
                                            \pLoopInv{\valPC, \pObBag}
                                            \slStar
                                            \valPC \neq 0
                                            \slStar
                                            \itPerm[\pAncPid, \pGenDeg]
                                    )
                                    \\
                                    \slStar\,
                                    (\nextValPC = 0
                                            \ \rightarrow\
                                            \noObs
                                    )
                            }
                                    &\proofHint{Reestablished loop invariant.}
                    \\
                    )\ \keyword{do}\ \cmdSkip
                    \\
                    \progProof{
                            \assRootForkedPhase
                            \slStar
                            \progProofCancel{
                                    \exists \valPC.\
                                    \pLoopInv{\valPC}
                                    \slStar
                                    \valPC \neq 0
                            }\,
                            \progProofNew{
                                    \noObs
                            }
                    }
                    \\
                    \dots&\referenceHintML{Continued in Figure~\ref{gsTR:appendix:fig:VerificationExampleBoundedFifoProducerOuterLoop}.}
            \end{array}
            $}

            \caption{Verification example bounded FIFO, producer loop.}
            \label{gsTR:appendix:fig:VerificationExampleBoundedFifoProducerInnerLoopNEW}
    \end{figure}

    \begin{figure}
            % line with approx. width of array
            % helps visualizing whether space to left and to right of array are equal            
%            \noindent\makebox[\linewidth]{\rule{1.35\textwidth}{1pt}} 

			\caseStudyFont
            \makebox[\textwidth][c]
            {$
            % box treats separator center between left and right column as "center of array"
            \hspace{0.14\textwidth}  % adjusting center
            \arxivShift

                                            \!\!\!\!\!\!\right\}
                                    $
                                    \vspace{0.1cm}
                                    \\
                                    \referenceHintML{For definition of \pPostIf 
	                                    	\\
	                                    	cf.\@ Figure~\ref{gsTR:appendix:fig:VerificationExampleBoundedFifoProducerInnerLoopNEW}.}
                            }
                    \\
                    \\
                    \ghostFont{\text{case:}\quad \valPC-1 > 0}
                            &\proofRuleHintML{\prSimpleViewShiftName \& \vsSigInitName}
%                    \\\vspace{-0.86cm}\\\hspace{5cm}\proofHint{Must create signal for next iteration.}
                    \\\vspace{-0.45cm}\\\hspace{5cm}\proofHint{Must create signal for next iteration.}
                    \\\quad\quad
                            \progProofML{
                                    \obsOf{
                                            \progProofNew{\idPush{\valPC-1}},
                                            \mut
                                    }
                                    \slStar
                                    \slBigStar_{i = 1, \dots, \valPC-2}
                                            \quadBack\quadBack\quadBack\!\!
                                            \sigUninit{\idPush{i}}
                                    \\
                                    \slStar\,
                                    \progProofCancel{
                                            \sigUninit{\idPush{\valPC-1}}
                                    }
                                    \,
                                    \progProofNew{
                                            \assSignal{\sigPush{\valPC-1}}{\slFalse}
                                    }
                                    \slStar \dots
                            }
                                    &\proofRuleHint{\prSimpleViewShiftName \& \vsSemImpName}
                    \\\quad\quad
                            \progProof{
                                    \assRootForkedPhase
                                    \slStar
                                    \exists \nextValPC, \pNextObBag.\
                                            \obs{\pNextObBag \msCup \multiset{\mut}}
                                            \slStar
                                            \assMutLocked{\mut}{\invM}{\frac{1}{2}}
                                            \slStar
                                            \pPostIf
                            }
                    \\
                    \\
                    \progProof{
                            \assRootForkedPhase
                            \slStar
                            \exists \nextValPC, \pNextObBag.\
                            \obs{\pNextObBag \msCup \multiset{\mut}}
                            \slStar
                            \assMutLocked{\mut}{\invM}{\frac{1}{2}}
                            \slStar
                            \pPostIf
                    }
                    \\
                    \dots&\referenceHintML{Continued in Figure~\ref{gsTR:appendix:fig:VerificationExampleBoundedFifoProducerInnerLoopNEW}.}
            \end{array}
            $}

            \caption{Verification example bounded FIFO, producer thread's production step.}
            \label{gsTR:appendix:fig:VerificationExampleBoundedFifoProductionStep}
    \end{figure}

    \begin{figure}
            % line with approx. width of array
            % helps visualizing whether space to left and to right of array are equal            
%            \noindent\makebox[\linewidth]{\rule{1.35\textwidth}{1pt}} 

			\caseStudyFont
            \makebox[\textwidth][c]
            {$
            % box treats separator center between left and right column as "center of array"
%            \hspace{0.025\textwidth}  % adjusting center
			\hspace{2cm}
			\arxivShift

            $}

            \caption{Verification example bounded FIFO, producer's wait step.}
            \label{gsTR:appendix:fig:VerificationExampleBoundedFifoProducerWaitStep}
    \end{figure}

    \begin{figure}
            % line with approx. width of array
            % helps visualizing whether space to left and to right of array are equal            
%            \noindent\makebox[\linewidth]{\rule{1.35\textwidth}{1pt}} 

			\caseStudyFont
            \makebox[\textwidth][c]
            {$
            % box treats separator center between left and right column as "center of array"
            \hspace{0.13\textwidth}  % adjusting center
            \arxivShift
            %
                                            \!\!\!\!\!\!\right\}
                                            \\
                                            {}
                                    }
                    \\\quad\quad
                            )\ \ghostFont{\keyword{else}\ (\ }
                    \\\quad\quad\quad\quad
                                    \progProof{
                                            \progProofNew{\lstSize{\valFifoInv} = 0}
                                            \slStar
                                            \assRootAfterForkPhase
                                            \slStar
                                            \cLoopInvLockedMut{\valCC, \cObBag}
                                            \slStar
                                            \invMNoExFifo{\valFifoInv}
                                            \slStar
                                            \valCC \neq 0
                                    }\
                                            &\proofRuleHint{\prSimpleViewShiftName \& \vsSemImpName}
                    \\\quad\quad\quad\quad
                            \progProofML{
                                    \progProofCancel{
                                            \lstSize{\valFifoInv} = 0
                                    }
                                    \slStar
                                    \assRootAfterForkPhase
                                    \,
                                    \progProofCancel{
                                            \slStar\,
                                            \cLoopInvLockedMut{\valCC, \cObBag}
                                            \slStar
                                            \invMNoExFifo{\valFifoInv}
                                            \,\slStar
                                    }\\
                                    \progProofCancel{
                                            \valCC \neq 0
                                    }\,
                                    \progProofNew{
                                            \exists \nextValCC, \cNextObBag.\
                                            \obs{\cNextObBag \msCup \multiset{\mut}}
                                            \slStar
                                            \assMutLocked{\mut}{\invM}{\frac{1}{2}}
                                            \slStar
                                            \cPostIf
                                    }
                            }
                    \\\quad\quad
                            \ghostFont{)};
                    \\\quad\quad
                    \progProofML{
                            \assRootAfterForkPhase
                            \slStar
                            \progProofCancel{
                                    \cLoopInvLockedMut{\valCC, \cObBag}
                                    \slStar
                                    \invMNoExFifo{\valFifoInv}
                                    \slStar
                                    \valCC \neq 0
                            }
                            \\
                            \progProofNew{
                                    \exists \nextValCC, \cNextObBag.\
                                    \obs{\cNextObBag \msCup \multiset{\mut}}
                                    \slStar
                                    \assMutLocked{\mut}{\invM}{\frac{1}{2}}
                                    \slStar
                                    \cPostIf
                            }
                    }
                            &\proofRuleHint{\prExistsName (2x)}
                    \\\quad\quad
                    \ghostFont{\forall \nextValCC, \cNextObBag.}
                    \\\vspace{-0.2cm}
                    \\\quad\quad
                    \cmdReleaseMut{\m};\ \
                            \quad\quad\quad\!\!\!
                            \referenceHintML{Wait step presented on Figure~\ref{gsTR:appendix:fig:MinimalVerificationExampleBoundedFifoConsumerWaitStep},
                                    i.e., view shift performed
                                    \\
                                    after releasing \mut but before consuming \invM.
                            }
                            &\proofRuleHint{\prReleaseName \& \prExistsName (2x)}
                    \\\vspace{-0.1cm}
                    \\\quad\quad
                            \ghostFont{\forall \cAncPid, \cGenDeg.}
                    \\\vspace{-0.3cm}
                    \\\quad\quad
                            \keyword{let}\ \ccLoaded \mathop{:=} \cmdReadHeapLoc{\cc}\ \keyword{in}\ \expNeq{\ccLoaded}{0}
                                    &\proofRuleHintML{\prLetName \& \prReadHeapLocName\\\& \prExpName}
                    \\\vspace{-0.3cm}
                    \\\quad\quad
                            \progProofML{
                                    \cAncPid \fpLeq \fpidRootAfterFork
                                    \slStar
                                    \assRootAfterForkPhase
                                    \\
                                    \slStar\,
                                    (\nextValCC \neq 0
                                            \ \rightarrow\
                                            \exists \valPC, \cObBag.\
                                            \cLoopInv{\valCC, \cObBag}
                                            \slStar
                                            \valCC \neq 0
                                            \slStar
                                            \itPerm[\cAncPid, \cGenDeg]
                                    )
                                    \\
                                    \slStar\,
                                    (\nextValCC = 0
                                            \ \rightarrow\
                                            \noObs
                                    )
                            }
                                    &\proofHint{Reestablished loop invariant.}
                    \\
                    )\ \keyword{do}\ \cmdSkip
                    \\
                    \progProof{
                            \assRootAfterForkPhase
                            \slStar
                            \progProofCancel{
                                    \exists \valPC.\
                                    \cLoopInv{\valPC}
                                    \slStar
                                    \valCC \neq 0
                            }\,
                            \progProofNew{
                                    \noObs
                            }
                    }
                    \\
                    \dots&\referenceHintML{Continued in Figure~\ref{gsTR:appendix:fig:VerificationExampleBoundedFifoProducerOuterLoop}.}
            \end{array}
            $}

            \caption{Verification example bounded FIFO, consumer loop.}
            \label{gsTR:appendix:fig:VerificationExampleBoundedFifoConsumerLoop}
    \end{figure}

    \begin{figure}
            % line with approx. width of array
            % helps visualizing whether space to left and to right of array are equal            
%            \noindent\makebox[\linewidth]{\rule{1.35\textwidth}{1pt}} 

			\caseStudyFont
            \makebox[\textwidth][c]
            {$
            % box treats separator center between left and right column as "center of array"
            \hspace{0.15\textwidth}  % adjusting center
            \arxivShift

                                            \!\!\!\!\!\!\right\}
                                    $
                                    \vspace{0.1cm}
                                    \\
                                    \referenceHintML{For definition of \cPostIf 
                                    	\\
                                    	cf.\@ Figure~\ref{gsTR:appendix:fig:VerificationExampleBoundedFifoConsumerLoop}.}
                            }
                    \\
                    \\
                    \ghostFont{\text{case:}\quad \valCC-1 > 0}
                            &\proofRuleHintML{\prSimpleViewShiftName \& \vsSigInitName}
                    \\\vspace{-0.45cm}\\\hspace{5cm}\proofHint{Must create signal for next iteration.}
                    \\\quad\quad
                            \progProofML{
                                    \obsOf{
                                            \progProofNew{\idPop{\valCC-1}},
                                            \mut
                                    }
                                    \slStar
                                    \slBigStar_{i = 1, ..., \valCC-1}
                                            \quadBack\quadBack\quadBack\!\!
                                            \sigUninit{\idPop{i}}
                                    \\
                                    \slStar\,
                                    \progProofCancel{
                                            \sigUninit{\idPop{\valCC-1}}
                                    }
                                    \,
                                    \progProofNew{
                                            \assSignal{\sigPop{\valCC-1}}{\slFalse}
                                    }
                                    \slStar \dots
                            }
                                    &\proofRuleHint{\prSimpleViewShiftName \& \vsSemImpName}
                    \\\quad\quad
                            \progProof{
                                    \assRootAfterForkPhase
                                    \slStar
                                    \exists \nextValCC, \cNextObBag.\
                                            \obs{\cNextObBag \msCup \multiset{\mut}}
                                            \slStar
                                            \assMutLocked{\mut}{\invM}{\frac{1}{2}}
                                            \slStar
                                            \cPostIf
                            }
                    \\
                    \\
                    \progProof{
                            \assRootAfterForkPhase
                            \slStar
                            \exists \nextValCC, \cNextObBag.\
                            \obs{\cNextObBag \msCup \multiset{\mut}}
                            \slStar
                            \assMutLocked{\mut}{\invM}{\frac{1}{2}}
                            \slStar
                            \cPostIf
                    }
                    \\
                    \dots&\referenceHintML{Continued in Figure~\ref{gsTR:appendix:fig:VerificationExampleBoundedFifoConsumerLoop}.}
            \end{array}
            $}

            \caption{Verification example bounded FIFO, consumer thread's consumption step.}
            \label{gsTR:appendix:fig:MinimalVerificationExampleBoundedFifoConsumptionStep}
    \end{figure}

    \begin{figure}
            % line with approx. width of array
            % helps visualizing whether space to left and to right of array are equal            
%            \noindent\makebox[\linewidth]{\rule{1.35\textwidth}{1pt}} 

			\caseStudyFont
            \makebox[\textwidth][c]
            {$
            % box treats separator center between left and right column as "center of array"
            \hspace{1.8cm}  % adjusting center
            \arxivShift

            $}

            \caption{Verification example bounded FIFO, consumer's wait step.}
            \label{gsTR:appendix:fig:MinimalVerificationExampleBoundedFifoConsumerWaitStep}
    \end{figure}

    }% end of local scope of verification figures

\end{document}

%%%%%%%%%%%%%%%%%%%%%%%%%%%%%%%%%%%%%%%%%%%%%%%%%%
% Keep the following \cleardoublepage at the end of this file, 
% otherwise \includeonly includes empty pages.
\cleardoublepage

% vim: tw=70 nocindent expandtab foldmethod=marker foldmarker={{{}{,}{}}}

% !TeX root = ../../thesis.tex
\chapter{Completeness Thresholds: Formalization}\label{ch:appendix-paper-3-ct4ms}

\NewDocumentCommand{\appCtSecDir}{}
	{chapters/appendix-paper-3-ct4ms/SOAP_TR}

This chapter formalizes the work described in chapter~\ref{ch:paper-3-ct4ms}.
The contents of this chapter were published as technical report on arXiv:
\citationBox
{
	%\cite{Reinhard2023ct4ms-TR}
	\textbf{T. Reinhard}. 
	\emph{Completeness Thresholds for Memory Safety of Array Traversing Programs: Early Technical Report}. 
	arXiv,
	CoRR, 
	abs/2211.11885, 2022. 
	https://doi.org/10.48550/arXiv.2211.11885
}

%{	% open local scope to avoid problems when we redefine commands

%\NewDocumentCommand{\keyword}{m}{\texttt{#1}}

% TODO: move to package and include package in main file

% Assertions
% ==============================================================================

\NewDocumentCommand{\assSet}{}{\metaVar{A}}
%\NewDocumentCommand{\indexVar}{}{\metaVar{i}}
\NewDocumentCommand{\indexSetVar}{}{\metaVar{I}}
%\NewDocumentCommand{\assFam}{o}
%	{\ensuremath{
%		\assVar
%		\IfValueT{#1}{(#1)}
%	}\xspace}

\RenewDocumentCommand{\preVar}{}{\metaVar{M}}

\NewDocumentCommand{\HeapPredSet}{}{\metaVar{P}}
\NewDocumentCommand{\hPredVar}{}{\metaVar{p}}

% program sets
% ----------------------------------------------------------------------------
\NewDocumentCommand{\setVar}{m}{\ensuremath{\mathcal{#1}}\xspace}

%\NewDocumentCommand{\HeapLocSet}{}{\setVar{L}}
%\NewDocumentCommand{\VarSet}{}{\setVar{X}}
\NewDocumentCommand{\dataValVarSet}{}{\setVar{X_D}}
\NewDocumentCommand{\ptrVarVarSet}{}{\setVar{X_L}}
\NewDocumentCommand{\ProcNameSet}{}{\setVar{X_P}}
%\NewDocumentCommand{\ValueSet}{}{\setVar{V}}
%\NewDocumentCommand{\OpSet}{}{\metaVar{Ops}}
\NewDocumentCommand{\BaseTypeSet}{}{\metaVar{BaseTypes}}
\NewDocumentCommand{\ComplexTypeSet}{}{\metaVar{ComplexTypes}}
\NewDocumentCommand{\TypeSet}{}{\metaVar{Types}}
\NewDocumentCommand{\DataTypeSet}{}{\metaVar{DataTypes}}
\NewDocumentCommand{\DataSet}{}{\metaVar{Data}}

%\NewDocumentCommand{\ExpSet}{}{\metaVar{Exps}}
\NewDocumentCommand{\LocExpSet}{}{\metaVar{LocExps}}
\NewDocumentCommand{\DataExpSet}{}{\metaVar{DataExps}}
%\NewDocumentCommand{\CmdSet}{}{\metaVar{Cmds}}
\NewDocumentCommand{\BasicCmdSet}{}{\metaVar{BCmds}}
\NewDocumentCommand{\HeapCmdSet}{}{\metaVar{HCmds}}
\NewDocumentCommand{\CmdPotErrSet}{}{\metaVar{Cmds^{+}}}
\NewDocumentCommand{\ProcSet}{}{\metaVar{Procs}}
\NewDocumentCommand{\TransUnitSet}{}{\metaVar{TUs}}

% program variables
% ----------------------------------------------------------------------------
%\NewDocumentCommand{\metaVar}{m}{\ensuremath{#1}\xspace}

%\NewDocumentCommand{\hlocVar}{}{\metaVar{l}}
%\NewDocumentCommand{\nVar}{}{\metaVar{n}}
\NewDocumentCommand{\nextNVar}{}{\metaVar{n'}}
%\NewDocumentCommand{\zVar}{}{\metaVar{z}}
\NewDocumentCommand{\bVar}{}{\metaVar{b}}
%\NewDocumentCommand{\varVar}{}{\metaVar{x}}
\NewDocumentCommand{\freshVarVar}{}{\metaVar{x_{\text{fresh}}}}
\NewDocumentCommand{\dataVarVar}{}{\metaVar{d}}
\NewDocumentCommand{\ptrVarVar}{}{\metaVar{ptr}}
\NewDocumentCommand{\procNameVar}{}{\metaVar{p}}
%\NewDocumentCommand{\valVar}{}{\metaVar{v}}
\NewDocumentCommand{\dataValVar}{o}
	{\metaVar{
		v^d
		\IfValueT{#1}{_{#1}}
	}}
\NewDocumentCommand{\nextDataVar}{o}
	{\metaVar{
		v^{d\,\prime}
		\IfValueT{#1}{_{#1}}
	}}
%\NewDocumentCommand{\nextValVar}{}{\metaVar{v^\prime}}
%\NewDocumentCommand{\expVar}{}{\metaVar{e}}
%\NewDocumentCommand{\nextExpVar}{}{\metaVar{e'}}
\NewDocumentCommand{\locExpVar}{}{\metaVar{e^L}}
\NewDocumentCommand{\nextLocExpVar}{}{\metaVar{e^{L\prime}}}
\NewDocumentCommand{\dataExpVar}{o}
	{\metaVar{
			e^d
			\IfValueT{#1}{_{#1}}
	}}
\NewDocumentCommand{\nextDataExpVar}{}{\metaVar{e^{d\prime}}}
%\NewDocumentCommand{\cmdVar}{}{\metaVar{c}}
%\NewDocumentCommand{\nextCmdVar}{}{\metaVar{c'}}
\NewDocumentCommand{\cmdPotErr}{}{\metaVar{\cmdVar^{+}}}
\NewDocumentCommand{\hCmdVar}{}{\metaVar{hc}}
\NewDocumentCommand{\bCmdVar}{}{\metaVar{bc}}
\NewDocumentCommand{\fctVar}{}{\metaVar{f}}
%\NewDocumentCommand{\opVar}{}{\metaVar{op}}
\NewDocumentCommand{\nextTypeVar}{}{\metaVar{T^\prime}}
\NewDocumentCommand{\dataTypeVar}{}{\metaVar{D}}
\NewDocumentCommand{\nextDataTypeVar}{}{\metaVar{D^\prime}}
\NewDocumentCommand{\procVar}{}{\metaVar{proc}}
\NewDocumentCommand{\tuVar}{}{\metaVar{tu}}
\NewDocumentCommand{\nextTuVar}{}{\metaVar{tu'}}

\NewDocumentCommand{\nSeqVar}{}{\metaVar{s}}
\NewDocumentCommand{\rangeVar}{}{\metaVar{R}}

%\NewDocumentCommand{\arrayVar}{}{\metaVar{a}}
\NewDocumentCommand{\listVar}{}{\metaVar{li}}
%\NewDocumentCommand{\sizeVar}{}{\metaVar{s}}

% types
% ----------------------------------------------------------------------------
\NewDocumentCommand{\UnitType}{}{\fixedFuncNameFont{Unit}}
\NewDocumentCommand{\PtrType}{O{\Z}}
	{\ensuremath{
			{}^\rightarrow
			#1
	}\xspace}

% operations
% ----------------------------------------------------------------------------
\NewDocumentCommand{\opPlusLocsZ}{}{\ensuremath{\operatorname{+_{\HeapLocSet\Z}}}}

% errors
% ----------------------------------------------------------------------------
\NewDocumentCommand{\memoryError}{}
	{\ensuremath{\fixedPredNameFont{error}}\xspace}

% values
% ----------------------------------------------------------------------------
\NewDocumentCommand{\unitVal}{}
	{\ensuremath{\fixedPredNameFont{()}}\xspace}

% expressions
% ----------------------------------------------------------------------------

% ranges
% ----------------------------------------------------------------------------
\NewDocumentCommand{\rangeFromTo}{m m}
	{\ensuremath{
		[#1\ \keyword{to}\ #2]
	}}

% commands
% ----------------------------------------------------------------------------
% #1 : heap cell to be read
\NewDocumentCommand{\cmdReadHloc}{m}
{\ensuremath{
		!#1
	}\xspace}

% #1 : heap cell to be set
% #2 : new value of #1
\NewDocumentCommand{\cmdAssignHloc}{m m}
	{\ensuremath{
		\cmdReadHloc{#1} := #2
	}}

% #1 : variable to be defined
% #2 : expression whose value will be bound to variable #1
% #3 : command 
%\NewDocumentCommand{\cmdLet}{m m m}
%    {\ensuremath{
%        \keyword{let}\ #1 := #2\ \keyword{in}\ #3
%    }}

% #1 : condition
% #2 : true-branch
% #3 : false-branch
\RenewDocumentCommand{\cmdIf}{m m m}
    {\ensuremath{
        \keyword{if}\ #1\ \keyword{then}\ #2\ \keyword{else}\ #3
    }}

% #1 : condition
% #2 : true-branch
\NewDocumentCommand{\cmdIfNoElse}{m m}
{\ensuremath{
		\keyword{if}\ #1\ \keyword{then}\ #2
}}

% #1 : loop condition
% #2 : loop body
%\NewDocumentCommand{\cmdWhile}{m m}
%{\ensuremath{
%		\keyword{while}\ #1\ \keyword{do}\ #2
%}}

% #1 : loop condition
% #2 : loop invariant
% #3 : loop body
\NewDocumentCommand{\cmdWhileInv}{m m m}
{\ensuremath{
		\keyword{while}\ #1\ \keyword{inv}\ #2\ \keyword{do}\ #3
}}

% #1 : loop variable
% #2 : range of #1
% #3 : loop body
\NewDocumentCommand{\cmdFor}{m m m}
{\ensuremath{
		\keyword{for}\ #1\ \keyword{in}\ #2\ \keyword{do}\ #3
}}

% #1 : loop variable
% #2 : range of #1
% #3 : loop invariant
% #4 : loop body
\NewDocumentCommand{\cmdForInv}{m m m m}
{\ensuremath{
		\keyword{for}\ #1\ \keyword{in}\ #2\ \keyword{inv}\ #3\ \keyword{do}\ #4
}}

\NewDocumentCommand{\cmdContextID}{}
{\contextIdFont{cmds}}

\NewDocumentCommand{\cmdGetProcNames}{o}
{\ensuremath{
		\fixedFuncNameFont{procNames}_\cmdContextID
		\IfValueT{#1}{(#1)}
	}\xspace}

% heap commands
% ----------------------------------------------------------------------------

\NewDocumentCommand{\heapCmdContextID}{}
	{\contextIdFont{hcmds}}

% memory predicates
% ----------------------------------------------------------------------------

\NewDocumentCommand{\heapPredContextID}{}
	{\contextIdFont{hpreds}}

% procedures
% ----------------------------------------------------------------------------
% #1 : procedure signature
% #2 : body
\NewDocumentCommand{\cmdLetProc}{m m}
{\ensuremath{
		\keyword{letproc}\ #1\ \{ #2 \}
}}

% #1 : variable name
% #2 : type
\NewDocumentCommand{\varDecl}{m m}
{\ensuremath{
		#1 : #2
}}

% programs
% ----------------------------------------------------------------------------
\NewDocumentCommand{\mainFct}{}{\keyword{main}}

% ==============================================================================
% Memory safety

\NewDocumentCommand{\safeCmdHeapName}{}
	{\ensuremath{\fixedPredNameFont{safe}}}

% #1 : physical heap
% #2 : command
\NewDocumentCommand{\safeCmdHeap}{m m}
	{\ensuremath{
			\safeCmdHeapName(#1, #2)
	}}

\NewDocumentCommand{\safeCmdAssName}{}
{\ensuremath{\fixedPredNameFont{safe}_\assertionContextID}}

% #1 : physical heap
% #2 : command
\NewDocumentCommand{\safeCmdAss}{m m}
{\ensuremath{
		\safeCmdAssName(#1, #2)
}}

% ==============================================================================
% Static semantics
\NewDocumentCommand{\TypingCtxtSet}{}
	{\ensuremath{TypingCtxts}\xspace}
\NewDocumentCommand{\typingCtxtVar}{}
	{\metaVar{\Gamma}\xspace}

% ==============================================================================
% Dynamic semantics
\NewDocumentCommand{\ProcCtxtSet}{}{\ensuremath{\metaVar{ProcCtxts}}\xspace}
\NewDocumentCommand{\procCtxtVar}{}{\metaVar{\Lambda}\xspace}
\NewDocumentCommand{\nextProcCtxtVar}{}{\metaVar{\Lambda'}\xspace}

\NewDocumentCommand{\procCtxtContextID}{}
	{\contextIdFont{procCtxt}}

\NewDocumentCommand{\procCtxtGetProcNames}{o}
	{\ensuremath{
		\fixedFuncNameFont{procNames}_\procCtxtContextID
		\IfValueT{#1}{(#1)}
	}\xspace}

% ==============================================================================
% Symbolic stuff
\NewDocumentCommand{\SymbLocSet}{}{\metaVar{\HeapLocSet^{\contextIdFont{symb}}}}
\NewDocumentCommand{\slocVar}{}{\metaVar{L}}

\NewDocumentCommand{\SymbValueSet}{}{\metaVar{\ValueSet^\contextIdFont{symb}}}
\NewDocumentCommand{\svalVar}{}{\metaVar{\valVar^\contextIdFont{s}}}
\NewDocumentCommand{\nextSvalVar}{}{\metaVar{\valVar^{\contextIdFont{s}\,\prime}}}

% #1 : precondition
% #2 : program
\NewDocumentCommand{\preProgTuple}{m m}
{\ensuremath{
		\{ #1 \}\ #2
	}\xspace}

\section{Introduction}

In this early technical report on an ongoing project, we present -- to the best of our knowledge -- the first study of completeness thresholds for memory safety proofs.
Specifically we consider heap-manipulating programs that iterate over arrays without allocating or freeing memory.
We present the first notion of completeness thresholds for program verification which reduce unbounded memory safety proofs to bounded ones.
Moreover, we present some preliminary ideas on how completeness thresholds can be computed for concrete programs.

\paragraph{Unbounded vs Bounded Proofs}
Memory safety is a very basic property we want to hold for every critical program, regardless of its nature or purpose.
Yet, it remains hard to prove and in general requires us to write tedious, inductive proofs.
One way to automate the verification process is to settle on bounded proofs and accept bounded guarantees.

Consider a program \cmdVar that searches through an array of size \sizeVar.
An unbounded memory safety proof for \cmdVar would yield that the program is safe for any possible input, in particular for any array size, i.e., 
$\forall \sizeVar.\ \texttt{safe}(\cmdVar)$.
A bounded proof that only considers input sizes smaller than 10 would only guarantees that the program is safe for any such bounded array, i.e., 
$\forall \sizeVar < 10.\ \texttt{safe}(\cmdVar)$.

\paragraph{Completeness Thresholds}
Approximating unbounded proofs by bounded ones is a technique often used in model checking.
Hence, the relationship between bounded and unbounded proofs about finite state transition systems has been studied extensively~\cite{Biere1999SymbolicMC, Clarke2004CompletenessAC, Kroening2003EfficientCO, Bundala2012OnTM, Abdulaziz2018FormallyVA, Heljanko2005IncrementalAC, Awedh2004ProvingMP, McMillan2003InterpolationAS}.
For a finite transition system $T$ and a property of interest $\phi$, a \emph{completeness threshold} is any number $k$ such that we can prove $\phi$ by only examining path prefixes of length $k$ in $T$, i.e., $T \models_k \phi \Rightarrow T \models \phi$~\cite{Clarke2004CompletenessAC}~\footnote{
	Note that the term completeness threshold is used inconsistently in literature.
	Some papers such as~\cite{Clarke2004CompletenessAC} use the definition above, according to which completeness thresholds are not unique.
	Others such as~\cite{Kroening2003EfficientCO} define them as the minimal number $k$ such that $T \models_k \phi \Rightarrow T \models \phi$, which makes them unique.
}.
Over the years, various works characterised over-approximations of least completeness thresholds for different types of properties~$\phi$.
These over-approximations are typically described in terms of key attributes of the transition system $T$, such as the recurrence diameter~\cite{Kroening2003EfficientCO}.

Heap-manipulating programs are essentially infinite state transition systems.
Hence, in general, these key attributes are infinite.
This vast structural difference between the programs we are interested in and the transition systems for which completeness thresholds have been studied prevents us from reusing any of the existing definitions or results.

In \S~\ref{ct4msTR:appendix:sec:NotationsAndDefinitions} we start by presenting basic definitions and notations that are used throughout this work.
In \S~\ref{ct4msTR:appendix:sec:Syntax} and \S~\ref{ct4msTR:appendix:sec:DynamicSemantics} we present the syntax and semantics of the programming language we consider.
In \S~\ref{ct4msTR:appendix:sec:Assertions} and \S~\ref{ct4msTR:appendix:sec:MemorySafety} we formalise our assertion language as well as the notion of memory safety we consider.
Our study of completeness thresholds relies on verification conditions, which we define in \S~\ref{ct4msTR:appendix:sec:VerificationConditions}.
We introduce completeness thresholds and study their properties in \S~\ref{ct4msTR:appendix:sec:CompletenessThresholds}.

\section{General Notation and Basic Definitions}\label{ct4msTR:appendix:sec:NotationsAndDefinitions}

The following definitions and notations will be used throughout this work.

\begin{definition}[Tuples]
	For any set $X$ we denote the set of tuples over $X$ as 
	$$
		X^* 
		\ := \
		\bigcup_{n\in\N}\ \setOf[(x_1, \dots x_n)]{x_1, \dots, x_n \in X}.
	$$
	We denote tuples by overlining the variable name, i.e., $\overline{x} \in X^*$, except if it is clear from the context.
\end{definition}

\begin{definition}[Non-Empty Tuples]
	For any set $X$ we denote the set of non-empty tuples over $X$ as 
	$$
	X^+
	\ := \
	\setOf[t\in X^*]{\tupleLen{t} > 0}.
	$$
\end{definition}

\begin{definition}[Disjoint Union]
	Let $A, B$ be sets.
	We define their disjoint union as
	$$
		A \disjCup B \ := \ A \cup B
	$$
	if $A \cap B = \emptyset$ and leave it undefined otherwise.
\end{definition}

\begin{notation}[Homomorphic \& Isomorphic]
	Let $A, B$ be algebraic structures.
	We define the following notations:
	\begin{itemize}
		\item $A \homomorphic B$ 
			expresses that $A$ and $B$ are homomorphic.
		\item $A \isomorphic B$
			expresses that $A$ and $B$ are isomorphic.
	\end{itemize}
\end{notation}

\begin{remark}[Canonical Homomorphism from Tuples to Sets]
	Let $X$ be a set.
	Then, $X^*$ with concatenation and $\powerSetOf{X}$ with union are both monoids and the canonical homomorphism from $X^*$ to $\powerSetOf{X}$ is
	$
		(x_1, \dots, x_n) \mapsto \setOf{x_1, \dots, x_n}.
	$
\end{remark}

\begin{definition}[Congruence Relation between Tuples and Sets]
	Let $X$ be a set and let $h_X : X^* \rightarrow \powerSetOf{X}$ be the canonical homomorphism.
	We define the congruence relation
	$\congruent_X\ \subseteq 
		(X^* \times \powerSetOf{X})
		\cup
		(\powerSetOf{X} \times X^*)
	$
	such that the following holds for all 
	$\overline{t} \in X^*$ and all $S \in \powerSetOf{X}$:
	$$
		S \congruent_X \overline{t}
		\quad\Leftrightarrow\quad
		\overline{t} \congruent_X S
		\quad\Leftrightarrow\quad
		h_X(\overline{t}) = S
	$$
	Whenever the base set $X$ is clear from the context, we write \congruent instead of $\congruent_X$.

\end{definition}

%\section{Universe}
%In the following we define assumptions on the sets that we use as a basis for our formalisation.
%As soon as the document structure is somewhat settled, these assumptions should be put in the place where the respective set is first introduced.

\NewDocumentCommand{\hcmdInterpret}{o}
{\ensuremath{
		\mathcal{I}_\heapCmdContextID
		\IfValueT{#1}{(#1)}
	}\xspace}

\NewDocumentCommand{\hpredInterpret}{o}
	{\ensuremath{
		\mathcal{I}_\heapPredContextID
		\IfValueT{#1}{(#1)}
	}\xspace}

\newpage
\section{Syntax}\label{ct4msTR:appendix:sec:Syntax}

In this section we define the syntax of the programming language that we use in the rest of this work.

\begin{definition}[Variables]
	We define \VarSet to be an infinite set of variable symbols.
\end{definition}

Our language allows for simple pointer arithmetic of the form $\hlocVar \opPlusLocsZ \zVar$ where \hlocVar is a heap location and \zVar is an offset.

\NewDocumentCommand{\HeapObjSet}{}{\metaVar{HObjs}}
\NewDocumentCommand{\hObjVar}{}{\metaVar{o}}
\NewDocumentCommand{\HeapIdxSet}{}{\metaVar{HIdxs}}
\NewDocumentCommand{\hIdxVar}{}{\metaVar{idx}}

\begin{definition}[Structured Heap Locations]
	We define the sets of heap objects \HeapObjSet and heap indices \HeapIdxSet to be infinite sets with $\N \subset \HeapIdxSet$.
	Further, we define the set of heap locations as
	$$
		\HeapLocSet \quad := \quad \HeapObjSet \times \HeapIdxSet.
	$$
	We denote heap locations by \hlocVar, heap objects by \hObjVar and heap indices by \hIdxVar.
\end{definition}

\begin{definition}[Unit Type]
	We define the unit type as $\UnitType := \setOf{\unitVal}$.
\end{definition}

\begin{definition}[Types]
	We define the set of types \TypeSet syntactically as follows:
	$$
	\begin{array}{l c l}
	\typeVar \in \TypeSet &:=
	&\HeapLocSet \alt \HeapObjSet \alt \Z \alt \B \alt \UnitType
	\end{array}
	$$
\end{definition}

Operations are pure functions that map inputs to output and cannot access the heap.

\begin{definition}[Operations]
	We define \OpSet to be a set of operations with 
	$\OpSet \subseteq \setOf[f: D\rightarrow C]{ D\in\TypeSet^*, C\in\TypeSet}$
	and with $<,\, + \in \OpSet$.
	
	For each $f: D \rightarrow C \in \OpSet$, we use the following notation
	$\dom{f} = D$ and
	$\codom{f} = C$.
\end{definition}

While this work currently only deals with arrays, our plan is to investigate completeness thresholds for programs that deal with arbitrary tree-like inductive data structures.
In order to keep the semantics of our language modular and to allow for easy extensions, we introduce an unspecified set of \emph{heap commands} that captures the APIs of the data structures we are interested in.

\begin{definition}[Heap Commands]
	We define \HeapCmdSet to be a set of symbols.
	Every $\hCmdVar \in \HeapCmdSet$ represents a command that accesses the heap.
\end{definition}

\begin{definition}[Program syntax]
	We define the set of commands \CmdSet, expressions \ExpSet and values \ValueSet syntactically by the grammar presented in Fig.~\ref{ct4msTR:appendix:fig:def:ProgSyntax}.
\end{definition}

\begin{figure}
	\begin{subfigure}{\textwidth}
		$$
		\begin{array}{lcl p{0.7cm}l}
			\hlocVar \in \HeapLocSet 
				&&&&\text{Heap locations}
			\\
			\hObjVar \in \HeapObjSet
				&&&&\text{Heap Objects}
			\\
			\zVar \in \Z
			\\
			\bVar \in \B
			\\
			\varVar  \in  \VarSet
			&&&&\text{Variables}
			\\
			\opVar \in \OpSet &
				&
				&&\text{Primitive Operations}
			\\
			\hCmdVar \in \HeapCmdSet
				&&&&\text{Heap Commands}
			\\
			\\
			% General
			% -------------------------------------------------------
			\valVar \in \ValueSet &::=
				&\hlocVar \alt \hObjVar \alt \zVar \alt \bVar \alt \unitVal
				&&\text{Values}
			\\
			\expVar \in \ExpSet &::=
				&\valVar \alt  
					\varVar \alt
					\opVar(\overline{\expVar})
				&&\text{Expressions}
			\\
			\\
			\cmdVar \in \CmdSet &::= 
				&\expVar \alt 
				\cmdLet{\varVar}{\cmdVar}{\cmdVar} \alt
				\\
				&&\cmdIf{\expVar}{\cmdVar}{\cmdVar} \alt
				\\
				&&\cmdWhile{\cmdReadHloc{\expVar}}{\cmdVar} \alt
				\\
				&&\cmdFor{\varVar}{\rangeFromTo{\expVar}{\expVar}}{\cmdVar} \alt
				\\
				&&\cmdReadHloc{\expVar} \alt 
				\cmdAssignHloc{\expVar}{\expVar} \alt
				\hCmdVar(\overline{\expVar})
				&&\text{Heap access}
		\end{array}
		$$
		\caption{Values, expressions and commands.}
	\end{subfigure}
	
	\begin{subfigure}{\textwidth}
		$$
		\begin{array}{c l l l}
			\cmdVar ; \nextCmdVar 
			&\ :=\
			&\cmdLet{\varVar}{\cmdVar}{\nextCmdVar}
			&\text{where \varVar is not free in \nextCmdVar}
			\\
			\cmdIfNoElse{\expVar}{\cmdVar}
			&\ :=\
			&
			\cmdIf{\expVar}{\cmdVar}{\unitVal}
		\end{array}
		$$
		\caption{Syntactic sugar.}
	\end{subfigure}

	\caption{Program syntax.}
	\label{ct4msTR:appendix:fig:def:ProgSyntax}
\end{figure}

\newpage
\newpage
\section{Dynamic Semantics}\label{ct4msTR:appendix:sec:DynamicSemantics}

\NewDocumentCommand{\MemTraceSet}{}{\metaVar{MemTraces}}
\NewDocumentCommand{\cmdMemTrace}{o}
{\ensuremath{
		\fixedFuncNameFont{memTrace}_\cmdContextID
		\IfValueT{#1}{(#1)}
	}\xspace}

\NewDocumentCommand{\memTraceVar}{}{\metaVar{tr}}
\NewDocumentCommand{\nextMemTraceVar}{o}
{\metaVar{
		tr^\prime
		\IfValueT{#1}{_{#1}}
}}

% ------------------------------------------------------------------------------
% Reduction of entire programs, i.e., translation units + commands
\NewDocumentCommand{\progRedStepSymb}{}
	{\ensuremath{\rightsquigarrow_\contextIdFont{prog}}\xspace}

% #1 : translation unit before reduction
% #2 : procedure context before reduction
% #3 : tuple (memory trace, heap, command) before reduction
% #4 : translation unit after reduction
% #5 : procedure context after reduction
% #6 : tuple (memory trace, heap, command) after reduction
\NewDocumentCommand{\progRedStep}{m m m m m m}
	{\ensuremath{
		#1, #2, #3
		\tuRedStepSymb
		#4, #5, #6
	}\xspace}

% ------------------------------------------------------------------------------
% Reduction of translation units

\NewDocumentCommand{\tuRedStepSymb}{}
	{\ensuremath{\rightsquigarrow_\contextIdFont{tu}}\xspace}
	
% #1 : translation unit before reduction
% #2 : procedure context before reduction
% #3 : translation unit after reduction
% #4 : procedure context after reduction
\NewDocumentCommand{\tuRedStep}{m m m m}
	{\ensuremath{
			#1, #2 
			\tuRedStepSymb
			#3, #4
		}\xspace}

% ------------------------------------------------------------------------------
% Reduction of commands
\NewDocumentCommand{\cmdRedStepSymb}{}
	{\ensuremath{\rightsquigarrow_\contextIdFont{cmd}}\xspace}
	
\NewDocumentCommand{\cmdRedStepStarSymb}{}
	{\ensuremath{\rightsquigarrow_\contextIdFont{cmd}^*}\xspace}

% #1 (opional) : heap before reduction
% #2 : command before reduction
% #3 (optional) : heap after reduction
% #4 : command after reduction
\NewDocumentCommand{\cmdRedStep}
	{O{\pheapVar} m O{\pheapVar} m}
	{\ensuremath{
		#1, #2
		\cmdRedStepSymb
		#3, #4
	}\xspace}

% #1 (opional) : heap before reduction
% #2 : command before reduction
% #3 (optional) : heap after reduction
% #4 : command after reduction
\NewDocumentCommand{\cmdRedStepStar}
	{O{\pheapVar} m O{\pheapVar} m}
	{\ensuremath{
		#1, #2,
		\cmdRedStepStarSymb
		#3, #4
	}\xspace}

% #1 (opional) : heap before reduction
% #2 : command before reduction
% #3 (optional) : heap after reduction
% #4 : command after reduction
\NewDocumentCommand{\cmdRedStepTwoLines}
	{O{\pheapVar} m O{\pheapVar} m}
	{\ensuremath{
		\begin{array}{l}
			#1, #2\\
			\cmdRedStepSymb\,
			#3, #4
		\end{array}
	}\xspace}

In order to keep things simple, our language uses a heap but no store.
Hence, variables are actually constants that can be bound to values via \keyword{let} commands.
As a consequence, the entire evaluation state of a program is represented by the heap and the program itself.
Further, expressions are pure, hence their evaluation does not depend on the heap.

\begin{definition}[Evaluation of Closed Expressions]\label{ct4msTR:appendix:def:EvaluationOfClosedExpressions}
	We define a partial evaluation function $\evalExp{\cdot}: \ExpSet \partialFunArrow \ValueSet$ on expressions by recursion on the structure of expressions as follows:
	$$
	\begin{array}{l l l l}
		\evalExp{\valVar} &:= &\valVar 
			&\text{if}\quad \valVar \in \ValueSet,
		\\
		\evalExp{\opVar(\expVar_1, \dots, \expVar_n)} 
			&:=
			&\opVar(\evalExp{\expVar_1}, \dots, \evalExp{\expVar_n})
			&\text{if}\quad \bot\not\in\setOf{\evalExp{\expVar_1}, \dots, \evalExp{\expVar_n}}
		\\
			&&&\text{and}\quad (\evalExp{\expVar_1}, \dots, \evalExp{\expVar_n}) \in \dom{\opVar},
		\\
		\evalExp{\expVar} &:= &\bot &\text{otherwise}.
	\end{array}
	$$
	We identify closed expressions \expVar with their ascribed value \evalExp{\expVar}.
\end{definition}

\begin{definition}[Evaluation Context]
	We define the set of evaluation contexts \EvalCtxtSet syntactically as follows:
	$$
	\begin{array}{l c l}
		\evalCtxt \in \EvalCtxtSet &\ ::=\
				\cmdLet{\varVar}{\evalCtxtHole}{\cmdVar} 
	\end{array}
	$$
	For any $\cmdVar \in \CmdSet$ and $\evalCtxt \in \EvalCtxtSet$, we define 
	$\evalCtxt[\cmdVar] := \subst{\evalCtxt}{\evalCtxtHole}{\cmdVar}$.
\end{definition}

% TODO: Remove
\RenewDocumentCommand{\freeVars}{o}
{\ensuremath{
		\fixedFuncNameFont{freeVars}
		\IfValueT{#1}{(#1)}
}\xspace}

\begin{definition}[Free Variables (Commands)]\label{ct4msTR:appendix:def:freeVariables_commands}
	We define free variables in the usual way.
	For any command \cmdVar we denote the set of variables that occur freely in \cmdVar by \freeVars[\cmdVar].
\end{definition}

\begin{definition}[Substitution (Commands)]\label{ct4msTR:appendix:def:substitution_commands}
	We define substitution in the usual way.
	For any command \cmdVar, variable \varVar and expression \expVar, we denote the result of substituting every free occurrence of \varVar in \cmdVar with \expVar by \subst{\cmdVar}{\varVar}{\expVar}.
	Further, we extend substitutions to tuples of variables and expressions in the canonical way.
\end{definition}

We explicitly model memory errors in our operational semantics.
This way we know that 
(i)~any execution which ends in a value does not involve memory errors and 
(ii)~any execution that does involve memory errors ends in the dedicated error state \memoryError.

\begin{definition}[Memory Errors]
	We denote the memory error state by \memoryError and the set of potentially erroneous commands by
	$\CmdPotErrSet := \CmdSet \cup \setOf{\memoryError}$.
	We denote potentially erroneous commands by \cmdPotErr.
\end{definition}

Heaps are finite collections of resources that can be manipulated by commands.

\begin{definition}[Physical Resources]
	We define the set of physical resources \PhysResSet syntactically as follows:
	$$
	\begin{array}{l c l}
		\presVar \in \PhysResSet 
		&\ ::=\
			&\slPointsTo{\hlocVar}{\valVar}
			\end{array}
	$$
	$$
	\begin{array}{c p{0.5cm} c}
		\hlocVar \in \HeapLocSet
		&&\valVar \in \ValueSet
	\end{array}
	$$
\end{definition}

\begin{definition}[Physical Heaps]
	We define the set of physical heaps as 
	$$\PhysHeapSet\ :=\ \finPowerSetOf{\PhysResSet}$$
	                and the function 
	$\phGetLocs: \PhysHeapSet \rightarrow \finPowerSetOf{\HeapLocSet}$
	mapping physical heaps to the sets of allocated heap locations as
	$$
	\begin{array}{l c l}
		\phGetLocs(\pheapVar) 
		& :=
		&\setOf
			[\hlocVar \in \HeapLocSet]
			{
				\exists \valVar \in \ValueSet.\
				\slPointsTo{\hlocVar}{\valVar} \in \pheapVar
			}.
	\end{array}
	$$
	We denote physical heaps by \pheapVar.
\end{definition}

\begin{definition}[Basic Commands]
	We define the set of basic commands \BasicCmdSet syntactically as follows:
	$$
	\begin{array}{l l l}
		\bCmdVar \in \BasicCmdSet &::= 
			&\expVar \alt 
				\cmdLet{\varVar}{\bCmdVar}{\bCmdVar} \alt
		\\
			&&\cmdIf{\expVar}{\bCmdVar}{\bCmdVar} \alt
		\\
			&&\cmdWhile{\cmdReadHloc{\expVar}}{\bCmdVar} \alt
		\\
			&&\cmdFor{\varVar}{\rangeFromTo{\expVar}{\expVar}}{\bCmdVar} \alt
		\\
			&&\cmdReadHloc{\expVar} \alt 
				\cmdAssignHloc{\expVar}{\expVar}
	\end{array}
	$$
\end{definition}

The set of basic commands is the subset of \CmdSet that consists exactly of those commands that do not involve any heap command call $\hCmdVar(\overline{\expVar})$.
Remember that the set of heap commands captures the APIs of data structures.
We do not want to change our operational semantics each time we want to consider a new data structure.
Hence, we assume that there exists an interpretation for each heap command that describes its behaviour in terms of basic commands.

\begin{assumption}[Heap Command Interpretation]
	We assume that there exists a function
	$\hcmdInterpret : \HeapCmdSet \rightarrow (\BasicCmdSet \times \VarSet^*)$
	that maps each heap command to a basic command and a vector of variables.
	Further, for every mapping of the form
	$\hcmdInterpret[\hCmdVar] = (\bCmdVar, (\varVar_1, \dots, \varVar_n))$ the following two properties hold:
	\begin{itemize}
		\item $\displaystyle \bigwedge_{i \neq j} \varVar_i \neq \varVar_j$
		\item $\freeVars[\bCmdVar] \subseteq \setOf{\varVar_1, \dots, \varVar_n}$.
	\end{itemize}
\end{assumption}

% ------------------------------------------------------------------------------
% Command reduction rules
\NewDocumentCommand{\cmdRedRuleName}{m}{CmdRed-#1}

\NewDocumentCommand{\cmdRedEvalCtxtName}{}{\cmdRedRuleName{EvalCtxt}}
\NewDocumentCommand{\cmdRedEvalCtxt}{}
{
	\inferrule[\cmdRedEvalCtxtName]
	{\cmdRedStep
		{\cmdVar}
		[\nextPheapVar]{\nextCmdVar}
	}
	{\cmdRedStep
		{\evalCtxt[\cmdVar]}
		[\nextPheapVar]{\evalCtxt[\nextCmdVar]}
	}
}

\NewDocumentCommand{\cmdRedEvalCtxtFailName}{}{\cmdRedRuleName{EvalCtxt-Fail}}
\NewDocumentCommand{\cmdRedEvalCtxtFail}{}
{
	\inferrule[\cmdRedEvalCtxtFailName]
	{\cmdRedStep
		{\cmdVar}
		[\nextPheapVar]{\memoryError}
	}
	{\cmdRedStep
		{\evalCtxt[\cmdVar]}
		[\nextPheapVar]{\memoryError}
	}
}

\NewDocumentCommand{\cmdRedIfTrueName}{}{\cmdRedRuleName{IfTrue}}
\NewDocumentCommand{\cmdRedIfTrue}{}
{
	\inferrule[\cmdRedIfTrueName]
	{}
	{\cmdRedStep
		{\cmdIf{\slTrue}{\cmdVar_t}{\cmdVar_f}}
		{\cmdVar_t}
	}
}

\NewDocumentCommand{\cmdRedIfFalseName}{}{\cmdRedRuleName{IfFalse}}
\NewDocumentCommand{\cmdRedIfFalse}{}
{
	\inferrule[\cmdRedIfFalseName]
	{}
	{\cmdRedStep
		{\cmdIf{\slFalse}{\cmdVar_t}{\cmdVar_f}}
		{\cmdVar_f}
	}
}

\NewDocumentCommand{\cmdRedWhileName}{}{\cmdRedRuleName{While}}
\NewDocumentCommand{\cmdRedWhile}{}
{
	\inferrule[\cmdRedWhileName]
	{\varVar \not\in \freeVars[\cmdVar]}
	{\cmdRedStep
		{\cmdWhile{\cmdReadHloc{\hlocVar}}{\cmdVar}}
		{
			\cmdLet
				{\varVar}
				{\cmdReadHloc{\hlocVar}}
				{\cmdIfNoElse
					{\varVar}
					{(
						\cmdVar;
						\cmdWhile{\cmdReadHloc{\hlocVar}}{\cmdVar}
					)}
				}
		}
	}
}

\NewDocumentCommand{\cmdRedForName}{}{\cmdRedRuleName{For}}
\NewDocumentCommand{\cmdRedFor}{}
{
	\inferrule[\cmdRedForName]
	{}
	{\cmdRedStepTwoLines
		{\cmdFor{\varVar}{\rangeFromTo{\nVar}{\nextNVar}}{\cmdVar}}
		{
			\cmdIfNoElse{\nVar \leq \nextNVar}
				{(
					\cmdVar;
					\cmdFor{\varVar}{\rangeFromTo{\nVar+1}{\nextNVar}}{\cmdVar}
				)}
		}
	}
}

\NewDocumentCommand{\cmdRedLetName}{}{\cmdRedRuleName{Let}}
\NewDocumentCommand{\cmdRedLet}{}
{
	\inferrule[\cmdRedLetName]
	{}
	{\cmdRedStep
		{\cmdLet{\varVar}{\valVar}{\cmdVar}}
		{\subst{\cmdVar}{\varVar}{\valVar}}
	}
}

\NewDocumentCommand{\cmdRedReadHeapLocName}{}{\cmdRedRuleName{ReadHeapLoc}}
\NewDocumentCommand{\cmdRedReadHeapLoc}{}
{
	\inferrule[\cmdRedReadHeapLocName]
	{
		\slPointsTo{\hlocVar}{\valVar} \in \pheapVar
	}
	{\cmdRedStep
		{\cmdReadHloc{\hlocVar}}
		[\pheapVar]{\valVar}
	}
}

\NewDocumentCommand{\cmdRedReadHeapLocFailName}{}{\cmdRedRuleName{ReadHeapLoc-Fail}}
\NewDocumentCommand{\cmdRedReadHeapLocFail}{}
{
	\inferrule[\cmdRedReadHeapLocFailName]
	{
		\hlocVar \not\in \phGetLocs[\pheapVar]
	}
	{\cmdRedStep
		{\cmdReadHloc{\hlocVar}}
		[\pheapVar]{\memoryError}
	}
}

\NewDocumentCommand{\cmdRedAssignHeapLocName}{}{\cmdRedRuleName{AssignHeapLoc}}
\NewDocumentCommand{\cmdRedAssignHeapLoc}{}
{
	\inferrule[\cmdRedAssignHeapLocName]
	{}
	{\cmdRedStep
		[\phAddResDisj
			{\slPointsTo
				{\hlocVar}
				{\slWildcard}
			}
		]
		{\cmdAssignHloc{\hlocVar}{\valVar}}
		[\phAddResDisj
			{\slPointsTo
				{\hlocVar}
				{\valVar}
			}
		]
		{\unitVal}
	}
}

\NewDocumentCommand{\cmdRedAssignHeapLocFailName}{}{\cmdRedRuleName{AssignHeapLoc-Fail}}
\NewDocumentCommand{\cmdRedAssignHeapLocFail}{}
{
	\inferrule[\cmdRedAssignHeapLocFailName]
	{
		\hlocVar \not\in \phGetLocs[\pheapVar]
	}
	{\cmdRedStep
		{\cmdAssignHloc{\hlocVar}{\valVar}}
		[\pheapVar]{\memoryError}
	}
}

\NewDocumentCommand{\cmdRedDesugarHeapCmdCallName}{}{\cmdRedRuleName{Desugar-HeapCmdCall}}
\NewDocumentCommand{\cmdRedDesugarHeapCmdCall}{}
{
	\inferrule[\cmdRedDesugarHeapCmdCallName]
	{
		\hcmdInterpret[\hCmdVar] = (\bCmdVar, (\varVar_1, \dots, \varVar_n))
	}
	{\cmdRedStep
		{\hCmdVar(\valVar_1, \dots, \valVar_n)}
		{\subst
			{\bCmdVar}
			{(\varVar_1, \dots, \varVar_n)}
			{(\valVar_1, \dots, \valVar_n)}
		}
	}
}

\begin{definition}[Command Reduction Relation]
	We define a command reduction relation \cmdRedStepSymb according to the
	rules presented in Fig.~\ref{ct4msTR:appendix:fig:def:CmdReductionRules}.
	A reduction step has the form 
	$$\cmdRedStep
		[\pheapVar]{\cmdVar}
		[\nextPheapVar]{\nextCmdVar}.
	$$
	
	We define \cmdRedStepStarSymb as the reflexive transitive closure of \cmdRedStepSymb.
\end{definition}

\begin{figure}
	\begin{mathpar}
		\cmdRedEvalCtxt
		\and
		\cmdRedEvalCtxtFail
		\and
		\cmdRedIfTrue
		\and
		\cmdRedIfFalse
		\and
		\cmdRedWhile
		\and
		\cmdRedFor
		\and
		\cmdRedLet
		\and
		\cmdRedReadHeapLoc
		\and
		\cmdRedReadHeapLocFail
		\and
		\cmdRedAssignHeapLoc
		\and
		\cmdRedAssignHeapLocFail
		\and
		\cmdRedDesugarHeapCmdCall
	\end{mathpar}
	
	\caption{Command reduction rules.}
	\label{ct4msTR:appendix:fig:def:CmdReductionRules}
\end{figure}

\section{Assertion Language}\label{ct4msTR:appendix:sec:Assertions}

In the previous sections we introduced heap commands that capture the APIs of data structures.
In a similar way, we introduce heap predicates that describe their memory layout.

\begin{assumption}[Heap Predicates]
	We assume that there is a set of symbols \HeapPredSet.
	Every $\hPredVar \in \HeapPredSet$ represents a predicate characterising the heap.
	Further, we assume that there is a function 
	$\hpredInterpret : 
		\HeapPredSet 
		\rightarrow 
		\powerSetOf{\PhysHeapSet \times \ValueSet^*}$
	that maps each heap predicate symbol to a predicate over heaps and value tuples.
\end{assumption}

\begin{definition}[Assertions]
	We define the set of \emph{assertions} \AssertionSet according to the syntax presented in Figure~\ref{ct4msTR:appendix:fig:def:Assertions}.
	
	We omit the index set \indexSetVar in quantifications when its choice becomes clear from the context and write 
	$\exists \indexVar.\, \assFam[\indexVar]$
	and
	$\forall \indexVar.\, \assFam[\indexVar]$
	instead of                
	$\exists \indexVar \in \indexSetVar.\, \assFam[\indexVar]$
	and
	$\forall \indexVar \in \indexSetVar.\, \assFam[\indexVar]$,
\end{definition}

\begin{figure}
	\begin{subfigure}{\textwidth}
		$$

		$$
		\caption{Assertion syntax.}
	\end{subfigure}

	\begin{subfigure}{\textwidth}
		$$
		%
		$$
		\caption{Syntactic sugar.}
	\end{subfigure}
	
	\caption{Assertions.}
	\label{ct4msTR:appendix:fig:def:Assertions}
\end{figure}

\begin{definition}{Assertion Model Relation}\label{ct4msTR:appendix:def:AssertionModelRelation}
	We define the assertion model relation $\assModelsSymb\ \subseteq \PhysHeapSet \times \AssertionSet$ by recursion over the structure of assertions according to the rules presented in Fig.~\ref{ct4msTR:appendix:fig:def:AssertionModelRelation}.
\end{definition}

\begin{figure}
	$$
	%
	$$
	\caption{
		Assertion model relation.
		We write \assNotModels{\pheapVar}{\assVar} if \assModels{\pheapVar}{\assVar} does not hold.
	}
	\label{ct4msTR:appendix:fig:def:AssertionModelRelation}
\end{figure}

\begin{definition}[Free Variables (Assertions)]\label{ct4msTR:appendix:def:freeVariables_assertions}
	We define the notion of free variables for assertions analogously to that of commands (cf.~\ref{ct4msTR:appendix:def:freeVariables_commands}).
\end{definition}

\begin{definition}[Substitution (Assertions)]\label{ct4msTR:appendix:def:substitution_assertions}
	We define substitution for assertions analogously to substitution for commands (cf.~\ref{ct4msTR:appendix:def:substitution_commands}).
\end{definition}

\begin{notation}[Free Variables (Tuples)]
	For convenience we define the following notation for any tuple of commands and assertions
	$(y_1, \dots, y_n) \in (\CmdSet \cup \AssertionSet)^*$:
	$$
		\freeVars[y_1, \dots, y_n] 
		\ :=\
		\freeVars[y_1] \cup \dots \cup \freeVars[y_n].
	$$
\end{notation}

\begin{definition}[Validity]
	Let $\assVar \in \AssertionSet$ be an assertion with
	$\freeVars[\assVar] 
		\congruent
		\overline{\varVar}
		= (\varVar_i)_i
	$.
	For each $i = 1, \dots, n$, let $\typeVar_i$ be the type of variables $\varVar_i$ and let
	$\overline{\typeVar} = (\typeVar_i)_i$.
	We call assertion \assVar \emph{valid} if the following holds:
	$$
		\forall \pheapVar.\
		\forall \overline{\valVar} \in \overline{\typeVar}.\
		\assModels
			{\pheapVar}
			{\subst
				{\assVar}
				{\overline{\varVar}}
				{\overline{\valVar}}
			}
	$$
	We denote validity of \assVar by writing \assValid{\assVar}.
\end{definition}

\section{Memory Safety}\label{ct4msTR:appendix:sec:MemorySafety}

\begin{definition}[Memory Safety of Commands and Heaps]
	We define the safety relation for commands 
	$\safeCmdHeapName \subseteq \PhysHeapSet \times \CmdSet$ 
	as follows:

	Let 
	$\overline{\varVar} = (\varVar_i)_i 
	\congruent
	\freeVars[\cmdVar]
	$
	be the variables occurring freely in \cmdVar.
	For each~$i$, let $\typeVar_i$ be the type of variable $\varVar_i$ and let $\overline{\typeVar} = (\typeVar_i)_i$.
	Then,
	$$
	\safeCmdHeap{\pheapVar}{\cmdVar}
	\ \Leftrightarrow \
	\forall \overline{\valVar} \in \overline{\typeVar}.\
	\neg\exists \nextPheapVar.\
		\cmdRedStepStar
			[\pheapVar]
			{\subst
				{\cmdVar}
				{\overline{\varVar}}
				{\overline{\valVar}}
			}
			[\nextPheapVar]
			{\memoryError}
	$$
	We say that a command \cmdVar is \emph{safe} under a physical heap \pheapVar if \safeCmdHeap{\pheapVar}{\cmdVar} holds.
\end{definition}

We consider a command safe under a heap if its execution does not lead to a memory error.
Note that a command's execution can get stuck without any memory error occurring.
Such cases arise for not-well-typed commands such as \cmdIfNoElse{13}{\dots}.
For this work, we only consider well-typed programs.
Hence, we do not care about cases in which a program gets stuck as long as no memory error occurs.

\begin{definition}[Memory Safety of Commands and Assertions]
	We define the safety relation for commands 
	$\safeCmdAssName \subseteq \AssertionSet \times \CmdSet$ 
	as follows:
	
	Let 
	$\overline{\varVar} = (\varVar_i)_i 
		\congruent
	 	\freeVars[\assVar, \cmdVar]
	 $
	be the variables occurring freely in \assVar and \cmdVar.
	For each~$i$, let $\typeVar_i$ be the type of variable $\varVar_i$ and let $\overline{\typeVar} = (\typeVar_i)_i$.
	Then,
	$$
		\safeCmdAss{\assVar}{\cmdVar}
		\quad \Longleftrightarrow \quad
		\forall \overline{\valVar} \in \overline{\typeVar}.\
		\forall \pheapVar.\
		\big(
			\assModels
				{\pheapVar}
				{\subst
						{\assVar}
						{\overline{\varVar}}
						{\overline{\valVar}}
				}
			\ \Rightarrow\
			\safeCmdHeap
				{\pheapVar}
				{\subst
						{\cmdVar}
						{\overline{\varVar}}
						{\overline{\valVar}}
				}
		\big)
	$$
	We say that a command \cmdVar is \emph{safe} under a assertion \assVar if \safeCmdAss{\assVar}{\cmdVar} holds.
\end{definition}

\begin{notation}
	We denote preconditions by \preVar.
	Further, we aggregate preconditions \preVar and programs \cmdVar into tuples \preProgTuple{\preVar}{\cmdVar}.
\end{notation}

\section{Verification Conditions}\label{ct4msTR:appendix:sec:VerificationConditions}

\NewDocumentCommand{\symbCmdContextID}{}
	{\contextIdFont{scmds}}

% #1 : set of used variable symbols
% #2 : symbolic store
% #3 : expression
\NewDocumentCommand{\symbEvalExp}{O{\cdot} O{\cdot} O{\cdot}}
	{\ensuremath{
		\evalExp{#1, #2, #3}
	}\xspace}

\NewDocumentCommand{\UsedVarSetVar}{}{\metaVar{U}}
\NewDocumentCommand{\symbStoreVar}{}{\metaVar{s}}
\NewDocumentCommand{\VcSetVar}{}{\metaVar{VC}}
%\NewDocumentCommand{\vcVar}{}{\metaVar{vc}}

A common approach in program verification is to derive a verification condition \vcVar from the program \cmdVar and correctness property $\phi$ in question~\cite{Flanagan2001AvoidingExpExplosionVC, Parthasarathy2021VCGenerator}.
Instead of verifying the program directly, we prove the verification condition.
In general, \vcVar describes an over-approximation of all possible program behaviours.
The process is sound iff truth of the verification condition indeed implies that our program is correct, i.e., $\models \vcVar \Rightarrow \cmdVar \models \phi$.
We proceed analogously during our study of completeness thresholds.

\begin{definition}[Verification Condition]\label{ct4msTR:appendix:def:verificationCondition}
	We call an assertion $\vcVar\in \AssertionSet$ a \emph{verification condition} for \preProgTuple{\preVar}{\cmdVar} if the following holds:
	$$
		\assValid{\vcVar}
		\quad\Rightarrow\quad
		\safeCmdAss{\preVar}{\cmdVar}
	$$
	We denote verification conditions by \vcVar.
\end{definition}

\begin{definition}[Precise Verification Conditions]\label{ct4msTR:appendix:def:preciseVerificationCondition}
	Let \vcVar be a verification condition for \preProgTuple{\preVar}{\cmdVar} and let $\varVar \in \freeVars[\preVar, \cmdVar]$ be a free variable of type \typeVar.
	We call \vcVar 
	\emph{precise in \varVar for \preProgTuple{\preVar}{\cmdVar}}
	if the following holds for every value $\valVar \in \typeVar$:
	$$
		\safeCmdAss
			{\subst{\preVar}{\varVar}{\valVar}}
			{\subst{\cmdVar}{\varVar}{\valVar}}
		\quad\Rightarrow\quad
		\assValid{\subst{\vcVar}{\varVar}{\valVar}}
	$$
	
\end{definition}

\RenewDocumentCommand{\arrayPred}{o}
{\ensuremath{
		\fixedPredNameFont{array}
		\IfValueT{#1}{(#1)}
	}\xspace}

\begin{assumption}[Array Predicate]
	We assume that there exists a predicate symbol $\arrayPred \in \HeapPredSet$.
	We further assume that 
	$\hpredInterpret[\arrayPred] \subseteq \PhysHeapSet \times \HeapObjSet \times \N$ 
	is the minimal relation for which the following holds:
	$$
	\begin{array}{l c l}
	\hpredInterpret[\arrayPred](\pheapVar, \arrayVar, \sizeVar)
	&\Longleftrightarrow
	&\displaystyle
	\assModels
	{\pheapVar}
	{
		\slBigStar_{0 \leq \indexVar < \sizeVar}
		\slPointsTo{(\arrayVar, \indexVar)}{\slWildcard}
	}
	\end{array}
	$$
	We identify the assertion \arrayPred[\arrayVar, \sizeVar] with the assertion 
	$\displaystyle
	\slBigStar_{0 \leq \indexVar < \sizeVar}
	\slPointsTo{(\arrayVar, \indexVar)}{\slWildcard}
	$.
\end{assumption}

\RenewDocumentCommand{\postVar}{}{\metaVar{Q}}
%\NewDocumentCommand{\resVar}{}{\metaVar{r}}
\NewDocumentCommand{\invVar}{}{\metaVar{I}}

\subsection{VC Generation}

\NewDocumentCommand{\AssertionLambdaSet}{}
{\metaVar{\AssertionSet_\lambda}}

% #1 : body of lambda term
\NewDocumentCommand{\lambdaIgnore}{m}
{\ensuremath{
		\lambda \_.\ #1
	}\xspace}

We use weakest liberal preconditions~\cite{Flanagan2001AvoidingExpExplosionVC, Dijsktra1976DisciplineOfProgramming} as verification conditions.

\begin{definition}{Assertion Lambdas}
	We define the set of assertion lambda terms as
	$\AssertionLambdaSet :=
	\setOf
	[\lambda \varVar.\ \assVar]
	{\varVar \in \VarSet, \assVar \in \AssertionSet}
	$.
	For convenience we define the notation
	$\lambdaIgnore{\assVar} := \lambda \varVar_{fresh}.\ \assVar$
	for $\varVar_{fresh} \not\in \freeVars[\assVar]$.
\end{definition}

We use assertion lambda terms $\lambda \resVar. \postVar$ to denote postconditions referring to a result value \resVar.

\begin{definition}{Weakest Liberal Precondition}
	We define the weakest liberal precondition function 
	$wlp : \CmdSet \times \AssertionLambdaSet \rightarrow \AssertionSet$
	by recursion over the structure of commands as follows:
	$$

	$$
\end{definition}

Note that our weakest precondition for \keyword{for} loops does not allow the postcondition to depend on the knowledge that any work was done.
This simplified precondition is sufficient to reason about the memory safety of array-traversing programs that do not allocate nor free memory.

\deleteBlock{
\begin{example}[WLP]
	$$
	%
	$$
\end{example}
}

\NewDocumentCommand{\hcmdVCs}{o}
{\ensuremath{
		vcs_\heapCmdContextID
		\IfValueT{#1}{(#1)}
	}\xspace}

\deleteBlock{
	\begin{assumption}[Verification Conditions for Heap Commands]
		We assume that there is a partial function 
		$\hcmdVCs : 
		\HeapCmdSet \times (\ValueSet \times \VarSet)^*
		\partialFunArrow
		\AssertionSet
		$
		that maps every well-typed heap command call to a verification condition.
	\end{assumption}
}

\deleteBlock{
	Note that according to the above definition \hcmdVCs is undefined for applications of the array access command to a wrong number of arguments.
}

\deleteBlock{
	\begin{definition}{Relaxed Liberal Precondition}
		We define the relaxed liberal precondition function 
		$rlp : \CmdSet \times \AssertionLambdaSet \rightarrow \AssertionSet$
		analogous to $wlp$ but with the adjustment that heap commands are handled by \hcmdVCs.
	\end{definition}
}

\section{Completeness Thresholds}\label{ct4msTR:appendix:sec:CompletenessThresholds}

%\NewDocumentCommand{\ctVar}{}{\metaVar{R}}
%\NewDocumentCommand{\ctElemVar}{}{\metaVar{q}}

Now that we have an intuition for what a completeness threshold should be and for how we want to use it, let's formalise this intuition.

\begin{definition}[Completeness Thresholds for Quantified Assertions]\label{ct4msTR:appendix:def:ass:completenessThreshold}
	Let $Q \in \setOf{\forall, \exists}$ be a quantifier and let 
	$Q\varVar \in \typeVar.\ \assVar$ be a quantified assertion.
	Further, let $\ctVar \subseteq \typeVar$ be a restriction of the domain of \varVar.
	We call \ctVar a \emph{completeness threshold} for \assVar if the following holds:
	$$
		\assValid{
			Q \varVar \in \ctVar.\ \assVar
		}
		\quad\Rightarrow\quad
		\assValid{
			Q \varVar \in \typeVar.\ \assVar
		}.
	$$
\end{definition}

\begin{definition}[Completeness Thresholds for Programs]\label{ct4msTR:appendix:def:prog:completenessThreshold}
	Let \typeVar be the type of \varVar in \preProgTuple{\preVar}{\cmdVar} and let $\ctVar \subseteq \typeVar$ be a restriction of this type.
	We call \ctVar a \emph{completeness threshold} for \varVar in \preProgTuple{\preVar}{\cmdVar} if the following holds:
	$$
		\forall \valVar \in \ctVar.\
		\safeCmdAss
			{\subst{\preVar}{\varVar}{\valVar}}
			{\subst{\cmdVar}{\varVar}{\valVar}}
		\quad\Rightarrow\quad
		\forall \valVar \in \typeVar.\
		\safeCmdAss
			{\subst{\preVar}{\varVar}{\valVar}}
			{\subst{\cmdVar}{\varVar}{\valVar}}
	$$
\end{definition}

\begin{theorem}[Soundness]\label{ct4msTR:appendix:soundness}
	Let $\forall \varVar \in \typeVar.\ \vcVar$ be a verification condition for \preProgTuple{\preVar}{\cmdVar} and let $\ctVar \subseteq \typeVar$ be a completeness threshold for \vcVar.
	Further, let \vcVar be precise in \varVar for \preProgTuple{\preVar}{\cmdVar}.
	Then \ctVar is a completeness threshold for \preProgTuple{\preVar}{\cmdVar}.
\end{theorem}
\begin{proof}
	Assume that
	$\forall \valVar \in \ctVar.\ 
		\safeCmdAss
			{\subst{\preVar}{\varVar}{\valVar}}
			{\subst{\cmdVar}{\varVar}{\valVar}}
	$
	holds.
	We have to prove that the program is safe for the unrestricted domain, i.e., \safeCmdAss{\preVar}{\cmdVar}.
	
	Together with the precision of \vcVar in \varVar, this assumption implies
	$\forall \valVar \in \ctVar.\ 
		\assValid{
			\subst{\vcVar}{\varVar}{\valVar}
		}
	$
	(cf. Def.~\ref{ct4msTR:appendix:def:preciseVerificationCondition}),
	which is equivalent to
	$\assValid{
			\forall \varVar \in \ctVar.\ 
			\vcVar
		}
	$.
	
	\ctVar is a completeness threshold for \vcVar.
	According to Def.~\ref{ct4msTR:appendix:def:ass:completenessThreshold}, this means
	$\assValid{
			\forall \varVar \in \typeVar.\ 
			\vcVar
		}
	$.
	Since the latter is a verification condition for \preProgTuple{\preVar}{\cmdVar}, 
	proposition
	\safeCmdAss{\preVar}{\cmdVar}
	holds by Def.~\ref{ct4msTR:appendix:def:verificationCondition}.
	Hence, \ctVar is a completeness threshold for \preProgTuple{\preVar}{\cmdVar}.
\end{proof}

Finding completeness threshold for a non-precise verification condition $\forall \varVar.\ \vcVar$ does not allow us to conclude that we found a completeness threshold for the actual program \preProgTuple{\preVar}{\cmdVar}.
However, as long as our ultimate goal is to verify the program by proving an assertion $A$ that is at least as strong as our verification condition, i.e., $A \Rightarrow \forall \varVar.\ \vcVar$,
we can leverage the completeness threshold.
Hence, it makes sense to first concentrate on completeness thresholds for verification conditions.
Later, we can try to relate our results to completeness thresholds for programs.

\begin{lemma}\label{ct4msTR:appendix:lemma:assWithRestrictedDomain}
	Let $\forall \varVar \in \typeVar.\ \assVar$ be an assertion.
	Let $\ctVar \subseteq \typeVar$ be such that
	$$
	\forall \valVar_1, \valVar_2 \in \ctVar.\quad
		(
			\assValid{\subst{\assVar}{\varVar}{\valVar_1}}
			\ \ \Leftrightarrow\ \
			\assValid{\subst{\assVar}{\varVar}{\valVar_2}}
		).
	$$
	Then, for every $\ctElemVar \in \ctVar$ it holds that
	$$
		\assValid{
			\forall \varVar \in 
				(\typeVar\setminus\ctVar) \cup \setOf{\ctElemVar}.\
				\assVar
		}
		\quad\Leftrightarrow\quad
		\assValid{
			\forall \varVar \in \typeVar.\
			\assVar
		}.
	$$
\end{lemma}
\begin{proof}
	Let $\ctElemVar \in \ctVar$.
	With the assumption from the lemma, we get
	$$
		\forall \valVar \in \ctVar.\quad
		(
			\assValid{\subst{\assVar}{\varVar}{\ctElemVar}}
			\ \ \Leftrightarrow\ \
			\assValid{\subst{\assVar}{\varVar}{\valVar}}
		)
	$$
	and hence
	$$
		\assValid{\subst{\assVar}{\varVar}{\ctElemVar}}
		\ \ \Leftrightarrow\ \
		\assValid{\forall \varVar \in \ctVar.\ \assVar}
	$$
	Further,
	$\forall \varVar \in \typeVar.\ \assVar$
	is valid iff both
	$\forall \varVar \in (\typeVar \setminus \ctVar).\ \assVar$
	and
	$\forall \varVar \in \ctVar.\ \assVar$
	are valid.
	As we can reduce validity of the latter to validity of \subst{\assVar}{\varVar}{\ctElemVar},
	we get
	$$
	\begin{array}{l l}
			&(\assValid
			{
				\forall \varVar \in 
					(\typeVar \setminus \ctVar).\ 
					\assVar
			})
			\ \wedge \
			(\assValid
			{
				\subst{\assVar}{\varVar}{\ctElemVar}
			})
		\\
		\Longleftrightarrow
			&(\assValid
			{
				\forall \varVar \in 
				(\typeVar \setminus \ctVar).\ 
				\assVar
			})
			\ \wedge \
			(\assValid{ \forall \varVar \in \ctVar.\ \assVar})
		\\
		\Longleftrightarrow
			&\assValid{\forall \varVar \in \typeVar.\ \assVar}
		.
	\end{array}
	$$

\end{proof}

\begin{corollary}\label{ct4msTR:appendix:corollary:ass:domainRestrictionCT}
	Let $\forall \varVar \in \typeVar.\ \assVar$ be an assertion.
	Let $\ctVar \subseteq \typeVar$ be such that
	$$
		\forall \valVar_1, \valVar_2 \in \ctVar.\quad
		(
			\assValid{\subst{\assVar}{\varVar}{\valVar_1}}
			\ \ \Leftrightarrow\ \
			\assValid{\subst{\assVar}{\varVar}{\valVar_2}}
		).
	$$
	Then, for every $\ctElemVar \in \ctVar$, the set 
	$(\typeVar \setminus \ctVar) \cup \setOf{\ctElemVar}$
	is a completeness threshold for assertion \assVar.
\end{corollary}
\begin{proof}
	Follows from Lem.~\ref{ct4msTR:appendix:lemma:assWithRestrictedDomain} and Def.~\ref{ct4msTR:appendix:def:ass:completenessThreshold}.
\end{proof}

Consider a verification condition $\forall \varVar \in \typeVar.\ \vcVar$ and suppose we are interested in a completeness threshold for \varVar.
By definition, the threshold is a restriction of \varVar's domain, i.e., $\ctVar \subseteq \typeVar$.
The lemmas and corollary above show us that one way forward is to identify a validity-preserving subset of \typeVar.
That is, we need to look for a subset $\ctVar \subseteq \typeVar$ of the domain within which the concrete choice for \varVar does not affect the validity of the verification condition.
Once we got this, we can collapse \ctVar to any representative $\ctElemVar \in \ctVar$ and we found our completeness threshold $(\typeVar \setminus \ctVar) \cup \setOf{\ctElemVar}$.

Notice that validity preservation of domain restrictions is a transitive property.
Hence, we can easily turn the search for completeness thresholds for a fixed variable into an iterative approach.
\begin{lemma}[Transitivity of Completeness Thresholds for Fixed Variable]\label{ct4msTR:appendix:lemma:CT_Transitivity_fixedVar}
	Let $\ctVar_0, \ctVar_1, \ctVar_2$ be sets with $\ctVar_2 \subseteq \ctVar_1 \subseteq \ctVar_0$.
	Let $\assVar_i = \forall \varVar \in \ctVar_i.\ \assVar$ be assertions.
	Let $\ctVar_1$ and $\ctVar_2$ be completeness thresholds for $\varVar$ in $\assVar_0$ and $\assVar_1$, respectively.
	Then, $\ctVar_2$ is also a completeness threshold for \varVar in $\assVar_0$.
\end{lemma}
\begin{proof}
	Since $\ctVar_1$ is a completeness threshold for \varVar in 
	$\assVar_0 = \forall \varVar \in \ctVar_0.\ \assVar$, 
	we get
	$$
		\assValid{
			\forall \varVar \in \ctVar_1.\ \assVar
		}
		\quad\Rightarrow\quad
		\assValid{
			\forall \varVar \in \ctVar_0.\ \assVar
		}.
	$$
	Since $\ctVar_2$ is a completeness threshold for \varVar in 
	$\assVar_1 = \forall \varVar \in \ctVar_1.\ \assVar$, 
	we get
	$$
		\assValid{
			\forall \varVar \in \ctVar_2.\ \assVar
		}
		\quad\Rightarrow\quad
		\assValid{
			\forall \varVar \in \ctVar_1.\ \assVar
		}.
	$$
	That is,
	$$
	\assValid{
		\forall \varVar \in \ctVar_2.\ \assVar
	}
	\quad\Rightarrow\quad
	\assValid{
		\forall \varVar \in \ctVar_1.\ \assVar
	}
	\quad\Rightarrow\quad
	\assValid{
		\forall \varVar \in \ctVar_0.\ \assVar
	}
	$$
	and hence $\ctVar_2 \subset \ctVar_0$ is a completeness threshold for \varVar in
	$\assVar_0 = \forall \varVar \in \ctVar_0.\ \assVar$
\end{proof}

\begin{corollary}
	Let \typeVar be a type and let $(\ctVar_i)_i$ be a family of sets with $\ctVar_0 = \typeVar$ and $\ctVar_{i+1} \subseteq \ctVar_i$.
	Let $(\assVar_i)_i = (\forall \varVar \in \ctVar_i.\ \assVar)_i$ be a family of assertions such that each $\ctVar_{i+1}$ is a completeness threshold for \varVar in $\assVar_i$.
	Then, each $\ctVar_i$ is a completeness threshold for \varVar in 
	$\assVar_0 = \forall \varVar \in \typeVar.\ \assVar$.
\end{corollary}
\begin{proof}
	Follows from Lem.~\ref{ct4msTR:appendix:lemma:CT_Transitivity_fixedVar} by induction.
\end{proof}

Consider a program that traverses an array $a$ of size $\sizeVar_a$ and an array $b$ of size $\sizeVar_b$.
When we analyse the verification condition of this program we find that it contains distinct parts that describe memory safety of the accesses to array $a$ and distinct parts for the accesses to $b$.
Since both arrays describe separate parts of the heap, we can bring the verification condition into a form that reflects this.
Thereby, we get a formula of the form
$\vcVar \equiv \vcVar_a \slStar \vcVar_b$
where $\vcVar_a$ and $\vcVar_b$ describe memory safety in respect to $a$ and $b$, respectively.

Suppose, we want to find a completeness threshold for $\sizeVar_a$.
In some cases, the manipulation of both arrays is entangled which means that $\sizeVar_a$ potentially affects the validity of $\vcVar_b$.
In such a case, we have no choice but to analyse the entire formula to find our completeness threshold.
However, often that's not the case and $\sizeVar_a$ only shows up in the subformula $\vcVar_a$ that actually concerns array $a$.
In this case, it is sufficient to analyse $\vcVar_a$ in order to find a completeness threshold for $\sizeVar_a$.

\begin{lemma}[Elimination]\label{ct4msTR:appendix:lemma:Elimination}
	Let $\assVar, \assVar_x, \assVar'$ be assertions with
	$\forall \varVar \in \typeVar.\ \assVar
		\equiv
		\forall \varVar \in \typeVar.\ \assVar_\varVar \slStar \assVar'
	$.
	Suppose the choice of \varVar does not affect the validity of $\assVar'$, i.e.,
	$$
		\forall \valVar \in \typeVar.\quad
		(
			\assValid{\assVar'}
			\quad\Leftrightarrow\quad
			\assValid{\subst{\assVar'}{\varVar}{\valVar}}
		).
	$$
	Let $\ctVar \subseteq \typeVar$ be a completeness threshold for \varVar in
	$\forall \varVar \in \typeVar.\ \assVar_x$.
	Then, \ctVar is also a completeness threshold for \varVar in 
	$\forall \varVar \in \typeVar.\ \assVar$.
\end{lemma}
\begin{proof}
	Since $\ctVar$ is a completeness threshold for \varVar in 
	$\forall \varVar \in \ctVar.\ \assVar_x$, 
	we get
	$$
	\begin{array}{l l}
		&\assValid{
			\forall \varVar \in \ctVar.\ \assVar_x \slStar \assVar'
		}
		\\
		\quad\Rightarrow\quad
		&\assValid{
			(\forall \varVar \in \ctVar.\ \assVar_x)
			\slStar
			(\forall \varVar \in \ctVar.\ \assVar')
		}
		\\
		\quad\Rightarrow\quad
		&\assValid{
			(\forall \varVar \in \typeVar.\ \assVar_x)
			\slStar
			(\forall \varVar \in \ctVar.\ \assVar')
		}.
	\end{array}
	$$
	By using the assumption that the choice of \varVar does not affect the validity of $\assVar'$ we can conclude
	$$
	\begin{array}{l l}
		&\assValid{
			(\forall \varVar \in \typeVar.\ \assVar_x)
			\slStar
			(\forall \varVar \in \ctVar.\ \assVar')
		}
		\\
		\quad\Rightarrow\quad
		&\assValid{
			(\forall \varVar \in \typeVar.\ \assVar_x)
			\slStar
			(\forall \varVar \in \typeVar.\ \assVar')
		}		
		\\
		\quad\Rightarrow\quad
		&\assValid{
			\forall \varVar \in \typeVar.\ \assVar_x \slStar \assVar'
		}.
	\end{array}
	$$
\end{proof}

\begin{corollary}[VC Slicing]
	Let $\assVar, \assVar_x, \assVar'$ be assertions with
	$\forall \varVar \in \typeVar.\ \assVar
	\equiv
	\forall \varVar \in \typeVar.\ \assVar_\varVar \slStar \assVar'
	$.
	Suppose \varVar is not free in $\assVar'$.
	Let $\ctVar \subseteq \typeVar$ be a completeness threshold for \varVar in
	$\forall \varVar \in \typeVar.\ \assVar_x$.
	Then, \ctVar is also a completeness threshold for \varVar in 
	$\forall \varVar \in \typeVar.\ \assVar$.
\end{corollary}
\begin{proof}
	Follows from Lem.~\ref{ct4msTR:appendix:lemma:Elimination}.
\end{proof}

\subsection{Iteratively Extracting Completeness Thresholds}

What we saw so far, gives us the tools to define an iterative process to extract completeness thresholds.

\begin{workflow}
	Let \preProgTuple{\preVar}{\cmdVar} be a program with variables $\varVar_1, \dots, \varVar_n$ for which we would like to extract completeness thresholds.
	\begin{enumerate}
		\item
			Compute a verification condition for \preProgTuple{\preVar}{\cmdVar}, e.g., by using weakest preconditions.
			The result has the form
			$\forall \overline{\varVar} \in \overline{\typeVar}.\ \vcVar$.
		\item 
			Iteratively extract completeness thresholds for each $\varVar_i$.\\
			For all $i \in \setOf{1, \dots, n}$:
			\begin{enumerate}
				\item 
					Let $\overline{\ctVar} \subseteq \overline{\typeVar}$ be the completeness thresholds extracted so far.
					(Initially $\overline{\ctVar} = \overline{\typeVar}$.)
				\item
					Bring the verification condition into the form\\
					$\forall \overline{\varVar} \in \overline{\ctVar}.\ \vcVar
						\quad\equiv\quad
						\forall \overline{\varVar} \in \overline{\ctVar}.\ 
							\slBigStar_{0 \leq j \leq m} \vcVar_j
					$.
				\item 
					Identify a subformula $\vcVar'$ whose validity is not affected by the choice of $\varVar_i$.
					Bring the verification condition into the form\\
					$\forall \overline{\varVar} \in \overline{\ctVar}.\ \vcVar
						\quad\equiv\quad
						\forall \overline{\varVar} \in \overline{\ctVar}.\ 
							\vcVar_i \slStar \vcVar'
					$.\\
					In the remaining steps it suffices to analyse
					$\forall \overline{\varVar}\ \in \overline{\ctVar}.\ \vcVar_i$.
				\item
					Examine $\vcVar_i$ and extract a completeness threshold.
					This can either be done purely manually or by identifying patterns for which we previously proved that we can extract completeness thresholds.\\
					This step yields a completeness threshold $\ctVar_i'$ which, in the worst case, does not yield an improvement, i.e., $\ctVar_i' = \ctVar_i$.
				\item 
					Repeat this process iteratively until the extracted completeness threshold does not improve in respect to the last iteration.
			\end{enumerate}
	\end{enumerate}
\end{workflow}

\subsection{Iterating over Arrays}

\NewDocumentCommand{\rangeLeft}{}{\metaVar{a}}
\NewDocumentCommand{\rangeRight}{}{\metaVar{b}}

In the following, we study patterns encountered in verification conditions of programs that iterate over arrays.
The goal of this section is to formulate reusable lemmas that allow us to automate the extraction of completeness thresholds.

\begin{lemma}[VC-CT for Bounded, Unconditional Array Access]
	\label{ct4msTR:appendix:lemma:vc:ct:boundedUnconditionalArrayAccess}
	Let $\zVar, \rangeLeft, \rangeRight \in \Z$ be constants and
	$\vcVar(\sizeVar) = 
		\forall \indexVar \in \Z.\
			\rangeLeft \leq \indexVar < \sizeVar+\rangeRight
			\rightarrow
			0 \leq \indexVar + \zVar < \sizeVar
	$.
	Then, for every $\ctElemVar \in \Z$ with $\ctElemVar > \rangeLeft - \rangeRight $ it holds that
	$$
	\assValid{
		\forall \sizeVar \in \Z.\  \vcVar(\sizeVar)
	}
	\quad\Leftrightarrow\quad
	\assValid{\vcVar(\ctElemVar)}
	$$
	That is, \setOf{\ctElemVar} is a completeness threshold for $\vcVar(s)$.
\end{lemma}
\begin{proof}
	For $\sizeVar \leq \rangeLeft-\rangeRight$ it holds
	$\rangeLeft \leq \indexVar < \sizeVar+\rangeRight \equiv \slFalse$
	and hence
	$\vcVar(\sizeVar) \equiv \slTrue$.
	
	For $\sizeVar > \rangeLeft - \rangeRight$ we get:
	$$
	\begin{array}{r l l}
		\vcVar(\sizeVar) \equiv
			&\forall \indexVar.\
				\rangeLeft \leq \indexVar < \sizeVar + \rangeRight
				\rightarrow
				0 \leq \indexVar + \zVar < \sizeVar
		\\
		\equiv
			&\forall \indexVar.\
				(
					\rangeLeft \leq \indexVar < \sizeVar + \rangeRight
					\rightarrow
					0 \leq \indexVar + \zVar
				)
				\wedge
				(
					\rangeLeft \leq \indexVar < \sizeVar + \rangeRight
					\rightarrow
					\indexVar + \zVar < \sizeVar			
				)				
		\\
		\equiv
			&\forall \indexVar.\
			(
				\rangeLeft \leq \indexVar
				\rightarrow
				0 \leq \indexVar + \zVar
			)
			\wedge
			(
				\indexVar < \sizeVar + \rangeRight
				\rightarrow
				\indexVar + \zVar < \sizeVar			
			)
		\\
		\equiv
			&\forall \indexVar.\
			(
				\rangeLeft \leq \indexVar
				\rightarrow
				0 \leq \indexVar + \zVar
			)
			\wedge
			\forall \indexVar.\
			(
				\indexVar < \sizeVar + \rangeRight
				\rightarrow
				\indexVar + \zVar < \sizeVar
			)
		\\
		\equiv
			&\forall \indexVar.\
			(
				\rangeLeft \leq \indexVar
				\rightarrow
				0 \leq \indexVar + \zVar
			)
			\wedge
			\forall \indexVar.\
			(
				\indexVar < \rangeRight
				\rightarrow
				\indexVar + \zVar < 0
			)
		\\
		=: 
			&\vcVar^+
	\end{array}
	$$
	\sizeVar does not occur in $\vcVar^+$ and, hence, the concrete value of \sizeVar does not impact the validity of $\vcVar^+$.
	By Lem.~\ref{ct4msTR:appendix:corollary:ass:domainRestrictionCT},
	for any choice of $\ctElemVar \in \Z$ with $\ctElemVar > l -r $, the set \setOf{\ctElemVar} is a completeness threshold for $\vcVar(s)$.
\end{proof}

%}

%%%%%%%%%%%%%%%%%%%%%%%%%%%%%%%%%%%%%%%%%%%%%%%%%%
% Keep the following \cleardoublepage at the end of this file, 
% otherwise \includeonly includes empty pages.
\cleardoublepage

% vim: tw=70 nocindent expandtab foldmethod=marker foldmarker={{{}{,}{}}}

% !TeX root = ../../thesis.tex
\chapter{Generalising Completeness Thresholds}\label{ch:appendix-genCTsTR}

\NewDocumentCommand{\appGenCtSecDir}{}
	{chapters/appendix-genCTsTR/TR-sections}

In the previous chapters \S~\ref{ch:paper-3-ct4ms} and \S~\ref{ch:appendix-paper-3-ct4ms} we studied completeness thresholds for memory safety proofs.
In this chapter we drop this limitation and show that we can straightforwardly generalise the theoretic foundations of our work to arbitrary correctness properties.

\section{Basic Definitions}

Since we are generalising our work formalised in appendix~\S~\ref{ch:appendix-paper-3-ct4ms}, we reuse most of that chapter's definitions throughout this chapter.
This includes but it not limited to the definitions presented in \S~\ref{ct4msTR:appendix:sec:NotationsAndDefinitions}.
Deviations from previous definitions are explicitly redefined.

\begin{definition}[Equivalence Classes]
	Let $R \subseteq X \times X$ be an equivalence relation.
	For any $x \in X$, we define the notation
	$$
	\eqClass{x}[R] :=
	\setOf[x^\prime \in X]{R(x, x^\prime)}
	\in\ \quotientSpace{X}{R}
	$$
	We omit $R$ whenever it is clear from the context and write \eqClass{x} instead of \eqClass{x}[R].
\end{definition}

\begin{definition}[Equivalence Relation Refinement]
	Let $R, S \subseteq X \times X$ be equivalence relations.
	We call $S$ a \emph{refinement} of $R$ if the following holds:
	$$
		\forall x \in X.\ \eqClass{x}[S] \subseteq \eqClass{x}[R].
	$$
\end{definition}

\begin{definition}[Cover]
	Let \domVar be a set.
	We call a family of subsets
	$(\subdomVar_i)_{i\in I}$
	a \emph{cover} of \domVar if
	$$
		\bigcup_{i\in I} \subdomVar_i
		\ =\
		\domVar.
	$$
\end{definition}

\begin{definition}[Equivalence Cover]
	Let 
	$R \subseteq \domVar \times \domVar$ 
	be an equivalence relation over a set \domVar.
	We call a subset
	$\subdomVar \subseteq \domVar$
	an \emph{$R$-cover} of \domVar if the family
	$(\eqClass{\subdomElemVar}[R])_{\subdomElemVar\in\subdomVar}$
	is a cover of \domVar.
	When $R$ is clear from the context or irrelevant, we also call \subdomVar an \emph{equivalence cover} of \domVar.
\end{definition}

\begin{notation}[Logical Term Operator Variables]
	We denote variables ranging over quantifiers~\setOf{\forall, \exists}
	by~\quantifierVar.
	Likewise, we denote variables ranging over the logical~connectors~\setOf{\wedge, \vee} by~\connectorVar.
\end{notation}

%\section{Universe}
%In the following we define assumptions on the sets that we use as a basis for our formalisation.
%As soon as the document structure is somewhat settled, these assumptions should be put in the place where the respective set is first introduced.

\section{Generalised Completeness Thresholds}\label{genCTsTR:appendix:sec:CompletenessThresholds}

In this chapter we treat correctness as an abstract property of programs.
Hence, the following definition of correctness encompasses both safety properties like memory safety and liveness properties such as termination.
Moreover, it also covers more complex functional correctness properties.

Without loss of generality, we can assume that any correctness property \correctnessPropVar is given in prenex normal form 
$\quantifierVar_1 \dots \quantifierVar_n. \correctnessPredVar$
for quantifiers $\quantifierVar \in \setOf{\forall, \exists}$.
To simplify the notation in subsequent definitions and lemmas, we call the quantifier-free core a \emph{correctness predicate} and denote it by \correctnessPredVar.
Meanwhile, we call the full correctness specification, where every variable except for the target program is explicitly quantified, a \emph{correctness property} and denote it by \correctnessPropVar.

\begin{definition}[Correctness Predicate]
    Let 
    $\domVar_1, 
        \dots, 
        \domVar_n
    $
    be variable domains.
    Let 
    $\correctnessPredVar(\cmdVar, \varVar_1, \dots, \varVar_n) 
        \subseteq
        \CmdSet \times \domVar_1 \times \dots \times \domVar_n
    $
    be a predicate over commands and typed values.
    We call \correctnessPredVar a \emph{correctness predicate}.
\end{definition}

\genericValid{\correctnessPredVar(\cmdVar, \valVar_1, \dots, \valVar_n)}
means that every run of 
\subst{\cmdVar}{\varVar_1, \dots, \varVar_n}{\valVar, \dots, \valVar_n}
that takes place under the conditions described by \correctnessPredVar is correct in some sense described by~\correctnessPredVar.

\begin{definition}[Correctness Property]\label{genCTsTR:appendix:def:correctnessProperty}
    Let
    $\correctnessPredVar(\cmdVar, \varVar_1, \dots, \varVar_n) 
        \subseteq
        \CmdSet \times \domVar_1 \times \dots \times \domVar_n
    $
    be a correctness predicate and let
    $\quantifierVar_1, \dots, \quantifierVar_n \in \setOf{\forall, \exists}$
    be quantifiers.
    We call a predicate $\correctnessPropVar(\cmdVar) \subseteq \CmdSet$ of the form
    $$
        \correctnessPropVar(\cmdVar)
        \quad:=\quad
        \quantifierVar_1 \varVar_1 \in \domVar_1.\
        \dots\
        \quantifierVar_n \varVar_n \in \domVar_n.\
        \correctnessPredVar(\cmdVar,\ \varVar_1, \dots, \varVar_n)
    $$
    a \emph{correctness property}.
    We write
    $\genericModels{\cmdVar}{\correctnessPropVar}$
    to express that $\correctnessPropVar(\cmdVar)$ holds.
\end{definition}

The following definitions are straightforward generalisations of 
Def.~\ref{ct4msTR:appendix:def:verificationCondition},
Def~\ref{ct4msTR:appendix:def:preciseVerificationCondition},
Def~\ref{ct4msTR:appendix:def:ass:completenessThreshold},
Def~\ref{ct4msTR:appendix:def:prog:completenessThreshold}.

\begin{definition}[Verification Condition]
	\label{genCTsTR:appendix:def:verificationCondition}
	We call an assertion $\vcVar \in \AssertionSet$ a \emph{verification condition} for a correctness property \correctnessPropVar if the following holds:
	$$
		\assValid{\vcVar}	
		\quad\Rightarrow\quad
		\genericModels{\cmdVar}{\correctnessPropVar}
	$$
\end{definition}

We call a VC precise in \varVar for some correctness property \correctnessPropVar if it does not over-approximate the influence of \varVar on \correctnessPropVar.
Note that precision is a significantly weaker property than stating that the VC as a whole does not over-approximate.

\begin{definition}[Precise Verification Conditions]
	\label{genCTsTR:appendix:def:preciseVerificationCondition}
	Let 
	$\quantifierVar \varVar \in \domVar.\ \vcVar$
	be a verification condition for a correctness property 
	$\quantifierVar \varVar \in \domVar.\ \correctnessPredVar$ and a command~\cmdVar.
	We call \vcVar \emph{precise} in \varVar for \correctnessPropVar and \cmdVar if the following holds for every value $\valVar \in \domVar$:
	$$
		\genericModels
			{\cmdVar}
			{\subst{\correctnessPredVar}{\varVar}{\valVar}}
		\quad\Rightarrow
		\quad
		\assValid{\subst{\vcVar}{\varVar}{\valVar}}
	$$
	We also say that \vcVar does not over-approximate \correctnessPredVar and \cmdVar with respect to \varVar.
\end{definition}

\begin{definition}[Completeness Thresholds for Quantified Assertions]\label{genCTsTR:appendix:def:ass:completenessThreshold}
	Let $\quantifierVar\varVar \in \domVar.\ \assVar$ be a quantified assertion.
	Further, let $\ctVar \subseteq \domVar$ be a subdomain.
	We call \ctVar a \emph{completeness threshold} for \varVar in \assVar if the following holds:
	$$
		\assValid{
			\quantifierVar \varVar \in \ctVar.\ \assVar
		}
		\quad\Rightarrow\quad
		\assValid{
			\quantifierVar \varVar \in \domVar.\ \assVar
		}.
	$$
\end{definition}

\begin{observation}[Trivial CTs for $\exists$-Assertions]
	\label{genCTsTR:appendix:obs:trivialCTexistsAss}
	For existentially quantified assertions of the form 
	$\exists\varVar \in \domVar.\ \assVar$
	any subdomain
	$\subdomVar \subseteq \domVar$
	is a completeness threshold for \varVar.
\end{observation}

\begin{definition}[Completeness Thresholds for Correctness Properties and Programs]
	\label{genCTsTR:appendix:def:prog:completenessThreshold}
	Let $\quantifierVar \varVar \in \domVar.\ \correctnessPredVar$ be a correctness property and \cmdVar a comand.
	We call a subdomain $\ctVar \subseteq \domVar$ a \emph{completeness threshold} for \varVar in \correctnessPredVar if the following holds:
	$$
		\genericModels
			{\cmdVar}
			{\quantifierVar \varVar \in \ctVar.\ \correctnessPredVar}
		\quad\Rightarrow\quad
		\genericModels
			{\cmdVar}
			{\quantifierVar \varVar \in \domVar.\ \correctnessPredVar}
	$$
\end{definition}

\begin{observation}[Trivial CTs for $\exists$-Correctness-Properties]
	\label{genCTsTR:appendix:obs:trivialCTexistsProp}
	For existentially quantified correctness properties of the form 
	$\exists\varVar \in \domVar.\ \correctnessPredVar$
	any subdomain
	$\subdomVar \subseteq \domVar$
	is a completeness threshold for \varVar.
\end{observation}

\begin{notation}
	We denote completeness thresholds by \ctVar and singleton ones by \setOf{\ctElemVar}.
\end{notation}

\begin{terminology}
	We call a completeness threshold \emph{minimal} if it is minimal wrt.\@~$\subseteq$.
\end{terminology}

The following theorem states that precision in some variable \varVar is a sufficiently strong requirement to ensure the soundness of any CT for \varVar extracted via the precise VC.

\begin{theorem}[Soundness]
	\label{genCTsTR:appendix:theo:precision}
	Let
	$\quantifierVar \varVar \in \domVar.\ \vcVar$
	be a verification condition
	for a correctness property
	$\quantifierVar \varVar \in \domVar.\ \correctnessPredVar$
	and a command \cmdVar.
	Let $\ctVar \subseteq \domVar$ be a completeness threshold for \varVar in \vcVar and let \vcVar be precise in \varVar.
	Then, \ctVar is also a completeness threshold for \varVar in \correctnessPredVar.
\end{theorem}

\begin{proof}
	We can summarize the proof with this commuting diagram:
	$$
	\hspace{-0.5cm}
	\begin{tikzcd}[row sep=1cm, column sep=6cm]
		% line 1, left
		\genericModels
			{\cmdVar}
			{\quantifierVar \varVar \in \ctVar.\ \correctnessPredVar}
		\arrow[d, Rightarrow,
			"\text{
				\vcVar is precise in \varVar, 
				cf. Def.~\ref{genCTsTR:appendix:def:preciseVerificationCondition}
			}"
		]
		\arrow[r, Rightarrow,
			"\text{
				\ctVar is CT for \varVar in \correctnessPredVar, 
				cf. Def.~\ref{genCTsTR:appendix:def:prog:completenessThreshold}
			}"
		]
			&%line 1, right
				\genericModels
					{\cmdVar}
					{\quantifierVar \varVar \in \domVar.\ \correctnessPredVar}
		\\
		% line 2, left
		\quantifierVar \valVar \in \ctVar.\
			\big(
			\assValid{\subst{\vcVar}{\varVar}{\valVar}}
			\,\big)
		\arrow[d, Leftrightarrow,
		]
		\\
		% line 3, left
		\assValid{
			\quantifierVar \varVar \in \ctVar.\
			\vcVar
		}
		\arrow[r, Rightarrow,
			"\text{
				\ctVar is CT for \varVar in \vcVar, 
				cf. Def.~\ref{genCTsTR:appendix:def:ass:completenessThreshold}
			}"
		]
			&% line 3, right
				\assValid{
					\quantifierVar \varVar \in \domVar.\
					\vcVar
				}
				\arrow[uu, Rightarrow,
					"\text{
						\vcVar is VC,
						Def.~\ref{genCTsTR:appendix:def:verificationCondition}
					}"
				]
	\end{tikzcd}
	$$
	Assume that
		$\genericModels
			{\cmdVar}
			{\quantifierVar \varVar \in \ctVar.\ \correctnessPredVar}
		$
	holds.
	We have to show that the correctness predicate \correctnessPredVar holds for the unrestricted domain \domVar, i.e.,
	$\genericModels
		{\cmdVar}
		{\quantifierVar \varVar \in \domVar.\ \correctnessPredVar}
	$.
	
	By our assumption and the precision of \vcVar in \varVar (cf. Def.~\ref{genCTsTR:appendix:def:preciseVerificationCondition}), we get
	$
		\quantifierVar \valVar \in \ctVar.\
		\big(
			\assValid{\subst{\vcVar}{\varVar}{\valVar}}
		\,\big)
	$,
	which is equivalent to
	$\assValid{
		\quantifierVar \varVar \in \ctVar.\
		\vcVar
	}$.
	Since \ctVar is a completeness threshold for \varVar in \vcVar
	(cf. Def.~\ref{genCTsTR:appendix:def:ass:completenessThreshold}), the latter implies
	$\assValid{
		\quantifierVar \varVar \in \domVar.\
		\vcVar
	}$.
	Further, 
	$\quantifierVar \varVar \in \domVar.\ \vcVar$ 
	is a verification condition for
	$\quantifierVar \varVar \in \domVar.\ \correctnessPredVar$.
	Per Def.~\ref{genCTsTR:appendix:def:verificationCondition}, this means that
	$\genericModels
		{\cmdVar}
		{\quantifierVar \varVar \in \domVar.\ \correctnessPredVar}
	$.
\end{proof}

\begin{lemma}[Subdomain Reduction]\label{genCTsTR:appendix:lemma:subdomainReduction}
	Let $\quantifierVar \varVar \in \domVar.\ \assVar$ be an assertion.
	Let $\subdomVar \subseteq \domVar$ be such that
	$$
		\quantifierVar \valVar_1, \valVar_2 \in \subdomVar.\quad
		(
			\assValid{\subst{\assVar}{\varVar}{\valVar_1}}
			\ \ \Leftrightarrow\ \
			\assValid{\subst{\assVar}{\varVar}{\valVar_2}}
		).
	$$
	Then, for every $\subdomElemVar \in \subdomVar$ it holds that
	$$
	\assValid{
		\quantifierVar \varVar \in 
		(\domVar\setminus\subdomVar) \cup \setOf{\subdomElemVar}.\
		\assVar
	}
	\quad\Leftrightarrow\quad
	\assValid{
		\quantifierVar \varVar \in \domVar.\
		\assVar
	}.
	$$
\end{lemma}
\begin{proof}
	Let $\subdomElemVar \in \subdomVar$.
	With the assumption from the lemma, we get
	$$
		\quantifierVar \valVar \in \subdomVar.\quad
		(
		\assValid{\subst{\assVar}{\varVar}{\subdomElemVar}}
		\ \ \Leftrightarrow\ \
		\assValid{\subst{\assVar}{\varVar}{\valVar}}
		)
	$$
	and hence
	$$
		\assValid{\subst{\assVar}{\varVar}{\subdomElemVar}}
		\ \ \Leftrightarrow\ \
		\assValid{\quantifierVar \varVar \in \subdomVar.\ \assVar}
	$$
	If
	$\quantifierVar = \forall$, we get that
	$\quantifierVar \varVar \in \domVar.\ \assVar$
	is valid iff both
	$\forall \varVar \in (\domVar \setminus \subdomVar).\ \assVar$
	and
	$\forall \varVar \in \subdomVar.\ \assVar$
	are valid.
	In this case we define
	$\connectorVar := \wedge$.
	If
	$\quantifierVar = \exists$, we get that
	$\quantifierVar \varVar \in \domVar.\ \assVar$
	is valid iff 
	$\exists \varVar \in (\domVar \setminus \subdomVar).\ \assVar$
	or
	$\exists \varVar \in \subdomVar.\ \assVar$
	is valid.
	In this case we define
	$\connectorVar := \vee$.
	
	As we can reduce validity of the $\quantifierVar \varVar \in \subdomVar.\ \assVar$ to validity of \subst{\assVar}{\varVar}{\subdomElemVar},
	we get
	$$
	\begin{array}{l l}
	&(\assValid
	{
		\quantifierVar \varVar \in 
		(\domVar \setminus \subdomVar).\ 
		\assVar
	})
	\ \connectorVar \
	(\assValid
	{
		\subst{\assVar}{\varVar}{\subdomElemVar}
	})
	\\
	\Longleftrightarrow
	&(\assValid
	{
		\quantifierVar \varVar \in 
		(\domVar \setminus \subdomVar).\ 
		\assVar
	})
	\ \connectorVar \
	(\assValid{ \quantifierVar \varVar \in \subdomVar.\ \assVar})
	\\
	\Longleftrightarrow
	&\assValid{\quantifierVar \varVar \in \domVar.\ \assVar}
	.
	\end{array}
	$$
\end{proof}

\begin{corollary}[CT Extraction]\label{genCTsTR:appendix:corollary:CTextraction}
	Let $\quantifierVar \varVar \in \domVar.\ \assVar$ be an assertion.
	Let $\subdomVar \subseteq \domVar$ be such that
	$$
	\forall \valVar_1, \valVar_2 \in \subdomVar.\quad
	(
	\assValid{\subst{\assVar}{\varVar}{\valVar_1}}
	\ \ \Leftrightarrow\ \
	\assValid{\subst{\assVar}{\varVar}{\valVar_2}}
	).
	$$
	Then, for every $\subdomElemVar \in \subdomVar$, the set 
	$(\domVar \setminus \subdomVar) \cup \setOf{\subdomElemVar}$
	is a completeness threshold for assertion \assVar.
\end{corollary}
\begin{proof}
	Follows from the Subdomain Reduction Lem.~\ref{genCTsTR:appendix:lemma:subdomainReduction} and Def.~\ref{genCTsTR:appendix:def:ass:completenessThreshold}.
\end{proof}

This corollary shows that we can extract CTs by factoring out validity preserving subdomains.
Later on we will see that it is indeed sufficient to factor out equivalence classes that capture all potential errors.
For now, it is sufficient to note that we can reduce CTs by dropping elements that provably do not cause any errors.

\begin{corollary}\label{genCTsTR:appendix:corollary:OmittingNoErrorElements}
	Let $\quantifierVar \varVar \in \domVar.\ \assVar$ be an assertion.
	Let $\subdomVar \subseteq \domVar$ be such that
	$$
	\forall \subdomElemVar \in \subdomVar.\quad
		\assValid{\subst{\assVar}{\varVar}{\subdomElemVar}}.
	$$
	Then, $(\domVar \setminus \subdomVar)$
	is a completeness threshold for assertion~\assVar.
\end{corollary}
\begin{proof}
	Let $\subdomElemVar \in \subdomVar$.
	By Cor.~\ref{genCTsTR:appendix:corollary:CTextraction}, 
	the set $(\domVar \setminus \subdomVar) \cup \setOf{\subdomElemVar}$ is a CT.
	By assumption, \assValid{\subst{\assVar}{\varVar}{\subdomElemVar}}.
	Hence we get
	$$
		\quantifierVar \varVar \in (\domVar \setminus \subdomVar).\ \assVar
		\quad\quad\Rightarrow\quad\quad
		\quantifierVar \varVar \in (\domVar \setminus \subdomVar \cup \setOf{\subdomElemVar}).\ \assVar
		\quad\quad\Rightarrow\quad\quad
		\quantifierVar \varVar \in \domVar.\ \assVar
		.
	$$
\end{proof}

Intuitively, CTs are subdomains that cover all possible types of potential errors.
In the following, we formalise this intuition by describing them as covers of (refinements of) a specific validity equivalence relation~\eqAssArg[\varVar]{\assVar}.

\begin{definition}[Equivalence Relation for Assertion Arguments]
	Let \assVar be an assertion, \varVar a variable with domain \domVar.
	We define the the relation
	$\eqAssArg[\varVar]{\assVar}\ \subseteq \domVar \times \domVar$
	as follows:
	$$
		\valVar_1 \eqAssArg[\varVar]{\assVar} \valVar_2
		\quad\Longleftrightarrow\quad
		\big(
			\assValid{
				\subst{\assVar}{\varVar}{\valVar_1}
			}
			\ \Leftrightarrow\ \
			\assValid{
				\subst{\assVar}{\varVar}{\valVar_2}
			}			
		\, \big)
	$$
	We omit the variable \varVar whenever it is clear from the context and write \eqAssArg{\assVar}.
\end{definition}

\begin{corollary}
	For any choice of \assVar and \varVar, the relation \eqAssArg[\varVar]{\assVar} is an equivalence relation.
\end{corollary}

\begin{theorem}[Equivalence Cover CT]
	\label{genCTsTR:appendix:theo:eqCoverCT}
	Let
	$\quantifierVar \varVar \in \domVar.\ \assVar$
	be an assertion and let
	$\subdomVar \subseteq \domVar$
	be a 
	\eqAssArg[\varVar]{\assVar}-cover 
	of \domVar.
	Then, \subdomVar is a completeness threshold for \varVar in \assVar.
\end{theorem}
\begin{proof}
	Suppose the claim does not hold.
	By Obs.~\ref{genCTsTR:appendix:obs:trivialCTexistsAss} we know that $\quantifierVar = \forall$.
	Further, it must hold that 
	\assValid{\forall \varVar \in \subdomVar.\ \assVar} 
	and there must exist a
	$\valVar_\domVar \in \domVar$ for which 
	\assNotValid{
		\subst{\assVar}{\varVar}{\valVar_\domVar}
	}.
	
	Since \subdomVar is a \eqAssArg[\varVar]{\assVar}-cover of \domVar, there exists a
	$\subdomElemVar \in \subdomVar$ with
	$\valVar_\domVar \in 
		\eqClass
			{\subdomElemVar}
			[\eqAssArg[\varVar]{\assVar}]
	$.
	By the definition of \eqAssArg[\varVar]{\assVar} we get 
	\assNotValid
		{\subst{\assVar}{\varVar}{\subdomElemVar}},
	which contradicts 
	\assValid{\forall \varVar \in \subdomVar.\ \assVar}.
\end{proof}

\begin{corollary}[Refinement Cover CT]
	Let
	$\quantifierVar \varVar \in \domVar.\ \assVar$
	be an assertion,
	let $\sim$ be a \eqAssArg[\varVar]{\assVar}-refinement
	and let
	$\subdomVar \subseteq \domVar$
	be a 
	$\sim$-cover 
	of \domVar.
	Then, \subdomVar is a completeness threshold for \varVar in \assVar.
\end{corollary}
\begin{proof}
	Since $\sim$ is a \eqAssArg[\varVar]{\assVar}-refinement,
	every $\sim$-cover of \domVar is also a \eqAssArg[\varVar]{\assVar}-cover of~\domVar.
	The rest follows by Theo.~\ref{genCTsTR:appendix:theo:eqCoverCT}.
\end{proof}

\begin{definition}[VC Error Equivalence Class]
	Let
	$\quantifierVar \varVar \in \domVar.\ \assVar$
	be an assertion,
	let $\sim$ be a \eqAssArg[\varVar]{\assVar}-refinement.
	For any $\valVar \in \domVar$, we call 
	\eqClass{\valVar}[\sim]
	a \emph{VC error equivalence class}
	if \assNotValid{\subst{\assVar}{\varVar}{\valVar}}.
\end{definition}

\begin{theorem}[Error Cover CT]\label{genCTsTR:appendix:theo:ErrorCoverCT}
	Let
	$\quantifierVar \varVar \in \domVar.\ \assVar$
	be an assertion,
	let $\sim$ be a \eqAssArg[\varVar]{\assVar}-refinement
	and let
	$\subdomVar \subseteq \domVar$
	cover all $\sim$-error equivalence classes.
	Then, \subdomVar is a completeness threshold for \varVar in \assVar.
\end{theorem}
\begin{proof}
	Let $\eqClass{e_1}[\sim], \dots, \eqClass{e_n}[\sim]$ be all the $\sim$-error equivalence classes.
	Without loss of generality, we can choose the representatives such that $e_i \in \subdomVar$ for all $i$.
	Since $\sim$ is a \eqAssArg[\varVar]{\assVar}-refinement, their union is exactly the \eqAssArg[\varVar]{\assVar}-error equivalence class $E$.
	(Note that \eqAssArg[\varVar]{\assVar} is an equivalence relation with only two equivalence classes.)
	Let $s \in \domVar \setminus E$ be an element for which \assVar holds.
	Then \setOf{e_1, \dots, e_n, s} is a \eqAssArg[\varVar]{\assVar}-cover.
	By Theo.~\ref{genCTsTR:appendix:theo:eqCoverCT} the set \setOf{e_1, \dots, e_n, s} is a CT.
	By Cor.~\ref{genCTsTR:appendix:corollary:OmittingNoErrorElements} we know that we can drop elements for which the VC holds.
	It follows that \setOf{e_1, \dots, e_n} is a CT.
	As $\setOf{e_1, \dots, e_n} \subseteq \subdomVar$, we also know that \subdomVar is a CT.
\end{proof}

\begin{lemma}[Transitivity of Completeness Thresholds for Fixed Variable]\label{genCTsTR:appendix:lemma:CT_Transitivity_fixedVar}
	Let $\domVar_0, \domVar_1, \domVar_2$ be sets with 
	$\domVar_0 \supseteq \domVar_1 \supseteq \domVar_2$.
	Let 
	$\assVar_i = \quantifierVar \varVar \in \domVar_i.\ \assVar$
	be assertions.
	Let $\domVar_1$ and $\domVar_2$ be completeness thresholds for $\varVar$ in $\assVar_0$ and $\assVar_1$, respectively.
	Then, $\domVar_2$ is also a completeness threshold for \varVar in $\assVar_0$.
\end{lemma}
\begin{proof}
	Since $\domVar_1$ is a completeness threshold for \varVar in 
	$\assVar_0 = \quantifierVar \varVar \in \domVar_0.\ \assVar$, 
	we get
	$$
		\assValid{
			\quantifierVar \varVar \in \domVar_1.\ \assVar
		}
		\quad\Rightarrow\quad
		\assValid{
			\quantifierVar \varVar \in \domVar_0.\ \assVar
		}.
	$$
	Since $\domVar_2$ is a completeness threshold for \varVar in 
	$\assVar_1 = \quantifierVar \varVar \in \domVar_1.\ \assVar$, 
	we get
	$$
		\assValid{
			\quantifierVar \varVar \in \domVar_2.\ \assVar
		}
		\quad\Rightarrow\quad
		\assValid{
			\quantifierVar \varVar \in \domVar_1.\ \assVar
		}.
	$$
	That is,
	$$
		\assValid{
			\quantifierVar \varVar \in \domVar_2.\ \assVar
		}
		\quad\Rightarrow\quad
		\assValid{
			\quantifierVar \varVar \in \domVar_1.\ \assVar
		}
		\quad\Rightarrow\quad
		\assValid{
			\quantifierVar \varVar \in \domVar_0.\ \assVar
		}
	$$
	and hence $\domVar_2 \subseteq \domVar_0$ is a completeness threshold for \varVar in
	$\assVar_0 = \quantifierVar \varVar \in \domVar_0.\ \assVar$
\end{proof}

\begin{corollary}[CT Refinement]
	Let \domVar be a set and let $(\subdomVar_i)_i$ be a family of sets with $\ctVar_0 = \domVar$ and $\ctVar_{i} \supseteq \ctVar_{i+1}$.
	Let 
	$(\assVar_i)_i = (\quantifierVar \varVar \in \ctVar_i.\ \assVar)_i$
	 be a family of assertions such that each $\ctVar_{i+1}$ is a completeness threshold for \varVar in $\assVar_i$.
	Then, each $\ctVar_i$ is a completeness threshold for \varVar in 
	$\assVar_0 = \forall \varVar \in \typeVar.\ \assVar$.
\end{corollary}
\begin{proof}
	Follows from the Transitivity Lem.~\ref{genCTsTR:appendix:lemma:CT_Transitivity_fixedVar} by induction.
\end{proof}

\begin{lemma}[Elimination]\label{genCTsTR:appendix:lemma:Elimination}
	Let $\assVar, \assVar_x, \assVar'$ be assertions with
	$\quantifierVar \varVar \in \domVar.\ \assVar
	\equiv
	\quantifierVar \varVar \in \domVar.\ \assVar_\varVar \slStar \assVar'
	$.
	Suppose the choice of \varVar does not affect the validity of $\assVar'$, i.e.,
	$$
	\forall \valVar \in \domVar.\quad
	(
	\assValid{\assVar'}
	\quad\Leftrightarrow\quad
	\assValid{\subst{\assVar'}{\varVar}{\valVar}}
	).
	$$
	Let $\ctVar \subseteq \domVar$ be a completeness threshold for \varVar in
	$\forall \varVar \in \domVar.\ \assVar_x$.
	Then, \ctVar is also a completeness threshold for \varVar in 
	$\forall \varVar \in \domVar.\ \assVar$.
\end{lemma}
\begin{proof}
	Since $\ctVar$ is a completeness threshold for \varVar in 
	$\quantifierVar \varVar \in \domVar.\ \assVar_x$, 
	we get
	$$
	\begin{array}{l l}
	&\assValid{
		\quantifierVar \varVar \in \ctVar.\ \assVar_x \slStar \assVar'
	}
	\\
	\quad\Rightarrow\quad
	&\assValid{
		(\quantifierVar \varVar \in \ctVar.\ \assVar_x)
		\slStar
		(\quantifierVar \varVar \in \ctVar.\ \assVar')
	}
	\\
	\quad\Rightarrow\quad
	&\assValid{
		(\quantifierVar \varVar \in \domVar.\ \assVar_x)
		\slStar
		(\quantifierVar \varVar \in \ctVar.\ \assVar')
	}.
	\end{array}
	$$
	By using the assumption that the choice of \varVar does not affect the validity of $\assVar'$ we can conclude
	$$
	\begin{array}{l l}
	&\assValid{
		(\quantifierVar \varVar \in \domVar.\ \assVar_x)
		\slStar
		(\quantifierVar \varVar \in \ctVar.\ \assVar')
	}
	\\
	\quad\Rightarrow\quad
	&\assValid{
		(\quantifierVar \varVar \in \domVar.\ \assVar_x)
		\slStar
		(\quantifierVar \varVar \in \domVar.\ \assVar')
	}		
	\\
	\quad\Rightarrow\quad
	&\assValid{
		\quantifierVar \varVar \in \domVar.\ \assVar_x \slStar \assVar'
	}.
	\end{array}
	$$
\end{proof}

\begin{corollary}[VC Slicing]
	Let $\assVar, \assVar_x, \assVar'$ be assertions with
	$\quantifierVar \varVar \in \domVar.\ \assVar
	\equiv
	\quantifierVar \varVar \in \domVar.\ \assVar_\varVar \slStar \assVar'
	$.
	Suppose \varVar is not free in $\assVar'$.
	Let $\ctVar \subseteq \domVar$ be a completeness threshold for \varVar in
	$\quantifierVar \varVar \in \domVar.\ \assVar_x$.
	Then, \ctVar is also a completeness threshold for \varVar in 
	$\quantifierVar \varVar \in \domVar.\ \assVar$.
\end{corollary}
\begin{proof}
	Follows from the Elimination Lem.~\ref{genCTsTR:appendix:lemma:Elimination}.
\end{proof}

%%%%%%%%%%%%%%%%%%%%%%%%%%%%%%%%%%%%%%%%%%%%%%%%%%
% Keep the following \cleardoublepage at the end of this file, 
% otherwise \includeonly includes empty pages.
\cleardoublepage

% vim: tw=70 nocindent expandtab foldmethod=marker foldmarker={{{}{,}{}}}

%%%%%%%%%%%%%%%%%%%%%%%%%%%%%%%%%%%%%%%%%%%%%%%%%%%%%%%%%%%%%%%%%%%%%%
\backmatter

\includebibliography
% BibTex
%\bibliographystyle{acm}
\bibliographystyle{plainnat}
\bibliography{bibliography}
% BibLatex (comment lines above and comment out biblatex lines in preamble)
%\printbibliography[title=\bibname]
%\instructionsbibliography

%\includecv{curriculum}

\makebackcoverXII

\end{document}